\documentclass[useAMS,usenatbib]{mn2e}
\usepackage{times}
\usepackage{multirow}
\usepackage[pdftex]{graphicx}
\usepackage{supertabular}
\usepackage{verbatim}

\def\gs{\mathrel{\raise0.35ex\hbox{$\scriptstyle >$}\kern-0.6em
\lower0.40ex\hbox{{$\scriptstyle \sim$}}}}
\def\ls{\mathrel{\raise0.35ex\hbox{$\scriptstyle <$}\kern-0.6em
\lower0.40ex\hbox{{$\scriptstyle \sim$}}}}

\title[Radio and mid-infrared counterparts to submm sources in the ECDFS]
      {The LABOCA survey of the Extended Chandra Deep Field South -- radio and mid-infrared counterparts to submillimetre galaxies}
\author[Biggs et al.]{A.~D.~Biggs,$^{\! 1,2}$\thanks{E--mail: abiggs@eso.org}
        R.~J.~Ivison,$^{\! 2,3}$ E.~Ibar,$^{\! 2,3}$ J.~L.~Wardlow,$^{4}$ H.~Dannerbauer,$^{\! 5,6}$ Ian~Smail,$^{\! 4}$
        \newauthor
        F.~Walter,$^{\! 5}$ A.~Wei{\ss},$^{\! 7}$ S.~C.~Chapman,$^{\! 8}$ K.~E.~K.~Coppin,$^4$ C.~De~Breuck,$^1$ M. Dickinson,$^{\! 9}$ 
        \newauthor
        K.~K.~Knudsen,$^{\! 10}$ V.~Mainieri,$^1$ K.~Menten$^7$ and C.~Papovich$^{11}$\\
        $^1$European Southern Observatory, Karl Schwarzschild Strasse 2,
            D-85748 Garching, Germany\\
        $^2$UK Astronomy Technology Centre, Royal Observatory, Blackford Hill,
            Edinburgh EH9 3HJ\\
        $^3$Institute for Astronomy, University of Edinburgh, Blackford Hill,
            Edinburgh EH9 3HJ\\
        $^4$Institute for Computational Cosmology, Durham University, Durham DH1 6LE\\
        $^5$Max--Planck--Institut f\"ur Astronomie, K\"onigstuhl 17,
            D-69117 Heidelberg, Germany\\
        $^6$Laboratoire AIM  Paris Saclay, CEA-CNRS-Universit{\' e}, Irfu/Service d'Astrophysique, CEA Saclay, Orme de Merisiers, 91191 Gif-sur-Yvette Cedex, France \\
        $^7$Max--Planck--Institut f\"ur Radioastronomie, Auf dem H\"ugel 69, D-53121 Bonn, Germany\\
        $^8$Institute of Astronomy, Madingley Road, Cambridge CB3 0HA\\
        $^9$National Optical Astronomy Observatory, 950 N. Cherry Ave., Tucson, AZ 85719, USA\\
        $^{10}$Argelander Institut f\"ur Astronomie, Auf dem H\"ugel 71, D-53121 Bonn, Germany\\
        $^{11}$Department of Physics, Texas A\&M University, College Station, TX, 77843-4242, USA}

\begin{document}

\maketitle

\begin{abstract} 
We present radio and infrared (3.6--24-$\umu$m) counterparts to
submillimetre galaxies (SMGs) detected in the Extended {\it Chandra}
Deep Field South with the LABOCA 870-$\umu$m bolometer camera on the
12-m Atacama Pathfinder Experiment. Using the Very Large Array at
1.4\,GHz and {\it Spitzer} we have identified secure counterparts to
79 of the 126 SMGs (SNR$>$3.7, $S_{870}> 4.4$~mJy) in the field, 62
via their radio and/or 24-$\umu$m emission, the remainder using a
colour-flux cut on IRAC 3.6- and 5.8-$\umu$m sources chosen to
maximise the number of secure, coincident radio and 24-$\umu$m
counterparts. In constructing our radio catalogue, we have corrected
for the effects of `flux boosting', then used the corrected flux
densities to estimate the redshifts of the SMGs based on the
radio/submm spectral indices. The effect of the boosting correction is
to increase the median redshift by 0.2 resulting in a value of
$\overline{z}=2.2^{+0.7}_{-0.8}$ (1-$\sigma$ errors) for the secure
radio counterparts, in agreement with other studies, both
spectroscopic and photometric.
\end{abstract}


\section{Introduction}
\label{introduction}

Although rare today, ultraluminous infrared galaxies (ULIRGs) --
galaxies with infrared (IR) luminosities exceeding
$10^{12}$\,L$_{\odot}$ -- were extremely common in the early Universe,
signposting systems undergoing intense, dust-obscured star
formation. Moreover, they contribute a significant fraction of the
submillimetre (submm) background \citep{fixsen98}. This important
high-redshift population was first discovered in the form of bright
submm sources behind massive, lensing clusters \citep*{smail97}, and
in blank fields \citep[e.g.][]{hughes98,barger98,eales99}, using the
Submm Common User Bolometer Array \citep[SCUBA;][]{holland99} on the
15-m James Clerk Maxwell Telescope (JCMT); a number of surveys with a
variety of instruments have now brought the number of known submm
galaxies (SMGs) to several hundred \citep[e.g.][]{coppin06,
  bertoldi07, greve08, scott08}.

Cross-identifying the submm sources with emission at other wavelengths
is made difficult by the poor spatial resolution of even the largest
submm telescopes. For example, the combination of JCMT and SCUBA
resulted in a resolution of 14\,arcsecond (arcsec; {\sc fwhm}) at
850\,$\umu$m. The best way to overcome this would be with mm/submm
interferometric observations -- capable of locating the submm emission
directly, with arcsec accuracy \citep[e.g.][]{downes99, gear00,
  iono06, wang07, younger07, ivison08, cowie09}. Such observations,
however, require a large investment of observing time with the few
existing facilities that are capable, although the advent of the
Atacama Large Millimeter/Submillimeter Array (ALMA) will make this
strategy much easier in the future.

In the meantime, attaining higher resolution is possible using radio
interferometric and IR observations, where the empirical correlations
between the far-IR and radio wavebands \citep{condon92} or the
bolometric IR/mid-IR \citep{elbaz02} make it much easier to identify
the submm emitter, particularly given the low source densities in the
radio \citep{ivison98, ivison00, ivison02, smail00,
  dannerbauer04}. This work has typically relied on data from the Very
Large Array (VLA) at 1.4\,GHz and {\it Spitzer} using the 24-$\umu$m
channel of the MIPS instrument \citep{werner04, rieke04}. In addition,
high-redshift SMGs can be identified through their IR colours as
measured by {\it Spitzer's} IRAC camera \citep[e.g.][]{pope06}.

Here we present radio, mid-IR (24-$\umu$m) and IRAC counterparts to
the 126 SMGs that have been detected in the Large APEX Bolometer
Camera (LABOCA) Extended Chandra Deep Field South [ECDFS] Submm Survey
(LESS), a deep blank-field 870-$\umu$m survey, down to a 3.7-$\sigma$
limit of 4.4\,mJy \citep{weiss09}. The ECDFS is an exceptional area
for multi-wavelength, wide-field studies of galaxy evolution due to
deep X-ray \citep{giacconi01,lehmer05,luo08}, optical
\citep{giavalisco04,beckwith06}, IR (Dickinson et al., in preparation)
and radio \citep{miller08,ivison10} data. The CDFS portion of the
field has also been surveyed \citep{scott10} with the AzTEC 1.1-mm
bolometric camera \citep{wilson08} on the Atacama Submillimeter
Telescope Experiment.

The paper is organised as follows: in Section~\ref{catalogues} we
describe the submm, radio, 24-$\umu$m and IRAC data that have been
used to identify counterparts to the submm sources, with particular
emphasis on the techniques used to extract source fluxes and positions
from the radio map. Section~\ref{strategy} contains details of our
counterpart identification strategy and in Sections~\ref{results} and
\ref{irac} we present lists of the likely counterparts and their
properties. Section~\ref{discussion} discusses these results in
detail, ascertaining the effectiveness of our strategy. We also derive
the redshift distribution of the radio-detected robust counterparts
using the radio-submm spectral index relation of
\citet{carilli99,carilli00} before drawing our conclusions in
Section~\ref{conclusions}. In an appendix we present detailed notes on
some of the sources as well as multi-wavelength maps with the
counterparts marked.

We assume a flat $\Lambda$CDM cosmology of
$\Omega_{\Lambda} = 0.73$, $\Omega_{m} = 0.27$ and $H_0 =
70.5$~km\,s$^{-1}$\,Mpc$^{-1}$ \citep{hinshaw09}.

\section{Observations, Reduction and analysis}
\label{catalogues}

\subsection{APEX 870-$\umu$m catalogue}

LABOCA \citep{siringo09} is a 295-element bolometer camera operating at
the 12-m Atacama Pathfinder Telescope (APEX\footnote{This publication
  is based on data acquired with the Atacama Pathfinder Experiment
  (APEX) under program IDs 078.F-9028(A), 079.F-9500(A), 080.A-3023(A)
  and 081.F-9500(A). APEX is a collaboration between the
  Max--Planck--Institut fur Radioastronomie, the European Southern
  Observatory and the Onsala Space Observatory.}) in the exceptionally
dry environment of the Atacama desert in Chile \citep{guesten06}. The
LESS map comprises 200\,hr of on-sky integration (excluding overheads)
and has extremely uniform noise coverage (average rms =
1.2\,mJy\,beam$^{-1}$) over the $30 \times 30$-arcmin$^2$ extent of
the ECDFS, with a resolution of 19\,arcsec ({\sc fwhm}). The catalogue
of submm sources identified by LESS is described in detail by
\citet{weiss09}. The full catalogue comprises 126 sources above
3.7-$\sigma$ with a false-detection expectation of $\approx$5. This is
based on extensive simulations as described in \citet{weiss09}.

\subsection{VLA 1.4-GHz catalogue}

To identify the radio counterparts to the LESS SMGs we use the VLA
1.4-GHz map of \citet{miller08} which we briefly describe here. The
map is constructed from six separate pointings arranged in a hexagonal
pattern, centred on the coordinates $\mathrm{03^h 32^m 28^s,
  -27^{\circ} 48^{\prime} 30^{\prime\prime}}$ (J2000). Each pointing
consists of approximately eight separate $\sim$5-hr observations. The
noise in the final $34 \times 34$-arcmin$^2$ mosaic is
$\sim$6.5$\umu$Jy\,beam$^{-1}$ at its deepest. All data were taken in
`A' configuration, resulting in a synthesised beam with dimensions
$2.8 \times 1.6$\,arcsec$^2$, aligned north--south. When looking for
radio counterparts to the SMGs, we do not use the \citet{miller08}
catalogue as this is truncated at a signal-to-noise ratio (SNR) of
seven; instead, we have created our own catalogue containing sources
down to a SNR of three.

Seven of the SMGs in the LESS catalogue lie outside the $34 \times
34$-arcmin$^2$ area of the radio map. For these, we use our own
reduction of the VLA data to search for counterparts. Our map was
created in a similar fashion to that of \cite{miller08} and achieves
an r.m.s.\ just below 7$\umu$Jy\,beam$^{-1}$. The flux density of the
brightest of the SMG counterparts has a flux density in the two maps
that differs by less than one~per~cent and thus we are confident
that the two maps are tied to the same flux scale.

\subsubsection{Source extraction}

The first step in producing a catalogue of radio sources is to create
a map of the noise across the field. Sources with a
peak-flux-density-to-noise ratio (PNR) greater than five are detected
and removed using the standard {\sc aips} source-extraction code, {\sc
  sad}. The residual image is then inverted and the source extraction
process repeated in order to remove `sources' with negative flux --
mainly prominent sidelobes caused by Gibbs ringing (associated with
high-SNR sources) which become increasingly prominent with distance
from the phase centre of each pointing. Once all significant sources
have been removed, a noise map is created for each pixel by fitting a
Gaussian to the histogram of pixel values contained within a
surrounding circle of diameter 50\,arcsec (using {\sc rmsd} with {\sc
  optype = `hist'}). Aided by the accurate noise map, we start the
source extraction again, this time restricting the fitting to positive
sources with a PNR equal to or greater than three.

\begin{figure*}
\begin{center}
\includegraphics[scale=0.4]{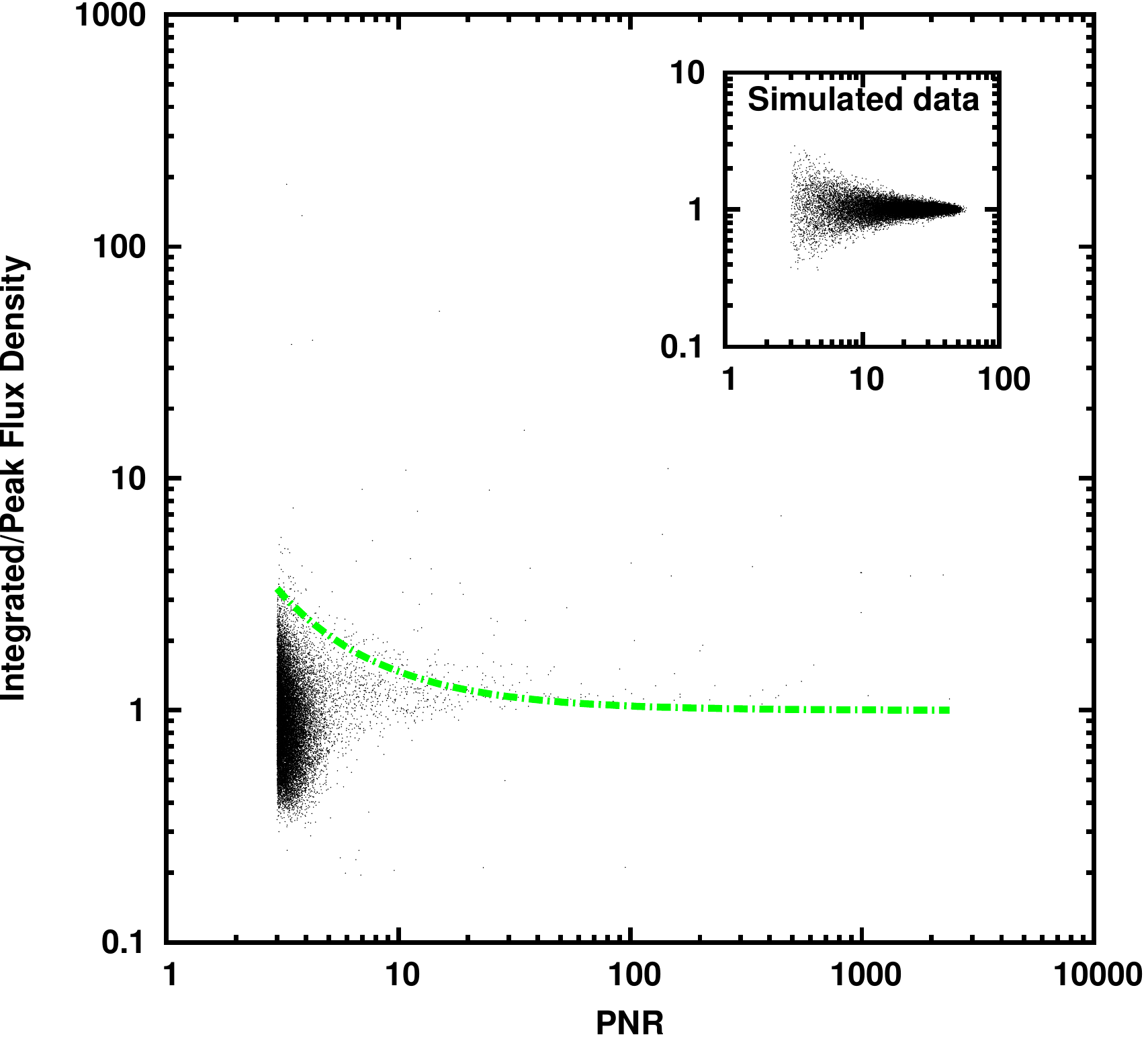}
\includegraphics[scale=0.4]{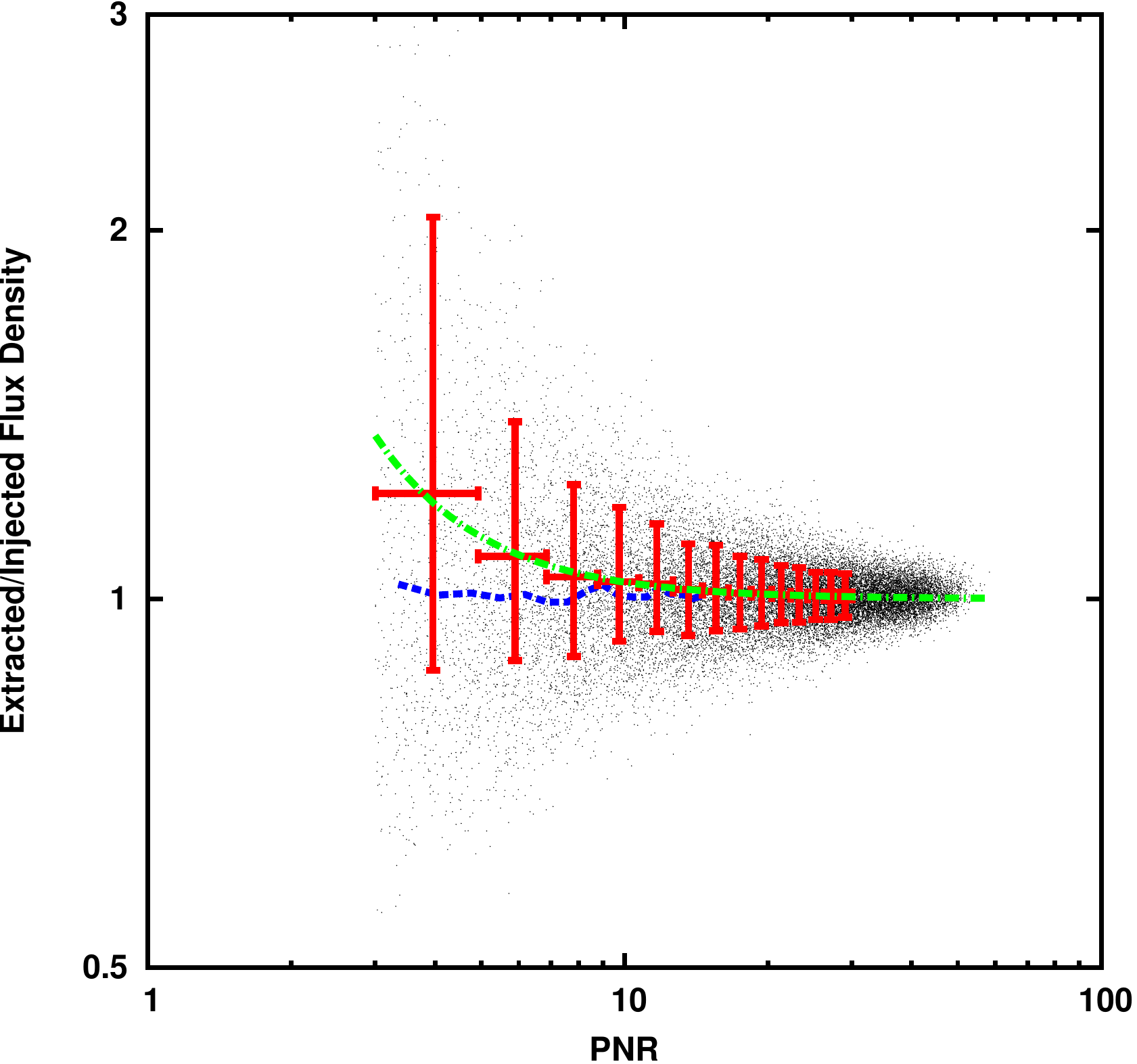}
\caption{{\it Left:} Plots of the ratio of the extracted integrated
  and peak flux densities for simulated sources as a function of
  PNR. The inset shows the results for simulated point sources and
  demonstrates how at low signal-to-noise the ratio deviates
  symmetrically from its initial value of unity. The main plot shows
  the same for the real data along with the upper envelope derived
  from the simulated data. All sources below the green line were
  fitted as point sources. {\it Right:} Ratio of the injected and
  extracted flux densities for simulated sources as a function of
  PNR. The points with error bars plot the median gain weighted by the
  differential source counts (see text) in consecutive bins. The
  vertical errorbars give the range including 68~per~cent of the
  sources in that bin. The dash-dotted line is a polynomial fit to the
  points and is used to correct the flux densities of the real sources
  for flux boosting. The lower line shows the median gain without
  weighting by the source counts -- its value is approximately equal
  to one, independent of PNR.}
\label{fig:areaplots}
\end{center}
\end{figure*}

To improve the accuracy of our extracted flux densities, we extract
sources in two ways. In the vast majority of cases we assume that the
source is unresolved and fix the size of the fitted Gaussian to that
of the restoring beam. For those sources that are significantly
resolved we instead allow the size of the Gaussian to vary. The reason
for this approach is that allowing the size of unresolved sources to
vary often produces cases where the peak flux density is greater than
the total, a consequence of its measured size being smaller than the
beam. The result is that the measured flux densities are less accurate
than if their sizes had been held fixed at the width of the restoring
beam. We have simulated this effect by injecting multiple point
sources into our residual map and extracting them, as with the real map,
first with the source size unconstrained, then again with the size
fixed to that of the beam. The scatter in the ratio of injected and
extracted flux densities was significantly reduced in the latter case
\citep[see also][]{ibar09}. In all the radio source simulations
described in this section, we created 50 fake maps, each containing
500 sources, i.e.\ a total of 25\,000 sources.

In order to use this approach it is obviously necessary to decide
which sources are unresolved and which are resolved. We do this in the
following way. When the source size is allowed to vary, the
uncertainties in the fitting process cause the ratio of peak to total
flux density to increase from unity as often (and by as much) as it
decreases; this symmetry is illustrated using our simulated data in
the inset of Fig.~\ref{fig:areaplots}. Each point represents a source
injected with the same size as the synthesised beam, but which has
increased or decreased in size upon being extracted. The envelope of
this plot locates sources in the real data which are inherently
unresolved and which should be fitted as such, yielding a more
accurate flux density. A similar approach was adopted by
\citet{bondi03}, but using the observed data only and not
simulations. Fig.~\ref{fig:areaplots} also shows the envelope
(containing 98~per~cent of the simulated sources) plotted over the
real data. Those sources lying above the upper envelope are fitted
using a variable width; all other sources are constrained to be point
sources.

\subsubsection{Bias correction}

We have also studied the effects of biases in the model fitting by
comparing the fluxes that we recover from our simulations to those
that were injected. A plot of this flux density ratio against PNR
(calculated based on the {\it recovered} peak flux, a measure against
which we can correct our data) is also shown in the right panel of
Fig.~\ref{fig:areaplots} (lower line) -- we find that the median
value is close to unity, independent of PNR i.e.\ there is no bias in
the measured flux densities. This is in contrast to the findings of
\citet{seymour04}, who find a significant positive bias. This is
because \citeauthor{seymour04} plot their flux ratios as a function of
{\it input} flux density, a quantity which is unknown in the real
radio data and which is biased towards sources whose flux densities
have increased due to the model-fitting uncertainties.

\subsubsection{Flux boosting}
\label{fluxboost}

`Flux boosting' is an effect regularly taken into account when
estimating the flux densities of SMGs
\citep[e.g.][]{coppin06,austermann09,weiss09}, but very rarely with
radio sources. The apparent flux density of such a source deviates
from its true value if it sits on/in a noise peak/trough. Because
faint sources are more numerous than bright ones, the measured flux
density of a catalogued source (i.e.\ a source lying above the chosen
SNR threshold) is more likely to have been boosted than reduced. The
most likely flux density is produced by `deboosting' the measured flux
densities by the appropriate factor.

We have measured the magnitude of the flux boosting as a function of
recovered peak flux density by using the same simulations that were
used to investigate the biases in the model fitting. These were
performed using equal numbers of sources per flux density bin and
therefore do not show the effects of flux boosting (the dots in the
right panel of Fig.~\ref{fig:areaplots}). The source counts can
however be added retrospectively by defining flux density bins and
again forming a median, but this time weighting each point (source) in
a bin by its differential source count ($dN/dS$). Using our catalogue,
we measured a Euclidean slope for the source counts, based on sources
above a flux density of 100\,$\umu$Jy. This should be valid for all
sources fainter than this limit as extremely deep radio observations
have shown that there is no change in the slope of the counts down to
flux densities as low as $\sim$15\,$\mu$Jy \citep{owen08}. The bins
and the value of the flux boosting correction in each are overplotted
on the unweighted data in the right panel of Fig.~\ref{fig:areaplots}
as red points with errorbars; the flux boosting at any value of PNR is
calculated by fitting a function to these points (also shown in the
figure). For a 3-$\sigma$ source the flux boosting is equal to
36~per~cent.

\subsection{{\it Spitzer} MIPS catalogues}

The 24-$\umu$m data are taken from FIDEL, the Far-Infrared Deep
Extragalactic Legacy Survey (Dickinson et al., in preparation), a
programme to map the ECDFS (as well as the Extended Groth Strip and
GOODS-N) at 24\,$\umu$m using the MIPS camera on board {\it
  Spitzer}. The FIDEL MIPS data were reduced following the procedures
given in \cite{chary04}, \citet{frayer06} and \citet{frayer09}. The
final image depth at 24\,$\umu$m varies across the field, with typical
exposure times ranging from 11\,000 to 30\,000\,s (with a maximum of
approximately 36\,000\,s). The 24-$\umu$m image almost completely
covers the area mapped by LABOCA and only one submm source (LESS046)
falls off its edge.

For the counterpart analysis, we have used a catalogue produced by the
{\sc daophot} package from {\sc iraf}; the source extraction was not
guided by information on positions from other wavelengths. Examination
of the differential number counts in the 24\,$\umu$m data show that
these turn over at $\sim$30\,$\umu$Jy due to incompleteness; thus we
have not considered sources with fluxes lower than this. The flux
errors reported by the {\sc daophot} software are gross
underestimates, but simulations have shown that the values of SNR
reported by the {\sc apex} point-source extraction software
specifically developed for {\it Spitzer} \citep{makovoz02} are
accurate. Although we have not used the {\sc apex} catalogue for our
counterpart analysis (it does not go as deep as that produced using
{\sc daophot}), matching sources from the two catalogues to within
1~arcsec shows that the {\sc apex} SNR and {\sc daophot}
flux/$\Delta$flux measurements are linearly related and that the
latter need to be multiplied by a factor of three; the simulations
also showed that the flux measurements from each catalogue were
consistent.

\subsection{{\it Spitzer} IRAC catalogues}

The {\it Spitzer} Infrared Array Camera \citep[IRAC;][]{fazio04}
images are taken from the {\it Spitzer} IRAC and MUSYC Public Legacy
in ECDFS (SIMPLE) survey (Damen et al., in preparation).  We use
SExtractor \citep{bertin96} to extract source positions on a summed
image of all four IRAC channels, weighted such that a source of a
given magnitude in each image is equally represented. The areas within
15\,arcsec of each LESS source were checked visually to ensure the
catalogues were complete. We then use {\sc apphot} in {\sc iraf} to
extract fluxes in 3.8-arcsec diameter apertures on the 3.6-$\umu$m and
5.8-$\umu$m images, and apply aperture corrections as derived by the
SWIRE team \citep{surace05} to obtain total source magnitudes.

\section{Identifying counterparts to SMGs}
\label{strategy}

Following several other authors \citep[e.g.][]{ivison02,ivison07,
  pope06,chapin09} we have identified the most likely radio and
24-$\umu$m counterparts to the LESS sources by calculating the {\it
  corrected Poissonian probability} \citep{browne78,downes86} of radio
and 24-$\umu$m sources that lie within a search radius, $r_{\mathrm s}$, of
each SMG. Given a potential counterpart at radius, $r$, with flux
density, $S$, we can calculate the {\it a priori} probability, $p$, of
finding at least one object within that radius of at least that flux
density from the expected number of events
\begin{equation}
\umu_r = \pi r^2 n_{\mathrm S}
\end{equation}
where $n_{\mathrm S}$ is the surface density of sources with fluxes $>S$.
The probability is
\begin{equation}
\label{eq:p}
p = 1 - \exp (-\umu_r).
\end{equation}
However, as the search is being conducted over the (generally) larger
radius $r_{\rm s}$, this is not the probability we require, i.e.\ searching
at random locations will find more sources as extreme as the one found
than would be expected given its measured probability, $p$. Having
found a source of probability $p$, at radius $r$, we need to know the
number of similar events that would be found in our random search out
to $r_{\rm s}$. Provided that $p \ll 1$, this is given by
\begin{equation}
\label{eq:mu_cor}
\umu_{\mathrm{cor}} = p \left[1 + \ln \left( \frac{p_{\rm c}}{p} \right) \right]
\end{equation}
where the so-called critical probability is defined as
\begin{equation}
p_{\rm c} = \pi r_{\rm s}^2 n_{\rm lim}
\end{equation}
and $n_{\rm lim}$ is the source surface density at our lowest
detectable flux density\footnote{The critical probability corresponds
  to finding the faintest possible source at the largest possible
  distance. By definition, there is no possibility of finding other
  sources that are at least as probable anywhere else within $r_{\rm
    s}$ and so such sources do not require a correction factor,
  i.e.\ the factor in square brackets in Equation~\ref{eq:mu_cor} is
  unity.}. The final probability of a counterpart being a chance
coincidence is calculated by inserting the corrected number of events
($\umu_{\mathrm{cor}}$) into Equation~\ref{eq:p} in place of
$\umu$. As has been typical in the literature
\citep{ivison02,ivison07,pope06,chapin09} we take a value of
$p\le0.05$ to indicate a secure association.

\begin{figure}
\begin{center}
\includegraphics[scale=0.45]{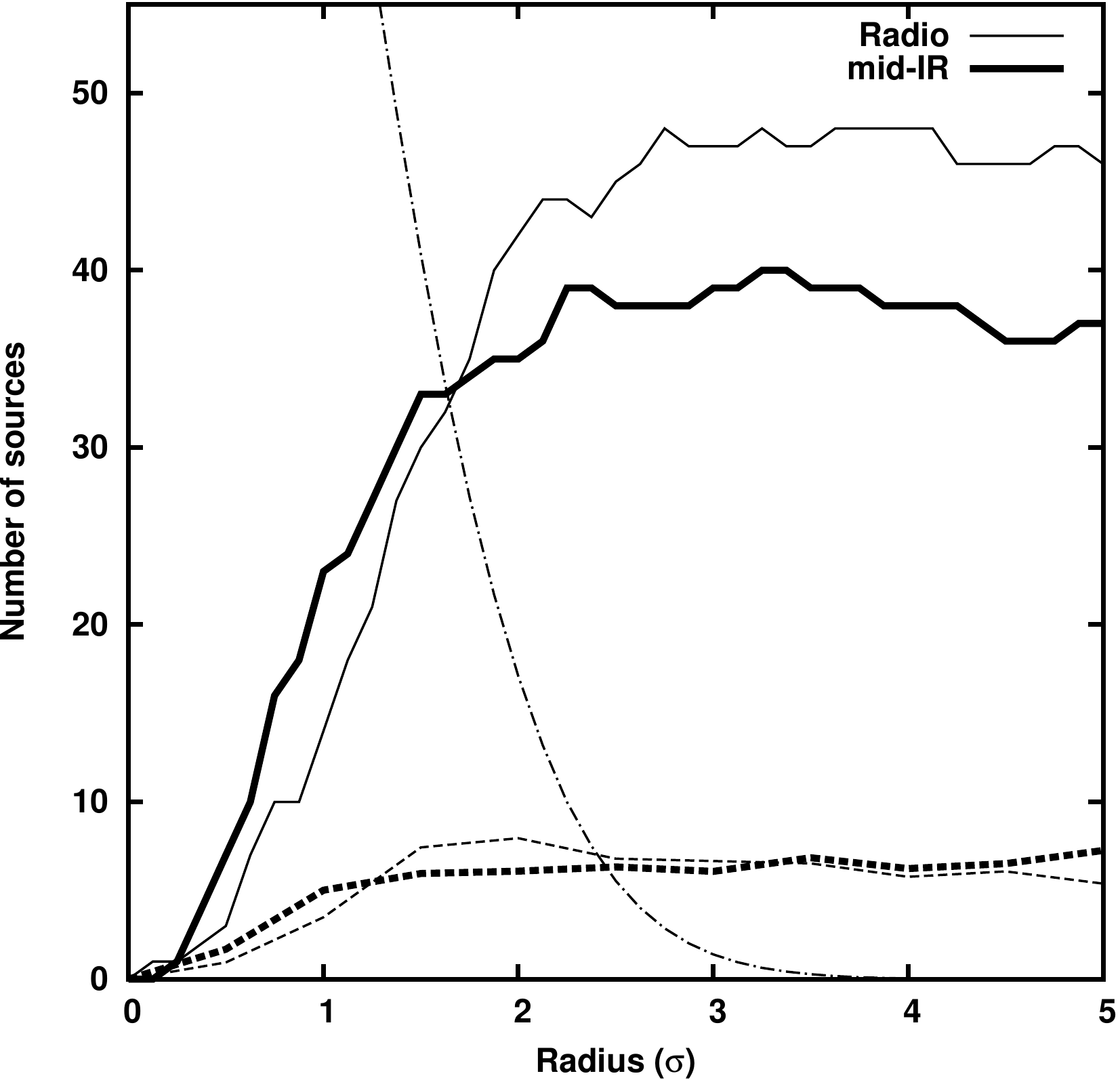}
\caption{The solid lines show the number of secure counterparts ($p
  \le 0.05$) as a function of radius (in units of the SMG 1-$\sigma$
  positional uncertainty) for the radio catalogue ({\it thin line})
  and 24-$\umu$m catalogue ({\it thick line}). Also shown is the
  number of SMGs for which the counterpart will not have been found,
  based on the cumulative distribution function of the Rayleigh
  distribution (dot-dashed line). The low dashed lines show the
  results of the Monte Carlo simulations, i.e.\ the number of secure
  counterparts found as a function of radius for randomly distributed
  SMGs. Choosing a radius of 3-$\sigma$ produces close to the maximum
  number of counterparts and results in only $\sim$1~per~cent of the
  SMGs not being searched out to a sufficient radius. Note that we do
  not show the IRAC IDs here as those were only searched for in error
  circles devoid of radio and MIPS counterparts, using a selection
  guided by the radio/MIPS IDs.}
\label{fig:summedp}
\end{center}
\end{figure}

Offsets between the SMG and radio/24-$\umu$m sources will be dominated
by the uncertainty in the SMG positions, this being a function of the
SNR of the submm detection. Therefore, in contrast to some studies of
this type that use a search radius based on some representative SNR,
we have chosen a different search radius for each SMG that is some
multiple of its 1-$\sigma$ positional uncertainty in Right
Ascension/Declination ($\approx$1--3~arcsec); \citet{smail00}
similarly used a SNR-dependent search radius. Effects such as
telescope pointing errors might conspire to produce systematic offsets
between the submm source and its counterpart, but as the final map is
an average of multiple observations taken at different times, any such
systematic offsets are minimal. \citet{greve10} come to the same
conclusion from a stacking analysis of the LESS data which confirms
the absolute astrometry of the submm map.

In choosing a value for $r_{\rm s}$, our over-riding concern has been
to make it large enough to avoid missing significant numbers of
counterparts, but small enough to avoid choosing counterparts from
unrelated, bright field sources; too large a radius also
over-estimates the value of $p$. A reasonable maximum value for
$r_{\rm s}$ is 3\,$\sigma$ as this will ensure that we only miss the
counterpart for one~per~cent of the SMGs\footnote{From the cumulative
  distribution function (CDF) of the Rayleigh distribution,
  $\mathrm{CDF} = 1 - e^{-r^2/2 \sigma^2}$, where $r_{\rm s}/\sigma = 3$.}
i.e.\ 1.3 sources. The number of SMGs with missed counterparts is
plotted in Fig.~\ref{fig:summedp} over a wide function of radius, 0 to
5\,$\sigma$. Also plotted in Fig.~\ref{fig:summedp} are the numbers of
secure counterparts found as a function of radius for both the radio
and 24-$\umu$m catalogues. Both rise steeply between 0 and 2\,$\sigma$
and gently decline above $\sim$3\,$\sigma$. We have therefore set our
search radius to $r_{\rm s} = 3\,\sigma$.

We have also used Monte Carlo simulations to investigate the effect of
varying the search radius, producing 100 realisations of the SMG
catalogue at radii between 0 and 5\,$\sigma$ in steps of 0.5\,$\sigma$
and searching for secure counterparts in the same way as with the real
data. Each simulated catalogue has the same distribution of flux
densities, and therefore search radii, as the real catalogue, but with
randomised positions. The results are again shown in
Fig.~\ref{fig:summedp} and illustrate that the number of false
detections is approximately constant beyond $r_{\rm s} \sim$1.5\,$\sigma$.

The positional errors for each SMG have been determined using the
simulated source extractions of \citet{weiss09}. This offers
advantages over analytical formulae such as Equation~B22 of
\citet{ivison07} in that it includes all sources of uncertainty,
including those originating in the data reduction and source
extraction processes. The empirical formula for the positional
uncertainties is
\begin{equation}
\sigma_{x,y} = a \exp (-b\,S_{\mathrm{in}}) + c
\end{equation}
where $a = 6.08$, $b=0.14$, $c=0.56$ and $S_{\mathrm{in}}$ is the
intrinsic flux of the source, i.e.\ the observed flux after
de-boosting.

Integrated source counts were calculated for both the radio and
24-$\umu$m data from the respective catalogues; these are used to
calculate the value of $n_{\rm S}$ at both the flux density of the potential
counterpart and the flux limit (for the radio catalogue we formed the
counts using the un-deboosted fluxes as these correspond to the actual
source densities in the radio map). In order to test these and the
entire $p$-statistic procedure, we have again performed Monte Carlo
simulations, producing 500 realisations of the 126-source submm
catalogue as described above. On average, five per~cent of the SMGs
should have a counterpart with $p\le0.05$. This corresponds to 6.3
sources on average and we indeed find values of 6.326 for the radio
and 6.302 for the 24-$\umu$m data. We are thus confident that we are
measuring the correct probablilities for each counterpart.

\section{The radio and MIPS counterparts}
\label{results}

\begin{table*}
\centering
{\small
\caption{Radio properties of potential counterparts to LESS
  870-$\umu$m sources in the ECDFS. SMGs are listed in order of
  decreasing SNR. SMG names appended with an $^*$ indicate that they
  are not on the \citet{miller08} map; radio counterparts have instead
  been searched for using our own reduction. Secure counterparts
  ($p\le0.05$) are in boldface and where $p$ lies between 0.05 and 1.0
  this is given in parentheses. Counterparts where $0.05 < p \le 0.1$
  is obtained at two out of radio, 24-$\umu$m (Table~\ref{tab:p24um})
  or 5.8-$\umu$m (Table~\ref{tab:pirac}) have their value of $p$ given
  in boldface within parentheses. Counterparts which formally have
  $p\le0.05$ but may be spurious are given in {\it square}
  parentheses. Although not used in the $p$-statistic procedure, all
  radio fluxes have been corrected for flux boosting (Section~\ref{fluxboost}.)}
\begin{tabular}{clcccccc} \hline
ID & SMG name & Submm position & $r_{\rm s}$ & Radio position & Radio flux & Offset & $p$ \\
   & & ($\alpha_{\mathrm{J2000}}$)\hspace{.9cm}($\delta_{\mathrm{J2000}}$) & (arcsec) & ($\alpha_{\mathrm{J2000}}$)\hspace{.9cm}($\delta_{\mathrm{J2000}}$) & ($\umu$Jy) & (arcsec) & \\ \hline
001 & LESS J033314.3$-$275611 & 03:33:14.26 $-$27:56:11.2 & 3.1 & $-$\hspace{1.5cm}$-$ & $-$ & $-$ & $-$\\
002 & LESS J033302.5$-$275643 & 03:33:02.50 $-$27:56:43.6 & 3.8 & 03:33:02.7150 $-$27:56:42.539 & $  234.6\pm   7.8$ &  3.0 & {\bf 0.004} \\
003 & LESS J033321.5$-$275520 & 03:33:21.51 $-$27:55:20.2 & 3.8 & $-$\hspace{1.5cm}$-$ & $-$ & $-$ & $-$\\
004 & LESS J033136.0$-$275439 & 03:31:36.01 $-$27:54:39.2 & 4.1 & $-$\hspace{1.5cm}$-$ & $-$ & $-$ & $-$\\
005 & LESS J033129.5$-$275907 & 03:31:29.46 $-$27:59:07.3 & 4.6 & $-$\hspace{1.5cm}$-$ & $-$ & $-$ & $-$\\
006 & LESS J033257.1$-$280102 & 03:32:57.14 $-$28:01:02.1 & 4.8 & 03:32:56.9734 $-$28:01:01.204 & $   42.7\pm   7.4$ &  2.4 & {\bf 0.013} \\
007 & LESS J033315.6$-$274523 & 03:33:15.55 $-$27:45:23.6 & 5.1 & 03:33:15.4267 $-$27:45:24.430 & $   75.8\pm   6.9$ &  1.8 & {\bf 0.006} \\
008 & LESS J033205.1$-$273108$^*$ & 03:32:05.07 $-$27:31:08.8 & 4.0 & $-$\hspace{1.5cm}$-$ & $-$ & $-$ & $-$\\
009 & LESS J033211.3$-$275210 & 03:32:11.29 $-$27:52:10.4 & 5.1 & 03:32:11.3737 $-$27:52:12.139 & $   31.0\pm   6.3$ &  2.1 & {\bf 0.025} \\
010 & LESS J033219.0$-$275219 & 03:32:19.02 $-$27:52:19.4 & 5.1 & 03:32:19.0632 $-$27:52:14.829 & $   54.9\pm   6.0$ &  4.6 & {\bf 0.035} \\
 & & & & 03:32:19.1370 $-$27:52:18.115 & $   51.1\pm   6.1$ &  2.0 & {\bf 0.011} \\
 & & & & 03:32:19.3086 $-$27:52:19.018 & $   50.1\pm   6.2$ &  3.8 & {\bf 0.029} \\
011 & LESS J033213.6$-$275602 & 03:32:13.58 $-$27:56:02.5 & 5.2 & 03:32:13.8475 $-$27:56:00.247 & $   55.1\pm   6.6$ &  4.2 & {\bf 0.029} \\
012 & LESS J033248.1$-$275414 & 03:32:48.12 $-$27:54:14.7 & 5.3 & 03:32:47.9995 $-$27:54:16.497 & $   39.9\pm   6.5$ &  2.4 & {\bf 0.020} \\
 & & & & 03:32:48.3987 $-$27:54:16.741 & $   21.4\pm   5.5$ &  4.2 & 0.113 \\
013 & LESS J033249.2$-$274246 & 03:32:49.23 $-$27:42:46.6 & 5.3 & $-$\hspace{1.5cm}$-$ & $-$ & $-$ & $-$\\
014 & LESS J033152.6$-$280320 & 03:31:52.64 $-$28:03:20.4 & 5.1 & 03:31:52.4870 $-$28:03:18.934 & $   89.4\pm   8.0$ &  2.5 & {\bf 0.007} \\
015 & LESS J033333.4$-$275930 & 03:33:33.36 $-$27:59:30.1 & 5.3 & $-$\hspace{1.5cm}$-$ & $-$ & $-$ & $-$\\
016 & LESS J033218.9$-$273738 & 03:32:18.89 $-$27:37:38.7 & 5.8 & 03:32:18.6870 $-$27:37:43.145 & $   49.0\pm   8.5$ &  5.2 & {\bf 0.039} \\
017 & LESS J033207.6$-$275123 & 03:32:07.59 $-$27:51:23.0 & 6.1 & 03:32:07.3105 $-$27:51:20.849 & $  120.3\pm  14.5$ &  4.3 & {\bf 0.015} \\
018 & LESS J033205.1$-$274652 & 03:32:05.12 $-$27:46:52.1 & 6.2 & 03:32:04.9033 $-$27:46:47.449 & $  130.1\pm  17.3$ &  5.5 & {\bf 0.020} \\
019 & LESS J033208.1$-$275818 & 03:32:08.10 $-$27:58:18.7 & 6.4 & 03:32:08.2721 $-$27:58:14.069 & $   30.8\pm   6.0$ &  5.2 & (0.090) \\
020 & LESS J033316.6$-$280018 & 03:33:16.56 $-$28:00:18.8 & 6.5 & 03:33:16.7726 $-$28:00:16.120 & $ 4251.9\pm  16.0$ &  3.9 & {\bf 0.001} \\
021 & LESS J033329.9$-$273441 & 03:33:29.93 $-$27:34:41.7 & 6.2 & $-$\hspace{1.5cm}$-$ & $-$ & $-$ & $-$\\
022 & LESS J033147.0$-$273243 & 03:31:47.02 $-$27:32:43.0 & 5.9 & 03:31:46.9496 $-$27:32:39.547 & $  111.3\pm  25.3$ &  3.6 & {\bf 0.009} \\
023 & LESS J033212.1$-$280508 & 03:32:12.11 $-$28:05:08.5 & 5.8 & $-$\hspace{1.5cm}$-$ & $-$ & $-$ & $-$\\
024 & LESS J033336.8$-$274401 & 03:33:36.79 $-$27:44:01.0 & 6.3 & 03:33:36.9881 $-$27:43:58.749 & $   60.0\pm   8.1$ &  3.5 & {\bf 0.019} \\
025 & LESS J033157.1$-$275940 & 03:31:57.05 $-$27:59:40.8 & 6.8 & 03:31:56.8845 $-$27:59:39.653 & $   61.3\pm   7.3$ &  2.5 & {\bf 0.012} \\
026 & LESS J033136.9$-$275456 & 03:31:36.90 $-$27:54:56.1 & 7.0 & $-$\hspace{1.5cm}$-$ & $-$ & $-$ & $-$\\
027 & LESS J033149.7$-$273432 & 03:31:49.73 $-$27:34:32.7 & 6.5 & $-$\hspace{1.5cm}$-$ & $-$ & $-$ & $-$\\
028 & LESS J033302.9$-$274432 & 03:33:02.92 $-$27:44:32.6 & 6.9 & $-$\hspace{1.5cm}$-$ & $-$ & $-$ & $-$\\
029 & LESS J033336.9$-$275813 & 03:33:36.90 $-$27:58:13.0 & 6.6 & 03:33:36.8866 $-$27:58:09.382 & $   44.7\pm   8.6$ &  3.6 & {\bf 0.024} \\
030 & LESS J033344.4$-$280346 & 03:33:44.37 $-$28:03:46.1 & 5.5 & $-$\hspace{1.5cm}$-$ & $-$ & $-$ & $-$\\
031 & LESS J033150.0$-$275743 & 03:31:49.96 $-$27:57:43.9 & 7.2 & 03:31:49.8280 $-$27:57:40.833 & $   25.9\pm   5.8$ &  3.5 & (0.085) \\
032 & LESS J033243.6$-$274644 & 03:32:43.57 $-$27:46:44.0 & 7.2 & $-$\hspace{1.5cm}$-$ & $-$ & $-$ & $-$\\
033 & LESS J033149.8$-$275332 & 03:31:49.78 $-$27:53:32.9 & 7.2 & $-$\hspace{1.5cm}$-$ & $-$ & $-$ & $-$\\
034 & LESS J033217.6$-$275230 & 03:32:17.64 $-$27:52:30.3 & 7.2 & $-$\hspace{1.5cm}$-$ & $-$ & $-$ & $-$\\
035 & LESS J033110.3$-$273714$^*$ & 03:31:10.35 $-$27:37:14.8 & 5.9 & $-$\hspace{1.5cm}$-$ & $-$ & $-$ & $-$\\
036 & LESS J033149.2$-$280208 & 03:31:49.15 $-$28:02:08.7 & 7.2 & 03:31:48.9740 $-$28:02:14.399 & $   47.5\pm   7.5$ &  6.2 & ({\bf 0.057 + MIPS}) \\
037 & LESS J033336.0$-$275347 & 03:33:36.04 $-$27:53:47.6 & 6.9 & $-$\hspace{1.5cm}$-$ & $-$ & $-$ & $-$\\
038 & LESS J033310.2$-$275641 & 03:33:10.20 $-$27:56:41.5 & 7.5 & $-$\hspace{1.5cm}$-$ & $-$ & $-$ & $-$\\
039 & LESS J033144.9$-$273435 & 03:31:44.90 $-$27:34:35.4 & 7.3 & 03:31:45.0493 $-$27:34:37.060 & $   45.9\pm   7.5$ &  2.6 & {\bf 0.017} \\
 & & & & 03:31:45.0634 $-$27:34:30.112 & $   27.9\pm   6.7$ &  5.7 & (0.105) \\
040 & LESS J033246.7$-$275120 & 03:32:46.74 $-$27:51:20.9 & 7.6 & 03:32:46.8465 $-$27:51:21.024 & $  119.1\pm  13.7$ &  1.4 & {\bf 0.003} \\
041 & LESS J033110.5$-$275233$^*$ & 03:31:10.47 $-$27:52:33.2 & 6.2 & $-$\hspace{1.5cm}$-$ & $-$ & $-$ & $-$\\
042 & LESS J033231.0$-$275858 & 03:32:31.02 $-$27:58:58.1 & 7.7 & $-$\hspace{1.5cm}$-$ & $-$ & $-$ & $-$\\
043 & LESS J033307.0$-$274801 & 03:33:07.00 $-$27:48:01.0 & 7.6 & 03:33:07.4844 $-$27:47:59.336 & $   25.2\pm   6.4$ &  6.6 & 0.168 \\
044 & LESS J033131.0$-$273238 & 03:31:30.96 $-$27:32:38.5 & 6.9 & 03:31:31.2272 $-$27:32:39.111 & $   90.3\pm   9.6$ &  3.6 & {\bf 0.012} \\
045 & LESS J033225.7$-$275228 & 03:32:25.71 $-$27:52:28.5 & 7.7 & 03:32:25.2714 $-$27:52:30.692 & $   31.1\pm   5.9$ &  6.2 & 0.135 \\
046 & LESS J033336.8$-$273247 & 03:33:36.80 $-$27:32:47.0 & 6.5 & 03:33:36.7533 $-$27:32:49.574 & $   73.2\pm  10.7$ &  2.6 & {\bf 0.008} \\
047 & LESS J033256.0$-$273317 & 03:32:56.00 $-$27:33:17.7 & 7.2 & $-$\hspace{1.5cm}$-$ & $-$ & $-$ & $-$\\
048 & LESS J033237.8$-$273202 & 03:32:37.77 $-$27:32:02.0 & 6.8 & 03:32:38.0090 $-$27:31:59.927 & $   84.5\pm   8.3$ &  3.8 & {\bf 0.015} \\
049 & LESS J033124.4$-$275040 & 03:31:24.45 $-$27:50:40.9 & 7.6 & 03:31:24.2001 $-$27:50:42.774 & $   31.5\pm   7.1$ &  3.8 & (0.056) \\
 & & & & 03:31:24.5046 $-$27:50:37.576 & $   36.0\pm   7.2$ &  3.4 & {\bf 0.038} \\
 & & & & 03:31:24.7140 $-$27:50:46.507 & $  115.9\pm  19.1$ &  6.6 & {\bf 0.029} \\
050 & LESS J033141.2$-$274441 & 03:31:41.15 $-$27:44:41.5 & 7.9 & 03:31:40.9917 $-$27:44:35.238 & $   77.3\pm   7.2$ &  6.6 & {\bf 0.047} \\
 & & & & 03:31:41.4170 $-$27:44:46.966 & $   38.8\pm   6.8$ &  6.5 & (0.090) \\
051 & LESS J033144.8$-$274425 & 03:31:44.81 $-$27:44:25.1 & 7.9 & 03:31:45.0647 $-$27:44:27.794 & $   29.5\pm   6.3$ &  4.3 & (0.093) \\
\end{tabular}
\label{tab:pradio}
}
\end{table*}

\begin{table*}
\centering
{\small
\contcaption{}
\begin{tabular}{clcccccc} \hline
ID & SMG name & Submm position & $r_{\rm s}$ & Radio position & Radio flux & Offset & $p$ \\
   & & ($\alpha_{\mathrm{J2000}}$)\hspace{.9cm}($\delta_{\mathrm{J2000}}$) & (arcsec) & ($\alpha_{\mathrm{J2000}}$)\hspace{.9cm}($\delta_{\mathrm{J2000}}$) & ($\umu$Jy) & (arcsec) & \\ \hline
052 & LESS J033128.5$-$275601 & 03:31:28.51 $-$27:56:01.3 & 7.9 & $-$\hspace{1.5cm}$-$ & $-$ & $-$ & $-$\\
053 & LESS J033159.1$-$275435 & 03:31:59.12 $-$27:54:35.5 & 8.0 & $-$\hspace{1.5cm}$-$ & $-$ & $-$ & $-$\\
054 & LESS J033243.6$-$273353 & 03:32:43.61 $-$27:33:53.6 & 7.6 & $-$\hspace{1.5cm}$-$ & $-$ & $-$ & $-$\\
055 & LESS J033302.2$-$274033 & 03:33:02.20 $-$27:40:33.6 & 8.0 & $-$\hspace{1.5cm}$-$ & $-$ & $-$ & $-$\\
056 & LESS J033153.2$-$273936 & 03:31:53.17 $-$27:39:36.1 & 8.1 & 03:31:53.1189 $-$27:39:38.555 & $   31.8\pm   6.5$ &  2.5 & {\bf 0.036} \\
057 & LESS J033152.0$-$275329 & 03:31:51.97 $-$27:53:29.7 & 8.0 & 03:31:51.9370 $-$27:53:27.179 & $   49.4\pm   6.7$ &  2.6 & {\bf 0.018} \\
058 & LESS J033225.8$-$273306 & 03:32:25.79 $-$27:33:06.7 & 7.6 & 03:32:25.5399 $-$27:33:06.953 & $   25.2\pm   6.7$ &  3.3 & (0.069) \\
059 & LESS J033303.9$-$274412 & 03:33:03.87 $-$27:44:12.2 & 8.2 & 03:33:03.8207 $-$27:44:14.497 & $   22.1\pm   5.9$ &  2.4 & (0.069) \\
 & & & & 03:33:03.5906 $-$27:44:13.586 & $   29.5\pm   6.3$ &  4.0 & (0.085) \\
060 & LESS J033317.5$-$275121 & 03:33:17.47 $-$27:51:21.5 & 8.3 & 03:33:17.4972 $-$27:51:28.796 & $   64.7\pm   6.9$ &  7.3 & ({\bf 0.067 + MIPS}) \\
061 & LESS J033245.6$-$280025 & 03:32:45.63 $-$28:00:25.3 & 8.3 & $-$\hspace{1.5cm}$-$ & $-$ & $-$ & $-$\\
062 & LESS J033236.4$-$273452 & 03:32:36.41 $-$27:34:52.5 & 8.2 & 03:32:36.5309 $-$27:34:53.363 & $  151.2\pm   7.3$ &  1.8 & {\bf 0.003} \\
 & & & & 03:32:36.6933 $-$27:34:47.261 & $   28.4\pm   6.4$ &  6.5 & 0.136 \\
063 & LESS J033308.5$-$280044 & 03:33:08.46 $-$28:00:44.3 & 8.3 & 03:33:08.5591 $-$28:00:44.866 & $   31.8\pm   7.2$ &  1.4 & [{\bf 0.013}] \\
064 & LESS J033201.0$-$280025 & 03:32:01.00 $-$28:00:25.6 & 8.4 & 03:32:00.9468 $-$28:00:26.467 & $   23.6\pm   6.2$ &  1.1 & {\bf 0.018} \\
065 & LESS J033252.4$-$273527 & 03:32:52.40 $-$27:35:27.7 & 8.3 & $-$\hspace{1.5cm}$-$ & $-$ & $-$ & $-$\\
066 & LESS J033331.7$-$275406 & 03:33:31.69 $-$27:54:06.1 & 8.2 & 03:33:31.9745 $-$27:54:10.257 & $   67.0\pm   7.8$ &  5.6 & {\bf 0.041} \\
067 & LESS J033243.3$-$275517 & 03:32:43.28 $-$27:55:17.9 & 8.4 & 03:32:43.2046 $-$27:55:14.289 & $   90.1\pm  14.8$ &  3.7 & {\bf 0.018} \\
 & & & & 03:32:43.8211 $-$27:55:15.380 & $   25.1\pm   5.8$ &  7.6 & 0.225 \\
068 & LESS J033233.4$-$273918 & 03:32:33.44 $-$27:39:18.5 & 8.4 & 03:32:33.9689 $-$27:39:14.491 & $   20.6\pm   5.5$ &  8.1 & 0.257 \\
069 & LESS J033134.3$-$275934 & 03:31:34.26 $-$27:59:34.3 & 8.5 & $-$\hspace{1.5cm}$-$ & $-$ & $-$ & $-$\\
070 & LESS J033144.0$-$273832 & 03:31:43.97 $-$27:38:32.5 & 8.5 & 03:31:44.0325 $-$27:38:35.859 & $  322.3\pm  14.6$ &  3.5 & {\bf 0.005} \\
071 & LESS J033306.3$-$273327 & 03:33:06.29 $-$27:33:27.7 & 8.0 & $-$\hspace{1.5cm}$-$ & $-$ & $-$ & $-$\\
072 & LESS J033240.4$-$273802 & 03:32:40.40 $-$27:38:02.5 & 8.5 & 03:32:40.0506 $-$27:38:09.235 & $   34.3\pm   7.1$ &  8.2 & 0.118 \\
073 & LESS J033229.3$-$275619 & 03:32:29.33 $-$27:56:19.3 & 8.5 & 03:32:29.3049 $-$27:56:19.404 & $   18.9\pm   5.1$ &  0.3 & {\bf 0.005} \\
 & & & & 03:32:29.3518 $-$27:56:23.802 & $   18.5\pm   5.2$ &  4.5 & 0.228 \\
074 & LESS J033309.3$-$274809 & 03:33:09.34 $-$27:48:09.9 & 8.4 & 03:33:09.1492 $-$27:48:16.833 & $   43.8\pm   7.5$ &  7.4 & ({\bf 0.085 + IRAC}) \\
 & & & & 03:33:09.3836 $-$27:48:15.887 & $   34.8\pm   7.0$ &  6.0 & (0.095) \\
075 & LESS J033126.8$-$275554 & 03:31:26.83 $-$27:55:54.6 & 8.4 & 03:31:27.1942 $-$27:55:51.287 & $   72.3\pm   8.2$ &  5.9 & {\bf 0.038} \\
076 & LESS J033332.7$-$275957 & 03:33:32.67 $-$27:59:57.2 & 8.4 & 03:33:32.3411 $-$27:59:54.831 & $   41.6\pm   8.4$ &  5.0 & {\bf 0.042} \\
077 & LESS J033157.2$-$275633 & 03:31:57.23 $-$27:56:33.2 & 8.8 & $-$\hspace{1.5cm}$-$ & $-$ & $-$ & $-$\\
078 & LESS J033340.3$-$273956 & 03:33:40.30 $-$27:39:56.9 & 8.4 & 03:33:40.1122 $-$27:39:49.684 & $   75.2\pm   9.8$ &  7.6 & {\bf 0.044} \\
079 & LESS J033221.2$-$275623 & 03:32:21.25 $-$27:56:23.5 & 8.8 & 03:32:21.6159 $-$27:56:23.755 & $   34.8\pm   6.3$ &  4.9 & (0.087) \\
080 & LESS J033142.2$-$274834 & 03:31:42.23 $-$27:48:34.4 & 8.9 & 03:31:41.8328 $-$27:48:36.131 & $   27.6\pm   6.2$ &  5.5 & 0.148 \\
 & & & & 03:31:42.8359 $-$27:48:36.936 & $   47.4\pm   6.6$ &  8.4 & 0.110 \\
081 & LESS J033127.4$-$274440 & 03:31:27.45 $-$27:44:40.4 & 8.8 & 03:31:27.5722 $-$27:44:39.651 & $  217.9\pm  15.3$ &  1.8 & {\bf 0.002} \\
082 & LESS J033253.8$-$273810 & 03:32:53.77 $-$27:38:10.9 & 9.0 & $-$\hspace{1.5cm}$-$ & $-$ & $-$ & $-$\\
083 & LESS J033308.9$-$280522 & 03:33:08.92 $-$28:05:22.0 & 8.3 & $-$\hspace{1.5cm}$-$ & $-$ & $-$ & $-$\\
084 & LESS J033154.2$-$275109 & 03:31:54.22 $-$27:51:09.8 & 8.9 & 03:31:54.5185 $-$27:51:05.700 & $   33.5\pm   6.1$ &  5.7 & 0.119 \\
 & & & & 03:31:54.8325 $-$27:51:10.973 & $   23.3\pm   5.7$ &  8.2 & 0.272 \\
085 & LESS J033110.3$-$274503$^*$ & 03:31:10.28 $-$27:45:03.1 & 7.7 & $-$\hspace{1.5cm}$-$ & $-$ & $-$ & $-$\\
086 & LESS J033114.9$-$274844 & 03:31:14.90 $-$27:48:44.3 & 8.5 & $-$\hspace{1.5cm}$-$ & $-$ & $-$ & $-$\\
087 & LESS J033251.1$-$273143 & 03:32:51.09 $-$27:31:43.0 & 8.4 & 03:32:50.8711 $-$27:31:41.762 & $  128.3\pm  25.8$ &  3.2 & {\bf 0.008} \\
 & & & & 03:32:51.0736 $-$27:31:45.730 & $   56.9\pm   8.7$ &  2.7 & {\bf 0.014} \\
088 & LESS J033155.2$-$275345 & 03:31:55.19 $-$27:53:45.3 & 9.0 & 03:31:54.7502 $-$27:53:41.012 & $   33.6\pm   6.1$ &  7.2 & 0.167 \\
 & & & & 03:31:54.8959 $-$27:53:41.303 & $   78.0\pm   6.6$ &  5.6 & {\bf 0.042} \\
 & & & & 03:31:55.7818 $-$27:53:48.183 & $   35.2\pm   6.0$ &  8.4 & 0.182 \\
089 & LESS J033248.4$-$280023 & 03:32:48.44 $-$28:00:23.8 & 9.1 & $-$\hspace{1.5cm}$-$ & $-$ & $-$ & $-$\\
090 & LESS J033243.7$-$273554 & 03:32:43.65 $-$27:35:54.1 & 9.1 & $-$\hspace{1.5cm}$-$ & $-$ & $-$ & $-$\\
091 & LESS J033135.2$-$274033 & 03:31:35.25 $-$27:40:33.7 & 9.1 & $-$\hspace{1.5cm}$-$ & $-$ & $-$ & $-$\\
092 & LESS J033138.4$-$274336 & 03:31:38.36 $-$27:43:36.0 & 9.2 & $-$\hspace{1.5cm}$-$ & $-$ & $-$ & $-$\\
093 & LESS J033110.8$-$275607 & 03:31:10.84 $-$27:56:07.2 & 8.4 & $-$\hspace{1.5cm}$-$ & $-$ & $-$ & $-$\\
094 & LESS J033307.3$-$275805 & 03:33:07.27 $-$27:58:05.0 & 9.1 & $-$\hspace{1.5cm}$-$ & $-$ & $-$ & $-$\\
095 & LESS J033241.7$-$275846 & 03:32:41.74 $-$27:58:46.1 & 9.2 & 03:32:41.2324 $-$27:58:41.752 & $   34.5\pm   6.6$ &  8.0 & 0.152 \\
096 & LESS J033313.0$-$275556 & 03:33:13.03 $-$27:55:56.8 & 9.2 & 03:33:12.6380 $-$27:55:51.515 & $   80.6\pm  17.4$ &  7.4 & (0.058) \\
097 & LESS J033313.7$-$273803 & 03:33:13.65 $-$27:38:03.4 & 9.2 & $-$\hspace{1.5cm}$-$ & $-$ & $-$ & $-$\\
098 & LESS J033130.2$-$275726 & 03:31:30.22 $-$27:57:26.0 & 9.3 & 03:31:29.8979 $-$27:57:22.733 & $  141.8\pm   8.1$ &  5.4 & {\bf 0.020} \\
099 & LESS J033251.4$-$275536 & 03:32:51.45 $-$27:55:36.0 & 9.2 & $-$\hspace{1.5cm}$-$ & $-$ & $-$ & $-$\\
100 & LESS J033111.3$-$280006 & 03:31:11.32 $-$28:00:06.2 & 8.8 & $-$\hspace{1.5cm}$-$ & $-$ & $-$ & $-$\\
101 & LESS J033151.5$-$274552 & 03:31:51.47 $-$27:45:52.1 & 9.3 & 03:31:51.6370 $-$27:45:52.262 & $   25.3\pm   6.1$ &  2.2 & {\bf 0.052} \\
102 & LESS J033335.6$-$274020 & 03:33:35.61 $-$27:40:20.1 & 9.2 & 03:33:36.1277 $-$27:40:18.677 & $   27.7\pm   7.6$ &  7.0 & 0.132 \\
103 & LESS J033325.3$-$273400 & 03:33:25.35 $-$27:34:00.4 & 9.2 & 03:33:25.8532 $-$27:33:57.591 & $   30.5\pm   8.0$ &  7.3 & 0.104 \\
104 & LESS J033258.5$-$273803 & 03:32:58.46 $-$27:38:03.0 & 9.4 & 03:32:57.8327 $-$27:37:59.389 & $   33.1\pm   6.8$ &  9.1 & 0.178 \\
\end{tabular}
}
\end{table*}

\begin{table*}
\centering
{\small
\contcaption{}
\begin{tabular}{clcccccc} \hline
ID & SMG name & Submm position & $r_{\rm s}$ & Radio position & Radio flux & Offset & $p$ \\
   & & ($\alpha_{\mathrm{J2000}}$)\hspace{.9cm}($\delta_{\mathrm{J2000}}$) & (arcsec) & ($\alpha_{\mathrm{J2000}}$)\hspace{.9cm}($\delta_{\mathrm{J2000}}$) & ($\umu$Jy) & (arcsec) & \\ \hline
105 & LESS J033115.8$-$275313 & 03:31:15.78 $-$27:53:13.1 & 9.0 & $-$\hspace{1.5cm}$-$ & $-$ & $-$ & $-$\\
106 & LESS J033140.1$-$275631 & 03:31:40.09 $-$27:56:31.4 & 9.4 & 03:31:40.1985 $-$27:56:23.051 & $   66.8\pm   7.1$ &  8.5 & (0.086) \\
107 & LESS J033130.8$-$275150 & 03:31:30.85 $-$27:51:50.9 & 9.4 & 03:31:30.5820 $-$27:51:45.062 & $   24.6\pm   6.3$ &  6.8 & 0.204 \\
 & & & & 03:31:31.3078 $-$27:51:44.774 & $   30.4\pm   6.8$ &  8.6 & 0.199 \\
108 & LESS J033316.4$-$275033 & 03:33:16.42 $-$27:50:33.1 & 9.5 & 03:33:16.5352 $-$27:50:39.704 & $  379.7\pm  36.4$ &  6.8 & {\bf 0.015} \\
109 & LESS J033328.1$-$274157 & 03:33:28.08 $-$27:41:57.0 & 9.5 & 03:33:28.0319 $-$27:42:03.554 & $   24.6\pm   6.6$ &  6.6 & 0.184 \\
110 & LESS J033122.6$-$275417 & 03:31:22.64 $-$27:54:17.2 & 9.4 & 03:31:22.7410 $-$27:54:12.315 & $   36.3\pm   7.9$ &  5.1 & (0.064) \\
111 & LESS J033325.6$-$273423 & 03:33:25.58 $-$27:34:23.0 & 9.4 & 03:33:25.1978 $-$27:34:25.322 & $   54.0\pm   9.4$ &  5.6 & {\bf 0.042} \\
112 & LESS J033249.3$-$273112$^*$ & 03:32:49.28 $-$27:31:12.3 & 9.0 & 03:32:48.8585 $-$27:31:13.054 & $   29.6\pm   8.0$ &  5.7 & (0.081) \\
 & & & & 03:32:49.4709 $-$27:31:19.667 & $   31.0\pm   7.9$ &  7.8 & (0.100) \\
113 & LESS J033236.4$-$275845 & 03:32:36.42 $-$27:58:45.9 & 9.5 & $-$\hspace{1.5cm}$-$ & $-$ & $-$ & $-$\\
114 & LESS J033150.8$-$274438 & 03:31:50.81 $-$27:44:38.5 & 9.7 & 03:31:51.1106 $-$27:44:37.552 & $   95.4\pm   6.7$ &  4.1 & {\bf 0.022} \\
115 & LESS J033349.7$-$274239$^*$ & 03:33:49.71 $-$27:42:39.2 & 8.9 & $-$\hspace{1.5cm}$-$ & $-$ & $-$ & $-$\\
116 & LESS J033154.4$-$274525 & 03:31:54.42 $-$27:45:25.5 & 9.7 & 03:31:54.2386 $-$27:45:27.809 & $   23.0\pm   5.7$ &  3.3 & 0.119 \\
 & & & & 03:31:54.4428 $-$27:45:31.605 & $   39.3\pm   6.2$ &  6.1 & (0.104) \\
117 & LESS J033128.0$-$273925 & 03:31:28.02 $-$27:39:25.2 & 9.7 & 03:31:27.5941 $-$27:39:27.989 & $   81.0\pm   8.4$ &  6.3 & {\bf 0.042} \\
118 & LESS J033121.8$-$274936 & 03:31:21.81 $-$27:49:36.8 & 9.7 & 03:31:21.9425 $-$27:49:41.894 & $   36.4\pm   6.9$ &  5.4 & (0.082) \\
 & & & & 03:31:21.8153 $-$27:49:35.179 & $   23.5\pm   6.3$ &  1.6 & [{\bf 0.032}] \\
119 & LESS J033256.5$-$280319 & 03:32:56.51 $-$28:03:19.1 & 9.7 & $-$\hspace{1.5cm}$-$ & $-$ & $-$ & $-$\\
120 & LESS J033328.4$-$275655 & 03:33:28.45 $-$27:56:55.9 & 9.8 & 03:33:28.5832 $-$27:56:54.376 & $   45.5\pm   8.4$ &  2.3 & {\bf 0.015} \\
 & & & & 03:33:28.5885 $-$27:56:58.901 & $   26.0\pm   7.5$ &  3.5 & (0.075) \\
121 & LESS J033333.3$-$273449 & 03:33:33.32 $-$27:34:49.3 & 9.7 & 03:33:33.0528 $-$27:34:51.686 & $   29.0\pm   8.2$ &  4.3 & (0.062) \\
 & & & & 03:33:33.0900 $-$27:34:42.616 & $   35.6\pm   8.6$ &  7.4 & (0.086) \\
122 & LESS J033139.6$-$274120 & 03:31:39.62 $-$27:41:20.4 & 9.9 & 03:31:39.5493 $-$27:41:19.658 & $  207.3\pm  14.5$ &  1.2 & {\bf 0.001} \\
123 & LESS J033330.9$-$275349 & 03:33:30.88 $-$27:53:49.3 & 9.8 & $-$\hspace{1.5cm}$-$ & $-$ & $-$ & $-$\\
124 & LESS J033203.6$-$273605 & 03:32:03.59 $-$27:36:05.0 & 10.0 & 03:32:03.1065 $-$27:36:01.967 & $   41.9\pm   7.5$ &  7.1 & (0.095) \\
 & & & & 03:32:03.4080 $-$27:36:08.877 & $   25.1\pm   6.4$ &  4.6 & 0.125 \\
125 & LESS J033146.0$-$274621 & 03:31:46.02 $-$27:46:21.2 & 9.9 & $-$\hspace{1.5cm}$-$ & $-$ & $-$ & $-$\\
126 & LESS J033209.8$-$274102 & 03:32:09.76 $-$27:41:02.0 & 9.9 & 03:32:09.5918 $-$27:41:07.368 & $   23.3\pm   5.5$ &  5.8 & 0.230 \\
\hline
\end{tabular}
}
\end{table*}

\begin{table*}
\centering
{\small
\caption{24-$\umu$m properties of potential counterparts to LESS
  870-$\umu$m sources in the ECDFS. SMGs are listed in order of
  decreasing SNR. Those SMG names that are appended with an $^*$ are
  not fully covered by the FIDEL map. Secure counterparts ($p\le0.05$)
  are in boldface and where $p$ lies between 0.05 and 1.0 this is
  given in parentheses. Counterparts where $0.05 < p \le 0.1$ is
  obtained at two out of radio (Table~\ref{tab:p24um}), 24-$\umu$m or
  5.8-$\umu$m (Table~\ref{tab:pirac}) have their value of $p$ given in
  boldface within parentheses.}
\begin{tabular}{clcccccc} \hline
ID & SMG name & Submm position & $r_{\rm s}$ & 24-$\umu$m position & 24-$\umu$m flux & Offset & $p$ \\
   &  & ($\alpha_{\mathrm{J2000}}$)\hspace{.9cm}($\delta_{\mathrm{J2000}}$) & (arcsec) & ($\alpha_{\mathrm{J2000}}$)\hspace{.9cm}($\delta_{\mathrm{J2000}}$) & ($\umu$Jy) & (arcsec) & \\ \hline
001 & LESS J033314.3$-$275611 & 03:33:14.26 $-$27:56:11.2 & 3.1 & 03:33:14.4124 $-$27:56:11.995 & $   38.0\pm   7.9$ &  2.2 & {\bf 0.053} \\
002 & LESS J033302.5$-$275643 & 03:33:02.50 $-$27:56:43.6 & 3.8 & 03:33:02.5305 $-$27:56:45.344 & $  186.5\pm  22.6$ &  1.8 & {\bf 0.014} \\
003 & LESS J033321.5$-$275520 & 03:33:21.51 $-$27:55:20.2 & 3.8 & 03:33:21.5113 $-$27:55:20.515 & $   33.6\pm   7.4$ &  0.3 & {\bf 0.004} \\
004 & LESS J033136.0$-$275439 & 03:31:36.01 $-$27:54:39.2 & 4.1 & $-$\hspace{1.5cm}$-$ & $-$ & $-$ & $-$\\
005 & LESS J033129.5$-$275907 & 03:31:29.46 $-$27:59:07.3 & 4.6 & $-$\hspace{1.5cm}$-$ & $-$ & $-$ & $-$\\
006 & LESS J033257.1$-$280102 & 03:32:57.14 $-$28:01:02.1 & 4.8 & 03:32:57.0774 $-$28:01:01.074 & $   35.7\pm   9.6$ &  1.3 & {\bf 0.041} \\
007 & LESS J033315.6$-$274523 & 03:33:15.55 $-$27:45:23.6 & 5.1 & 03:33:15.3999 $-$27:45:24.008 & $  368.3\pm   8.1$ &  2.0 & {\bf 0.008} \\
008 & LESS J033205.1$-$273108 & 03:32:05.07 $-$27:31:08.8 & 4.0 & $-$\hspace{1.5cm}$-$ & $-$ & $-$ & $-$\\
009 & LESS J033211.3$-$275210 & 03:32:11.29 $-$27:52:10.4 & 5.1 & 03:32:11.3060 $-$27:52:13.235 & $  110.6\pm  10.6$ &  2.8 & (0.063) \\
010 & LESS J033219.0$-$275219 & 03:32:19.02 $-$27:52:19.4 & 5.1 & 03:32:19.0566 $-$27:52:14.801 & $  119.8\pm  18.0$ &  4.6 & (0.090) \\
011 & LESS J033213.6$-$275602 & 03:32:13.58 $-$27:56:02.5 & 5.2 & 03:32:13.8449 $-$27:55:59.965 & $  103.1\pm   8.4$ &  4.3 & 0.113 \\
012 & LESS J033248.1$-$275414 & 03:32:48.12 $-$27:54:14.7 & 5.3 & 03:32:47.7383 $-$27:54:13.569 & $   33.7\pm   9.3$ &  5.2 & 0.178 \\
 & & & & 03:32:48.0689 $-$27:54:16.266 & $   43.2\pm   8.8$ &  1.7 & (0.058) \\
 & & & & 03:32:48.5059 $-$27:54:15.795 & $  174.3\pm   9.5$ &  5.2 & (0.095) \\
013 & LESS J033249.2$-$274246 & 03:32:49.23 $-$27:42:46.6 & 5.3 & $-$\hspace{1.5cm}$-$ & $-$ & $-$ & $-$\\
014 & LESS J033152.6$-$280320 & 03:31:52.64 $-$28:03:20.4 & 5.1 & 03:31:52.4265 $-$28:03:18.033 & $   95.5\pm   9.5$ &  3.7 & (0.096) \\
015 & LESS J033333.4$-$275930 & 03:33:33.36 $-$27:59:30.1 & 5.3 & 03:33:33.3439 $-$27:59:29.407 & $  108.6\pm  10.1$ &  0.7 & {\bf 0.008} \\
016 & LESS J033218.9$-$273738 & 03:32:18.89 $-$27:37:38.7 & 5.8 & $-$\hspace{1.5cm}$-$ & $-$ & $-$ & $-$\\
017 & LESS J033207.6$-$275123 & 03:32:07.59 $-$27:51:23.0 & 6.1 & 03:32:07.2947 $-$27:51:20.431 & $  219.3\pm   7.7$ &  4.7 & (0.071) \\
018 & LESS J033205.1$-$274652 & 03:32:05.12 $-$27:46:52.1 & 6.2 & 03:32:05.0405 $-$27:46:55.728 & $   39.9\pm   6.6$ &  3.8 & 0.174 \\
 & & & & 03:32:04.8558 $-$27:46:47.248 & $  560.5\pm   8.2$ &  6.0 & {\bf 0.029} \\
019 & LESS J033208.1$-$275818 & 03:32:08.10 $-$27:58:18.7 & 6.4 & 03:32:07.8975 $-$27:58:23.595 & $   40.0\pm   8.6$ &  5.6 & 0.239 \\
 & & & & 03:32:08.2306 $-$27:58:14.218 & $   79.8\pm   7.5$ &  4.8 & 0.172 \\
020 & LESS J033316.6$-$280018 & 03:33:16.56 $-$28:00:18.8 & 6.5 & 03:33:16.7545 $-$28:00:15.608 & $  176.6\pm   7.6$ &  4.1 & (0.078) \\
021 & LESS J033329.9$-$273441 & 03:33:29.93 $-$27:34:41.7 & 6.2 & 03:33:29.7579 $-$27:34:46.266 & $  217.1\pm  21.8$ &  5.1 & (0.067) \\
022 & LESS J033147.0$-$273243 & 03:31:47.02 $-$27:32:43.0 & 5.9 & 03:31:46.9134 $-$27:32:38.841 & $  409.8\pm  12.7$ &  4.4 & {\bf 0.025} \\
023 & LESS J033212.1$-$280508 & 03:32:12.11 $-$28:05:08.5 & 5.8 & 03:32:11.9457 $-$28:05:06.229 & $   35.7\pm  10.2$ &  3.1 & 0.135 \\
024 & LESS J033336.8$-$274401 & 03:33:36.79 $-$27:44:01.0 & 6.3 & 03:33:36.9853 $-$27:43:58.522 & $  130.2\pm   9.3$ &  3.6 & (0.086) \\
025 & LESS J033157.1$-$275940 & 03:31:57.05 $-$27:59:40.8 & 6.8 & 03:31:56.8419 $-$27:59:38.856 & $  233.2\pm   8.0$ &  3.4 & {\bf 0.043} \\
026 & LESS J033136.9$-$275456 & 03:31:36.90 $-$27:54:56.1 & 7.0 & $-$\hspace{1.5cm}$-$ & $-$ & $-$ & $-$\\
027 & LESS J033149.7$-$273432 & 03:31:49.73 $-$27:34:32.7 & 6.5 & 03:31:49.8900 $-$27:34:36.658 & $  171.9\pm  16.4$ &  4.5 & (0.082) \\
 & & & & 03:31:50.2090 $-$27:34:32.901 & $  277.3\pm  18.8$ &  6.4 & (0.072) \\
028 & LESS J033302.9$-$274432 & 03:33:02.92 $-$27:44:32.6 & 6.9 & $-$\hspace{1.5cm}$-$ & $-$ & $-$ & $-$\\
029 & LESS J033336.9$-$275813 & 03:33:36.90 $-$27:58:13.0 & 6.6 & 03:33:36.8692 $-$27:58:08.874 & $  136.0\pm   9.2$ &  4.1 & (0.103) \\
030 & LESS J033344.4$-$280346 & 03:33:44.37 $-$28:03:46.1 & 5.5 & $-$\hspace{1.5cm}$-$ & $-$ & $-$ & $-$\\
031 & LESS J033150.0$-$275743 & 03:31:49.96 $-$27:57:43.9 & 7.2 & 03:31:49.7330 $-$27:57:39.858 & $   66.2\pm   9.0$ &  5.0 & 0.214 \\
032 & LESS J033243.6$-$274644 & 03:32:43.57 $-$27:46:44.0 & 7.2 & 03:32:43.5139 $-$27:46:39.630 & $   81.5\pm  13.4$ &  4.4 & 0.157 \\
033 & LESS J033149.8$-$275332 & 03:31:49.78 $-$27:53:32.9 & 7.2 & $-$\hspace{1.5cm}$-$ & $-$ & $-$ & $-$\\
034 & LESS J033217.6$-$275230 & 03:32:17.64 $-$27:52:30.3 & 7.2 & 03:32:17.5943 $-$27:52:28.656 & $  223.3\pm   9.4$ &  1.8 & {\bf 0.016} \\
035 & LESS J033110.3$-$273714 & 03:31:10.35 $-$27:37:14.8 & 5.9 & 03:31:10.4778 $-$27:37:15.134 & $   73.3\pm   9.6$ &  1.7 & {\bf 0.046} \\
036 & LESS J033149.2$-$280208 & 03:31:49.15 $-$28:02:08.7 & 7.2 & 03:31:48.9432 $-$28:02:13.486 & $  274.5\pm   9.6$ &  5.5 & ({\bf 0.071 + radio}) \\
037 & LESS J033336.0$-$275347 & 03:33:36.04 $-$27:53:47.6 & 6.9 & 03:33:36.0581 $-$27:53:49.812 & $  201.8\pm  19.7$ &  2.2 & {\bf 0.025} \\
 & & & & 03:33:36.2881 $-$27:53:47.112 & $   92.4\pm  19.4$ &  3.3 & (0.092) \\
038 & LESS J033310.2$-$275641 & 03:33:10.20 $-$27:56:41.5 & 7.5 & 03:33:10.0986 $-$27:56:45.026 & $  154.9\pm   6.8$ &  3.8 & (0.086) \\
 & & & & 03:33:10.5182 $-$27:56:44.627 & $   78.4\pm   7.5$ &  5.2 & 0.216 \\
039 & LESS J033144.9$-$273435 & 03:31:44.90 $-$27:34:35.4 & 7.3 & 03:31:45.0100 $-$27:34:36.567 & $  131.0\pm   7.5$ &  1.9 & {\bf 0.036} \\
040 & LESS J033246.7$-$275120 & 03:32:46.74 $-$27:51:20.9 & 7.6 & 03:32:46.8016 $-$27:51:20.648 & $  119.8\pm   7.2$ &  0.9 & {\bf 0.011} \\
041 & LESS J033110.5$-$275233 & 03:31:10.47 $-$27:52:33.2 & 6.2 & $-$\hspace{1.5cm}$-$ & $-$ & $-$ & $-$\\
042 & LESS J033231.0$-$275858 & 03:32:31.02 $-$27:58:58.1 & 7.7 & 03:32:30.9901 $-$27:59:02.928 & $   70.5\pm  17.5$ &  4.8 & 0.180 \\
043 & LESS J033307.0$-$274801 & 03:33:07.00 $-$27:48:01.0 & 7.6 & 03:33:06.6209 $-$27:48:02.051 & $  229.6\pm   7.2$ &  5.1 & (0.086) \\
 & & & & 03:33:07.4686 $-$27:47:59.241 & $  190.1\pm  10.8$ &  6.5 & 0.140 \\
 & & & & 03:33:07.1581 $-$27:47:55.911 & $   77.7\pm   9.9$ &  5.5 & 0.224 \\
044 & LESS J033131.0$-$273238 & 03:31:30.96 $-$27:32:38.5 & 6.9 & 03:31:31.2040 $-$27:32:38.469 & $  400.9\pm  13.0$ &  3.2 & {\bf 0.017} \\
045 & LESS J033225.7$-$275228 & 03:32:25.71 $-$27:52:28.5 & 7.7 & 03:32:25.2320 $-$27:52:30.520 & $  116.1\pm   8.1$ &  6.7 & 0.226 \\
046 & LESS J033336.8$-$273247$^*$ & 03:33:36.80 $-$27:32:47.0 & 6.5 & $-$\hspace{1.5cm}$-$ & $-$ & $-$ & $-$\\
047 & LESS J033256.0$-$273317 & 03:32:56.00 $-$27:33:17.7 & 7.2 & 03:32:55.9048 $-$27:33:19.557 & $   55.9\pm   9.7$ &  2.2 & (0.089) \\
048 & LESS J033237.8$-$273202 & 03:32:37.77 $-$27:32:02.0 & 6.8 & 03:32:37.9882 $-$27:31:59.611 & $  406.8\pm  11.9$ &  3.8 & {\bf 0.021} \\
049 & LESS J033124.4$-$275040 & 03:31:24.45 $-$27:50:40.9 & 7.6 & 03:31:24.4752 $-$27:50:37.619 & $  113.5\pm  14.6$ &  3.3 & (0.090) \\
 & & & & 03:31:24.7116 $-$27:50:46.277 & $  122.9\pm  11.6$ &  6.4 & 0.198 \\
 & & & & 03:31:24.2346 $-$27:50:43.663 & $  120.3\pm   9.3$ &  4.0 & 0.118 \\
\end{tabular}
\label{tab:p24um}
}
\end{table*}

\begin{table*}
\centering
{\small
\contcaption{}
\begin{tabular}{clcccccccc} \hline
ID & SMG name & Submm position & $r_{\rm s}$ & 24-$\umu$m position & 24-$\umu$m flux & Offset & $p$ \\
   &  & ($\alpha_{\mathrm{J2000}}$)\hspace{.9cm}($\delta_{\mathrm{J2000}}$) & (arcsec) & ($\alpha_{\mathrm{J2000}}$)\hspace{.9cm}($\delta_{\mathrm{J2000}}$) & ($\umu$Jy) & (arcsec) & \\ \hline
050 & LESS J033141.2$-$274441 & 03:31:41.15 $-$27:44:41.5 & 7.9 & 03:31:41.3603 $-$27:44:47.005 & $  197.4\pm  12.5$ &  6.2 & 0.127 \\
 & & & & 03:31:41.1223 $-$27:44:42.531 & $   64.4\pm  12.3$ &  1.1 & {\bf 0.028} \\
 & & & & 03:31:41.5587 $-$27:44:40.948 & $  101.9\pm  10.2$ &  5.5 & 0.196 \\
 & & & & 03:31:40.6007 $-$27:44:40.929 & $  175.0\pm   9.5$ &  7.3 & 0.183 \\
 & & & & 03:31:40.9980 $-$27:44:34.928 & $  306.6\pm  24.1$ &  6.9 & (0.072) \\
051 & LESS J033144.8$-$274425 & 03:31:44.81 $-$27:44:25.1 & 7.9 & 03:31:45.0279 $-$27:44:27.859 & $  117.1\pm   8.5$ &  4.0 & 0.125 \\
 & & & & 03:31:44.4130 $-$27:44:20.236 & $   93.0\pm   9.6$ &  7.2 & 0.271 \\
052 & LESS J033128.5$-$275601 & 03:31:28.51 $-$27:56:01.3 & 7.9 & 03:31:28.3837 $-$27:56:07.987 & $  127.8\pm  14.0$ &  6.9 & 0.203 \\
053 & LESS J033159.1$-$275435 & 03:31:59.12 $-$27:54:35.5 & 8.0 & 03:31:58.9803 $-$27:54:38.097 & $  137.5\pm   8.3$ &  3.2 & (0.079) \\
054 & LESS J033243.6$-$273353 & 03:32:43.61 $-$27:33:53.6 & 7.6 & 03:32:43.6418 $-$27:33:56.925 & $  222.1\pm   8.8$ &  3.4 & {\bf 0.048} \\
055 & LESS J033302.2$-$274033 & 03:33:02.20 $-$27:40:33.6 & 8.0 & $-$\hspace{1.5cm}$-$ & $-$ & $-$ & $-$\\
056 & LESS J033153.2$-$273936 & 03:31:53.17 $-$27:39:36.1 & 8.1 & 03:31:53.1272 $-$27:39:37.490 & $  270.3\pm  11.4$ &  1.5 & {\bf 0.010} \\
057 & LESS J033152.0$-$275329 & 03:31:51.97 $-$27:53:29.7 & 8.0 & 03:31:51.9121 $-$27:53:26.733 & $  297.2\pm   8.4$ &  3.1 & {\bf 0.027} \\
058 & LESS J033225.8$-$273306 & 03:32:25.79 $-$27:33:06.7 & 7.6 & $-$\hspace{1.5cm}$-$ & $-$ & $-$ & $-$\\
059 & LESS J033303.9$-$274412 & 03:33:03.87 $-$27:44:12.2 & 8.2 & 03:33:03.6615 $-$27:44:11.811 & $  172.1\pm  10.6$ &  2.8 & {\bf 0.050} \\
060 & LESS J033317.5$-$275121 & 03:33:17.47 $-$27:51:21.5 & 8.3 & 03:33:17.4868 $-$27:51:28.081 & $  292.8\pm   9.8$ &  6.6 & ({\bf 0.089 + radio}) \\
061 & LESS J033245.6$-$280025 & 03:32:45.63 $-$28:00:25.3 & 8.3 & 03:32:45.9477 $-$28:00:22.155 & $   58.6\pm   7.9$ &  5.3 & 0.264 \\
062 & LESS J033236.4$-$273452 & 03:32:36.41 $-$27:34:52.5 & 8.2 & 03:32:36.5400 $-$27:34:53.319 & $  243.5\pm  35.6$ &  1.9 & {\bf 0.014} \\
 & & & & 03:32:36.1180 $-$27:34:53.299 & $   65.1\pm  14.6$ &  4.0 & 0.167 \\
063 & LESS J033308.5$-$280044 & 03:33:08.46 $-$28:00:44.3 & 8.3 & 03:33:08.4096 $-$28:00:42.440 & $   62.4\pm   9.8$ &  2.0 & (0.073) \\
064 & LESS J033201.0$-$280025 & 03:32:01.00 $-$28:00:25.6 & 8.4 & 03:32:00.5600 $-$28:00:25.733 & $   83.1\pm  17.1$ &  5.8 & 0.215 \\
 & & & & 03:32:00.9399 $-$28:00:25.316 & $  338.4\pm  17.2$ &  0.8 & {\bf 0.002} \\
065 & LESS J033252.4$-$273527 & 03:32:52.40 $-$27:35:27.7 & 8.3 & $-$\hspace{1.5cm}$-$ & $-$ & $-$ & $-$\\
066 & LESS J033331.7$-$275406 & 03:33:31.69 $-$27:54:06.1 & 8.2 & 03:33:31.9057 $-$27:54:10.024 & $  543.7\pm  12.3$ &  4.9 & {\bf 0.023} \\
067 & LESS J033243.3$-$275517 & 03:32:43.28 $-$27:55:17.9 & 8.4 & 03:32:43.7519 $-$27:55:16.397 & $  109.6\pm   6.4$ &  6.4 & 0.240 \\
 & & & & 03:32:43.0426 $-$27:55:24.757 & $  209.4\pm   7.1$ &  7.5 & 0.167 \\
 & & & & 03:32:43.1879 $-$27:55:14.295 & $  516.4\pm   8.0$ &  3.8 & {\bf 0.017} \\
068 & LESS J033233.4$-$273918 & 03:32:33.44 $-$27:39:18.5 & 8.4 & 03:32:33.3296 $-$27:39:13.479 & $   41.9\pm   9.4$ &  5.2 & 0.291 \\
 & & & & 03:32:33.9234 $-$27:39:14.713 & $  191.7\pm   8.2$ &  7.5 & 0.180 \\
069 & LESS J033134.3$-$275934 & 03:31:34.26 $-$27:59:34.3 & 8.5 & 03:31:33.8107 $-$27:59:32.287 & $   97.0\pm  11.7$ &  6.3 & 0.241 \\
 & & & & 03:31:34.6801 $-$27:59:34.629 & $  134.1\pm  17.2$ &  5.6 & 0.155 \\
070 & LESS J033144.0$-$273832 & 03:31:43.97 $-$27:38:32.5 & 8.5 & 03:31:44.0299 $-$27:38:35.076 & $  365.2\pm  17.4$ &  2.7 & {\bf 0.015} \\
 & & & & 03:31:43.9688 $-$27:38:30.430 & $   85.5\pm  18.7$ &  2.1 & (0.058) \\
071 & LESS J033306.3$-$273327 & 03:33:06.29 $-$27:33:27.7 & 8.0 & $-$\hspace{1.5cm}$-$ & $-$ & $-$ & $-$\\
072 & LESS J033240.4$-$273802 & 03:32:40.40 $-$27:38:02.5 & 8.5 & 03:32:40.0462 $-$27:38:08.479 & $  471.4\pm  22.9$ &  7.6 & {\bf 0.051} \\
073 & LESS J033229.3$-$275619 & 03:32:29.33 $-$27:56:19.3 & 8.5 & $-$\hspace{1.5cm}$-$ & $-$ & $-$ & $-$\\
074 & LESS J033309.3$-$274809 & 03:33:09.34 $-$27:48:09.9 & 8.4 & 03:33:09.1309 $-$27:48:16.747 & $  202.0\pm  23.4$ &  7.4 & 0.135 \\
 & & & & 03:33:09.3973 $-$27:48:14.431 & $   55.4\pm   9.5$ &  4.6 & 0.234 \\
 & & & & 03:33:09.5602 $-$27:48:03.494 & $  291.2\pm  25.0$ &  7.0 & (0.082) \\
 & & & & 03:33:09.0479 $-$27:48:07.069 & $   37.1\pm   9.1$ &  4.8 & 0.278 \\
075 & LESS J033126.8$-$275554 & 03:31:26.83 $-$27:55:54.6 & 8.4 & 03:31:27.1769 $-$27:55:50.848 & $ 1018.1\pm  38.3$ &  5.9 & {\bf 0.011} \\
076 & LESS J033332.7$-$275957 & 03:33:32.67 $-$27:59:57.2 & 8.4 & $-$\hspace{1.5cm}$-$ & $-$ & $-$ & $-$\\
077 & LESS J033157.2$-$275633 & 03:31:57.23 $-$27:56:33.2 & 8.8 & 03:31:56.7468 $-$27:56:37.786 & $   48.0\pm  10.2$ &  7.9 & 0.382 \\
 & & & & 03:31:56.8109 $-$27:56:32.398 & $   32.3\pm   8.9$ &  5.6 & 0.339 \\
 & & & & 03:31:57.2381 $-$27:56:40.234 & $  149.6\pm   7.9$ &  7.0 & 0.217 \\
 & & & & 03:31:57.6947 $-$27:56:28.842 & $  129.8\pm   6.7$ &  7.5 & 0.261 \\
078 & LESS J033340.3$-$273956$^*$ & 03:33:40.30 $-$27:39:56.9 & 8.4 & $-$\hspace{1.5cm}$-$ & $-$ & $-$ & $-$\\
079 & LESS J033221.2$-$275623 & 03:32:21.25 $-$27:56:23.5 & 8.8 & 03:32:21.5939 $-$27:56:23.782 & $  520.7\pm   9.2$ &  4.6 & {\bf 0.023} \\
 & & & & 03:32:21.1259 $-$27:56:26.704 & $   64.8\pm   7.3$ &  3.6 & 0.171 \\
 & & & & 03:32:20.8843 $-$27:56:18.455 & $   42.5\pm   6.7$ &  7.0 & 0.381 \\
 & & & & 03:32:21.4284 $-$27:56:16.710 & $  140.5\pm   7.9$ &  7.2 & 0.234 \\
080 & LESS J033142.2$-$274834 & 03:31:42.23 $-$27:48:34.4 & 8.9 & 03:31:42.6032 $-$27:48:41.050 & $  138.2\pm   8.1$ &  8.3 & 0.277 \\
 & & & & 03:31:42.7846 $-$27:48:36.479 & $   79.8\pm   5.8$ &  7.6 & 0.344 \\
081 & LESS J033127.4$-$274440 & 03:31:27.45 $-$27:44:40.4 & 8.8 & 03:31:27.5539 $-$27:44:39.264 & $  523.6\pm  11.0$ &  1.8 & {\bf 0.005} \\
082 & LESS J033253.8$-$273810 & 03:32:53.77 $-$27:38:10.9 & 9.0 & 03:32:53.5601 $-$27:38:14.872 & $   64.4\pm  10.3$ &  4.9 & 0.241 \\
083 & LESS J033308.9$-$280522 & 03:33:08.92 $-$28:05:22.0 & 8.3 & 03:33:08.8774 $-$28:05:14.298 & $   59.5\pm   8.8$ &  7.7 & 0.352 \\
084 & LESS J033154.2$-$275109 & 03:31:54.22 $-$27:51:09.8 & 8.9 & 03:31:54.8059 $-$27:51:10.132 & $   98.7\pm   7.6$ &  7.8 & 0.318 \\
 & & & & 03:31:54.5080 $-$27:51:05.155 & $  133.5\pm   7.3$ &  6.0 & 0.200 \\
 & & & & 03:31:53.8175 $-$27:51:03.834 & $  141.6\pm   7.6$ &  8.0 & 0.263 \\
085 & LESS J033110.3$-$274503 & 03:31:10.28 $-$27:45:03.1 & 7.7 & $-$\hspace{1.5cm}$-$ & $-$ & $-$ & $-$\\
086 & LESS J033114.9$-$274844 & 03:31:14.90 $-$27:48:44.3 & 8.5 & $-$\hspace{1.5cm}$-$ & $-$ & $-$ & $-$\\
087 & LESS J033251.1$-$273143 & 03:32:51.09 $-$27:31:43.0 & 8.4 & 03:32:51.6586 $-$27:31:40.858 & $   55.8\pm  10.6$ &  7.9 & 0.349 \\
 & & & & 03:32:50.8267 $-$27:31:41.290 & $  419.4\pm  13.2$ &  3.9 & {\bf 0.023} \\
\end{tabular}
}
\end{table*}

\begin{table*}
\centering
{\small
\contcaption{}
\begin{tabular}{clcccccccc} \hline
ID & SMG name & Submm position & $r_{\rm s}$ & 24-$\umu$m position & 24-$\umu$m flux & Offset & $p$ \\
   &  & ($\alpha_{\mathrm{J2000}}$)\hspace{.9cm}($\delta_{\mathrm{J2000}}$) & (arcsec) & ($\alpha_{\mathrm{J2000}}$)\hspace{.9cm}($\delta_{\mathrm{J2000}}$) & ($\umu$Jy) & (arcsec) & \\ \hline
088 & LESS J033155.2$-$275345 & 03:31:55.19 $-$27:53:45.3 & 9.0 & 03:31:55.7552 $-$27:53:47.714 & $  195.4\pm   7.2$ &  7.9 & 0.197 \\
 & & & & 03:31:54.7914 $-$27:53:41.259 & $  269.5\pm  19.4$ &  6.7 & (0.094) \\
089 & LESS J033248.4$-$280023 & 03:32:48.44 $-$28:00:23.8 & 9.1 & 03:32:48.6548 $-$28:00:21.217 & $   68.1\pm   8.8$ &  3.8 & 0.183 \\
090 & LESS J033243.7$-$273554 & 03:32:43.65 $-$27:35:54.1 & 9.1 & $-$\hspace{1.5cm}$-$ & $-$ & $-$ & $-$\\
091 & LESS J033135.2$-$274033 & 03:31:35.25 $-$27:40:33.7 & 9.1 & 03:31:35.0211 $-$27:40:38.070 & $   44.8\pm   9.5$ &  5.3 & 0.309 \\
092 & LESS J033138.4$-$274336 & 03:31:38.36 $-$27:43:36.0 & 9.2 & 03:31:38.2703 $-$27:43:39.387 & $  101.3\pm   9.2$ &  3.6 & 0.130 \\
 & & & & 03:31:38.2101 $-$27:43:28.525 & $   66.7\pm  13.5$ &  7.7 & 0.342 \\
093 & LESS J033110.8$-$275607 & 03:31:10.84 $-$27:56:07.2 & 8.4 & $-$\hspace{1.5cm}$-$ & $-$ & $-$ & $-$\\
094 & LESS J033307.3$-$275805 & 03:33:07.27 $-$27:58:05.0 & 9.1 & 03:33:07.6169 $-$27:58:06.076 & $  106.2\pm   7.3$ &  4.7 & 0.180 \\
 & & & & 03:33:06.6787 $-$27:58:06.259 & $   70.6\pm   6.7$ &  7.9 & 0.376 \\
095 & LESS J033241.7$-$275846 & 03:32:41.74 $-$27:58:46.1 & 9.2 & 03:32:41.2420 $-$27:58:41.239 & $  328.8\pm   6.5$ &  8.2 & 0.110 \\
096 & LESS J033313.0$-$275556 & 03:33:13.03 $-$27:55:56.8 & 9.2 & 03:33:12.6193 $-$27:55:51.500 & $  961.6\pm   9.0$ &  7.6 & {\bf 0.023} \\
 & & & & 03:33:13.0725 $-$27:55:55.873 & $   33.1\pm   7.6$ &  1.1 & {\bf 0.042} \\
097 & LESS J033313.7$-$273803 & 03:33:13.65 $-$27:38:03.4 & 9.2 & $-$\hspace{1.5cm}$-$ & $-$ & $-$ & $-$\\
098 & LESS J033130.2$-$275726 & 03:31:30.22 $-$27:57:26.0 & 9.3 & 03:31:29.9230 $-$27:57:22.432 & $  268.8\pm  15.7$ &  5.3 & (0.073) \\
099 & LESS J033251.4$-$275536 & 03:32:51.45 $-$27:55:36.0 & 9.2 & 03:32:51.3366 $-$27:55:43.489 & $  120.9\pm   6.2$ &  7.6 & 0.287 \\
100 & LESS J033111.3$-$280006 & 03:31:11.32 $-$28:00:06.2 & 8.8 & $-$\hspace{1.5cm}$-$ & $-$ & $-$ & $-$\\
101 & LESS J033151.5$-$274552 & 03:31:51.47 $-$27:45:52.1 & 9.3 & 03:31:51.3802 $-$27:46:00.784 & $   43.9\pm   6.6$ &  8.8 & 0.448 \\
102 & LESS J033335.6$-$274020 & 03:33:35.61 $-$27:40:20.1 & 9.2 & 03:33:35.5719 $-$27:40:22.963 & $  318.4\pm  16.5$ &  2.9 & {\bf 0.022} \\
103 & LESS J033325.3$-$273400 & 03:33:25.35 $-$27:34:00.4 & 9.2 & 03:33:25.3661 $-$27:33:58.314 & $  113.3\pm  13.5$ &  2.1 & {\bf 0.052} \\
104 & LESS J033258.5$-$273803 & 03:32:58.46 $-$27:38:03.0 & 9.4 & 03:32:57.8377 $-$27:37:59.368 & $  108.2\pm  10.0$ &  9.0 & 0.354 \\
105 & LESS J033115.8$-$275313 & 03:31:15.78 $-$27:53:13.1 & 9.0 & $-$\hspace{1.5cm}$-$ & $-$ & $-$ & $-$\\
106 & LESS J033140.1$-$275631 & 03:31:40.09 $-$27:56:31.4 & 9.4 & 03:31:40.4427 $-$27:56:34.335 & $   65.7\pm  10.2$ &  5.5 & 0.282 \\
 & & & & 03:31:40.1806 $-$27:56:22.325 & $  363.6\pm  10.2$ &  9.2 & 0.113 \\
107 & LESS J033130.8$-$275150 & 03:31:30.85 $-$27:51:50.9 & 9.4 & 03:31:31.3083 $-$27:51:53.866 & $   89.4\pm   9.6$ &  6.8 & 0.301 \\
 & & & & 03:31:30.5400 $-$27:51:58.629 & $   96.9\pm   9.4$ &  8.8 & 0.365 \\
 & & & & 03:31:30.6286 $-$27:51:45.285 & $  124.2\pm  16.5$ &  6.3 & 0.208 \\
 & & & & 03:31:31.2827 $-$27:51:43.895 & $  321.7\pm   9.2$ &  9.1 & 0.131 \\
 & & & & 03:31:30.3348 $-$27:51:48.424 & $  102.8\pm  10.9$ &  7.3 & 0.295 \\
108 & LESS J033316.4$-$275033 & 03:33:16.42 $-$27:50:33.1 & 9.5 & 03:33:16.4864 $-$27:50:39.550 & $ 3722.6\pm  73.7$ &  6.5 & {\bf 0.002} \\
109 & LESS J033328.1$-$274157 & 03:33:28.08 $-$27:41:57.0 & 9.5 & 03:33:27.9952 $-$27:42:02.797 & $  169.0\pm  15.5$ &  5.9 & 0.151 \\
 & & & & 03:33:28.5385 $-$27:41:51.142 & $  128.8\pm  13.3$ &  8.4 & 0.289 \\
110 & LESS J033122.6$-$275417 & 03:31:22.64 $-$27:54:17.2 & 9.4 & 03:31:22.6458 $-$27:54:21.881 & $  104.3\pm  16.6$ &  4.7 & 0.165 \\
111 & LESS J033325.6$-$273423 & 03:33:25.58 $-$27:34:23.0 & 9.4 & 03:33:25.2095 $-$27:34:23.302 & $  353.7\pm  53.3$ &  4.9 & {\bf 0.031} \\
112 & LESS J033249.3$-$273112 & 03:32:49.28 $-$27:31:12.3 & 9.0 & 03:32:48.8281 $-$27:31:12.959 & $  190.8\pm  13.2$ &  6.0 & 0.136 \\
113 & LESS J033236.4$-$275845 & 03:32:36.42 $-$27:58:45.9 & 9.5 & $-$\hspace{1.5cm}$-$ & $-$ & $-$ & $-$\\
114 & LESS J033150.8$-$274438 & 03:31:50.81 $-$27:44:38.5 & 9.7 & 03:31:50.7294 $-$27:44:40.607 & $   73.6\pm  18.5$ &  2.4 & (0.083) \\
 & & & & 03:31:51.0920 $-$27:44:37.132 & $  515.0\pm   7.8$ &  4.0 & {\bf 0.019} \\
 & & & & 03:31:50.9679 $-$27:44:43.872 & $   59.8\pm   6.5$ &  5.8 & 0.320 \\
115 & LESS J033349.7$-$274239$^*$ & 03:33:49.71 $-$27:42:39.2 & 8.9 & $-$\hspace{1.5cm}$-$ & $-$ & $-$ & $-$\\
116 & LESS J033154.4$-$274525 & 03:31:54.42 $-$27:45:25.5 & 9.7 & $-$\hspace{1.5cm}$-$ & $-$ & $-$ & $-$\\
117 & LESS J033128.0$-$273925 & 03:31:28.02 $-$27:39:25.2 & 9.7 & 03:31:27.5895 $-$27:39:27.598 & $  203.1\pm  13.1$ &  6.2 & 0.137 \\
118 & LESS J033121.8$-$274936 & 03:31:21.81 $-$27:49:36.8 & 9.7 & 03:31:21.7699 $-$27:49:41.451 & $   53.8\pm   8.6$ &  4.7 & 0.269 \\
119 & LESS J033256.5$-$280319 & 03:32:56.51 $-$28:03:19.1 & 9.7 & 03:32:56.5806 $-$28:03:11.789 & $  168.6\pm   7.0$ &  7.4 & 0.219 \\
120 & LESS J033328.4$-$275655 & 03:33:28.45 $-$27:56:55.9 & 9.8 & 03:33:29.0593 $-$27:56:57.375 & $   47.8\pm   9.6$ &  8.2 & 0.442 \\
 & & & & 03:33:28.5322 $-$27:56:54.295 & $  345.6\pm   9.5$ &  1.9 & {\bf 0.010} \\
121 & LESS J033333.3$-$273449$^*$ & 03:33:33.32 $-$27:34:49.3 & 9.7 & $-$\hspace{1.5cm}$-$ & $-$ & $-$ & $-$\\
122 & LESS J033139.6$-$274120 & 03:31:39.62 $-$27:41:20.4 & 9.9 & 03:31:39.5353 $-$27:41:19.449 & $ 1392.5\pm  15.8$ &  1.5 & {\bf 0.001} \\
123 & LESS J033330.9$-$275349 & 03:33:30.88 $-$27:53:49.3 & 9.8 & $-$\hspace{1.5cm}$-$ & $-$ & $-$ & $-$\\
124 & LESS J033203.6$-$273605 & 03:32:03.59 $-$27:36:05.0 & 10.0 & 03:32:03.8787 $-$27:36:06.030 & $  123.4\pm  10.6$ &  4.0 & 0.131 \\
 & & & & 03:32:03.0844 $-$27:36:01.864 & $   71.2\pm  10.0$ &  7.4 & 0.375 \\
125 & LESS J033146.0$-$274621 & 03:31:46.02 $-$27:46:21.2 & 9.9 & 03:31:46.3415 $-$27:46:23.329 & $   36.4\pm   7.1$ &  4.8 & 0.316 \\
 & & & & 03:31:45.5405 $-$27:46:15.445 & $   50.7\pm   8.6$ &  8.6 & 0.462 \\
126 & LESS J033209.8$-$274102 & 03:32:09.76 $-$27:41:02.0 & 9.9 & 03:32:09.5687 $-$27:41:06.810 & $  263.4\pm   8.7$ &  5.4 & (0.086) \\
 & & & & 03:32:10.1562 $-$27:40:56.226 & $   44.3\pm   8.0$ &  7.8 & 0.451 \\
\hline
\end{tabular}
}
\end{table*}

\begin{table*}
\centering
{\small
\caption{IRAC properties of potential counterparts to LESS 870-$\umu$m
  sources in the ECDFS that do not have robust counterparts identified
  in either the radio or at 24-$\umu$m and fall within the colour-flux
  cut shown in Fig.~\ref{fig:iraccolourmag}. SMGs are listed in order
  of decreasing SNR. Secure counterparts ($p\le0.05$) are in boldface
  and where $p$ lies between 0.05 and 0.1 this is given in
  parentheses. One counterpart where $0.05 < p \le 0.1$ is also
  obtained in the radio (Table~\ref{tab:pradio})
  (Table~\ref{tab:p24um}) is also considered robust and has its value
  of $p$ given in boldface within parentheses. Those SMG names that
  are appended with an $^*$ are not fully covered by the SIMPLE data
  at both 3.6- and 5.8-$\umu$m and so counterparts for them could not
  be found with this method.}
\begin{tabular}{clcccccc} \hline
ID & SMG name & Submm position & $r_{\rm s}$ & 5.8-$\umu$m position & 5.8-$\umu$m flux & Offset & $p$ \\
   &  & ($\alpha_{\mathrm{J2000}}$)\hspace{.9cm}($\delta_{\mathrm{J2000}}$) & (arcsec) & ($\alpha_{\mathrm{J2000}}$)\hspace{.9cm}($\delta_{\mathrm{J2000}}$) & ($\umu$Jy) & (arcsec) & \\ \hline
004 & LESS J033136.0$-$275439 & 03:31:36.01 $-$27:54:39.2 & 4.1 & $-$ & $-$ & $-$ & $-$ \\
005 & LESS J033129.5$-$275907 & 03:31:29.46 $-$27:59:07.3 & 4.6 & $-$ & $-$ & $-$ & $-$ \\
008 & LESS J033205.1$-$273108 & 03:32:05.07 $-$27:31:08.8 & 4.0 & $-$ & $-$ & $-$ & $-$ \\
013 & LESS J033249.2$-$274246 & 03:32:49.23 $-$27:42:46.6 & 5.3 & $-$ & $-$ & $-$ & $-$ \\
019 & LESS J033208.1$-$275818 & 03:32:08.10 $-$27:58:18.7 & 6.4 & 03:32:07.9138 $-$27:58:23.279 & $    6.7\pm   1.2$ &  5.2 & {\bf 0.053} \\
 & & & & 03:32:08.2382 $-$27:58:13.717 & $    6.7\pm   1.2$ &  5.3 & {\bf 0.053} \\
021 & LESS J033329.9$-$273441 & 03:33:29.93 $-$27:34:41.7 & 6.2 & $-$ & $-$ & $-$ & $-$ \\
023 & LESS J033212.1$-$280508 & 03:32:12.11 $-$28:05:08.5 & 5.8 & $-$ & $-$ & $-$ & $-$ \\
026 & LESS J033136.9$-$275456 & 03:31:36.90 $-$27:54:56.1 & 7.0 & $-$ & $-$ & $-$ & $-$ \\
027 & LESS J033149.7$-$273432 & 03:31:49.73 $-$27:34:32.7 & 6.5 & 03:31:49.9238 $-$27:34:36.790 & $    9.6\pm   1.5$ &  4.8 & {\bf 0.040} \\
 & & & & 03:31:49.8854 $-$27:34:30.428 & $    6.1\pm   1.2$ &  3.1 & {\bf 0.033} \\
028 & LESS J033302.9$-$274432 & 03:33:02.92 $-$27:44:32.6 & 6.9 & $-$ & $-$ & $-$ & $-$ \\
030 & LESS J033344.4$-$280346 & 03:33:44.37 $-$28:03:46.1 & 5.5 & $-$ & $-$ & $-$ & $-$ \\
031 & LESS J033150.0$-$275743 & 03:31:49.96 $-$27:57:43.9 & 7.2 & 03:31:49.7364 $-$27:57:39.280 & $    7.4\pm   1.3$ &  5.5 & (0.059) \\
 & & & & 03:31:49.7741 $-$27:57:40.439 & $    7.7\pm   1.3$ &  4.2 & {\bf 0.044} \\
032 & LESS J033243.6$-$274644 & 03:32:43.57 $-$27:46:44.0 & 7.2 & 03:32:43.5170 $-$27:46:38.978 & $    6.5\pm   1.2$ &  5.1 & (0.059) \\
033 & LESS J033149.8$-$275332 & 03:31:49.78 $-$27:53:32.9 & 7.2 & $-$ & $-$ & $-$ & $-$ \\
038 & LESS J033310.2$-$275641 & 03:33:10.20 $-$27:56:41.5 & 7.5 & $-$ & $-$ & $-$ & $-$ \\
041 & LESS J033110.5$-$275233 & 03:31:10.47 $-$27:52:33.2 & 6.2 & 03:31:10.0942 $-$27:52:36.347 & $   39.6\pm   3.1$ &  5.9 & {\bf 0.010} \\
042 & LESS J033231.0$-$275858 & 03:32:31.02 $-$27:58:58.1 & 7.7 & $-$ & $-$ & $-$ & $-$ \\
043 & LESS J033307.0$-$274801 & 03:33:07.00 $-$27:48:01.0 & 7.6 & 03:33:07.4822 $-$27:47:59.172 & $    9.2\pm   1.4$ &  6.7 & (0.065) \\
 & & & & 03:33:06.6365 $-$27:48:01.919 & $   13.3\pm   1.7$ &  4.9 & {\bf 0.035} \\
045 & LESS J033225.7$-$275228 & 03:32:25.71 $-$27:52:28.5 & 7.7 & 03:32:25.2458 $-$27:52:30.162 & $   15.1\pm   1.8$ &  6.4 & {\bf 0.042} \\
047 & LESS J033256.0$-$273317 & 03:32:56.00 $-$27:33:17.7 & 7.2 & 03:32:55.9356 $-$27:33:19.678 & $    7.3\pm   1.3$ &  2.2 & {\bf 0.019} \\
 & & & & 03:32:55.9910 $-$27:33:18.900 & $    6.9\pm   1.2$ &  1.2 & {\bf 0.008} \\
051 & LESS J033144.8$-$274425 & 03:31:44.81 $-$27:44:25.1 & 7.9 & $-$ & $-$ & $-$ & $-$ \\
052 & LESS J033128.5$-$275601 & 03:31:28.51 $-$27:56:01.3 & 7.9 & $-$ & $-$ & $-$ & $-$ \\
053 & LESS J033159.1$-$275435 & 03:31:59.12 $-$27:54:35.5 & 8.0 & $-$ & $-$ & $-$ & $-$ \\
055 & LESS J033302.2$-$274033 & 03:33:02.20 $-$27:40:33.6 & 8.0 & $-$ & $-$ & $-$ & $-$ \\
058 & LESS J033225.8$-$273306 & 03:32:25.79 $-$27:33:06.7 & 7.6 & $-$ & $-$ & $-$ & $-$ \\
061 & LESS J033245.6$-$280025 & 03:32:45.63 $-$28:00:25.3 & 8.3 & $-$ & $-$ & $-$ & $-$ \\
065 & LESS J033252.4$-$273527 & 03:32:52.40 $-$27:35:27.7 & 8.3 & $-$ & $-$ & $-$ & $-$ \\
068 & LESS J033233.4$-$273918 & 03:32:33.44 $-$27:39:18.5 & 8.4 & $-$ & $-$ & $-$ & $-$ \\
069 & LESS J033134.3$-$275934 & 03:31:34.26 $-$27:59:34.3 & 8.5 & 03:31:33.7745 $-$27:59:32.150 & $   13.0\pm   1.7$ &  6.8 & (0.057) \\
 & & & & 03:31:34.6841 $-$27:59:33.029 & $    7.3\pm   1.3$ &  5.8 & (0.074) \\
071 & LESS J033306.3$-$273327 & 03:33:06.29 $-$27:33:27.7 & 8.0 & $-$ & $-$ & $-$ & $-$ \\
074 & LESS J033309.3$-$274809 & 03:33:09.34 $-$27:48:09.9 & 8.4 & 03:33:09.1416 $-$27:48:16.650 & $   14.0\pm   1.7$ &  7.2 & ({\bf 0.057 + radio}) \\
 & & & & 03:33:09.3454 $-$27:48:15.998 & $   12.6\pm   1.7$ &  6.1 & {\bf 0.051} \\
077 & LESS J033157.2$-$275633 & 03:31:57.23 $-$27:56:33.2 & 8.8 & 03:31:57.2544 $-$27:56:39.815 & $   11.6\pm   1.6$ &  6.6 & (0.062) \\
080 & LESS J033142.2$-$274834 & 03:31:42.23 $-$27:48:34.4 & 8.9 & 03:31:42.5995 $-$27:48:41.155 & $    8.0\pm   1.3$ &  8.3 & 0.101 \\
 & & & & 03:31:42.8066 $-$27:48:36.659 & $   11.9\pm   1.6$ &  8.0 & (0.076) \\
 & & & & 03:31:41.6700 $-$27:48:30.031 & $   10.4\pm   1.5$ &  8.6 & (0.090) \\
082 & LESS J033253.8$-$273810 & 03:32:53.77 $-$27:38:10.9 & 9.0 & 03:32:53.9789 $-$27:38:14.633 & $    5.7\pm   1.1$ &  4.7 & (0.072) \\
083 & LESS J033308.9$-$280522 & 03:33:08.92 $-$28:05:22.0 & 8.3 & $-$ & $-$ & $-$ & $-$ \\
084 & LESS J033154.2$-$275109 & 03:31:54.22 $-$27:51:09.8 & 8.9 & 03:31:54.4937 $-$27:51:05.382 & $   15.6\pm   1.8$ &  5.7 & {\bf 0.039} \\
 & & & & 03:31:53.8248 $-$27:51:03.805 & $   12.5\pm   1.6$ &  8.0 & (0.073) \\
085 & LESS J033110.3$-$274503$^*$ & 03:31:10.28 $-$27:45:03.1 & 7.7 & $-$ & $-$ & $-$ & $-$ \\
086 & LESS J033114.9$-$274844 & 03:31:14.90 $-$27:48:44.3 & 8.5 & $-$ & $-$ & $-$ & $-$ \\
089 & LESS J033248.4$-$280023 & 03:32:48.44 $-$28:00:23.8 & 9.1 & $-$ & $-$ & $-$ & $-$ \\
090 & LESS J033243.7$-$273554 & 03:32:43.65 $-$27:35:54.1 & 9.1 & $-$ & $-$ & $-$ & $-$ \\
091 & LESS J033135.2$-$274033 & 03:31:35.25 $-$27:40:33.7 & 9.1 & $-$ & $-$ & $-$ & $-$ \\
092 & LESS J033138.4$-$274336 & 03:31:38.36 $-$27:43:36.0 & 9.2 & $-$ & $-$ & $-$ & $-$ \\
093 & LESS J033110.8$-$275607$^*$ & 03:31:10.84 $-$27:56:07.2 & 8.4 & $-$ & $-$ & $-$ & $-$ \\
094 & LESS J033307.3$-$275805 & 03:33:07.27 $-$27:58:05.0 & 9.1 & 03:33:07.5902 $-$27:58:05.840 & $    7.9\pm   1.3$ &  4.3 & {\bf 0.053} \\
095 & LESS J033241.7$-$275846 & 03:32:41.74 $-$27:58:46.1 & 9.2 & 03:32:41.2330 $-$27:58:41.239 & $   16.0\pm   1.9$ &  8.3 & (0.062) \\
097 & LESS J033313.7$-$273803 & 03:33:13.65 $-$27:38:03.4 & 9.2 & $-$ & $-$ & $-$ & $-$ \\
099 & LESS J033251.4$-$275536 & 03:32:51.45 $-$27:55:36.0 & 9.2 & $-$ & $-$ & $-$ & $-$ \\
100 & LESS J033111.3$-$280006$^*$ & 03:31:11.32 $-$28:00:06.2 & 8.8 & $-$ & $-$ & $-$ & $-$ \\
\end{tabular}
\label{tab:pirac}
}
\end{table*}

\begin{table*}
\centering
{\small
\contcaption{}
\begin{tabular}{clcccccccc} \hline
ID & SMG name & Submm position & $r_{\rm s}$ & 5.8-$\umu$m position & 5.8-$\umu$m flux & Offset & $p$ \\
   &  & ($\alpha_{\mathrm{J2000}}$)\hspace{.9cm}($\delta_{\mathrm{J2000}}$) & (arcsec) & ($\alpha_{\mathrm{J2000}}$)\hspace{.9cm}($\delta_{\mathrm{J2000}}$) & ($\umu$Jy) & (arcsec) & \\ \hline
104 & LESS J033258.5$-$273803 & 03:32:58.46 $-$27:38:03.0 & 9.4 & 03:32:58.2588 $-$27:38:11.522 & $    8.9\pm   1.4$ &  8.9 & 0.108 \\
 & & & & 03:32:57.8215 $-$27:37:59.117 & $    7.6\pm   1.3$ &  9.3 & 0.119 \\
105 & LESS J033115.8$-$275313 & 03:31:15.78 $-$27:53:13.1 & 9.0 & 03:31:15.4680 $-$27:53:11.450 & $    9.2\pm   2.5$ &  4.5 & {\bf 0.047} \\
106 & LESS J033140.1$-$275631 & 03:31:40.09 $-$27:56:31.4 & 9.4 & 03:31:40.1741 $-$27:56:22.412 & $   39.3\pm   2.9$ &  9.1 & {\bf 0.023} \\
107 & LESS J033130.8$-$275150 & 03:31:30.85 $-$27:51:50.9 & 9.4 & 03:31:30.5506 $-$27:51:58.716 & $   10.8\pm   1.6$ &  8.8 & (0.095) \\
109 & LESS J033328.1$-$274157 & 03:33:28.08 $-$27:41:57.0 & 9.5 & 03:33:28.0061 $-$27:42:02.408 & $   11.5\pm   1.6$ &  5.5 & {\bf 0.053} \\
 & & & & 03:33:28.5086 $-$27:41:50.420 & $    7.7\pm   1.3$ &  8.7 & 0.116 \\
110 & LESS J033122.6$-$275417 & 03:31:22.64 $-$27:54:17.2 & 9.4 & 03:31:22.6330 $-$27:54:17.014 & $    6.8\pm   1.3$ &  0.2 & {\bf 0.000} \\
112 & LESS J033249.3$-$273112 & 03:32:49.28 $-$27:31:12.3 & 9.0 & 03:32:48.8558 $-$27:31:12.868 & $   24.5\pm   2.3$ &  5.7 & {\bf 0.021} \\
113 & LESS J033236.4$-$275845 & 03:32:36.42 $-$27:58:45.9 & 9.5 & $-$ & $-$ & $-$ & $-$ \\
115 & LESS J033349.7$-$274239 & 03:33:49.71 $-$27:42:39.2 & 8.9 & 03:33:49.6663 $-$27:42:34.067 & $    9.9\pm   1.6$ &  5.2 & {\bf 0.053} \\
116 & LESS J033154.4$-$274525 & 03:31:54.42 $-$27:45:25.5 & 9.7 & $-$ & $-$ & $-$ & $-$ \\
119 & LESS J033256.5$-$280319 & 03:32:56.51 $-$28:03:19.1 & 9.7 & $-$ & $-$ & $-$ & $-$ \\
121 & LESS J033333.3$-$273449 & 03:33:33.32 $-$27:34:49.3 & 9.7 & $-$ & $-$ & $-$ & $-$ \\
123 & LESS J033330.9$-$275349 & 03:33:30.88 $-$27:53:49.3 & 9.8 & $-$ & $-$ & $-$ & $-$ \\
124 & LESS J033203.6$-$273605 & 03:32:03.59 $-$27:36:05.0 & 10.0 & 03:32:03.0814 $-$27:36:01.278 & $    7.9\pm   1.3$ &  7.7 & 0.109 \\
 & & & & 03:32:04.0070 $-$27:36:05.810 & $    6.5\pm   1.2$ &  5.6 & (0.090) \\
125 & LESS J033146.0$-$274621 & 03:31:46.02 $-$27:46:21.2 & 9.9 & $-$ & $-$ & $-$ & $-$ \\
126 & LESS J033209.8$-$274102 & 03:32:09.76 $-$27:41:02.0 & 9.9 & 03:32:09.6084 $-$27:41:06.983 & $   12.0\pm   1.6$ &  5.4 & {\bf 0.051} \\
\hline
\end{tabular}
}
\end{table*}

In the following we try and identify counterparts from the radio and
MIPS catalogues. We then use the properties of these counterparts to
select the parameter space in IRAC colour and flux to identify
potential counterparts to the radio and MIPS-undetected sources.

The results of our counterpart search are given in
Tables~\ref{tab:pradio} (radio) and ~\ref{tab:p24um} (24\,$\umu$m)
where we include the position and flux of the counterpart as well as
its radial offset from the submm source and the value of the search
radius. The first column also gives the `ID' of each SMG, an integer
describing the position of each in a ranked list of decreasing SNR (as
they appear in each table). For brevity, we will often refer to an
individual source using this integer, e.g.\ LESS001; the integers also
refer to the SNR-ranked list in \citet{weiss09}. Postage stamp maps
with a size of $36\times36$\,arcsec$^2$ centred on the submm position
are shown in Fig.~\ref{fig:mainplots}. We show radio contours
superimposed on IRAC 3.6-$\umu$m greyscales. The IRAC images are taken
from the SIMPLE Legacy Program (P.I.: P.\ van Dokkum) and where the
SMGs do not lie fully within the SIMPLE coverage we replace the images
with ones from SWIRE; LESS046 also has its FIDEL image replaced with
one from SWIRE. We also show the LESS contours, overplotted on
24-$\umu$m greyscales, in a separate panel.

The criteria for considering a potential counterpart as a secure
identification is that $p\le0.05$ (Tables~\ref{tab:pradio} and
~\ref{tab:p24um}). These sources have their values of $p$ given in
boldface in Tables~\ref{tab:pradio} and ~\ref{tab:p24um}, and their
positions are marked in Fig.~\ref{fig:mainplots}. Of the 126 submm
sources, 47 have at least one radio and 39 at least one 24-$\umu$m
counterpart with $p\le0.05$; together they produce 60 robust
counterparts. Of these, two (LESS063 and LESS118) are extremely weak
($\la$30\,$\umu$Jy, deboosted), have low values of $p$ (by virtue of
lying very close to the submm position), but do not appear to have
associated MIPS or IRAC emission. It is impossible to rule them out as
genuine counterparts, but we have highlighted them in
Table~\ref{tab:pradio} ($p$ contained within {\it square}
parentheses).

As we have both radio and 24-$\umu$m data for most sources, the
combination of the results for each SMG enables us to identify
additional reliable counterparts. For weak sources, the presence of
coincident emission in both wavebands makes it more likely that the
source is real, but for any source the presence of emission at radio
wavelengths {\it and} 24-$\umu$m makes it more likely that the galaxy
is the correct identification. Individual sources with $0.05 < p \le
0.1$ (i.e.\ still low enough to indicate a likely counterpart) have
their value of $p$ within parentheses, but where coincident radio and
24-$\umu$m components have $0.05 < p \le 0.1$ we consider this to be a
secure identification and present the value of $p$ in parenthesised
boldface. Two more SMGs gain robust counterparts in this way, LESS036
and LESS060.

In summary, we find statistically robust counterparts to 62
(49~per~cent) of the SMGs using the radio and 24-$\umu$m data
(Table~\ref{tab:summary}). We now go on to extend our sample of
identified SMGs by exploiting the very deep IRAC observations of this
field.

\section{IRAC counterparts}
\label{irac}

\begin{figure}
\includegraphics[width=8.5cm]{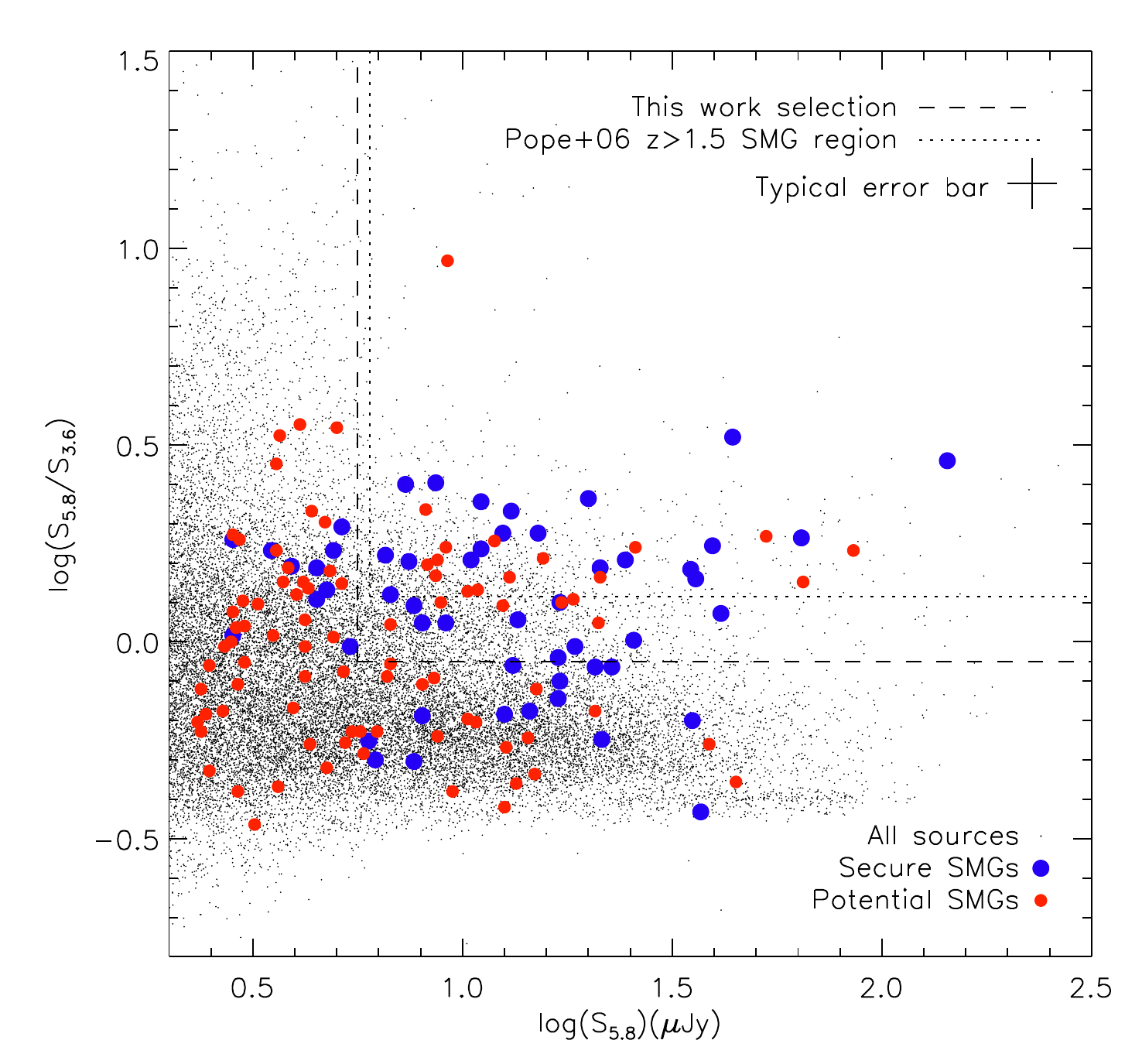}
\caption{3.6- and 5.8-$\umu$m colour-flux diagram for IRAC-selected
  sources in the ECDFS. Secure radio- and MIPS-identified SMG
  counterparts, and potential SMG counterparts are highlighted. The
  secure radio- and MIPS-identified SMG counterparts are typically
  redder than the field population, a property which we use to define
  the marked selection criteria for SMGs (dashed rectangle). Within
  this region there is a 10~per~cent contamination rate from field
  galaxies and we recover 78~per~cent of the secure radio- and
  MIPS-identified SMG counterparts. The dotted rectangle shows the
  region defined by \citet{pope06}.}
\label{fig:iraccolourmag}
\end{figure}

To identify counterparts to submillimetre sources without secure radio
or MIPS identifications we employ 3.6- and 5.8-$\umu$m {\it Spitzer}
IRAC data. Fig.~\ref{fig:iraccolourmag} shows the 3.6- and
5.8-$\umu$m colour-flux diagram for sources in the ECDFS, with secure
radio- or MIPS-identified SMG counterparts highlighted. An IRAC
counterpart to every MIPS and radio robust identification was found by
examining each by eye, taking into account any radio emission and
optical sources in the region under consideration (from MUSYC
imaging). It is apparent from Fig.~\ref{fig:iraccolourmag} that SMGs
are typically redder than the field population and it is this property
we will exploit to identify counterparts to some unidentified SMGs.

To decide on our colour and flux selection limits we exploit the
radio- and MIPS-identified SMGs to determine limits which balance
completeness and purity. We arbitrarily chose a limit of 10~per~cent
contamination (within the 3-$\sigma$ search radii of SMGs) and
maximise the completeness of recovery. Based upon these requirements
we select SMG counterparts with ${\rm log(S_{5.8\umu m}(\umu Jy))}\ge
0.75$ and ${\rm log(S_{5.8\umu m}/S_{3.6\umu m})}\ge -0.05$, returning
78~per~cent of the secure radio and MIPS SMG counterparts. Also shown
on Fig.~\ref{fig:iraccolourmag} are the colour-magnitude cuts of
\citet{pope06}, the colour division of which separates SMGs into low
and high redshift samples (SMGs redder than the colour limit
correspond to $z>1.5$).

We apply this colour selection to the 64 error circles of SMGs lacking
secure radio or MIPS counterparts and calculate a value of $p$ in a
similar way to the radio and 24-$\umu$m sources described in
Section~\ref{strategy} (Table ~\ref{tab:pirac}) -- Monte Carlo
simulations again show that the values of $p$ are reliable. In total,
we identify 17 additional SMGs with robust counterparts. Four have
multiple robust identifications and in one case (LESS074) the second
counterpart results from there being a probable ($0.05<p\le0.1$) in
both the IRAC and radio data. With the addition of the IRAC
counterparts, a total of 79 (63~per~cent) of the SMGs have at least
one robust counterpart. The details of the 79 robust counterparts are
shown in Table~\ref{tab:summary}, including the deboosted submm,
24-$\umu$m and deboosted radio flux densities.

\section{Analysis and Discussion}
\label{discussion}

Before the addition of the IRAC results, compared to some other
studies \citep{ivison02, ivison05, pope06, ivison07} the fraction of
SMGs with a secure identification (49~per~cent) is a little low. For
example, in the SCUBA HAlf-Degree Extragalactic Survey
\citep[SHADES;][]{coppin06} 79 out of 120 SMGs (66~per~cent) were
found to have secure counterparts \citep{ivison07} and in the {\it
  Hubble Deep Field}--North (HDF--N) \citet{pope06} claim secure
counterparts for 60~per~cent of their sample. However, the fraction of
SMGs with robust counterparts is clearly a function of the sensitivity
of the radio and mid-IR data and we note that in the case of the
HDF--N (5.3\,$\umu$Jy\,beam$^{-1}$ -- \citealt{biggs06}) and the
Lockman Hole portion of SHADES (4.2\,$\umu$Jy\,beam$^{-1}$ --
\citealt{biggs06}), the radio maps were more sensitive than that in
the ECDFS (6.5\,$\umu$Jy\,beam$^{-1}$ -- \citealt{miller08}). Assuming
no cosmic variance and an integral source count slope of $-1.5$, the
increase in sensitivity of the Lockman Hole radio map compared to that
of the ECDFS produces a density of sources that is higher by almost a
factor of two at the parts of the map corresponding to that
sensitivity. However, we also note that the radio map of the Subaru
{\em XMM-Newton} Deep Field (SXDF) that was used to find counterparts
to that portion of SHADES had a similar depth to the ECDFS map
(6.3\,$\umu$Jy\,beam$^{-1}$), but produced a significantly greater
fraction of SMGs with robust IDs, 58~per~cent \citep{ivison07}.

We now go on to consider other reasons why true counterparts might be
missed, beginning by noting that a small number ($\approx$5) of the
SMG detections are likely to be spurious \citep{weiss09}, as is common
in surveys of this kind. Other possible reasons include:
\begin{enumerate}
\item The counterpart lies outside the search radius
\item Multiple SMGs have become blended due to the low resolution of the submm data.
\end{enumerate}

\subsection{Search radius and radio astrometry}
\label{sec:astrometry}

As it is necessary to impose a limit to how far you search from a SMG
position, and because choosing too large a radius increases the
chances of finding unrelated counterparts whilst reducing the
significance of genuine associations -- it is likely that a small
number of SMGs will not have been searched out to a sufficient radius
to locate their counterpart. We have estimated that this will amount
to 1--2~counterparts, but if we have under-estimated the size of the
submm positional errors then this number will be larger. Taking SHADES
as an example, \citet{ivison07} estimated that 5~per~cent of
counterparts would be missed in this manner and recent Submillimeter
Array (SMA) observations with sub-arcsec accuracy have identified one
example of this, SXDF850.06 lying just outside the \citet{ivison07}
8-arcsec search radius \citep{hatsukade10}.

\begin{figure}
\begin{center}
\includegraphics[scale=0.4]{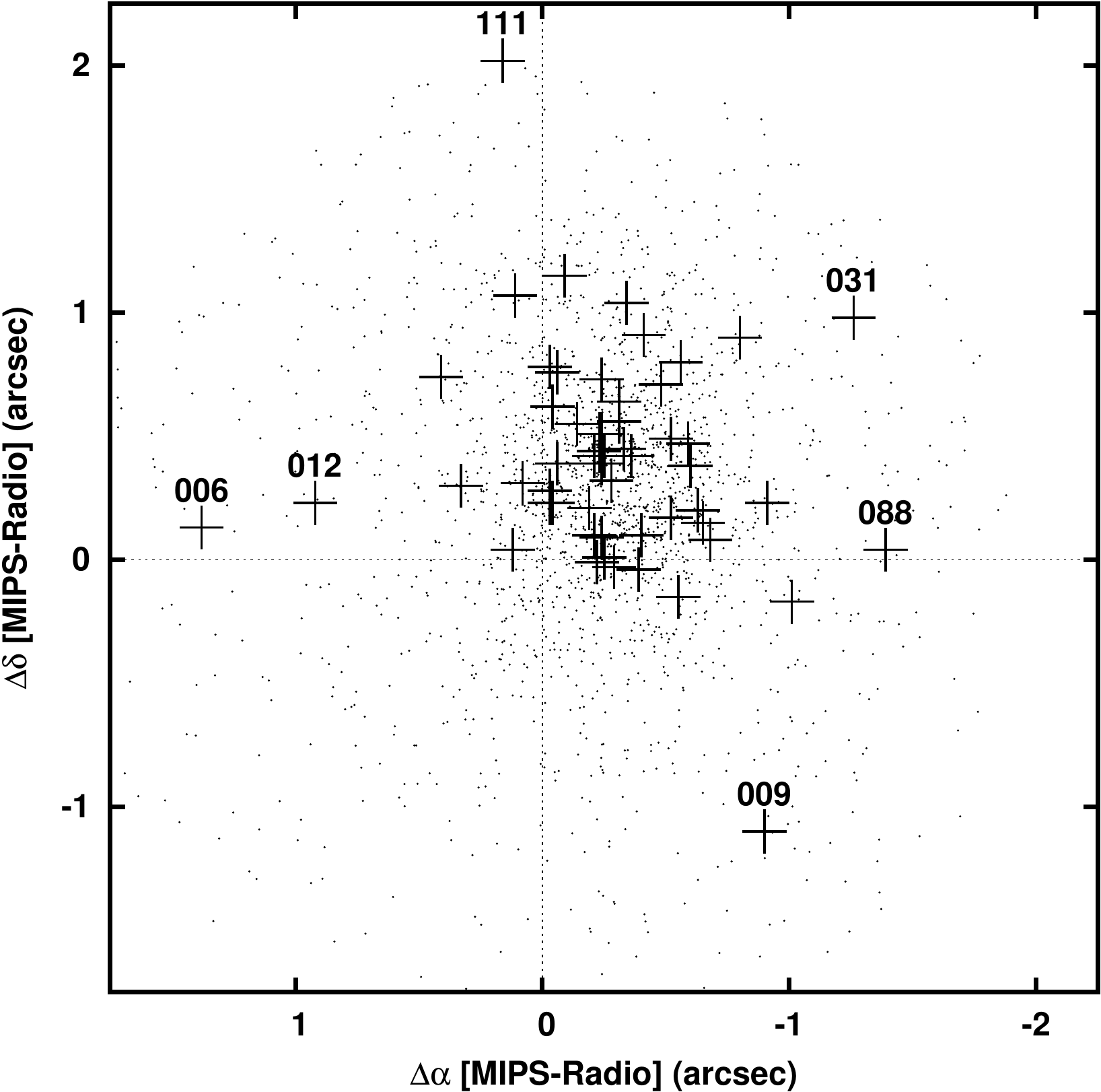}
\caption{Offsets (arcsec) in Right Ascension and Declination between
  the MIPS and radio counterpart positions (large crosses) with the
  largest offsets marked with the relevant SMG. Also plotted are the
  offsets for 2019 matches between the radio and MIPS catalogues --
  the median offset is equal to $-0.25$~arcsec in Right Ascension and
  $+0.29$~arcsec in Declination (MIPS $-$ radio) and the plot is
  centered on these coordinates.}
\label{fig:mips-radio}
\end{center}
\end{figure}

Errors in the absolute astrometric accuracy of the MIPS and radio
catalogues will also affect our ability to reliably determine
counterparts. In Fig.~\ref{fig:mips-radio} we show the differences
between the Right Ascension and Declination for each counterpart that
has emission at both 24-$\umu$m and 21~cm. There is a clear offset
between the two which has a median value of $-0.25$~arcsec in Right
Ascension and $+0.39$~arcsec in Declination (MIPS $-$ radio). In order
to improve the determination of this offset, we have used {\sc topcat}
\citep{taylor05} to measure the median offset between the radio and
MIPS catalogues by finding all unique matches within 2~arcsec. These
2019 matches are also plotted in Fig.~\ref{fig:mips-radio} and have
similar median offsets of $-0.25$~arcsec in Right Ascension and
$+0.29$~arcsec in Declination.

Applying these more accurate offsets during the $p$-statistic
procedure as if they purely originated from the MIPS data reduces the
number of robust counterparts by one, whereas assuming that the origin
lies with the radio data increases the number of robust counterparts
by one. Hence, the effect of the offsets is actually rather small and
the additional radio counterpart is found anyway without the offsets
applied due to it having $p<0.1$ in both the MIPS and radio. As a
result, we have not taken the offsets into account when calculating
the values of $p$ and the radio positions given in
Table~\ref{tab:pradio} have not been corrected for them.

\begin{figure*}
\begin{center}
\includegraphics[scale=0.4]{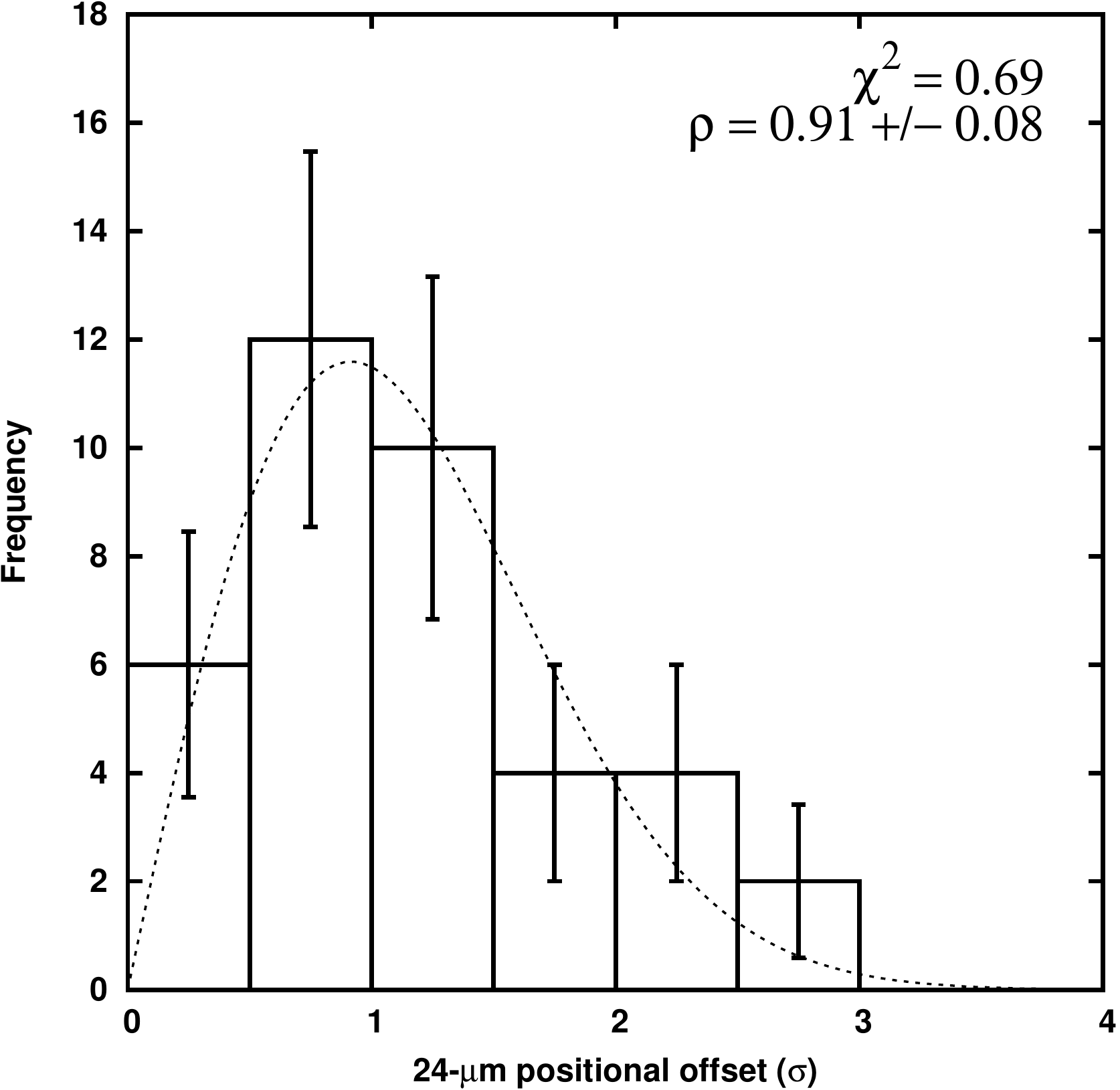}
\includegraphics[scale=0.4]{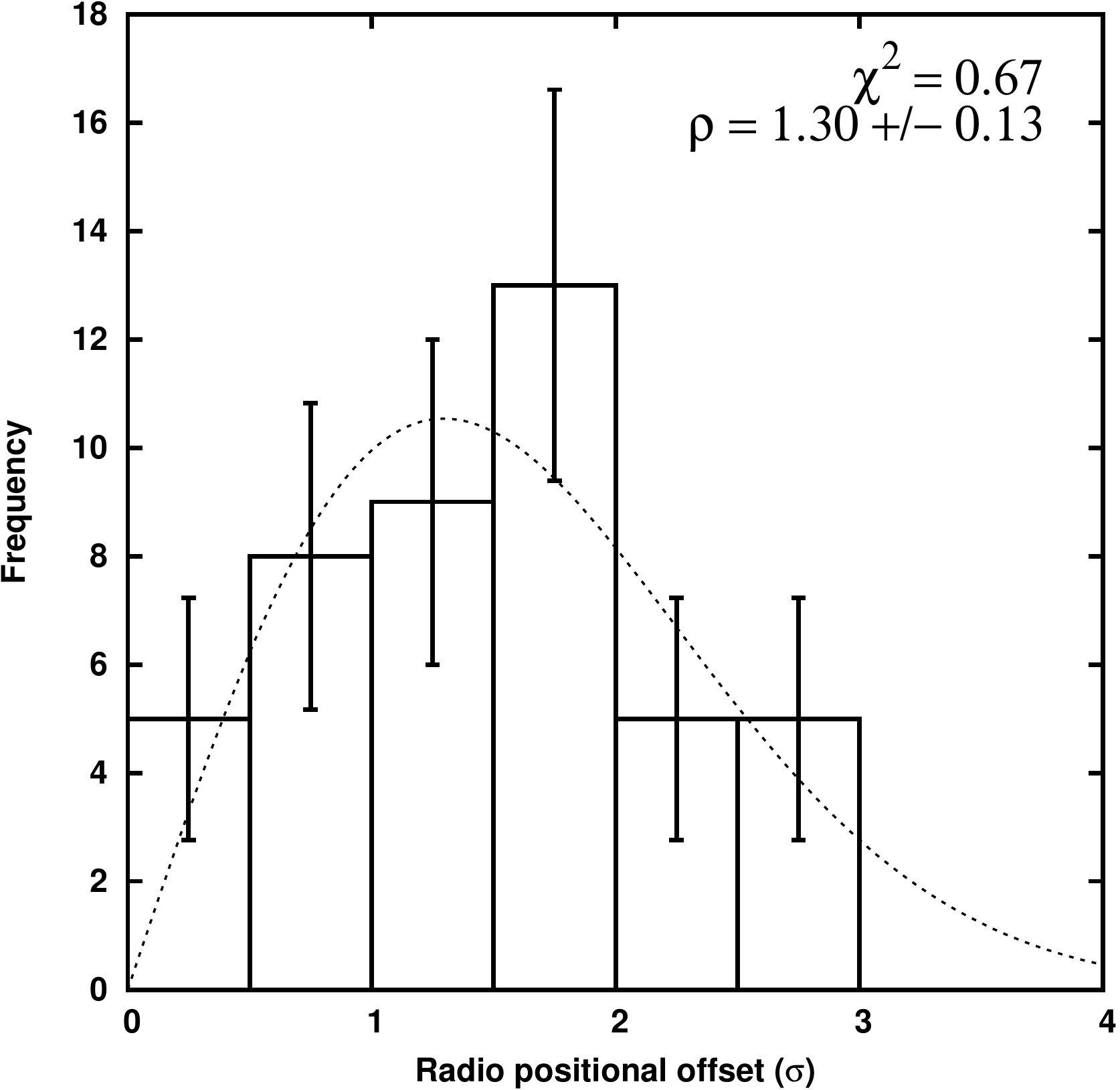}
\caption{Distribution of positional offsets (in units of $\sigma$)
  from the submm position for the radio (right) and 24-$\umu$m (left)
  sources with $p\le0.05$; error bars are Poissonian. The dashed line
  shows a fitted Rayleigh distribution ($R(r) \propto
  r\,e^{-r^2/2\rho^2}$) where $r$ is the radial offset and $\rho$ the
  standard deviation of the positional errors in either R.A.\ (or
  Dec., the two being equal). The reduced chi-squared of the fit
  ($\chi^2$) and the value of $\rho$ are given in the top-right-hand
  corner. For the MIPS counterparts, the offsets conform closely to
  the expected distribution and $\rho$ is close to the expected
  value. For the radio, despite removing the astrometric offset seen
  in Fig.~\ref{fig:mips-radio}, it is possible that there are more
  radio counterparts at larger radii than expected.}
\label{fig:posoff}
\end{center}
\end{figure*}

As a test of our search radius strategy, we have examined the
distribution of the radial offsets between the radio and 24-$\umu$m
secure counterparts and their SMG. If these truly originate from
random uncertainties in the positions of the SMGs then they should
conform to a Rayleigh distribution. In addition, we note that this
should only be the case if the offsets are calculated as a multiple of
the 1-$\sigma$ SMG position uncertainty. Fig.~\ref{fig:posoff} shows
the positional offsets for both the radio and MIPS data in bins of
0.5~$\sigma$, with a fitted Rayleigh distribution overplotted. Writing
this as $R(r) \propto r\,e^{-r^2/2\rho^2}$, the width parameter $\rho$
is equal to the one-dimensional standard deviation of the positional
errors. As we are plotting these in units of standard deviations then
by definition this parameter (which is also equal to the mode of the
distribution) should be equal to unity.

In the case of the MIPS data, the fit to the data is excellent with a
reduced chi-square of 0.7 and a value for the width parameter, $\rho$
of $0.91\pm0.08$. This strongly suggests that we have correctly
calculated the magnitude of the positional errors and that our
assumption that the offsets originate predominantly in the submm
source position is also correct. The fit in the case of the radio data
is formally very similar, but has a larger value of $\rho =
1.30\pm0.13$, a 2.3-$\sigma$ deviation from the expected value of
1.0. Given that the 24-$\umu$m data give the expected answer, this
points to something special about the radio positional offsets,
separate from the astrometric error already identified (which has
been removed when calculating the offsets displayed in
Fig.~\ref{fig:posoff}). If the larger offset in the radio is not a
statistical fluke, then it points to an additional, systematic offset
between the radio and the MIPS/submm position, perhaps due to emission
from more extended or structured radio emission e.g.\ radio jets.


\subsection{Blending of SMGs}

The second potential cause of the low identification rate is that the
SMG, instead of being a single unresolved source, may actually be a
blend of several sources. This is more likely to be the case here than
many other submm surveys due to the slightly larger beamsize,
19~arcsec compared to the 14~arcsec of the JCMT/SCUBA and the
11~arcsec of the IRAM 30-m/MAMBO (we note, however, that the lack of
chopping for LABOCA reduces the possibility of confusion from that
source in these maps). The problem is that the $p$-statistic
implicitly assumes that the submm position corresponds to a single
source; if multiple sources are responsible for the submm emission,
the SMG centroid will be offset from the genuine counterparts which
are less likely to fall within the search radius.

A visual inspection of the plots in Fig.~\ref{fig:mainplots} makes it
clear that this is often the case. One noteworthy case is the chain of
three MIPS galaxies visible under LESS004. This source is clearly
elongated along the axis of the galaxy chain and as two of the
galaxies are detected in the radio map they presumably both contribute
to the submm emission. As the resolution of the submm map is not
sufficient to separate the individual sources it has instead been
classified as a single source with its centroid lying in between the
two radio sources. This places it close, but not close enough, to the
third galaxy which is also radio quiet. We further note that this
chain of galaxies continues to the south where it blends into another
submm source, LESS026.

The use of a SNR-dependent search radius makes it more likely that
sources that are blended are not robustly identified as the summed
submm flux density causes a smaller value of $r_s$ to be used than if
a constant radius were adopted. This alternative approach often
chooses a search radius that is large enough to maximise the
probability of detecting counterparts to the weaker SMGs (of which the
median submm flux density is often representative) that have larger
positional errors. However, the cost of this is that higher values of
$p$ are measured for unblended, brighter SMGs which results in
potentially less counterparts being identified.

Finally, in forming the $p$-statistic, we have only included radio and
MIPS sources with $\mathrm{SNR} \ge 3.5$ (the IRAC sources were all
much more significantly detected due to all sources being brighter
than 5.6~$\umu$Jy). Below these SNR thresholds the detections become
less reliable (`sources' are increasingly likely to be noise spikes,
residual sidelobes, map artefacts, etc.) and automated source-finding
algorithms such as {\sc sad}, used to form the catalogues, have
greater difficulty in producing reliable fits to genuine sources,
which are therefore rejected. A detailed examination of the radio and
mid-IR sources shown in Fig.~\ref{fig:mainplots} can help to identify
such cases, especially where there is coincident emission in both
wavebands, and in Appendix~\ref{appendix} we give a brief description
of any noteworthy SMGs. This includes likely blends of individual
submm sources and possible counterparts that lie outside the search
radius (along with their fluxes and positions).

\begin{figure}
\begin{center}
\includegraphics[scale=0.45,type=pdf,ext=.pdf,read=.pdf]{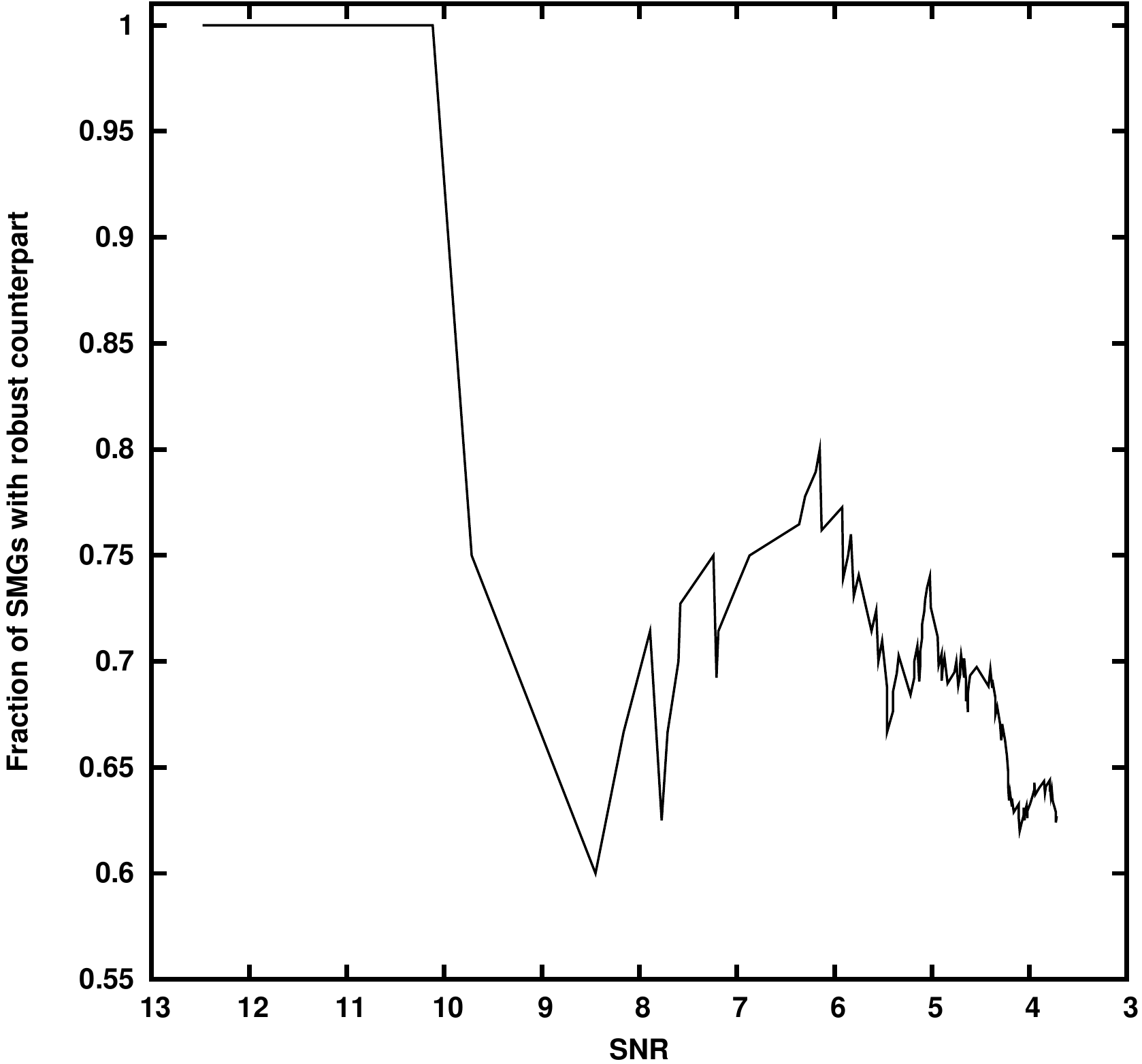}
\caption{Cumulative recovery fraction of secure radio and MIPS
  counterparts per SMG as a function of SNR. Below a SNR of about six,
  there is a steady decline in the fraction of SMGs with robust
  counterparts, as expected. At higher SNR, the trend is noisier and
  two of the five brightest SMGs do not have identifications.}
\label{fig:smgsnr}
\end{center}
\end{figure}

Fainter SMGs should, broadly speaking, also be fainter in the radio
and mid-IR (different spectral properties or their redshift dependency
will of course weaken the correlation). Indeed, it is very apparent
from Tables~\ref{tab:pradio} and \ref{tab:p24um} that fewer SMGs have
secure counterparts as you move to the bottom of the tables,
i.e.\ towards decreasing SNR. This is illustrated in
Fig.~\ref{fig:smgsnr} where we plot the cumulative recovery fraction
as a function of SNR (which, given the uniform noise in the LABOCA
map, is strongly correlated with submm flux). At high SNR there are
few sources and the trend is very noisy, but as the SNR declines below
six there is a clear trend towards a declining fraction of SMGs with
secure counterparts, as expected.

Of the five SMGs with the highest SNR, only one has a radio source
within the search radius. One of these, LESS004, can be seen in
Fig.~\ref{fig:mainplots} to clearly be a blend of multiple
sources. For the other sources, no such blend is obvious. Most
striking is the brightest SMG in the LESS sample, LESS001, for which
there is no radio emission at all and only extremely faint mid-IR
emission within the search radius (although this is classed as the
counterpart to the SMG). Blends do, of course, contribute doubly to
this effect -- their flux density is over-estimated and the additional
positional offsets render the identification of any counterpart more
difficult. However, it is also possible that the reason for the lack
of counterpart emission is due to the source lying at very high
redshift \citep[e.g.][]{ivison02,younger07,dannerbauer08} or the dust
in the galaxy being colder than average \citep{chapman05}.

\subsection{Redshift distribution}
\label{redshift}

Whilst the flux density of an SMG is essentially independent of
redshift (up to $z\sim8$), both the radio and IR emission will fade
with increasing distance. The median spectroscopic redshift of
radio-identified SMGs is 2.2 \citep{chapman05} although the
requirement that an accurate radio position be available (in order to
place the slit accurately for the spectroscopic observations -- see
\citealt{ivison05}) may skew the redshift distribution towards a lower
range. An increasing number of SMGs have been identified at $z>4$
\citep{schinnerer08,daddi09a,daddi09b,knudsen10}, including one from
this survey with a spectroscopically-determined redshift of 4.76
\citep{coppin09,coppin10}: LESS073 (with $p=0.003$ in
Table~\ref{tab:pradio}).

\begin{figure}
\begin{center}
\includegraphics[scale=0.5,type=pdf,ext=.pdf,read=.pdf]{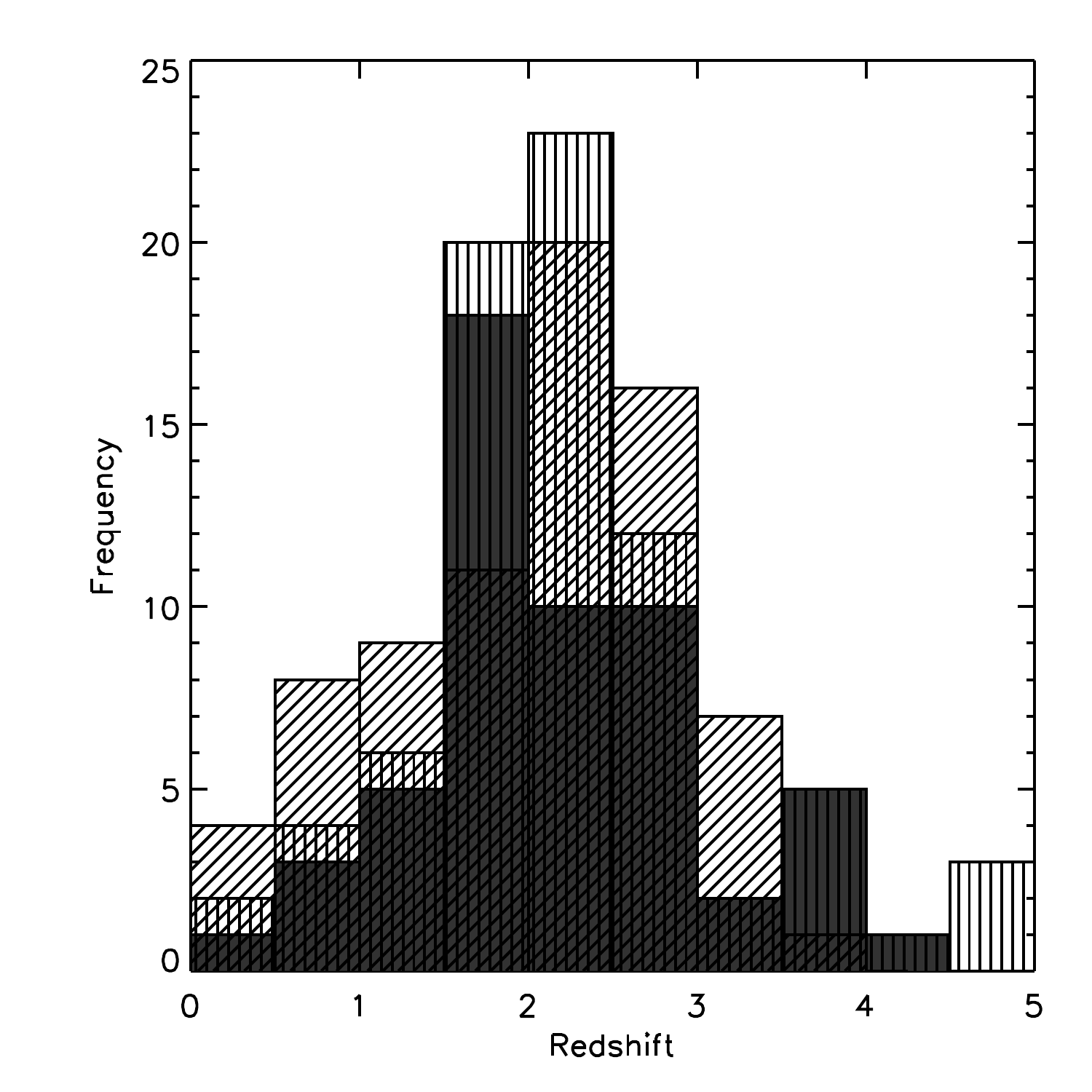}
\caption{Left: The solid filled histogram shows the redshift
  distribution calculated using the radio/submm spectral index
  relation of \citet{carilli00}; this includes all 55 robust
  counterparts from Tables~\ref{tab:pradio}, \ref{tab:p24um} and
  \ref{tab:pirac} that have a single, robust radio detection. The
  median of the distribution is 2.2 with an interquartile range of
  1.6--2.6. Also shown are the redshift distributions of (diaganol
  hatch) \citet{chapman05} and (vertical hatch)
  \citet{wardlow10}. None of the distributions have been scaled -- the
  $y$-axis shows the actual number of redshifts in each case.}
\label{fig:redshift}
\end{center}
\end{figure}

We can investigate the redshift distribution of the radio-detected
secure counterparts by utilising the radio-to-submm spectral index
relation (1.4:350\,GHz, $\alpha^{350}_{1.4}$) of \citet[][hereafter
  CY00]{carilli00} who characterise the variation of
$\alpha^{350}_{1.4}$ with $z$ using the average of 17 template
SEDs. The resulting redshifts are therefore averaged over various
source properties including radio spectral index,
$\alpha_{\mathrm{radio}}$, submm spectral index,
$\alpha_{\mathrm{submm}}$ and dust temperature. In order to make our
redshift estimates as reliable as possible, we have only included
sources where the robust counterpart consists of only a single radio
component. The resulting 55 redshifts are plotted in
Fig.~\ref{fig:redshift} and listed in Table~\ref{tab:summary}; the
1-$\sigma$ errors include the uncertainties in the fluxes and the
spread in the submm spectral index.

\begin{figure}
\begin{center}
\includegraphics[scale=0.43,type=pdf,ext=.pdf,read=.pdf]{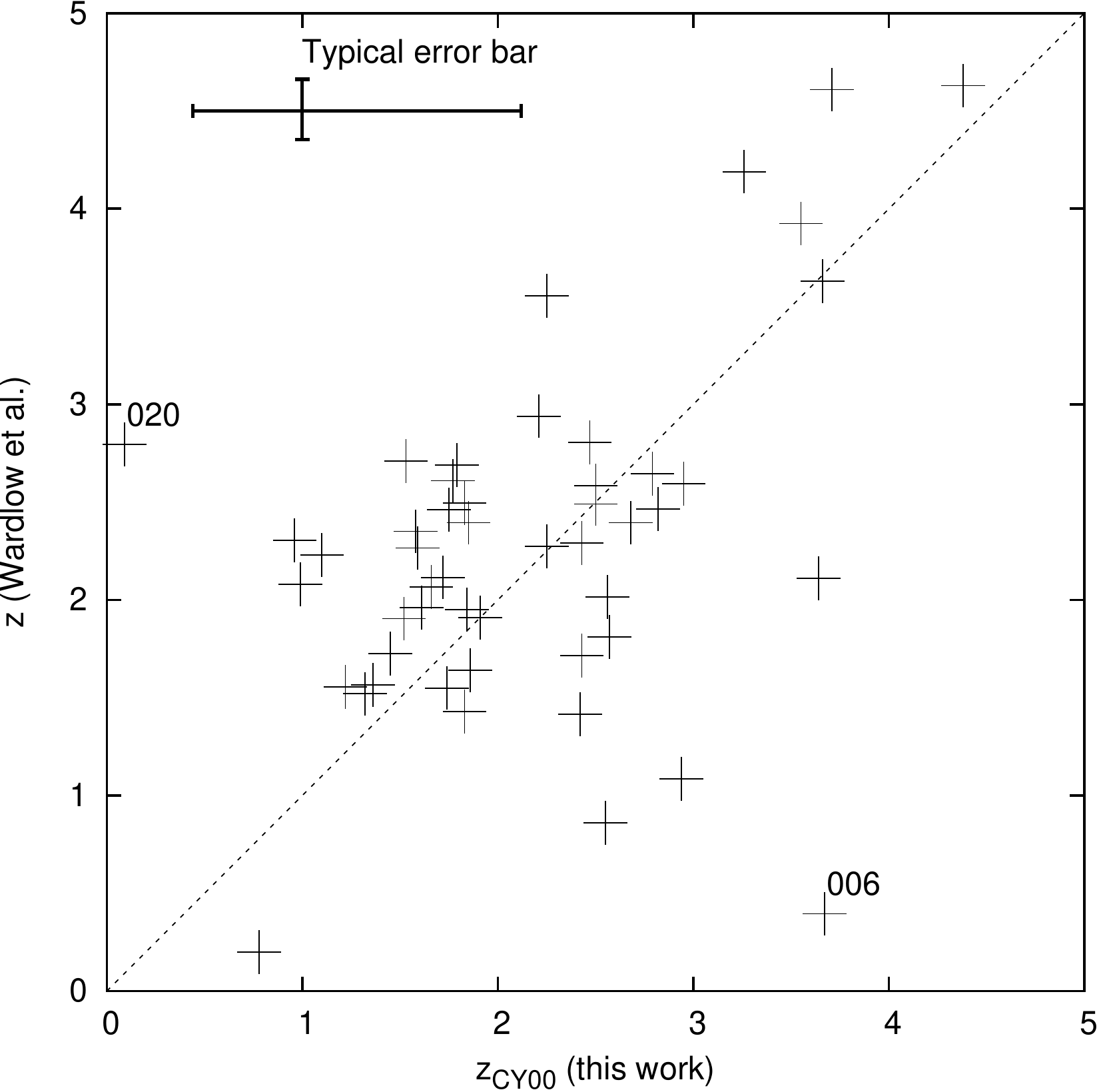}
\caption{A comparison between the redshifts measured for sources
  common to \citet{wardlow10} and this work. Error bars have been
  omitted for clarity, but a typical errorbar is shown in the top-left
  corner. The Spearman rank correlation coefficient is equal to 0.43
  (significant at $>$95~per~cent confidence) which confirms that there
  is a positive correlation between the two redshift measures.}
\label{fig:mejuliecomp}
\end{center}
\end{figure}

Our distribution is similar to those found by \citet{chapman05} and
\citet{wardlow10} which are also plotted in
Fig.~\ref{fig:redshift}. The median of our redshift distribution is
$\overline{z} = 2.2^{+0.8}_{-0.7}$ (1-$\sigma$ errors), identical to
both the spectroscopically-derived median for the radio-identified
SMGs of \citet{chapman05} and that measured by \citet{wardlow10} using
17-band optical to MIR photometry; \citet{aretxaga07} measure median
redshifts of between 2.2 and 2.7 for the two SHADES fields. The
\citet{wardlow10} study is particularly relevant as it uses the sample
of robust counterparts identified in this work, although it is not
confined to those with a radio detection. A comparison of the
redshifts measured here and by \citet{wardlow10} is shown in
Fig.~\ref{fig:mejuliecomp}.

There are a number of significant outliers, perhaps the most obvious
of which is LESS020, for which the CY00 technique gives a much lower
value of $z = 0.09$ compared to 2.8 from the full photometric
analysis. This is by far the brightest of the radio counterparts
($S_{1.4} > 4$~mJy) and its radio flux is most likely boosted by a
radio-loud AGN for which the SED templates of CY00 do not apply. The
other most prominent outlier is LESS006 for which there is a large
offset between the 24-$\umu$m and radio positions and where the radio
emission lies predominantly between two peaks in the 3.6-$\umu$m image
(Fig.~\ref{fig:mainplots}). Based on the large ($>1$~arcsec)
positional offsets, \citet{wardlow10} suggest that this SMG is being
gravitationally lensed by the low-redshift ($z=0.4$) optical/MIR
galaxy and the much larger redshift ($z=3.7$) measured based on the
radio/submm flux would support this conclusion. Excluding these two
outliers results in a Spearman rank correlation coefficient of 0.43
which, for 48 common redshifts, easily exceeds the critical value
(0.24) for 95~per~cent significance and allows us to reject the
hypothesis that there is no correlation.

The excellent agreement in the measurement of the median redshift of
the SMG population, using three different techniques, is very
encouraging and strongly argues that the peak in SMG activity was at
or close to $z = 2.2$. However, in all three cases, the majority of
the redshifts were made possible due to the presence of radio
emission, in the \citet{chapman05} case as an indicator of where to
place the spectroscopic slit. Because of the fading of radio emission
with increased distance (positive $k$-correction) this means that many
of the SMGs in these samples which are not detected in the radio, the
majority of which remain unidentified, are likely to be biased towards
higher redshifts. Indeed, the majority of the LESS sample are
undetected in the radio and although some of these will be undetected
because they are unusually cold, overall we expect this radio-faint
sub-sample to have a redshift distribution skewed to larger values
than the radio-detected SMGs. It is thus possible that the overall
median redshift of our SMG sample is higher than 2.2 although we note
that there is an additional bias in the other direction -- the IRAC
counterparts are preferentially located at high redshift as they have
weak radio emission and lie in the high-redshift quadrant of the
\citet{pope06} diagram (Fig.~\ref{fig:iraccolourmag}).

\begin{figure}
\begin{center}
\includegraphics[scale=0.43,type=pdf,ext=.pdf,read=.pdf]{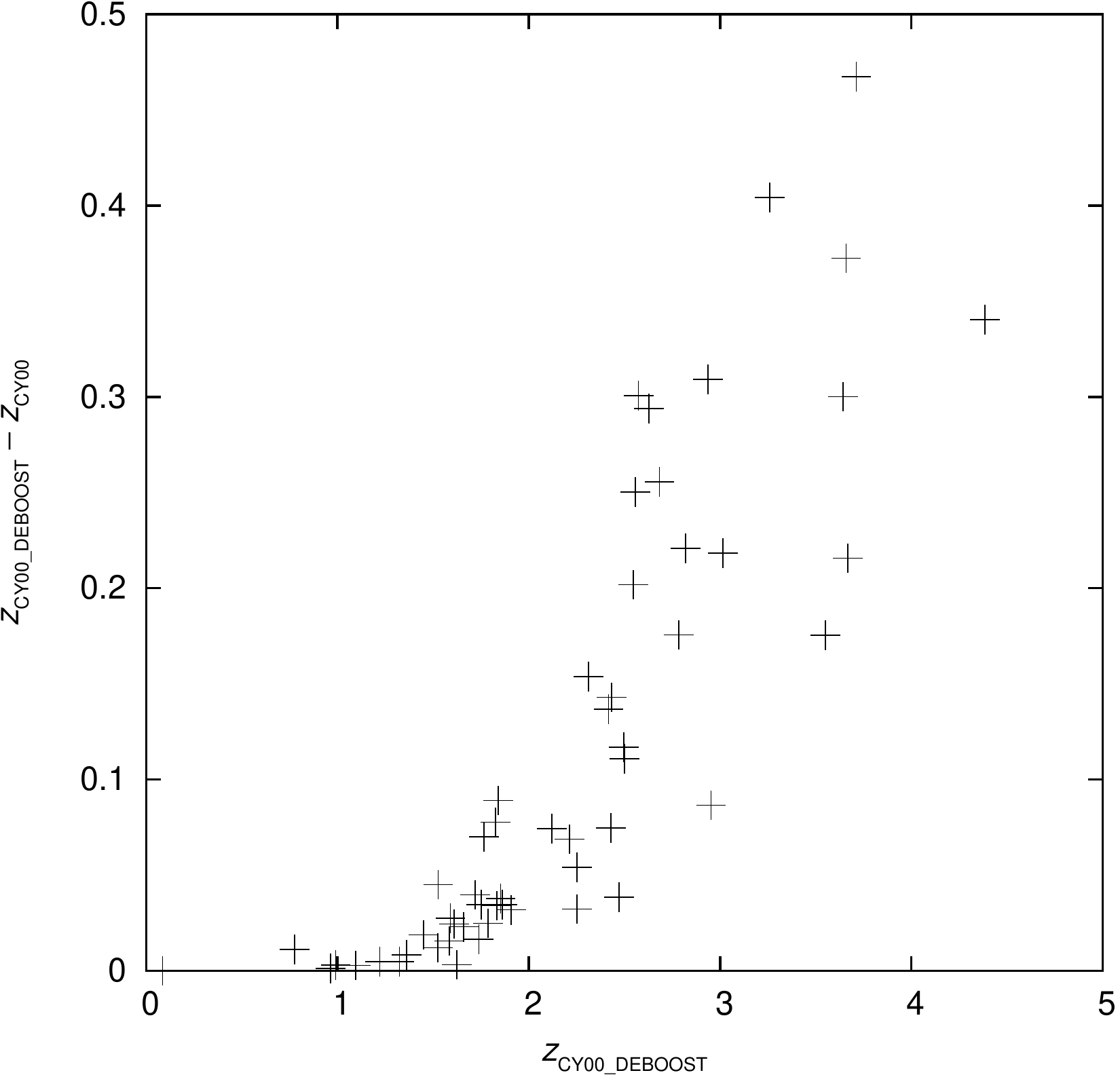}
\caption{Increase in the measured CY00 redshift caused by using
  deboosted flux densities. The increase can be significant ($\le0.5$)
  and is most pronounced at high redshifts (high redshifts are biased
  towards weaker radio flux densities which are in turn most affected
  by flux boosting).}
\label{fig:redshift_boost}
\end{center}
\end{figure}

The use of deboosted radio flux densities also has the effect of
increasing the redshifts measured using the CY00 technique. This is
demonstrated in Fig.~\ref{fig:redshift_boost} where we plot the CY00
redshifts measured using both deboosted and un-deboosted radio flux
densities. The increase due to the deboosting is most pronounced at
the highest redshifts which, for a given submm flux density,
correspond to the weaker radio sources that are most affected by flux
boosting. Whilst the majority of the SMGs have $\Delta z < 0.1$, the
maximum increase is $\la 0.5$, this corresponding to the only source
for which we have a spectroscopic redshift (LESS073 at $z =
4.76$). The CY00 redshift is 3.7, a considerable improvement on the
value of 3.2 that would otherwise have been measured without the use
of deboosted radio fluxes. Overall, the combined effect is to increase
the median redshift of the SMGs by 0.2 i.e. from 2.0 to 2.2.

\begin{figure}
\begin{center}
\includegraphics[scale=0.45,type=pdf,ext=.pdf,read=.pdf]{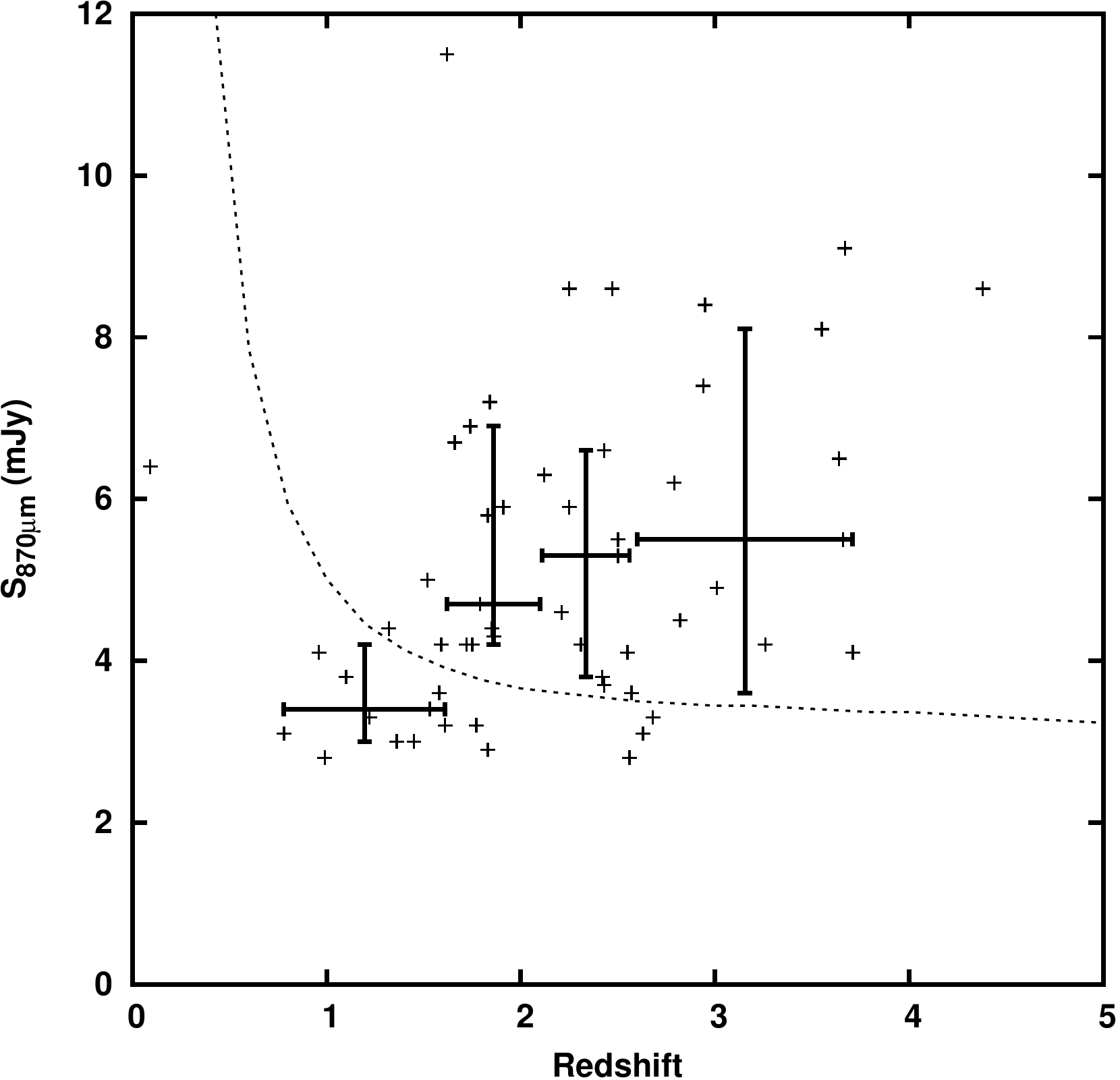}
\caption{Deboosted submm flux density versus redshifts calculated
  using the Carilli--Yun radio/submm spectral index
  relation. Superimposed are the median flux densities in four
  redshift bins; each bin contains 13 or 14 sources. Error bars are
  1-$\sigma$. Also shown is the 870-$\umu$m flux density of Arp~220 as
  a function of redshift (dashed line) normalised to the lowest
  redshift bin's average. As can be seen, there is a weak trend of
  increasing flux at higher redshifts -- the Spearman rank correlation
  coefficient for the unbinned data is 0.39 which is significant at
  $>$95~per~cent confidence.}
\label{fig:z_submm}
\end{center}
\end{figure}

In Fig.~\ref{fig:z_submm} we have plotted deboosted submm flux density
as a function of Carilli--Yun redshift. Also shown are the median
redshifts of the data in four separate flux-density bins. This reveals
an increase in the average flux density with redshift, an effect
previously noted by \citet{ivison02}, \citet{pope05} and
\citet{younger08}. The 870-$\umu$m flux density of Arp\,220 is also
plotted in Fig.~\ref{fig:z_submm}, as a function of redshift, to
illustrate that the apparent evolution of SMG luminosity with redshift
(note that \citet{wardlow10} do not find evidence for such an effect)
is unlikely to be caused by their spectral energy distributions, or by
cosmology. The Spearman rank correlation coefficient, with LESS020 and
LESS006 again excluded, is equal to 0.39 which allows us to reject the
null hypothesis with $>$95~per~cent confidence (the critical value for
53 pairs is equal to 0.23).


The exact form of the evolution of SMGs with redshift remains unclear
\citep*[e.g.][]{chapin09b}, particularly whether it is predominantly
in luminosity, density or both, but there are signs that luminosity
evolution plays a role. \citet{wall08} have even suggested that there are
in fact two populations of SMGs, separated by luminosity, and that
these evolve differently. It is therefore tempting to ascribe the
lack of detections of the brightest SMGs as being due to their high
redshifts. However, it is equally possible that the brightest
galaxies arise due to confusion from clustering in the brightest
sources, or from temperature and luminosity evolution which results in
them being high-luminosity, low-temperature galaxies at low redshift.

\begin{table*}
\centering
\caption{Summary of counterparts to SMGs. We show the deboosted submm, MIPS 24-$\umu$m and deboosted radio flux densities for each robust counterpart. Also included are the redshifts estimated from the submm/radio spectral index. LESS063 and 118 have not had redshifts calculated as the radio emission may be spurious. We have also not calculated the redshifts when there are multiple robust radio counterparts.}
\begin{tabular}{cccccccccc} \hline
SMG & 870~$\umu$m & 24~$\umu$m & 21~cm & $z$ & SMG & 870~$\umu$m & 24~$\umu$m & 21~cm & $z$ \\
    & (mJy)       & ($\umu$Jy) & ($\umu$Jy) & &    & (mJy)       & ($\umu$Jy) & ($\umu$Jy) & \\ \hline
{\bf 001} & $13.8\pm1.1$ & $   38.0\pm   7.9$ & $-$                &  $-$                  & {\bf 056} & $ 4.5\pm1.4$ & $  270.3\pm  11.4$ & $   31.8\pm   6.5$ & $2.82^{+1.69}_{-0.92}$ \\
{\bf 002} & $11.5\pm1.2$ & $   61.7\pm  18.3$ & $  234.6\pm   7.8$ &  $1.62^{+0.68}_{-0.39}$ & {\bf 057} & $ 4.6\pm1.5$ & $  297.2\pm   8.4$ & $   49.4\pm   6.7$ & $2.21^{+1.20}_{-0.64}$ \\
          & $-$          & $  186.5\pm  22.6$ & $-$                &                       & {\bf 059} & $ 4.4\pm1.5$ & $  172.1\pm  10.6$ & $-$                & $-$                  \\
{\bf 003} & $11.3\pm1.2$ & $   33.6\pm   7.4$ & $-$                &  $-$                  & {\bf 060} & $ 4.3\pm1.4$ & $  292.8\pm   9.8$ & $   64.7\pm   6.9$ & $1.86^{+0.89}_{-0.48}$ \\
{\bf 006} & $ 9.1\pm1.2$ & $   35.7\pm   9.6$ & $   42.7\pm   7.4$ &  $3.67^{+2.33}_{-1.32}$ & {\bf 062} & $ 4.4\pm1.5$ & $  243.5\pm  35.6$ & $  151.2\pm   7.3$ & $1.32^{+0.45}_{-0.31}$ \\
{\bf 007} & $ 8.6\pm1.2$ & $  368.3\pm   8.1$ & $   75.8\pm   6.9$ &  $2.47^{+1.41}_{-0.76}$ & {\bf 063} & $ 4.3\pm1.5$ & $-$                & $   31.8\pm   7.2$ & $-$                  \\
{\bf 009} & $ 8.6\pm1.3$ & $  110.6\pm  10.6$ & $   31.0\pm   6.3$ &  $4.38^{+2.84}_{-1.63}$ & {\bf 064} & $ 4.2\pm1.4$ & $  338.4\pm  17.2$ & $   23.6\pm   6.2$ & $3.26^{+2.03}_{-1.13}$ \\
{\bf 010} & $ 8.5\pm1.3$ & $  119.8\pm  18.0$ & $   54.9\pm   6.0$ &  $-$                  & {\bf 066} & $ 4.4\pm1.6$ & $  543.7\pm  12.3$ & $   67.0\pm   7.8$ & $1.85^{+0.88}_{-0.48}$ \\
          &              & $-$                & $   51.1\pm   6.1$ &                       & {\bf 067} & $ 4.2\pm1.5$ & $  516.4\pm   8.0$ & $   90.1\pm  14.8$ & $1.59^{+0.65}_{-0.38}$ \\
          &              & $-$                & $   50.1\pm   6.2$ &                       & {\bf 070} & $ 4.1\pm1.4$ & $  365.2\pm  17.4$ & $  322.3\pm  14.6$ & $0.96^{+0.31}_{-0.31}$ \\
{\bf 011} & $ 8.4\pm1.3$ & $  103.1\pm   8.4$ & $   55.1\pm   6.6$ &  $2.95^{+1.80}_{-0.98}$ & {\bf 072} & $ 4.1\pm1.4$ & $  471.4\pm  22.9$ & $   34.3\pm   7.1$ & $2.55^{+1.47}_{-0.80}$ \\
{\bf 012} & $ 8.1\pm1.3$ & $   43.2\pm   8.8$ & $   39.9\pm   6.5$ &  $3.55^{+2.25}_{-1.26}$ & {\bf 073} & $ 4.1\pm1.4$ & $   15.2\pm   6.9$ & $   18.9\pm   5.1$ & $3.71^{+2.37}_{-1.33}$ \\
{\bf 014} & $ 8.6\pm1.4$ & $   95.5\pm   9.5$ & $   89.4\pm   8.0$ &  $2.25^{+1.23}_{-0.65}$ & {\bf 074} & $ 4.2\pm1.5$ & $  202.0\pm  23.4$ & $   43.8\pm   7.5$ & $-$                  \\
{\bf 015} & $ 8.1\pm1.4$ & $  108.6\pm  10.1$ & $-$                &  $-$                  &           &              & $-$                & $   34.8\pm   7.0$ &                      \\
{\bf 016} & $ 7.4\pm1.3$ & $  457.3\pm  15.9$ & $   49.0\pm   8.5$ &  $2.94^{+1.78}_{-0.98}$ & {\bf 075} & $ 4.2\pm1.5$ & $ 1018.1\pm  38.3$ & $   72.3\pm   8.2$ & $1.75^{+0.79}_{-0.44}$ \\
{\bf 017} & $ 6.9\pm1.3$ & $  219.3\pm   7.7$ & $  120.3\pm  14.5$ &  $1.74^{+0.78}_{-0.43}$ & {\bf 076} & $ 4.2\pm1.5$ & $-$                & $   41.6\pm   8.4$ & $2.31^{+1.28}_{-0.68}$ \\
{\bf 018} & $ 6.7\pm1.3$ & $  560.5\pm   8.2$ & $  130.1\pm  17.3$ &  $1.66^{+0.71}_{-0.40}$ & {\bf 078} & $ 4.2\pm1.7$ & $  369.3\pm  43.0$ & $   75.2\pm   9.8$ & $1.72^{+0.76}_{-0.43}$ \\
{\bf 019} & $ 6.5\pm1.3$ & $   79.8\pm   7.5$ & $   30.8\pm   6.0$ &  $3.64^{+2.32}_{-1.30}$ & {\bf 079} & $ 3.8\pm1.4$ & $  520.7\pm   9.2$ & $   34.8\pm   6.3$ & $2.42^{+1.36}_{-0.73}$ \\
          &              & $   40.0\pm   8.6$ & $-$                &                       & {\bf 081} & $ 3.8\pm1.5$ & $  523.6\pm  11.0$ & $  217.9\pm  15.3$ & $1.10^{+0.33}_{-0.31}$ \\
{\bf 020} & $ 6.4\pm1.3$ & $  176.6\pm   7.6$ & $ 4251.9\pm  16.0$ &  $0.09^{+0.24}_{-0.09}$ & {\bf 084} & $ 3.7\pm1.4$ & $  133.5\pm   7.3$ & $   33.5\pm   6.1$ & $2.43^{+1.38}_{-0.74}$ \\
{\bf 022} & $ 7.2\pm1.6$ & $  409.8\pm  12.7$ & $  111.3\pm  25.3$ &  $1.84^{+0.87}_{-0.47}$ & {\bf 087} & $ 4.2\pm1.9$ & $  419.4\pm  13.2$ & $  128.3\pm  25.8$ & $-$                  \\
{\bf 024} & $ 6.6\pm1.5$ & $  130.2\pm   9.3$ & $   60.0\pm   8.1$ &  $2.43^{+1.37}_{-0.74}$ &           &              &                    & $   56.9\pm   8.7$ &                      \\
{\bf 025} & $ 5.9\pm1.3$ & $  233.2\pm   8.0$ & $   61.3\pm   7.3$ &  $2.25^{+1.23}_{-0.65}$ & {\bf 088} & $ 3.6\pm1.4$ & $  269.5\pm  19.4$ & $   78.0\pm   6.6$ & $1.58^{+0.65}_{-0.37}$ \\
{\bf 027} & $ 6.3\pm1.5$ & $  171.9\pm  16.4$ & $-$                &  $-$                  & {\bf 094} & $ 3.5\pm1.4$ & $  106.2\pm   7.3$ & $-$                & $-$                  \\
          &              & $  277.3\pm  18.8$ & $-$                &                       & {\bf 096} & $ 3.4\pm1.4$ & $  961.6\pm   9.0$ & $   80.6\pm  17.4$ & $1.53^{+0.59}_{-0.36}$ \\
{\bf 029} & $ 6.2\pm1.6$ & $  136.0\pm   9.2$ & $   44.7\pm   8.6$ &  $2.79^{+1.66}_{-0.91}$ &           &              & $   33.1\pm   7.6$ & $-$                &                      \\
{\bf 031} & $ 5.5\pm1.4$ & $   66.2\pm   9.0$ & $   25.9\pm   5.8$ &  $3.66^{+2.33}_{-1.31}$ & {\bf 098} & $ 3.3\pm1.4$ & $  268.8\pm  15.7$ & $  141.8\pm   8.1$ & $1.22^{+0.38}_{-0.30}$ \\
{\bf 034} & $ 5.5\pm1.4$ & $  223.3\pm   9.4$ & $-$                &  $-$                  & {\bf 101} & $ 3.3\pm1.4$ & $   43.9\pm   6.6$ & $   25.3\pm   6.1$ & $2.68^{+1.59}_{-0.85}$ \\
{\bf 035} & $ 7.2\pm2.0$ & $   73.3\pm   9.6$ & $-$                &  $-$                  & {\bf 102} & $ 3.4\pm1.5$ & $  318.4\pm  16.5$ & $-$                & $-$                  \\
{\bf 036} & $ 5.5\pm1.5$ & $  274.5\pm   9.6$ & $   47.5\pm   7.5$ &  $2.50^{+1.44}_{-0.77}$ & {\bf 103} & $ 3.4\pm1.5$ & $  113.3\pm  13.5$ & $-$                & $-$                  \\
{\bf 037} & $ 5.8\pm1.6$ & $  201.8\pm  19.7$ & $-$                &  $-$                  & {\bf 105} & $ 3.6\pm1.7$ & $-$                & $-$                & $-$                  \\
{\bf 039} & $ 5.3\pm1.5$ & $  131.0\pm   7.5$ & $   45.9\pm   7.5$ &  $2.50^{+1.43}_{-0.77}$ & {\bf 106} & $ 3.2\pm1.3$ & $  363.6\pm  10.2$ & $   66.8\pm   7.1$ & $1.61^{+0.67}_{-0.39}$ \\
{\bf 040} & $ 5.0\pm1.4$ & $  119.8\pm   7.2$ & $  119.1\pm  13.7$ &  $1.52^{+0.60}_{-0.35}$ & {\bf 108} & $ 3.1\pm1.3$ & $ 3722.6\pm  73.7$ & $  379.7\pm  36.4$ & $0.78^{+0.30}_{-0.33}$ \\
{\bf 041} & $ 6.7\pm2.0$ & $  241.8\pm  20.8$ & $-$                &  $-$                  & {\bf 109} & $ 3.1\pm1.4$ & $  169.0\pm  15.5$ & $   24.6\pm   6.6$ & $2.63^{+1.54}_{-0.83}$ \\
{\bf 043} & $ 5.0\pm1.5$ & $  229.6\pm   7.2$ & $-$                &  $-$                  & {\bf 110} & $ 3.2\pm1.5$ & $-$                & $-$                & $-$                  \\
{\bf 044} & $ 5.8\pm1.8$ & $  400.9\pm  13.0$ & $   90.3\pm   9.6$ &  $1.83^{+0.87}_{-0.47}$ & {\bf 111} & $ 3.2\pm1.4$ & $  353.7\pm  53.3$ & $   54.0\pm   9.4$ & $1.77^{+0.80}_{-0.45}$ \\
{\bf 045} & $ 4.9\pm1.4$ & $  116.1\pm   8.1$ & $   31.1\pm   5.9$ &  $3.01^{+1.85}_{-1.01}$ & {\bf 112} & $ 3.6\pm1.9$ & $  190.8\pm  13.2$ & $   29.6\pm   8.0$ & $2.57^{+1.50}_{-0.80}$ \\
{\bf 046} & $ 6.3\pm1.9$ & $-$                & $   73.2\pm  10.7$ &  $2.12^{+1.12}_{-0.59}$ & {\bf 114} & $ 3.0\pm1.3$ & $  515.0\pm   7.8$ & $   95.4\pm   6.7$ & $1.36^{+0.47}_{-0.32}$ \\
{\bf 047} & $ 5.4\pm1.6$ & $-$                & $-$                &  $-$                  & {\bf 115} & $ 3.7\pm2.2$ & $-$                & $-$                & $-$                  \\
{\bf 048} & $ 5.9\pm1.8$ & $  406.8\pm  11.9$ & $   84.5\pm   8.3$ &  $1.91^{+0.93}_{-0.50}$ & {\bf 117} & $ 3.0\pm1.4$ & $  203.1\pm  13.1$ & $   81.0\pm   8.4$ & $1.45^{+0.54}_{-0.34}$ \\
{\bf 049} & $ 5.0\pm1.5$ & $  122.9\pm  11.6$ & $   36.0\pm   7.2$ &  $-$                  & {\bf 118} & $ 3.0\pm1.4$ & $-$                & $   23.5\pm   6.3$ & $-$                  \\
          &              & $  113.5\pm  14.6$ & $  115.9\pm  19.1$ &                       & {\bf 120} & $ 2.9\pm1.4$ & $  345.6\pm   9.5$ & $   45.5\pm   8.4$ & $1.83^{+0.85}_{-0.47}$ \\
{\bf 050} & $ 4.7\pm1.4$ & $  306.6\pm  24.1$ & $   77.3\pm   7.2$ &  $1.79^{+0.82}_{-0.46}$ & {\bf 122} & $ 2.8\pm1.3$ & $ 1392.5\pm  15.8$ & $  207.3\pm  14.5$ & $0.99^{+0.31}_{-0.31}$ \\
          &              & $   64.4\pm  12.3$ & $-$                &                       & {\bf 126} & $ 2.8\pm1.3$ & $  263.4\pm   8.7$ & $   23.3\pm   5.5$ & $2.56^{+1.48}_{-0.80}$ \\
{\bf 054} & $ 5.0\pm1.6$ & $  222.1\pm   8.8$ & $-$                &  $-$                  &           &              &                    &                    &                      \\
\hline
\end{tabular}
\label{tab:summary}
\end{table*}

\section{Conclusions}
\label{conclusions}

Using a probabilistic approach, we have attempted to identify reliable
counterparts to the 126 SMGs recently discovered at a wavelength of
870\,$\umu$m in the LESS survey of the ECDFS using the LABOCA camera
on the APEX telescope \citep{weiss09}. Taking values of the corrected
Poissonian probability (the so-called $p$-statistic, $p$) that are
less than or equal to 0.05 to indicate a secure identification,
i.e.\ a highly unlikely chance coincidence, we have found reliable
radio and/or 24-$\umu$m counterparts to 62 SMGs. A further 17 SMGs were
identified using IRAC sources that fell within a
colour-flux cut that was constructed from the results of the radio and
MIPS analysis. In contrast to most previous work of a similar nature,
we have based our identifications on rigorously constructed catalogues
of 1.4-GHz and MIPS/IRAC sources.

In total we find that 79 out of the 126 SMGs have secure counterparts,
an identification fraction of 63~per~cent. This is not as high as some
other studies, partly due to the relatively shallow radio map and
somewhat larger submm beam. In several cases it is obvious that
multiple submm emitters are blended and consequently difficult to
identify.

Finally, in creating our radio catalogue we have performed simulations
in order to correct the flux densities for `flux boosting'. This has
particular relevance to the calculation of source redshifts based on
the radio-submm spectral index, a technique which often uses
de-boosted submm fluxes, but ignores the corresponding effect in the
radio. With the systematic shift towards lower redshifts removed, the
median redshift of the radio-detected SMGs in our sample is
$\overline{z}=2.2^{+0.8}_{-0.7}$ (1-$\sigma$ errors). This is
identical to that found by both \citet{chapman05} and
\citet{wardlow10}, the latter using the sample of SMGs identified in
this paper, but using a different technique (optical to mid-IR
multi-band photometry) for measuring the source redshifts. The median
redshift of the full sample is likely to be rather higher as the
unidentified SMGs by definition have weak radio emission.

The current generation of submm cameras produce maps with such poor
resolution that a probabilistic approach to identifying submm galaxies
is inevitable. Ideally, identification work such as that presented in
this paper would be done with telescopes offering sub-arcsecond
resolution, such as the IRAM Plateau de Bure Interferometer and the
Submillimeter Array (SMA). However, due to their limited sensitivity
(small numbers of antennas and relatively poor atmospheric
transmission), many hours are required for a reliable detection of a
typical submm galaxy. In the future, the Atacama Large
Millimeter/Submillimeter Array (ALMA) will revolutionise the study of
high-redshift star formation with its order of magnitude increase in
sensitivity and imaging fidelity which will make pinpointing the
origin of the submm emission in surveys such as LESS a relatively
trivial exercise, requiring only minutes per source to achieve a high
dynamic range image.

\section*{Acknowledgments}

The authors would like to acknowledge the comments provided by the
anonymous referee that significantly improved the manuscript. The
National Radio Astronomy Observatory is a facility of the National
Science Foundation operated under cooperative agreement by Associated
Universities, Inc. This work is based in part on observations made
with the {\em Spitzer Space Telescope}, which is operated by the Jet
Propulsion Laboratory, California Institute of Technology under a
contract with NASA. We thank Jacqueline Monkiewicz and the FIDEL team
for the {\em Spitzer} 24-$\umu$m data reductions. JLW acknowledges the
receipt of a STFC studentship. IRS acknowledges support from the
STFC. This research has made use of the NASA/IPAC Extragalactic
Database (NED) which is operated by the Jet Propulsion Laboratory,
California Institute of Technology, under contract with the National
Aeronautics and Space Administration. ADB would like to thank Phillip
Helbig for the use of his cosmology code.

\bibliographystyle{mnras}
\bibliography{deep}

\appendix

\section{Detailed description of SMGs}
\label{appendix}

\begin{figure*}
\centering
\includegraphics[scale=0.295]{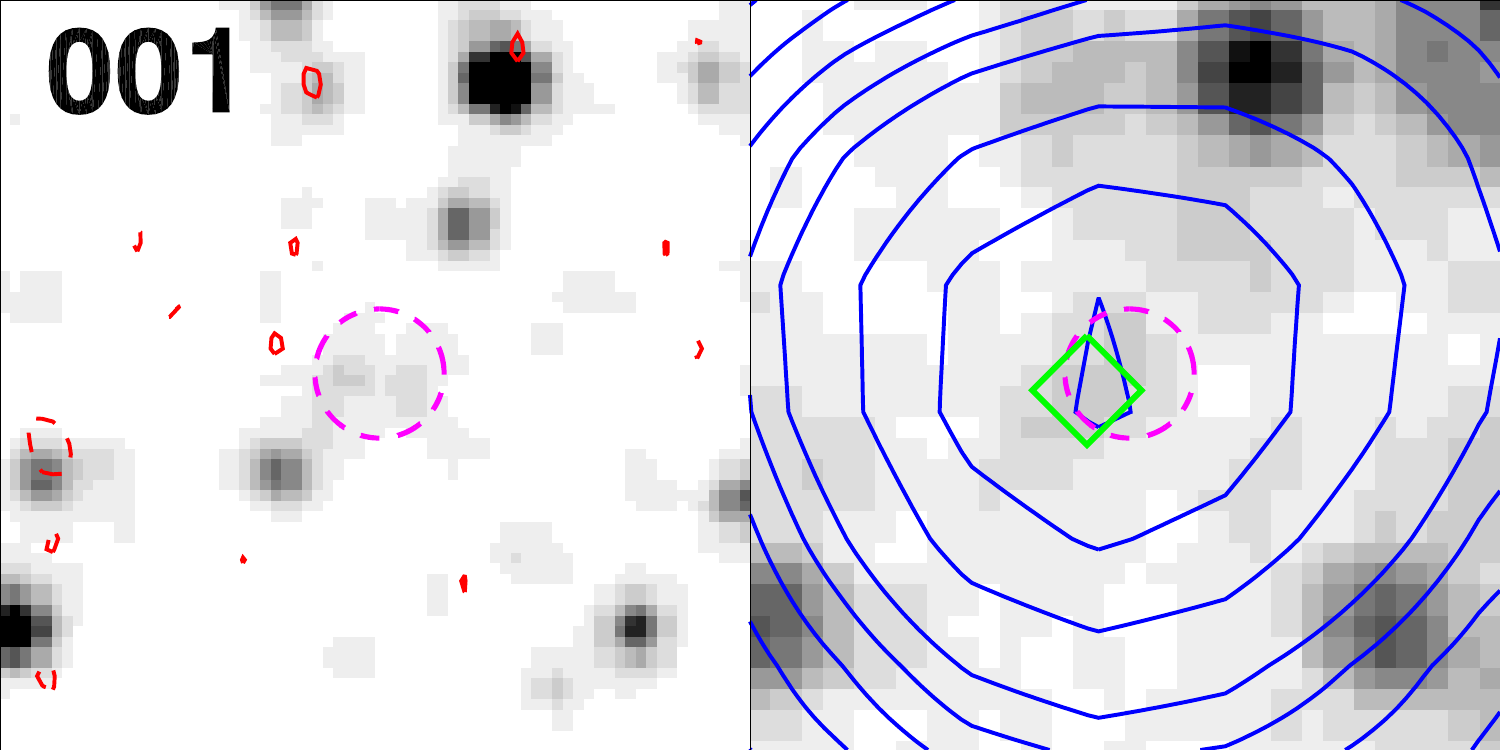}%
\hspace{1cm}%
\includegraphics[scale=0.295]{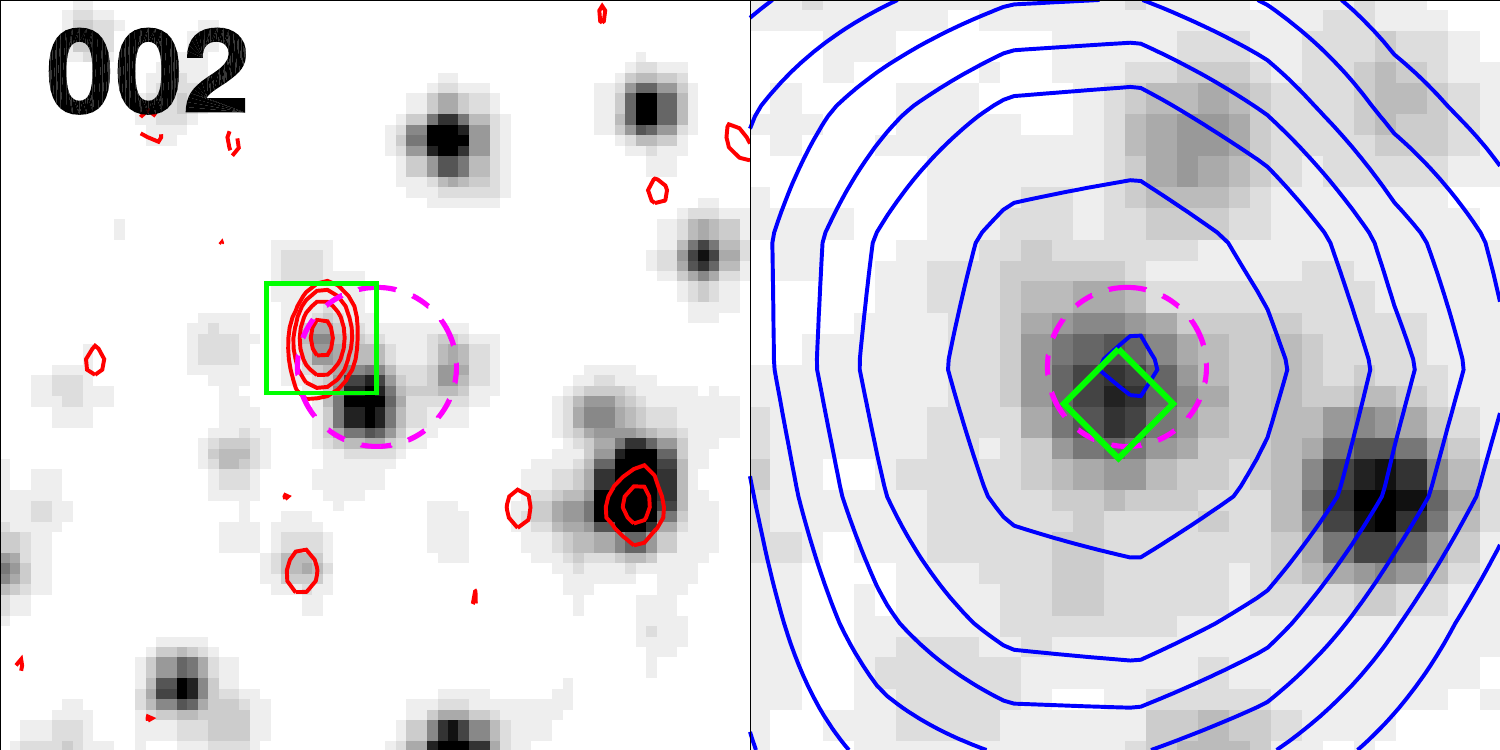}%
\hspace{1cm}%
\includegraphics[scale=0.295]{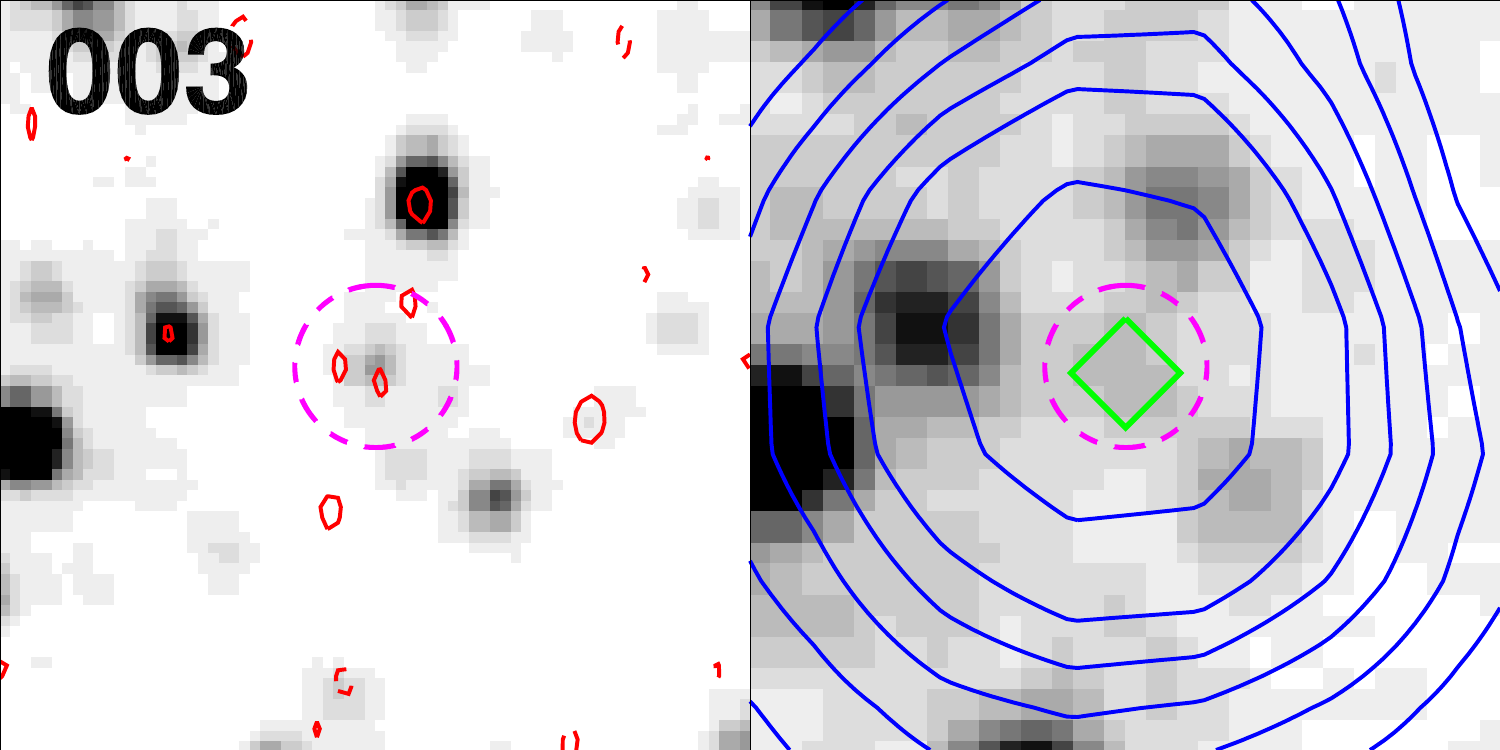}
\includegraphics[scale=0.295]{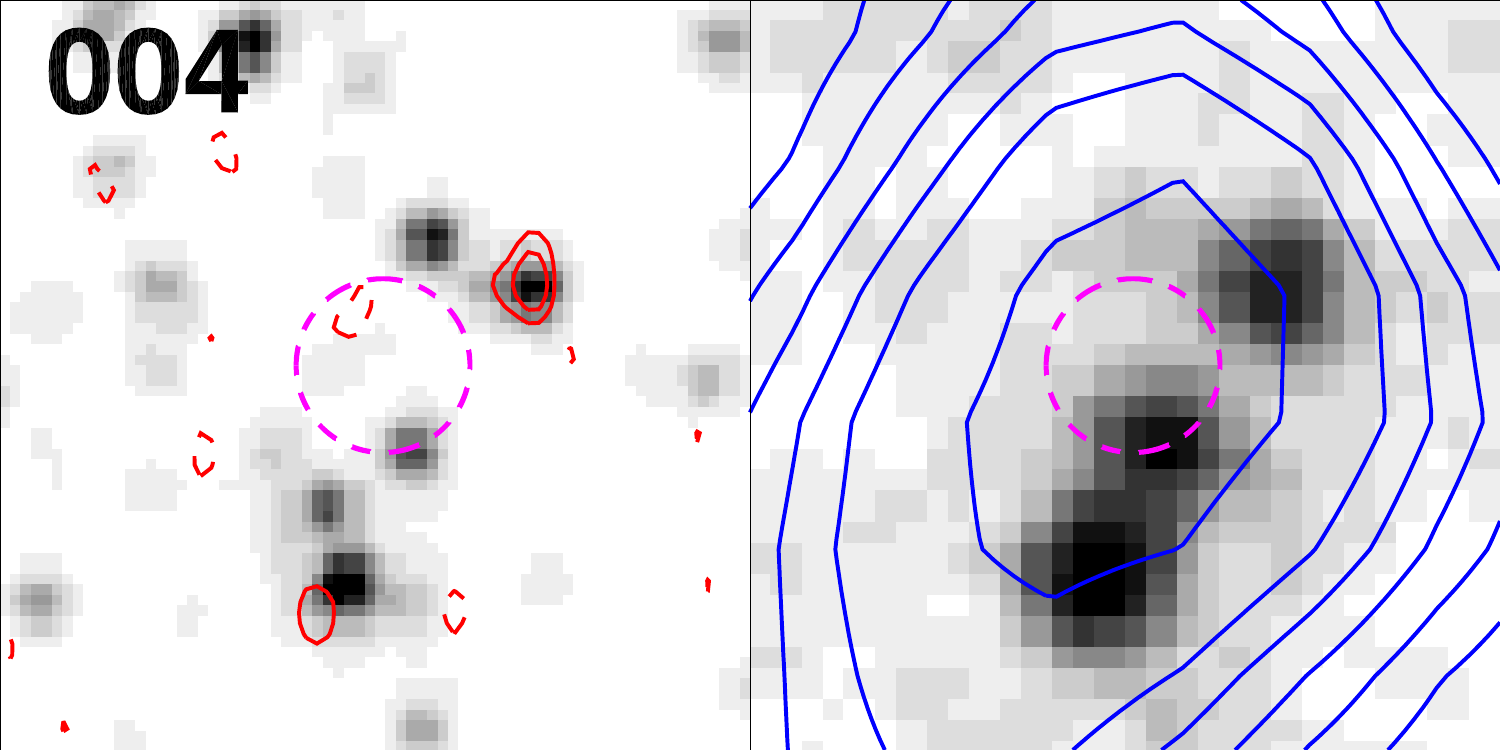}%
\hspace{1cm}%
\includegraphics[scale=0.295]{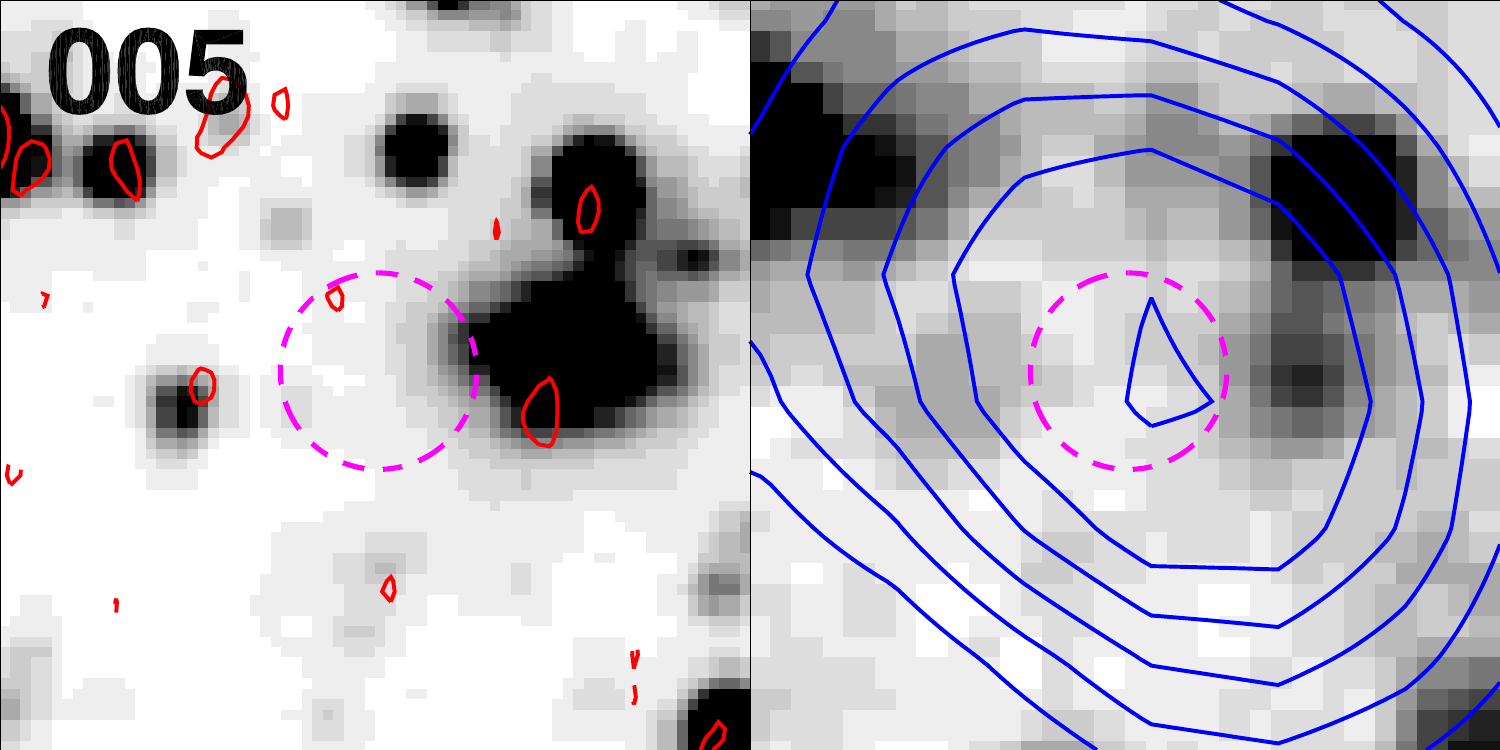}%
\hspace{1cm}%
\includegraphics[scale=0.295]{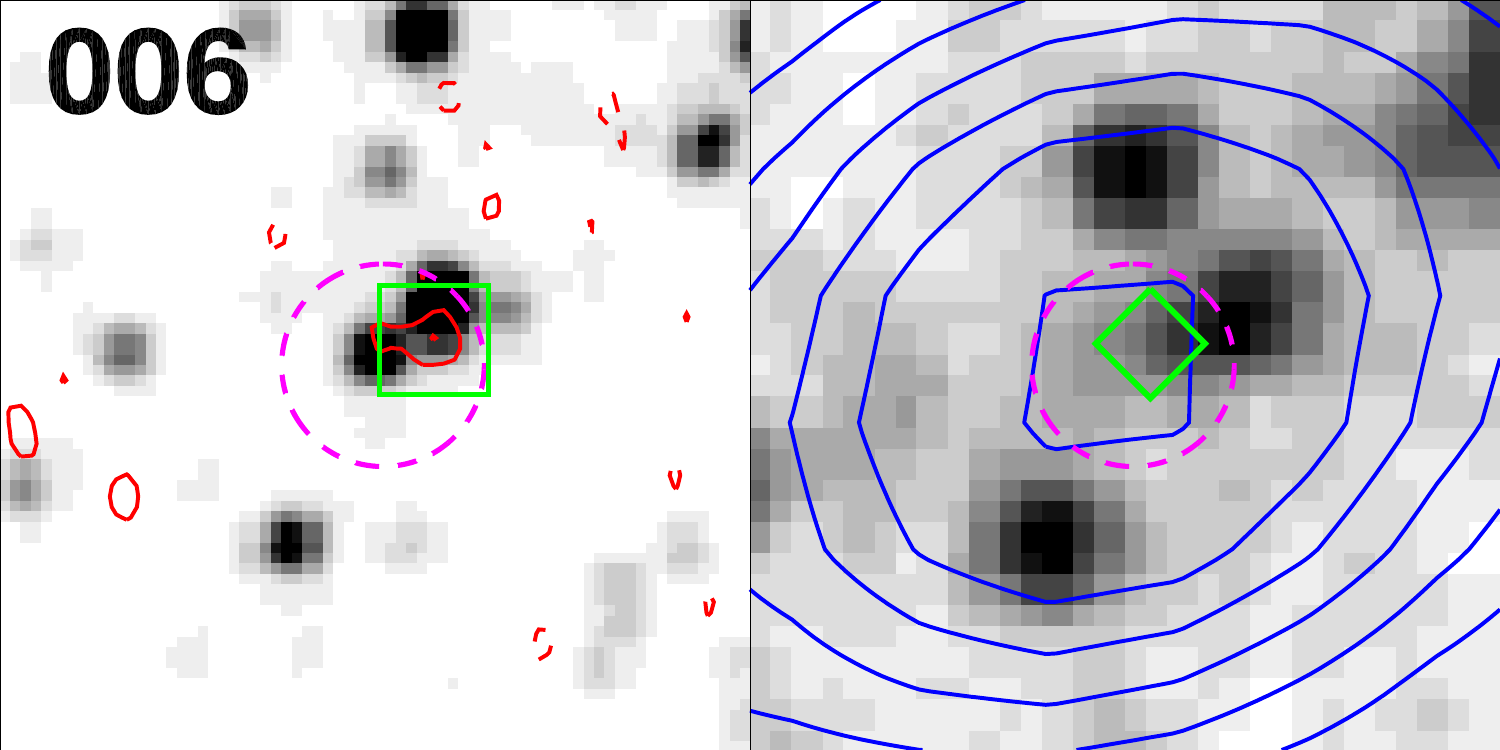}
\includegraphics[scale=0.295]{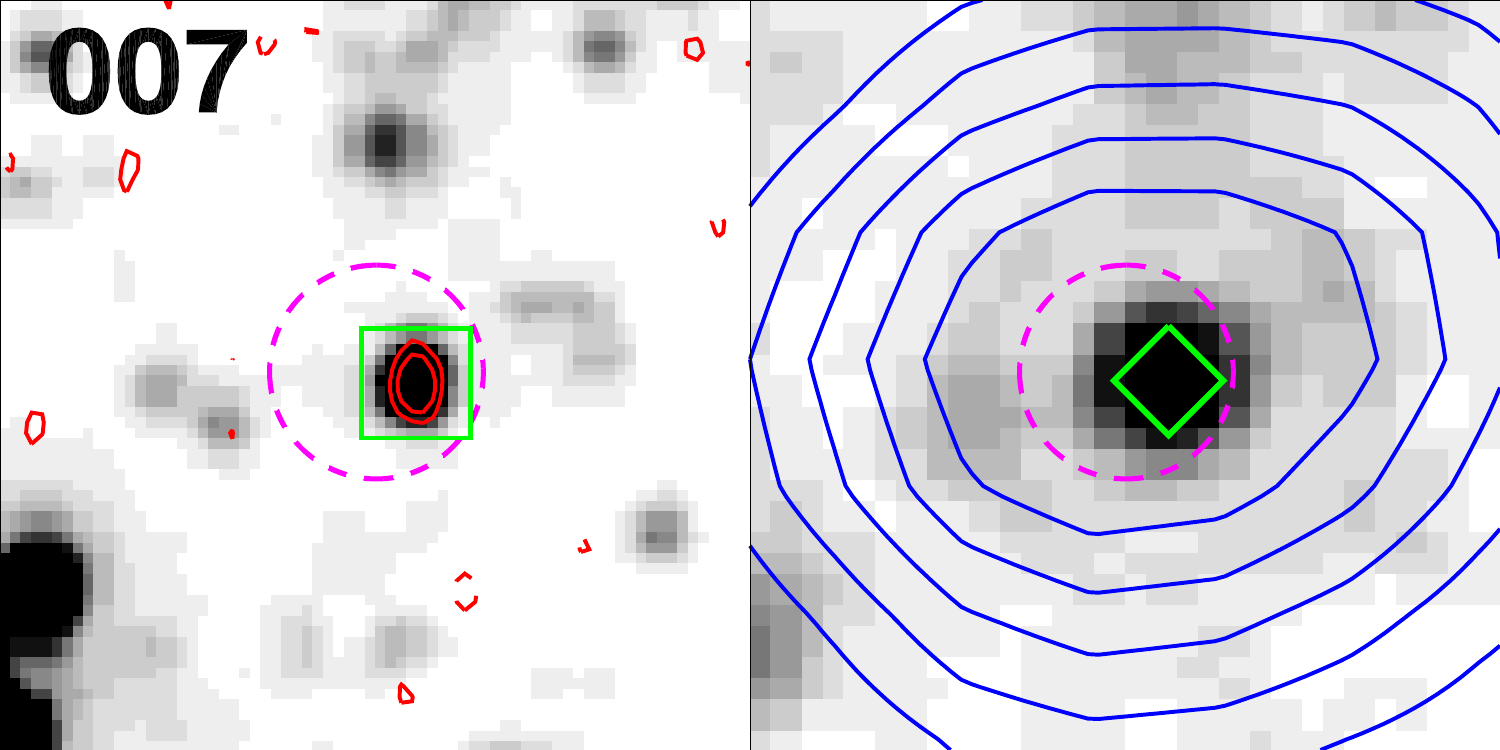}%
\hspace{1cm}%
\includegraphics[scale=0.295]{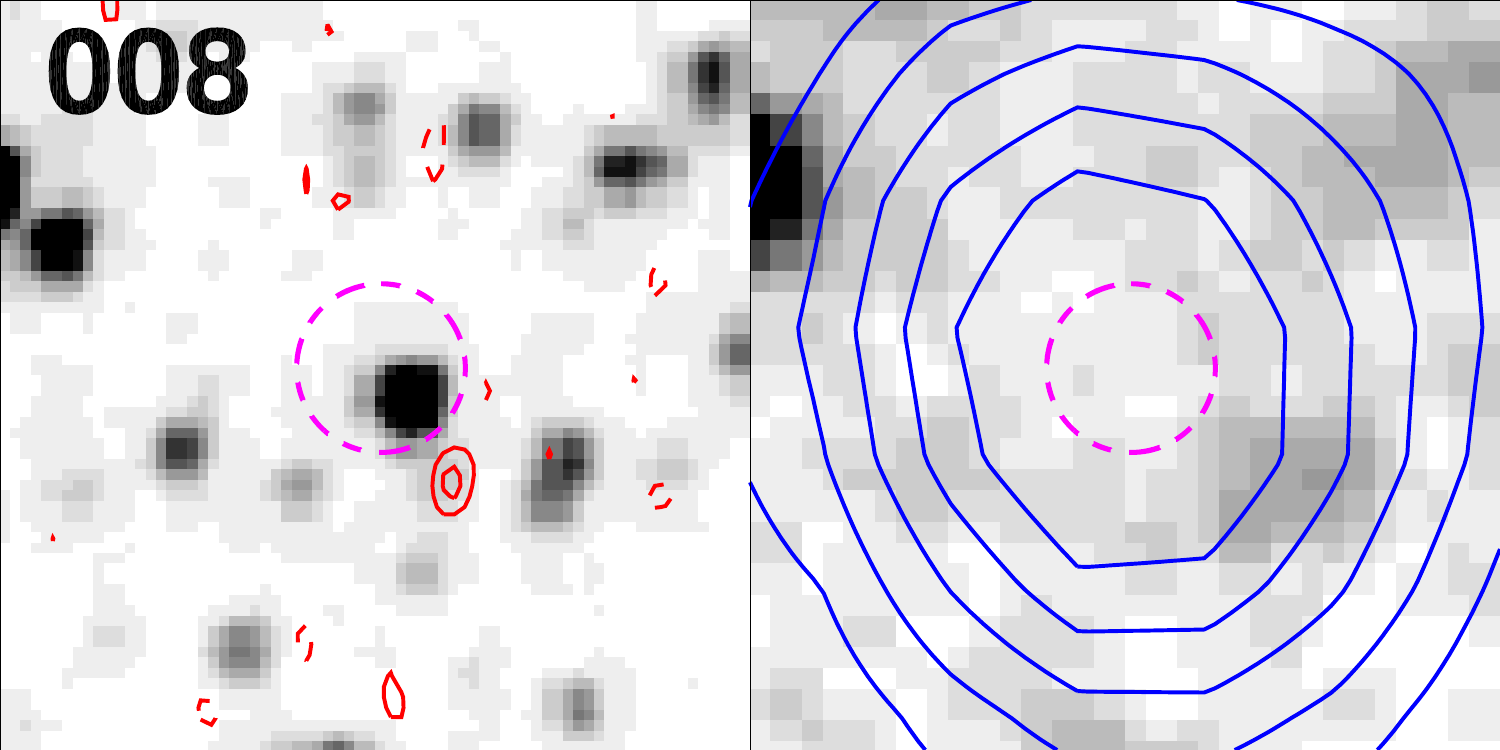}%
\hspace{1cm}%
\includegraphics[scale=0.295]{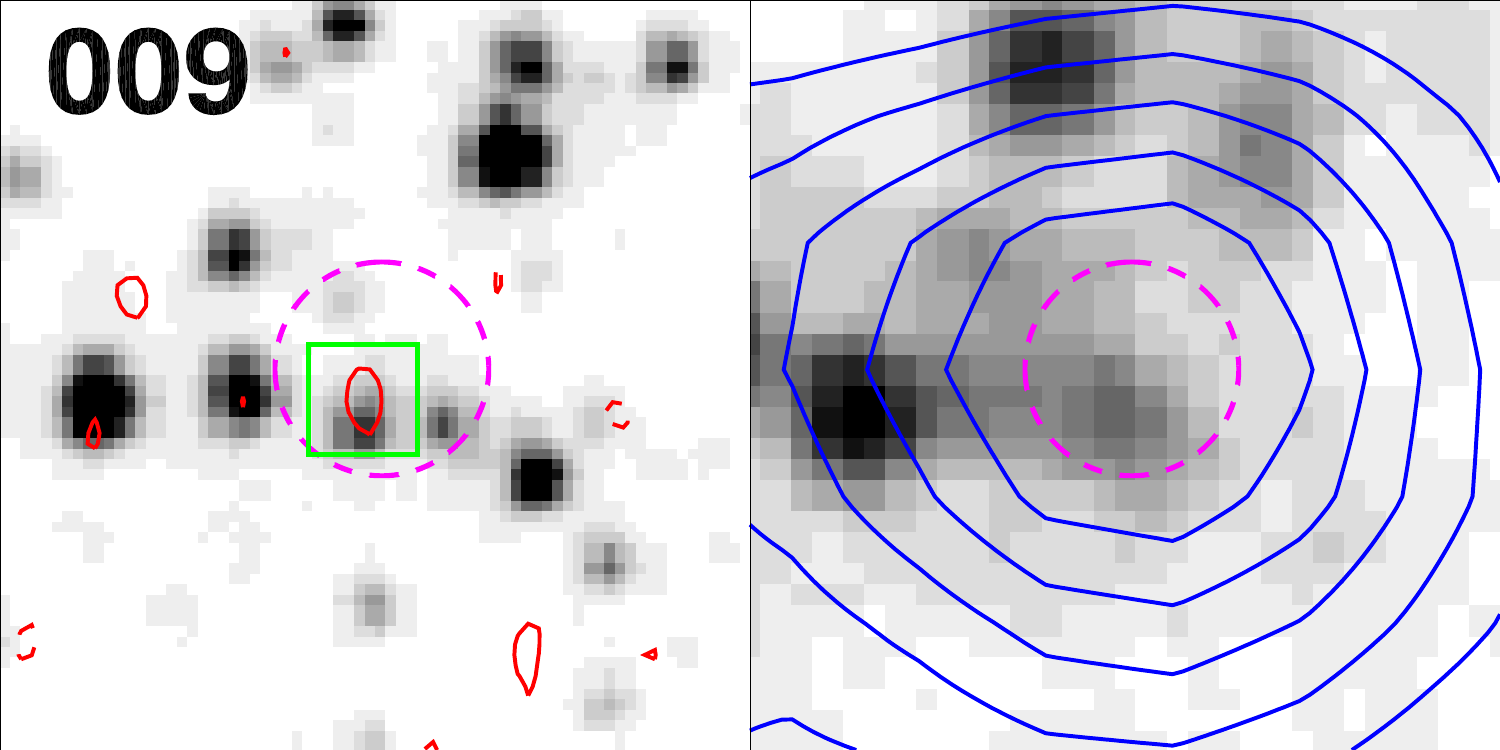}
\includegraphics[scale=0.295]{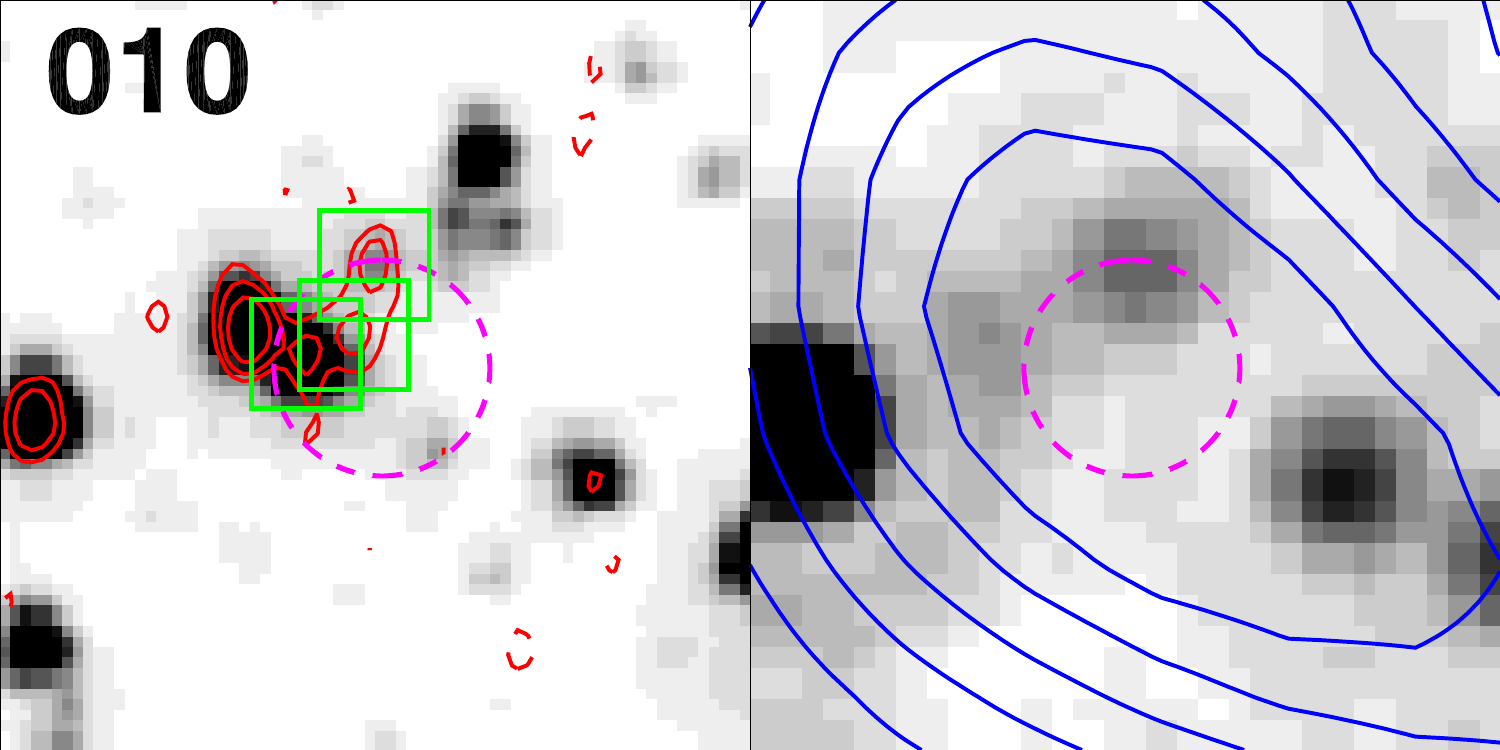}%
\hspace{1cm}%
\includegraphics[scale=0.295]{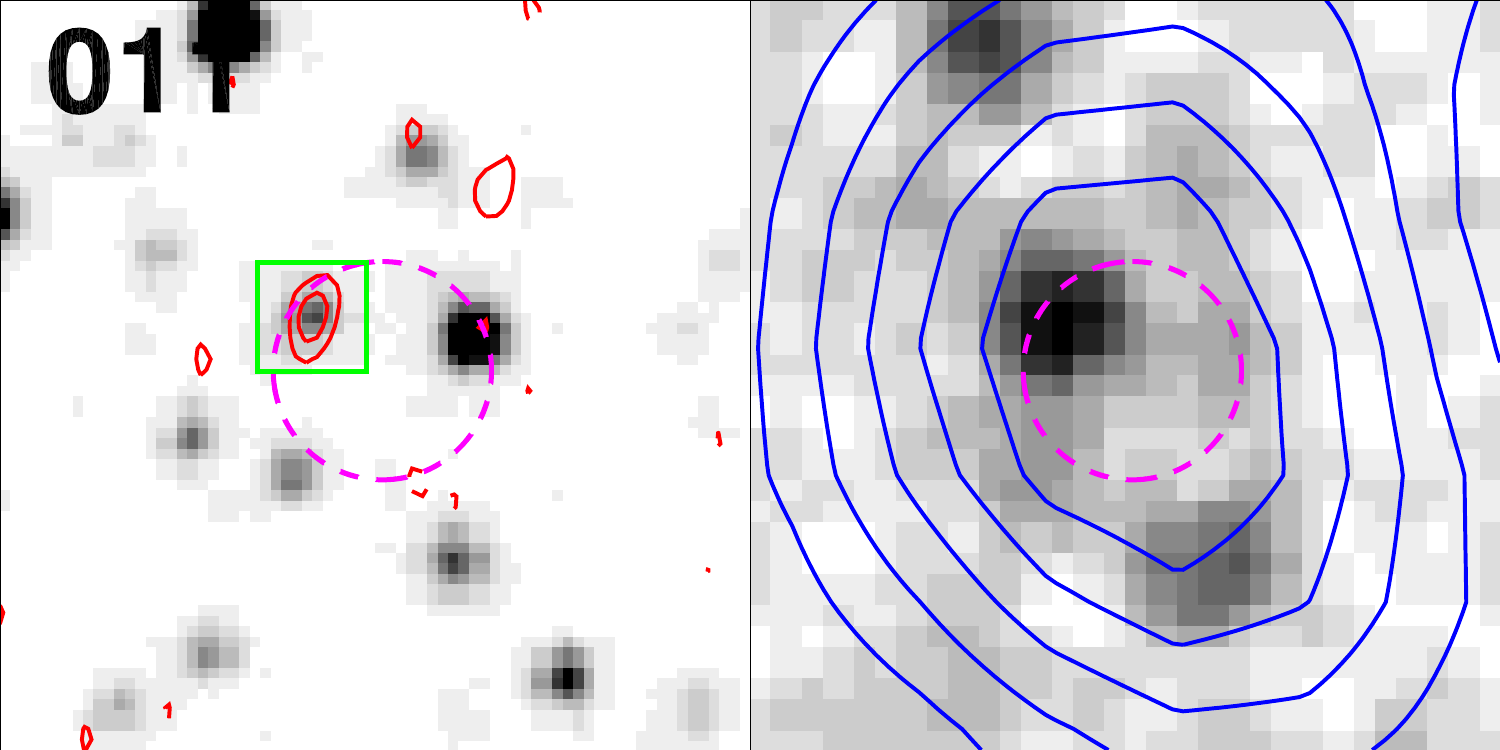}%
\hspace{1cm}%
\includegraphics[scale=0.295]{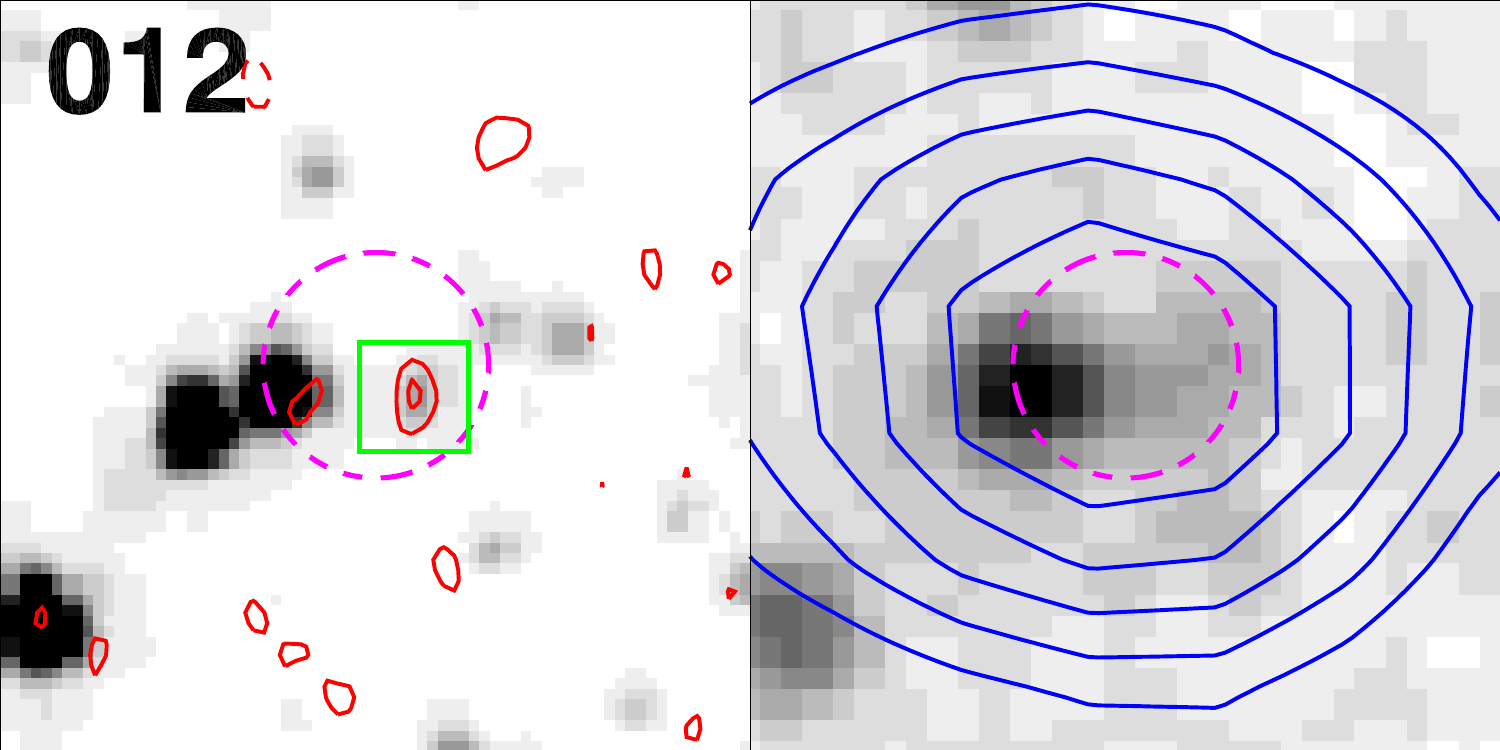}
\includegraphics[scale=0.295]{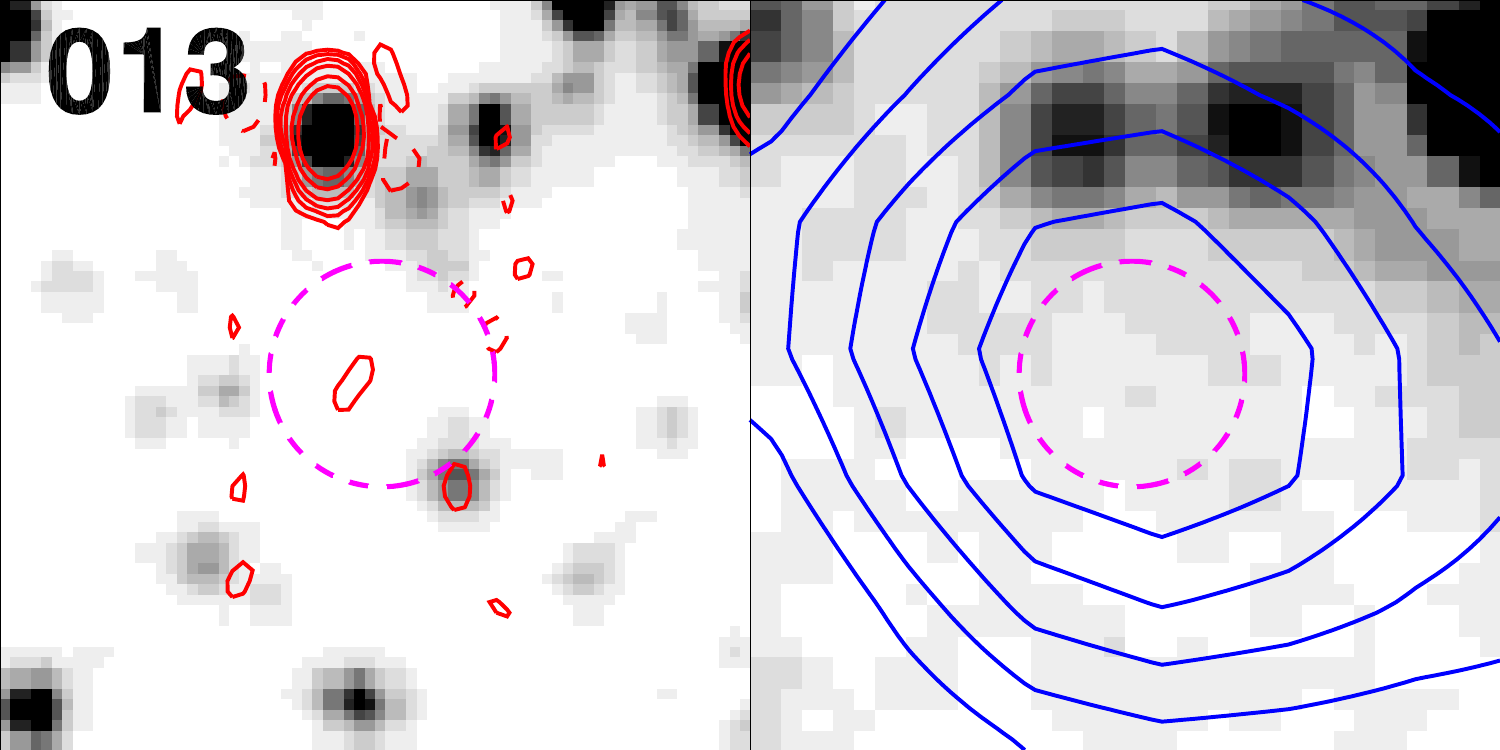}%
\hspace{1cm}%
\includegraphics[scale=0.295]{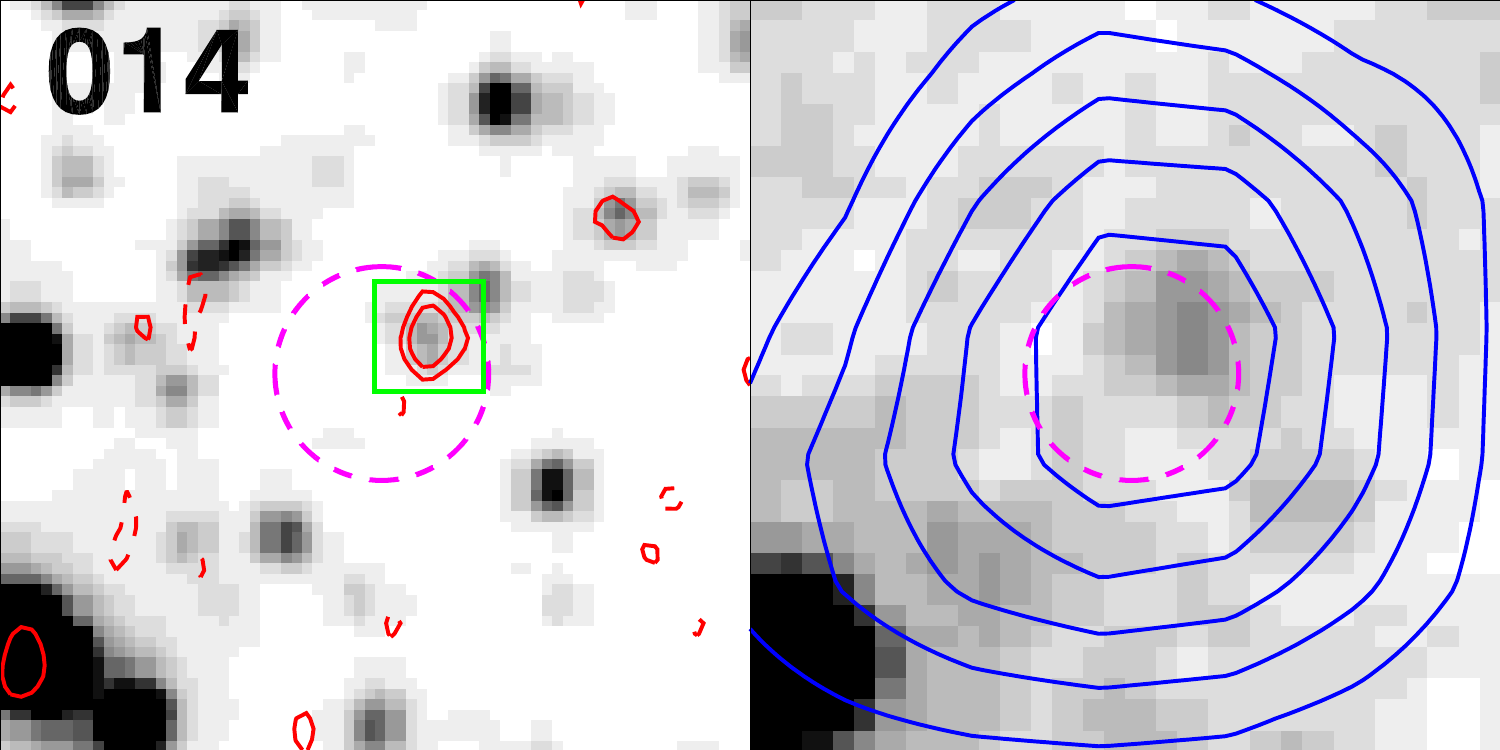}%
\hspace{1cm}%
\includegraphics[scale=0.295]{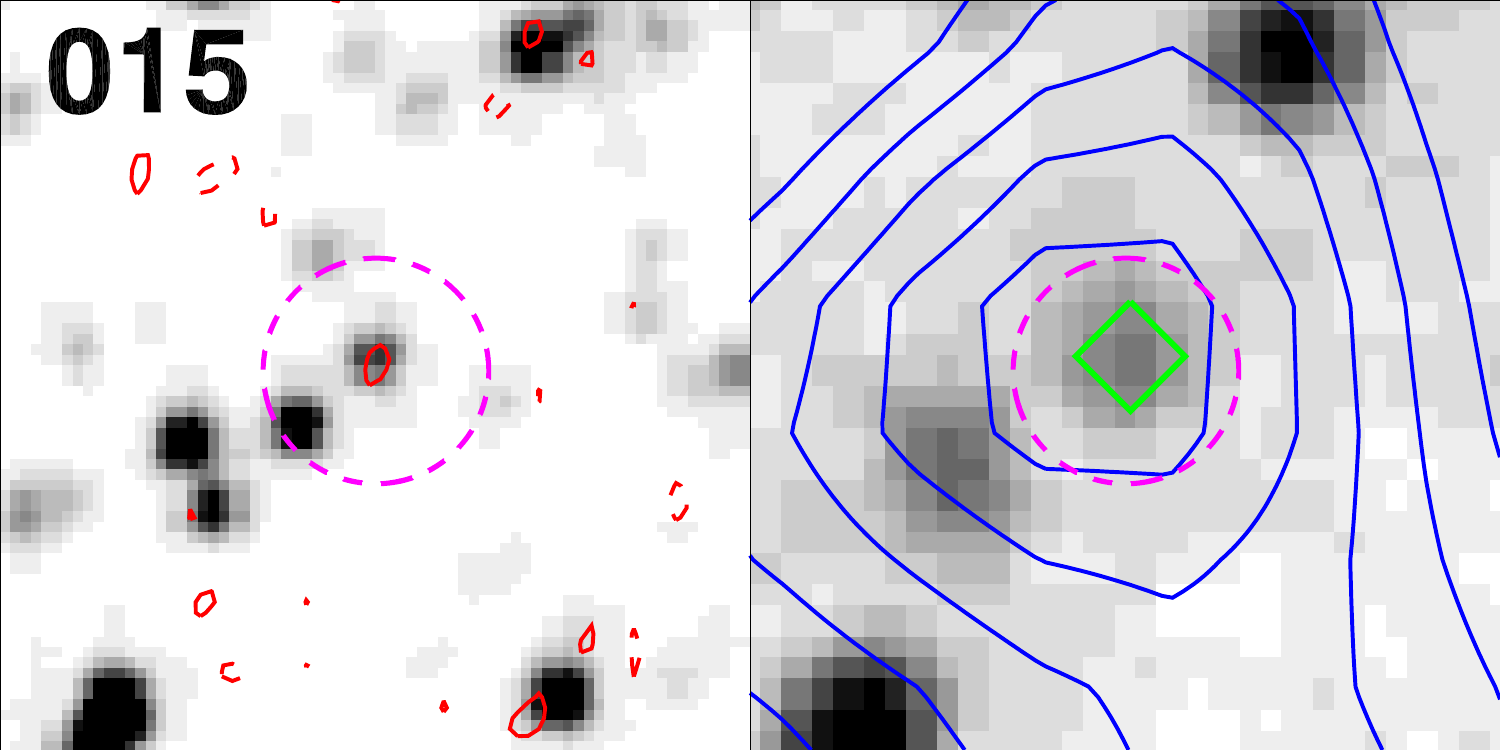}
\includegraphics[scale=0.295]{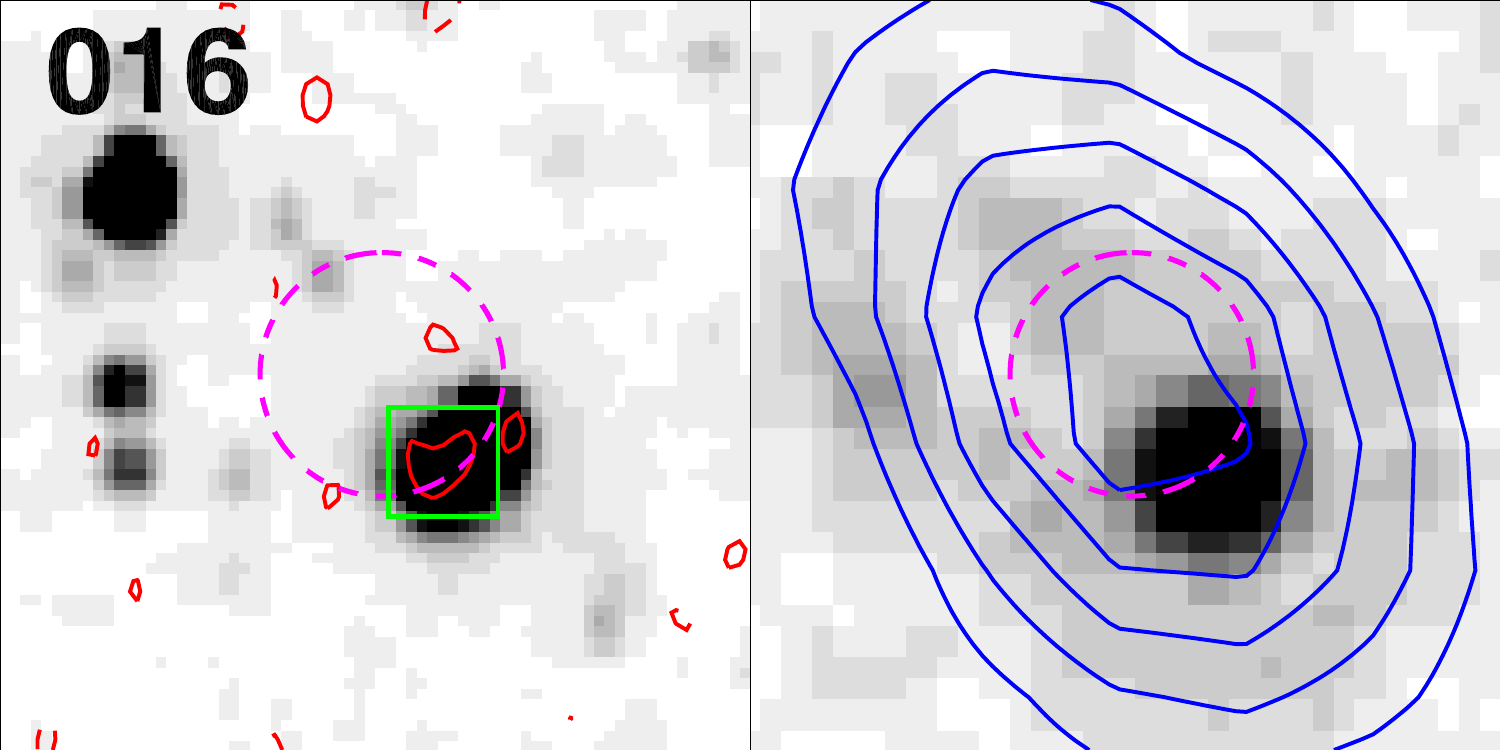}%
\hspace{1cm}%
\includegraphics[scale=0.295]{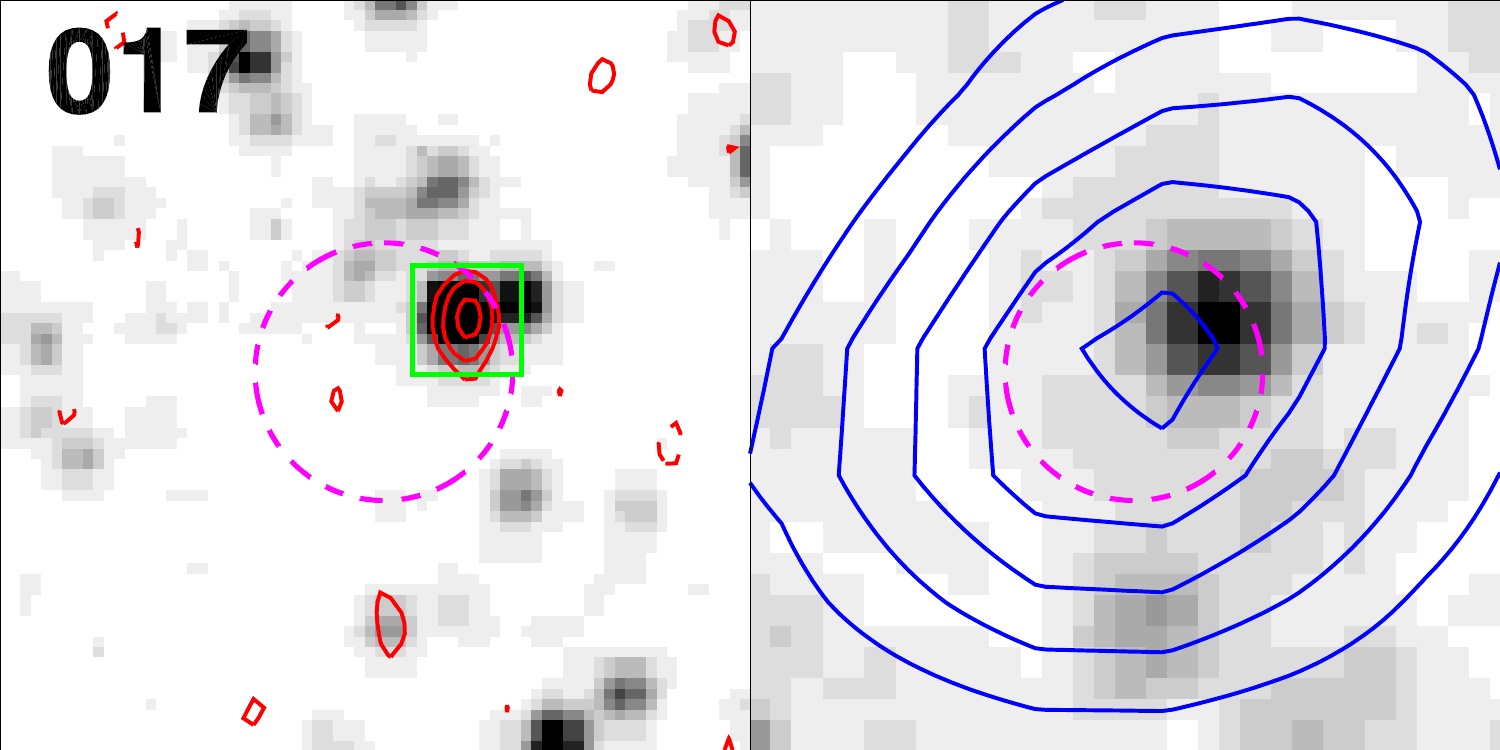}%
\hspace{1cm}%
\includegraphics[scale=0.295]{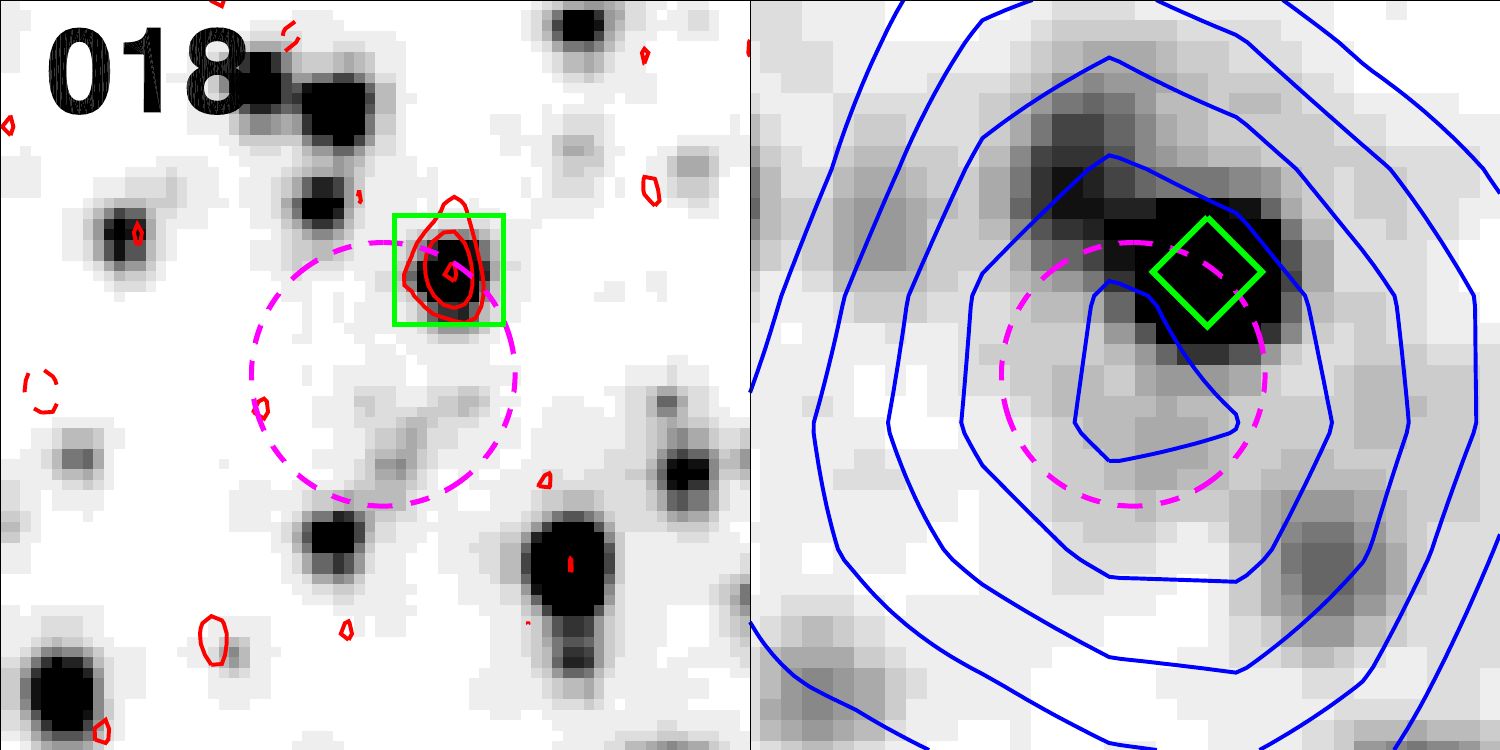}
\includegraphics[scale=0.295]{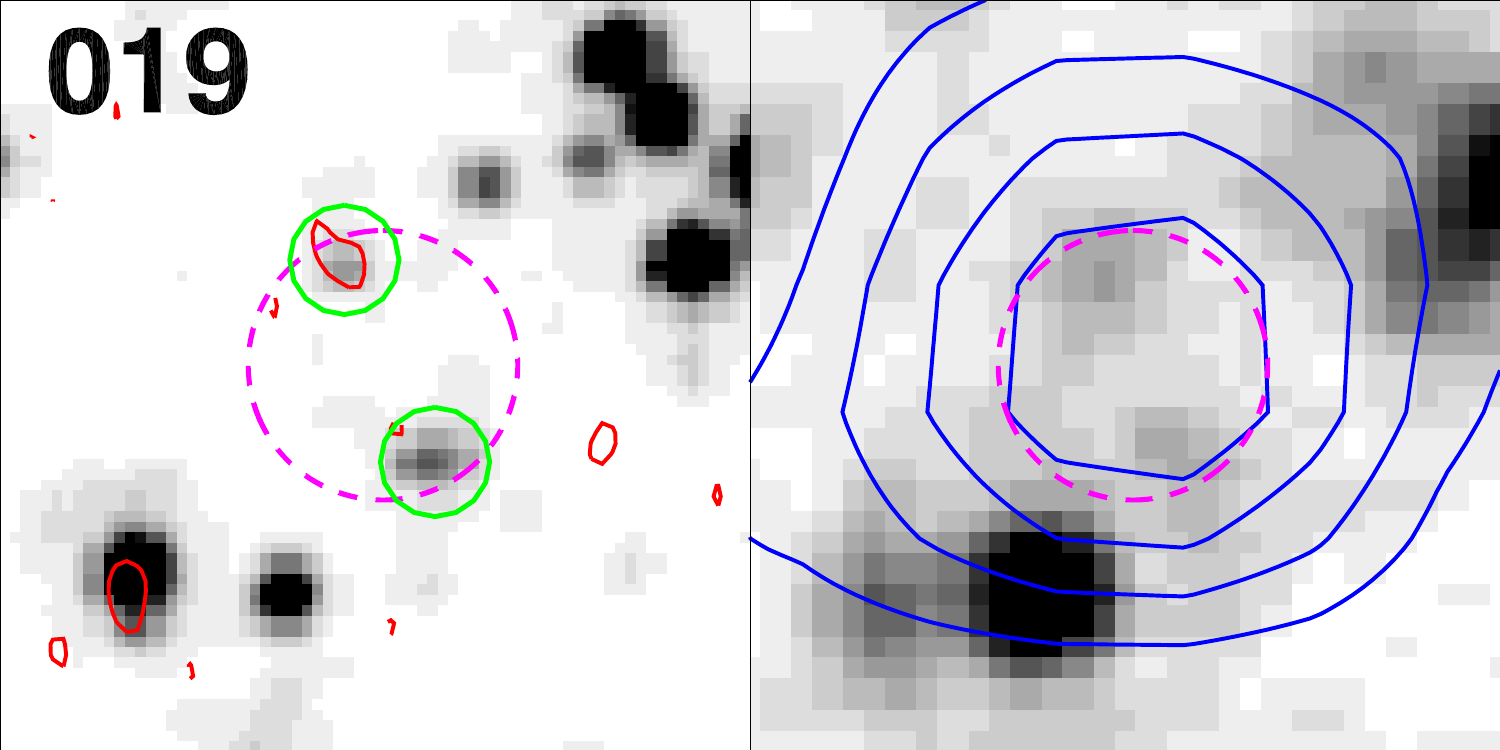}%
\hspace{1cm}%
\includegraphics[scale=0.295]{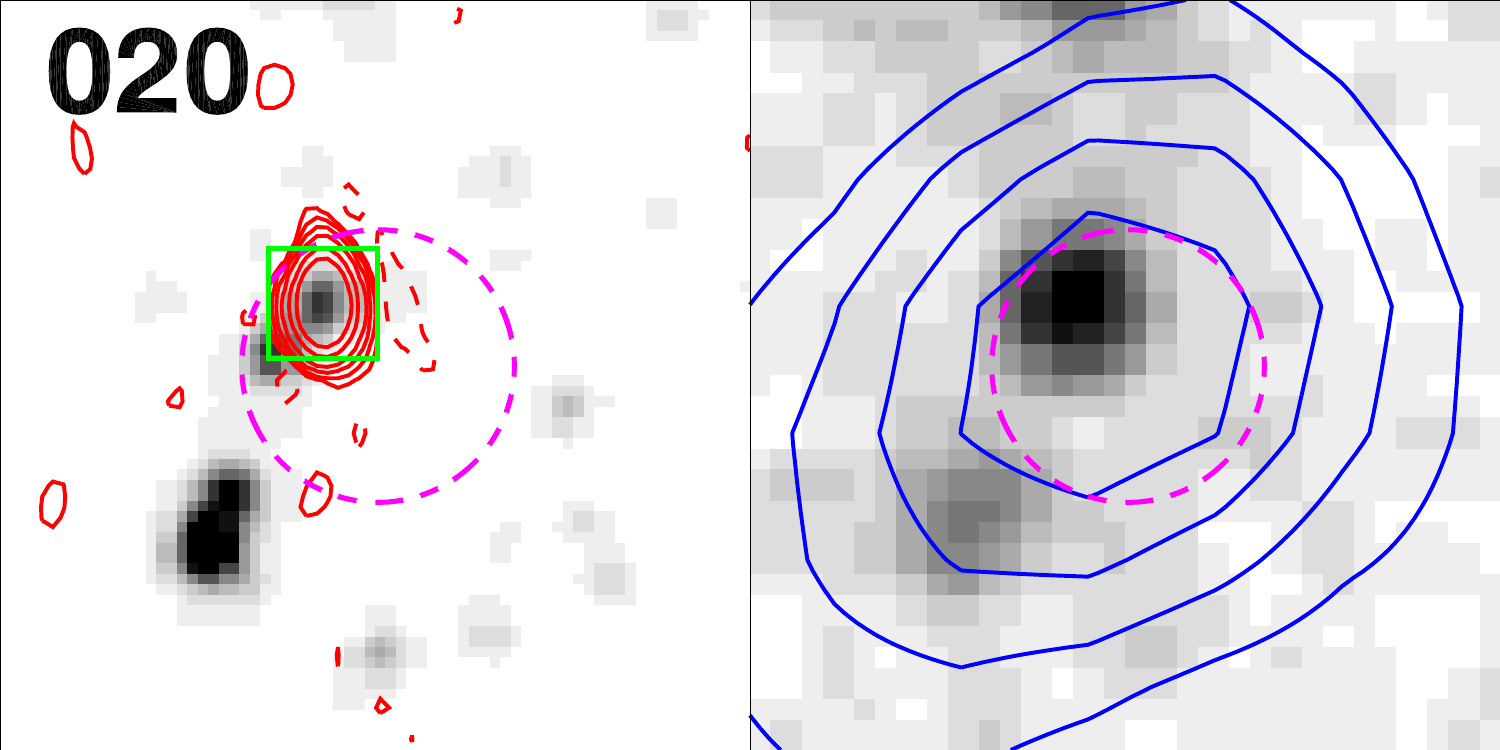}%
\hspace{1cm}%
\includegraphics[scale=0.295]{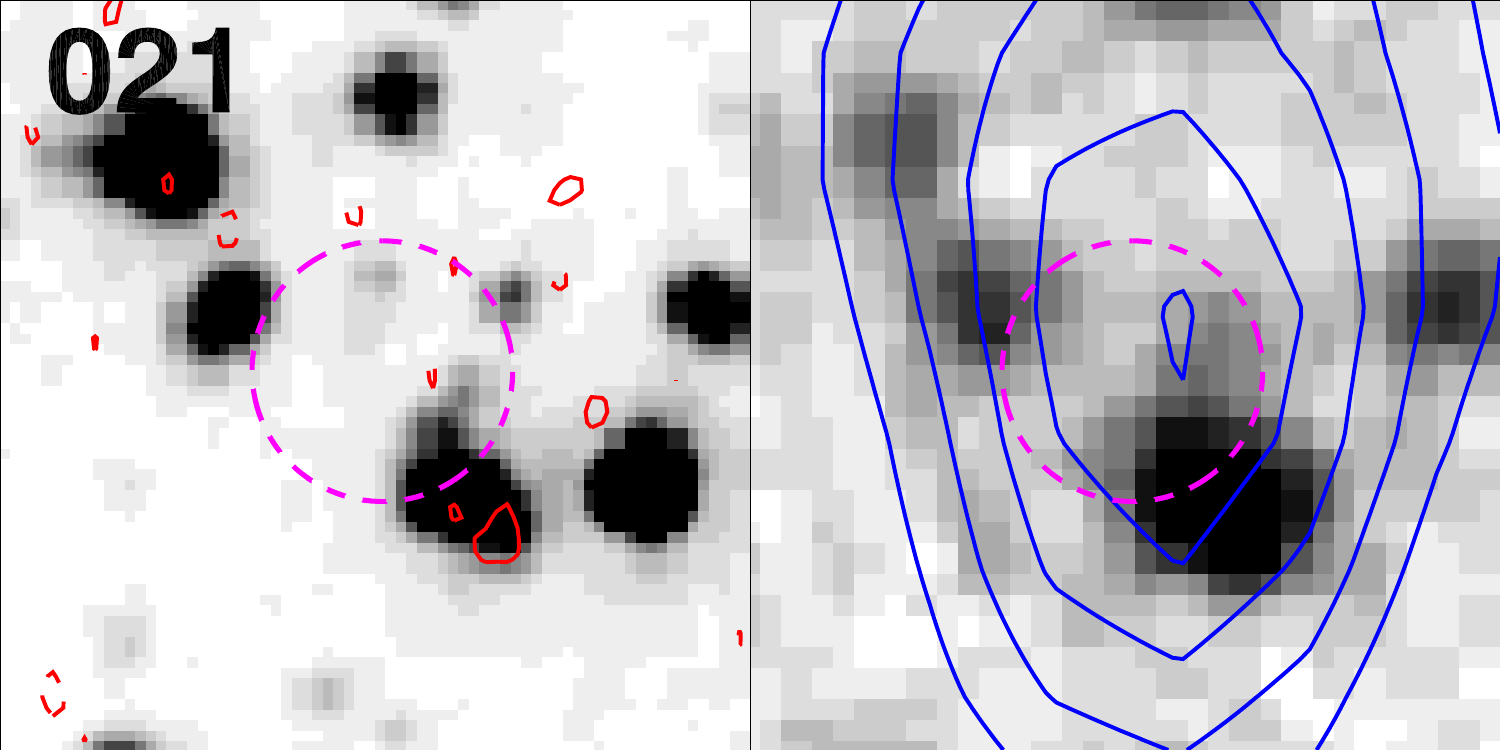}
\includegraphics[scale=0.295]{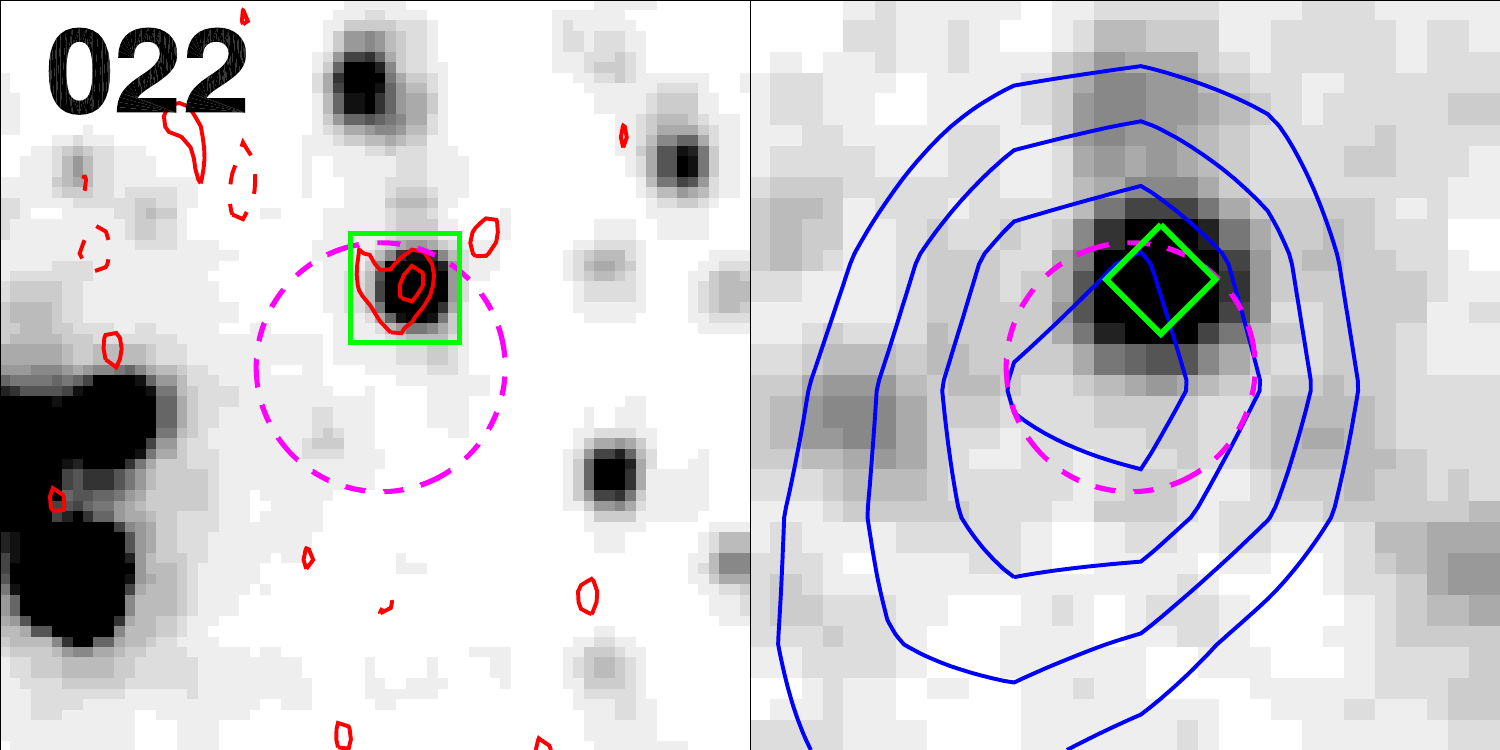}%
\hspace{1cm}%
\includegraphics[scale=0.295]{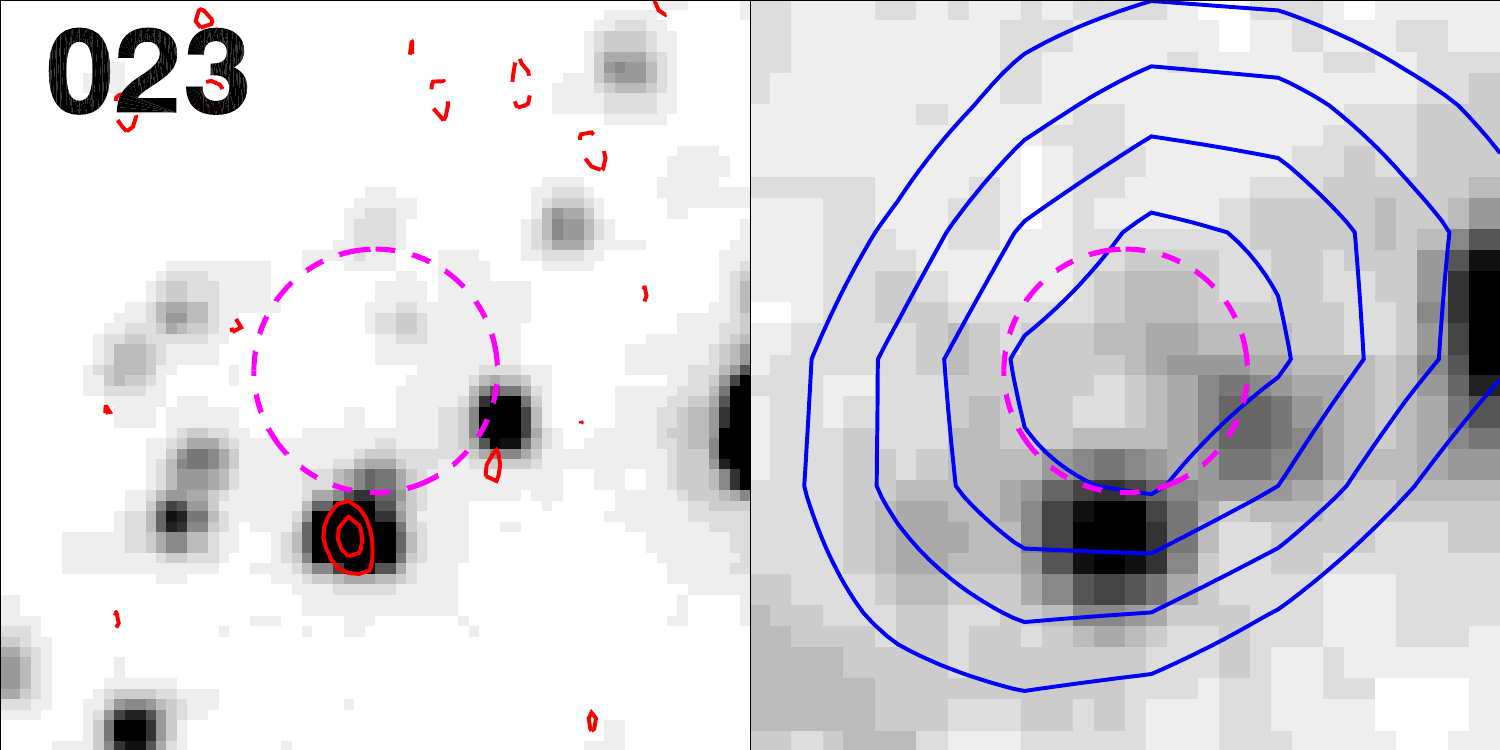}%
\hspace{1cm}%
\includegraphics[scale=0.295]{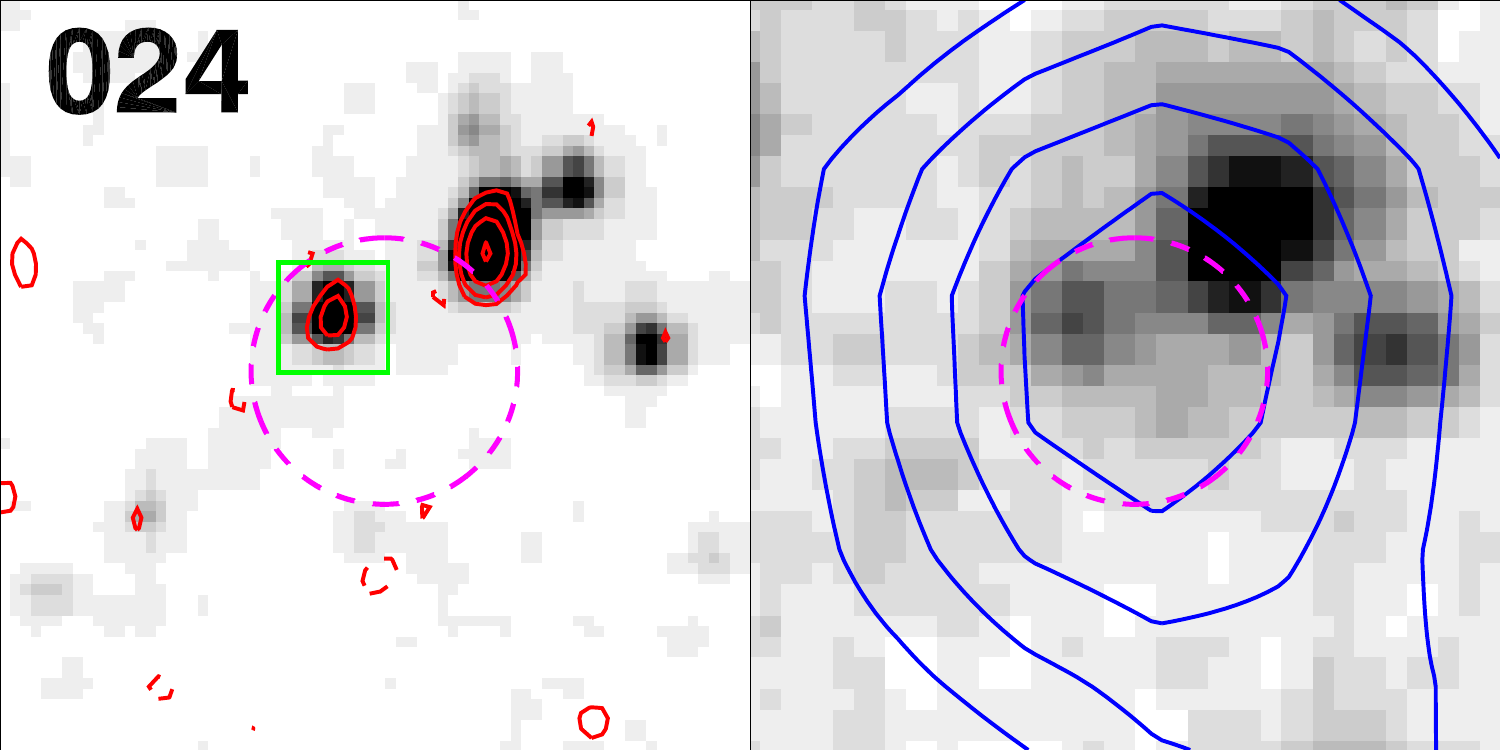}
\includegraphics[scale=0.295]{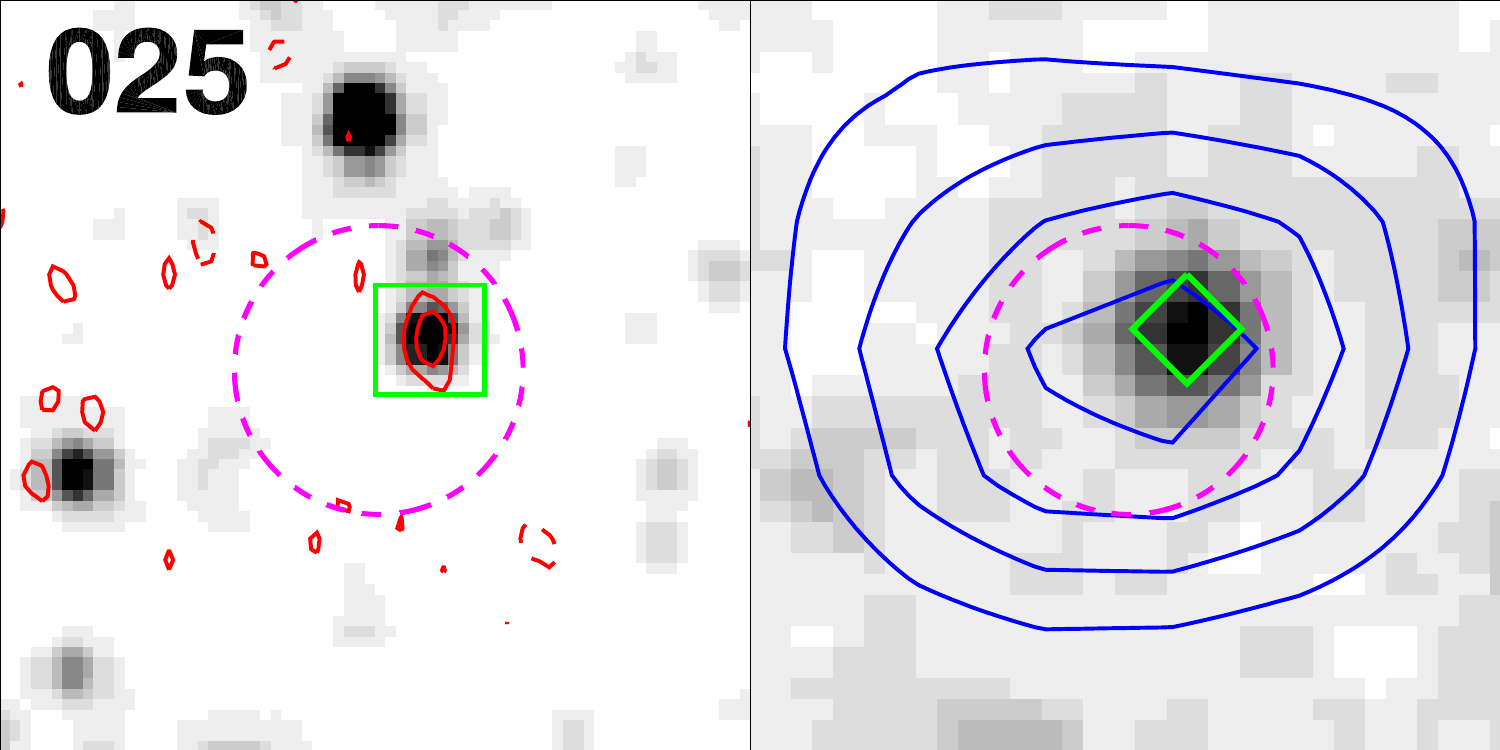}%
\hspace{1cm}%
\includegraphics[scale=0.295]{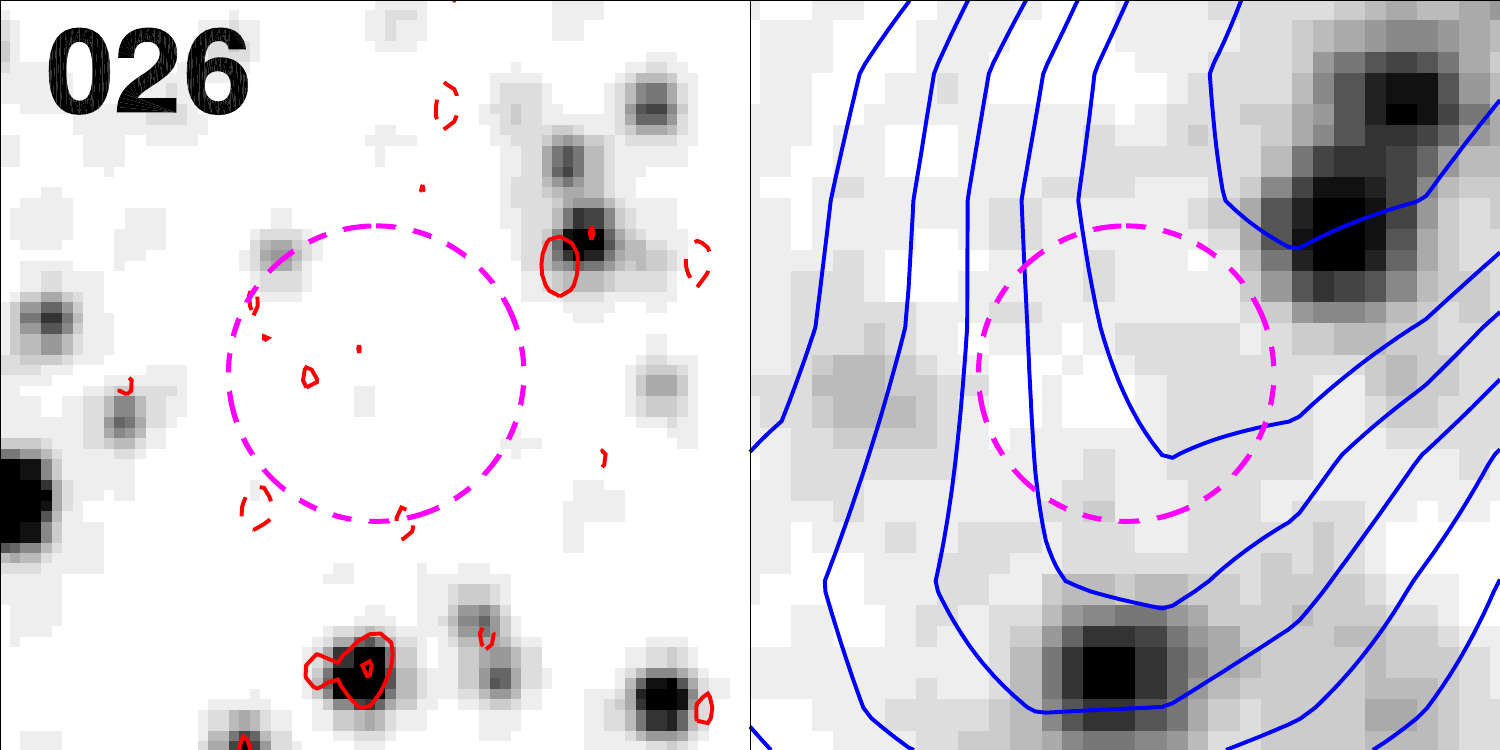}%
\hspace{1cm}%
\includegraphics[scale=0.295]{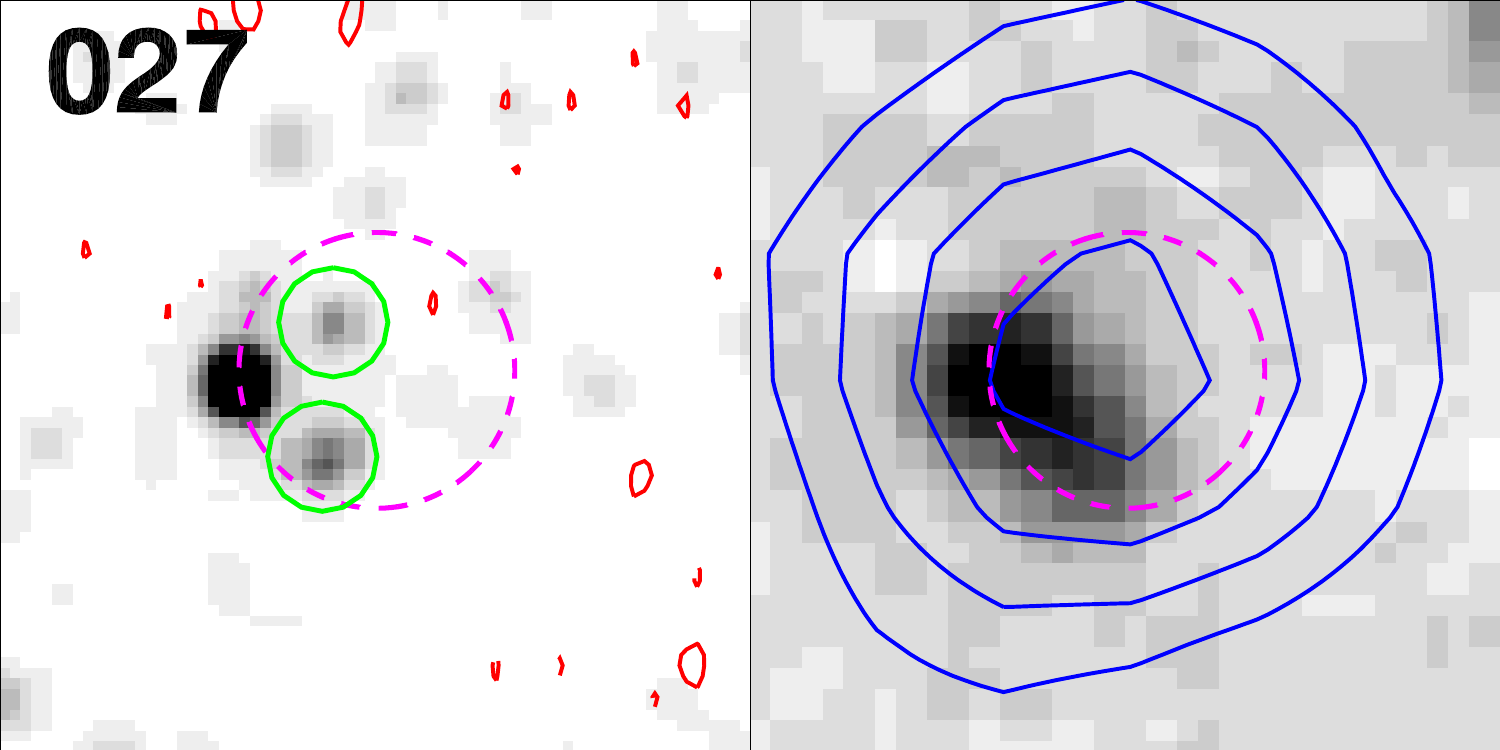}
\caption{Plots centered on the location of each LABOCA-detected source in the ECDFS; each is $36\times36$\,arcsec$^2$. Left panel: 3.6-$\umu$m IRAC greyscale image with radio 21-cm contours overlaid, right panel: 24-$\umu$m MIPS greyscale with submm contours (SNR) overlaid. The radio images have all been shifted by 0.25~arcsec to the East and by 0.29~arcsec to the North (Section~\ref{sec:astrometry}). The circle shows the search radius used to search for counterparts. Secure identifications ($p\le0.05$) are indicated by green squares (radio), diamonds (24-$\umu$m) and circles (5.8-$\umu$m). Paired yellow symbols represent those counterparts that are considered robust based on coincident emission having $0.05 < p \le 0.1$ in two separate wavebands. Radio contours are plotted at $-$3, 3, 5, 10, 20, 50 and 100 times the 1-$\sigma$ rms noise. 870-$\umu$m contours are plotted at 2, 3, 4, 5, 6, 8, 10, 12, 14 times the 1-$\sigma$ rms noise. Please note that the submm contours correspond to the beam-smoothed map that was used to identify the SMGs: see \citet{weiss09} for details. LESS046 is not located on the FIDEL 24-$\umu$m or the SIMPLE 3.6-$\umu$m images and so we have instead plotted the shallower Spitzer Wide-area Infrared Survey \citep[SWIRE;][]{lonsdale03} data for both. The SWIRE data has also been substituted for LESS035, 085, 093 and 100 at 3.6~$\umu$m.}
\label{fig:mainplots}
\end{figure*}

\begin{figure*}
\centering
\includegraphics[scale=0.295]{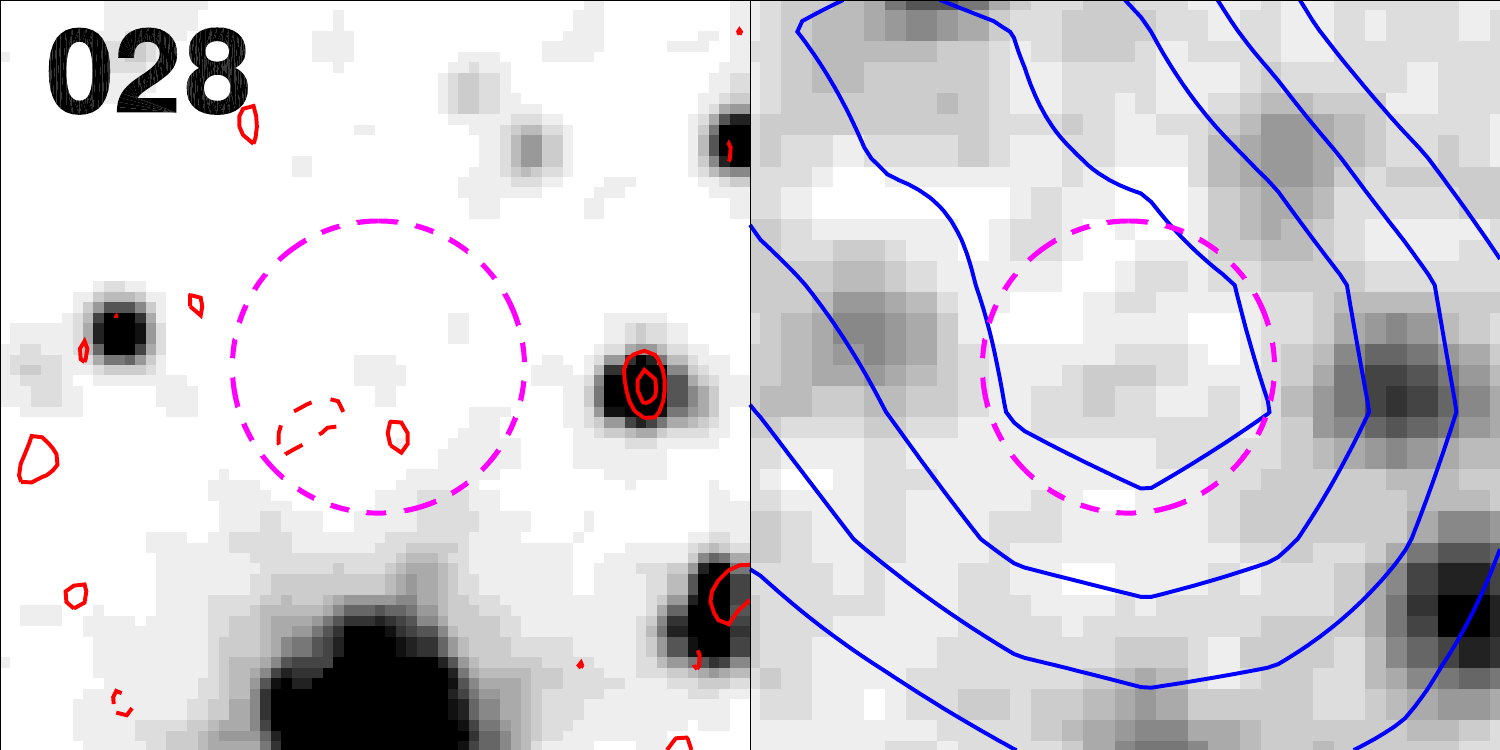}%
\hspace{1cm}%
\includegraphics[scale=0.295]{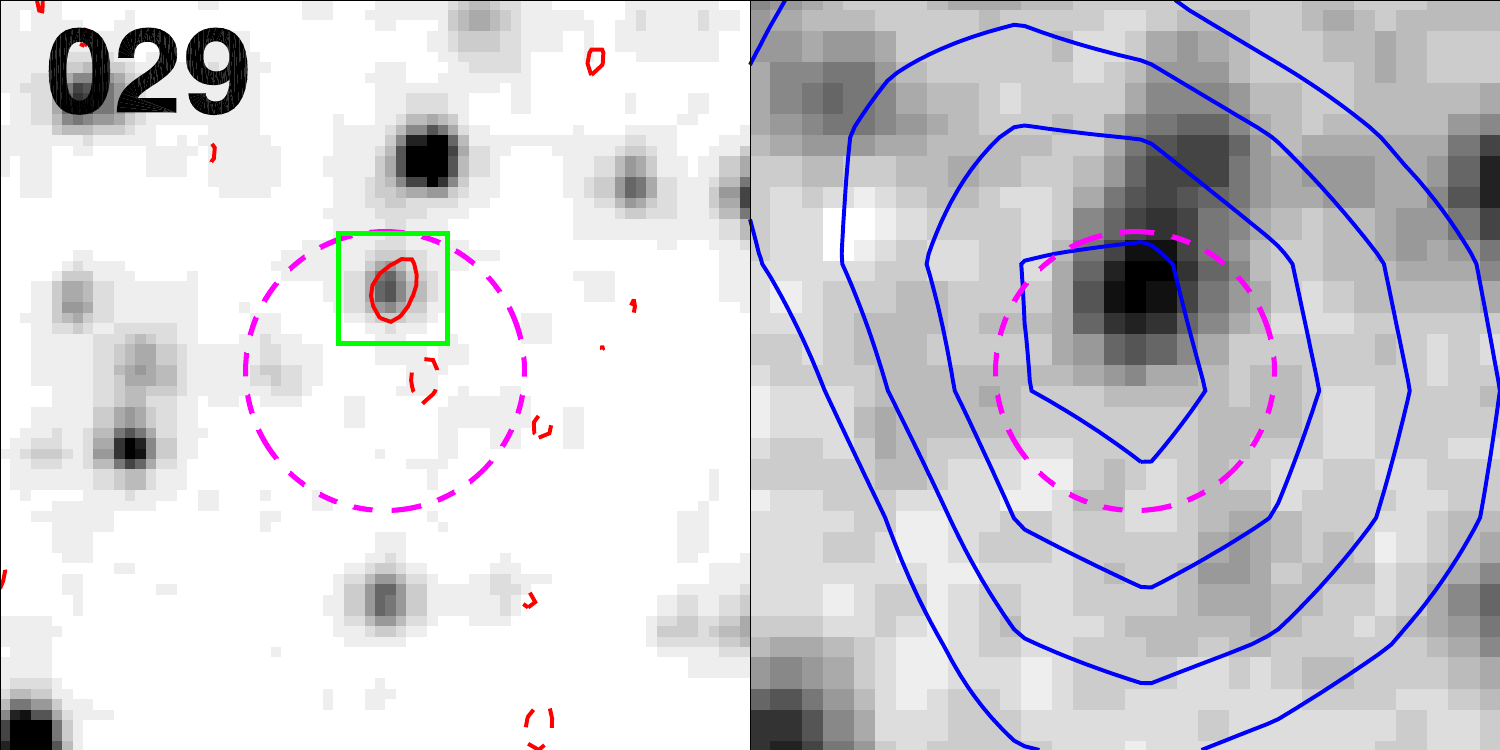}%
\hspace{1cm}%
\includegraphics[scale=0.295]{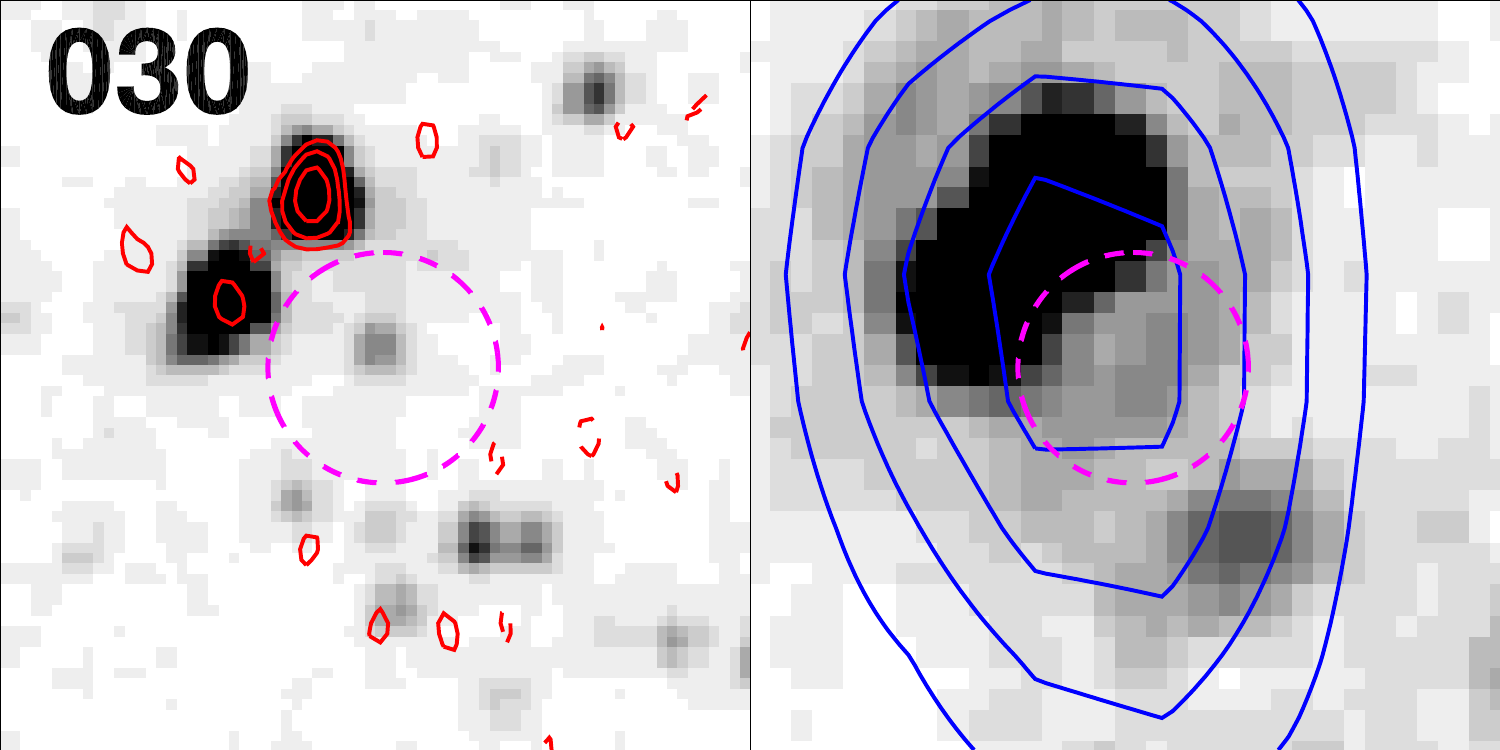}
\includegraphics[scale=0.295]{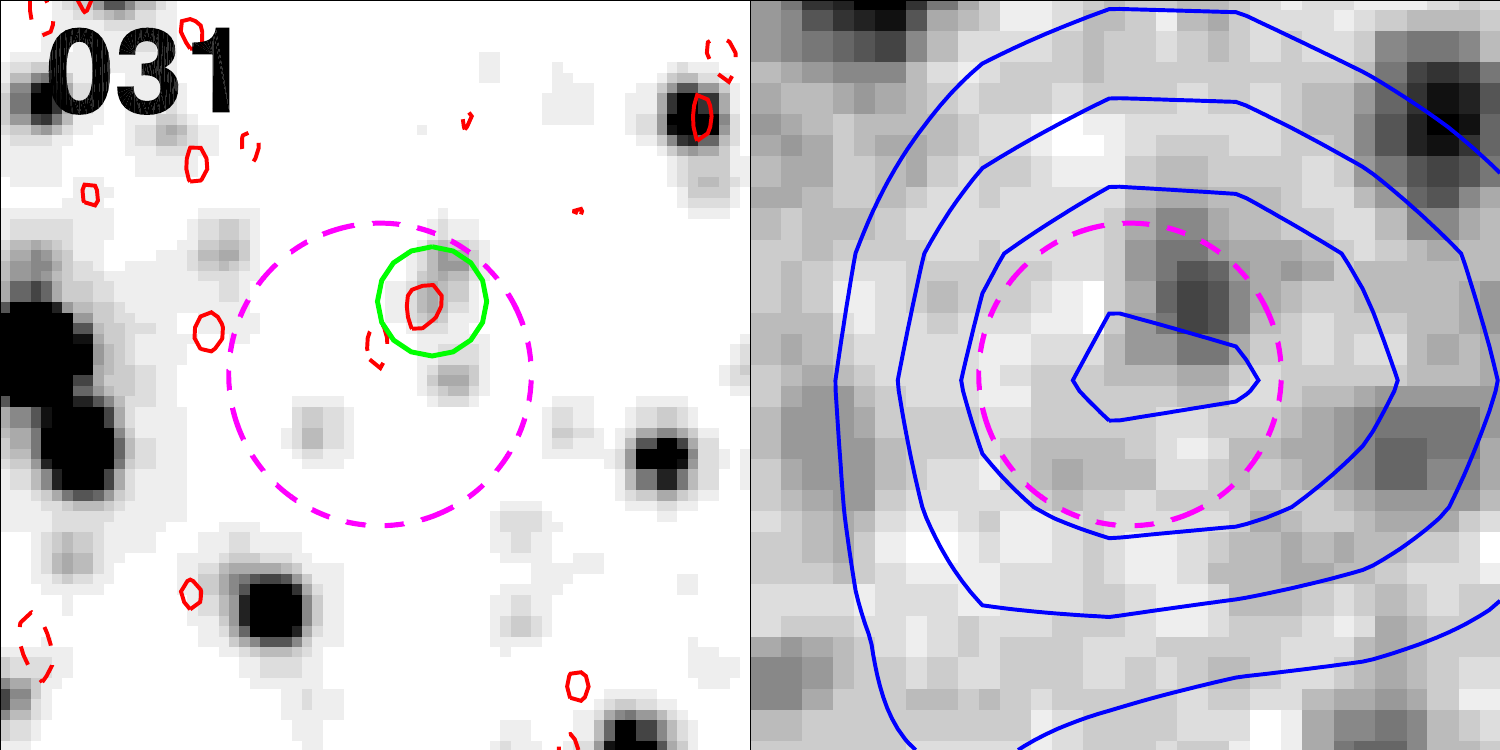}%
\hspace{1cm}%
\includegraphics[scale=0.295]{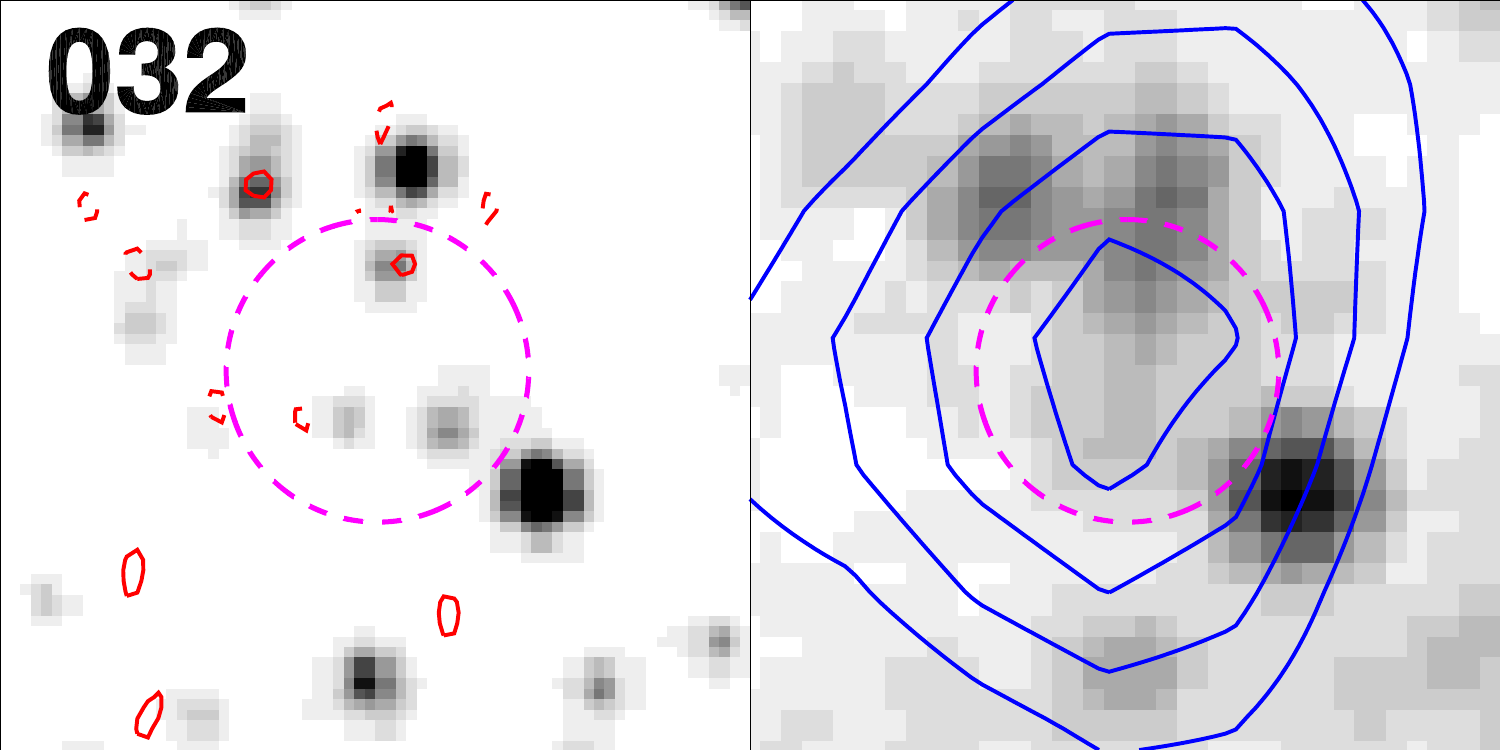}%
\hspace{1cm}%
\includegraphics[scale=0.295]{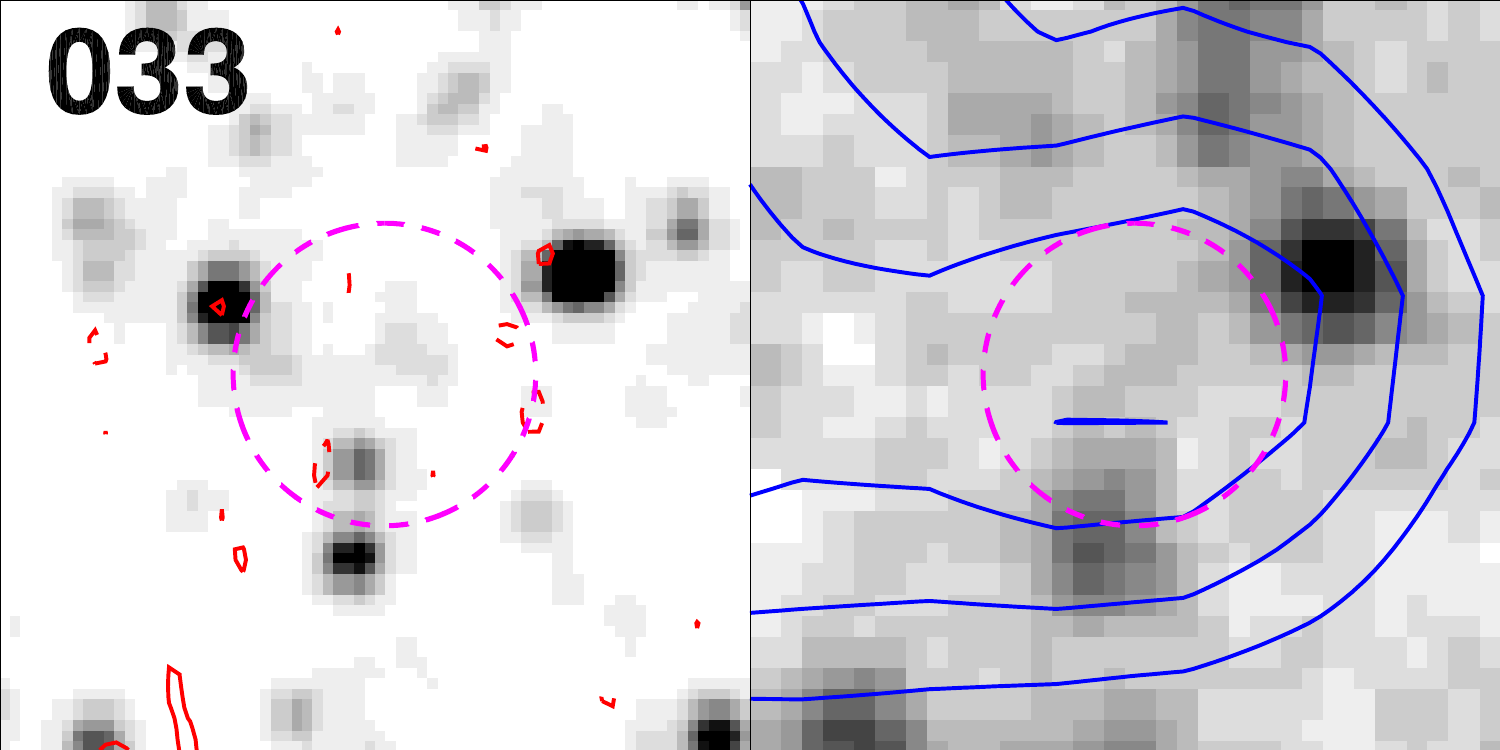}
\includegraphics[scale=0.295]{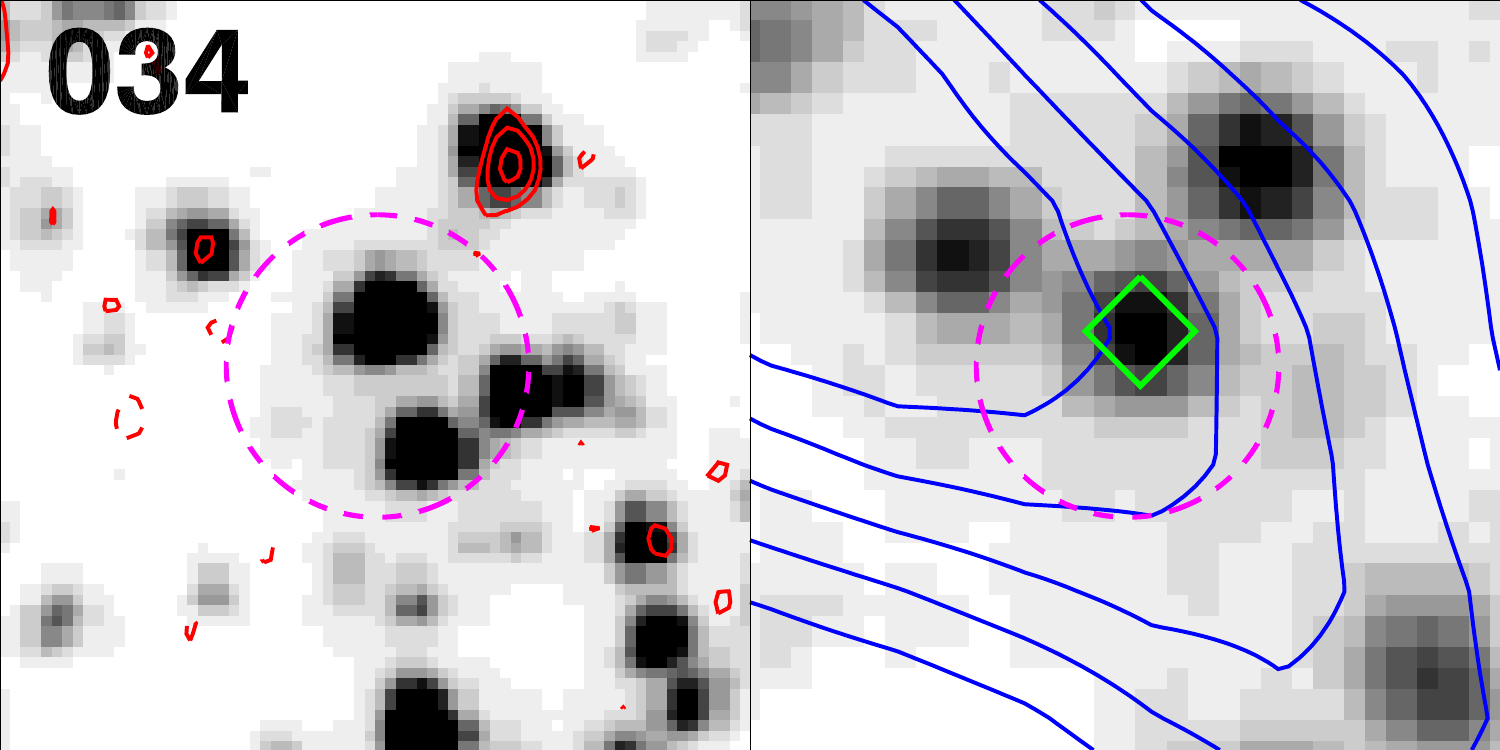}%
\hspace{1cm}%
\includegraphics[scale=0.295]{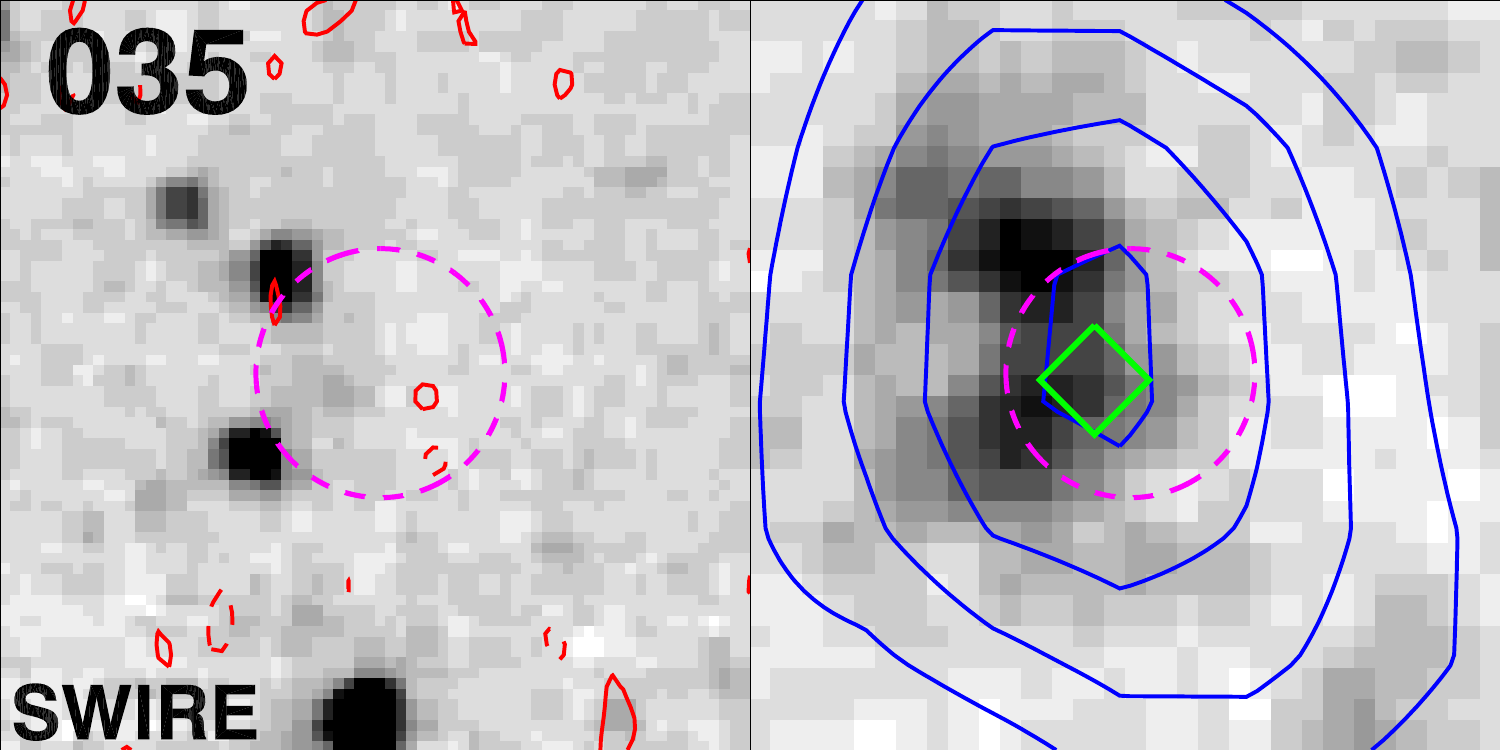}%
\hspace{1cm}%
\includegraphics[scale=0.295]{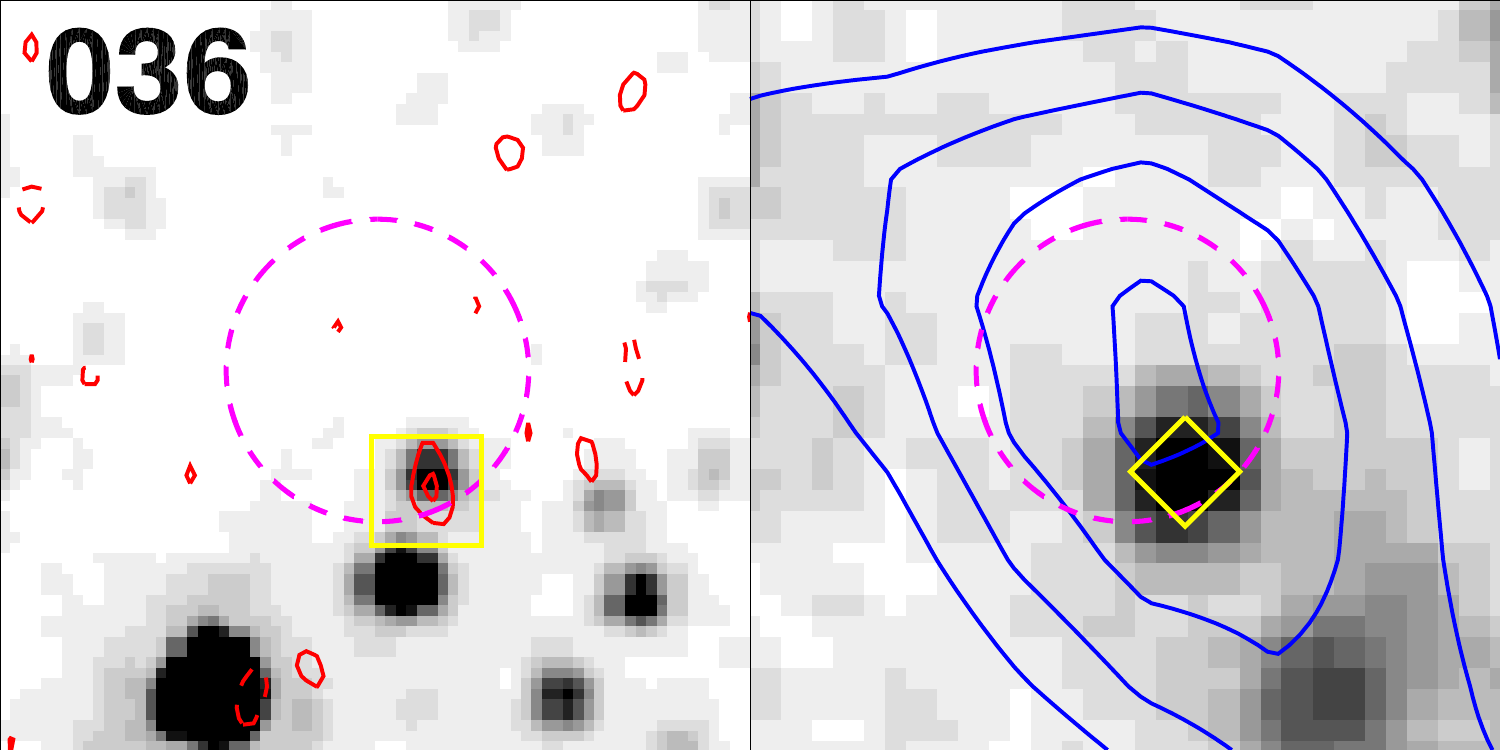}
\includegraphics[scale=0.295]{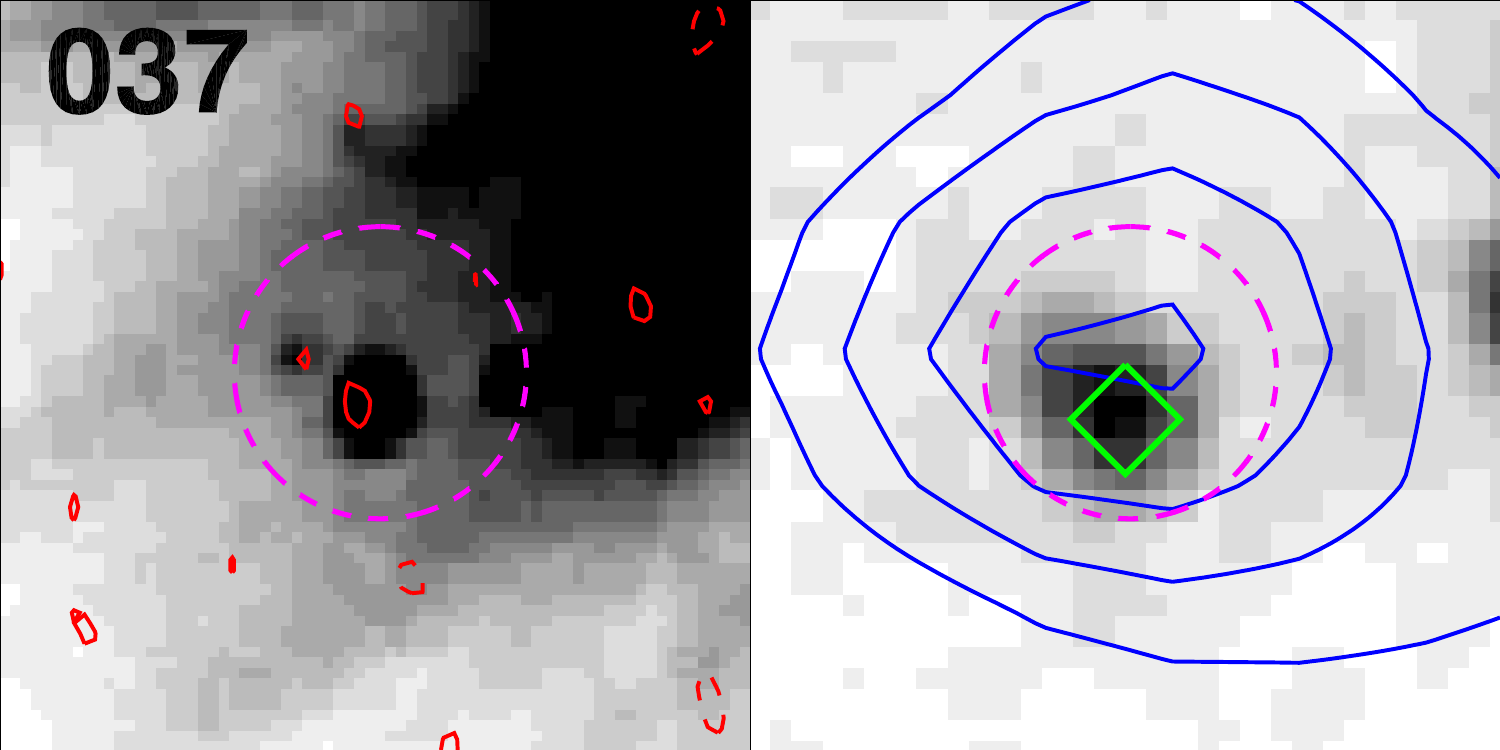}%
\hspace{1cm}%
\includegraphics[scale=0.295]{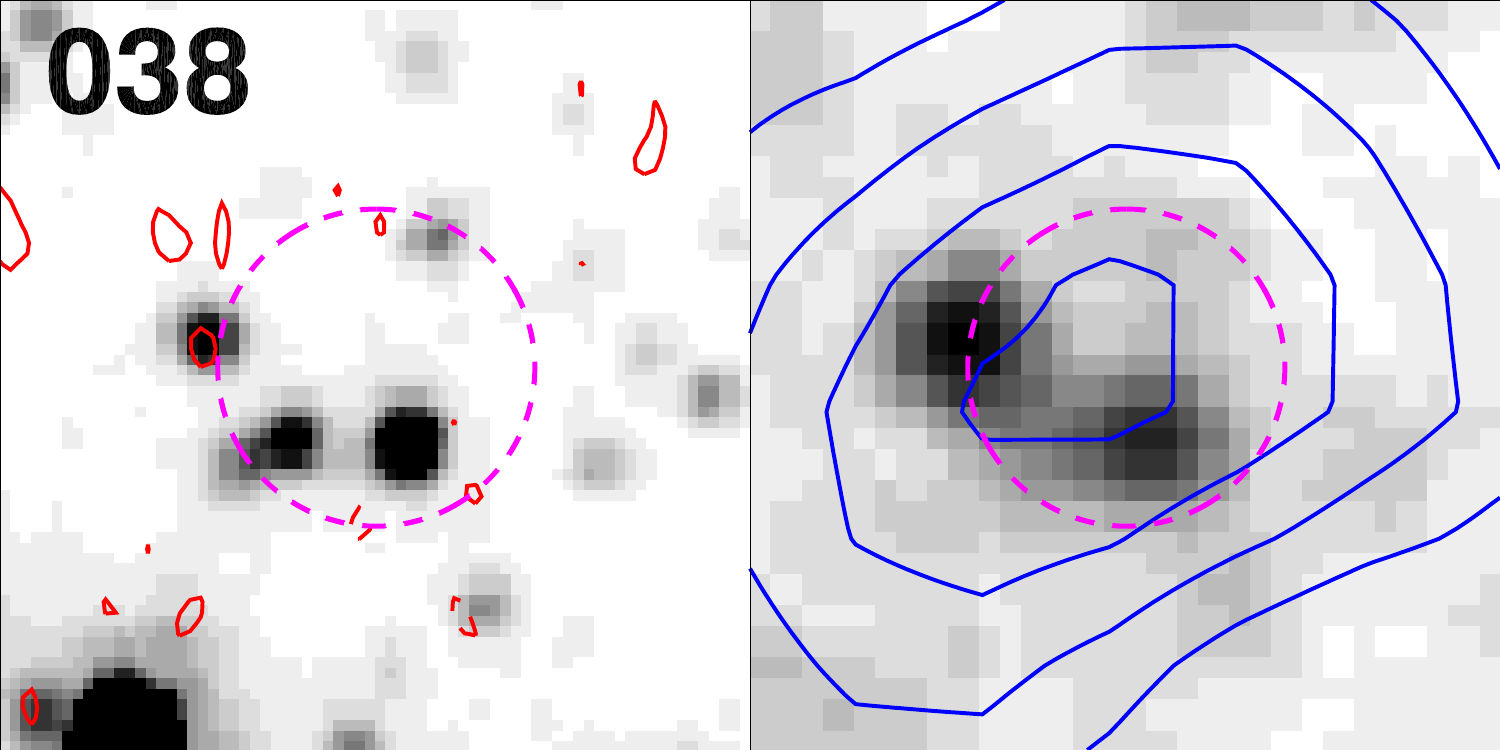}%
\hspace{1cm}%
\includegraphics[scale=0.295]{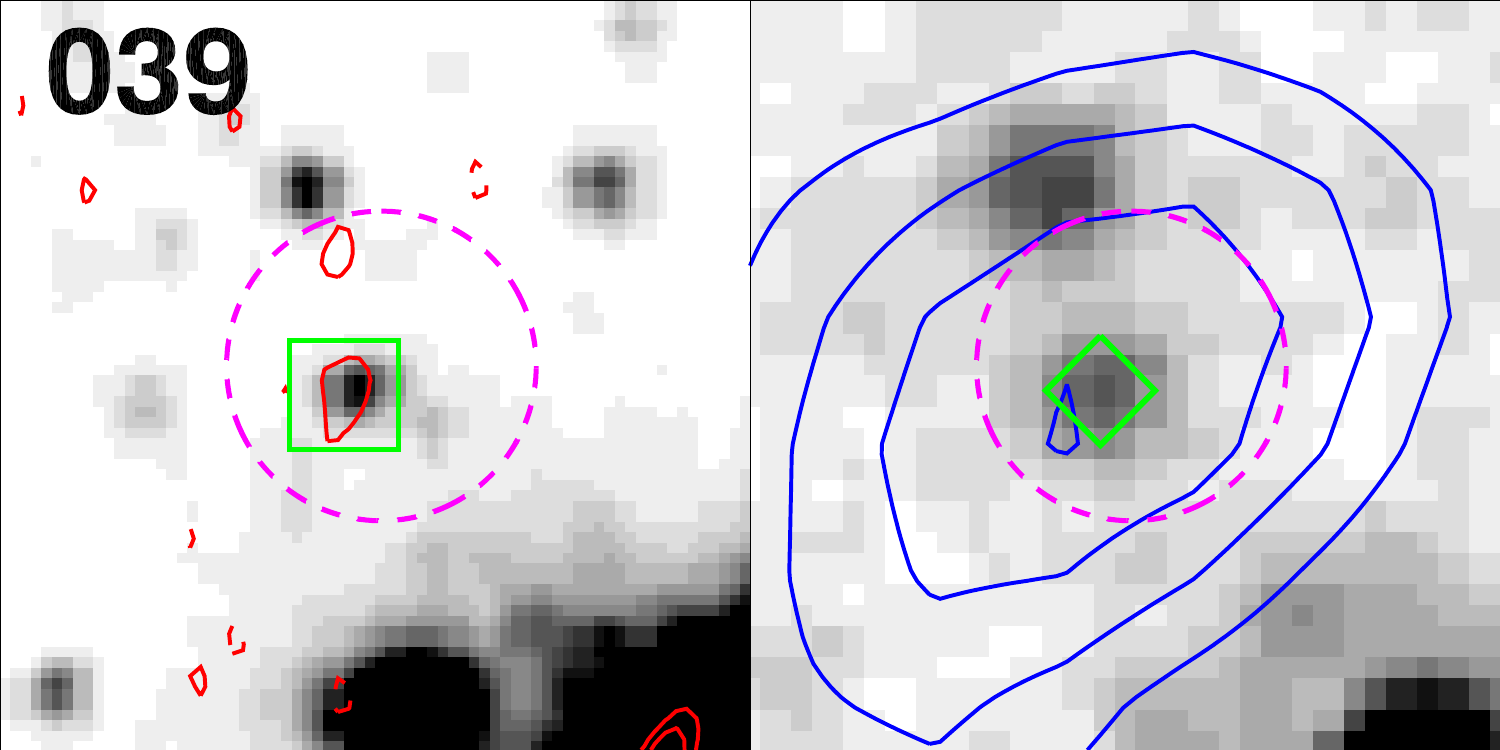}
\includegraphics[scale=0.295]{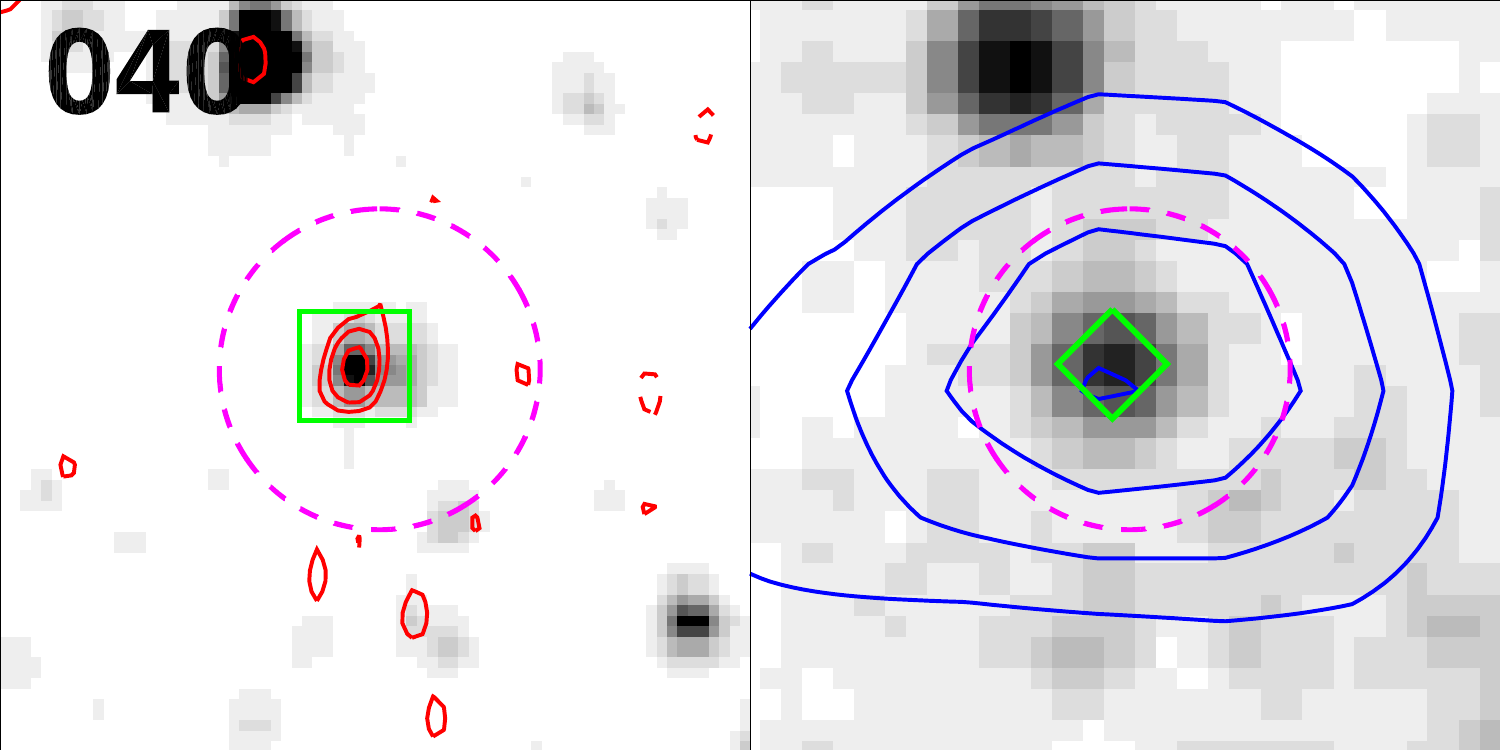}%
\hspace{1cm}%
\includegraphics[scale=0.295]{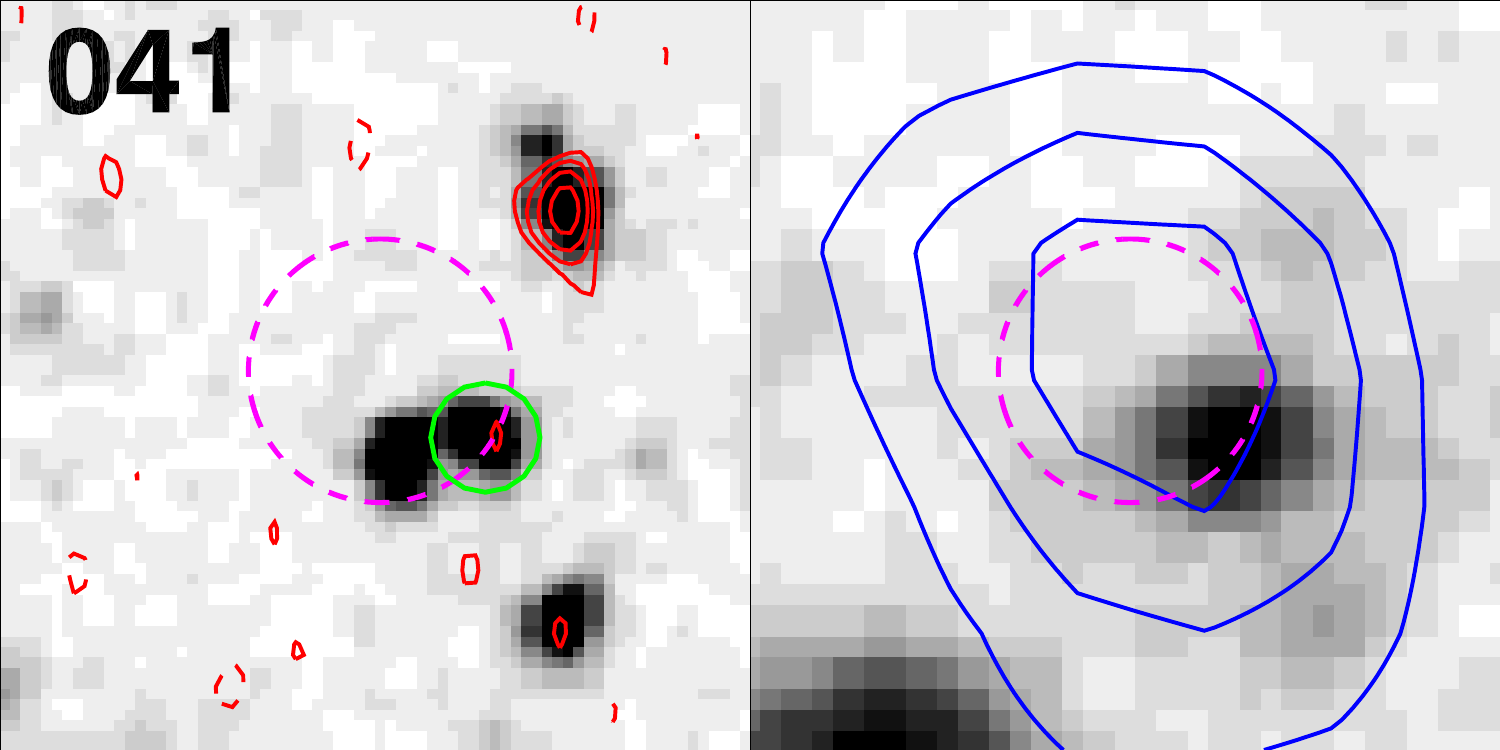}%
\hspace{1cm}%
\includegraphics[scale=0.295]{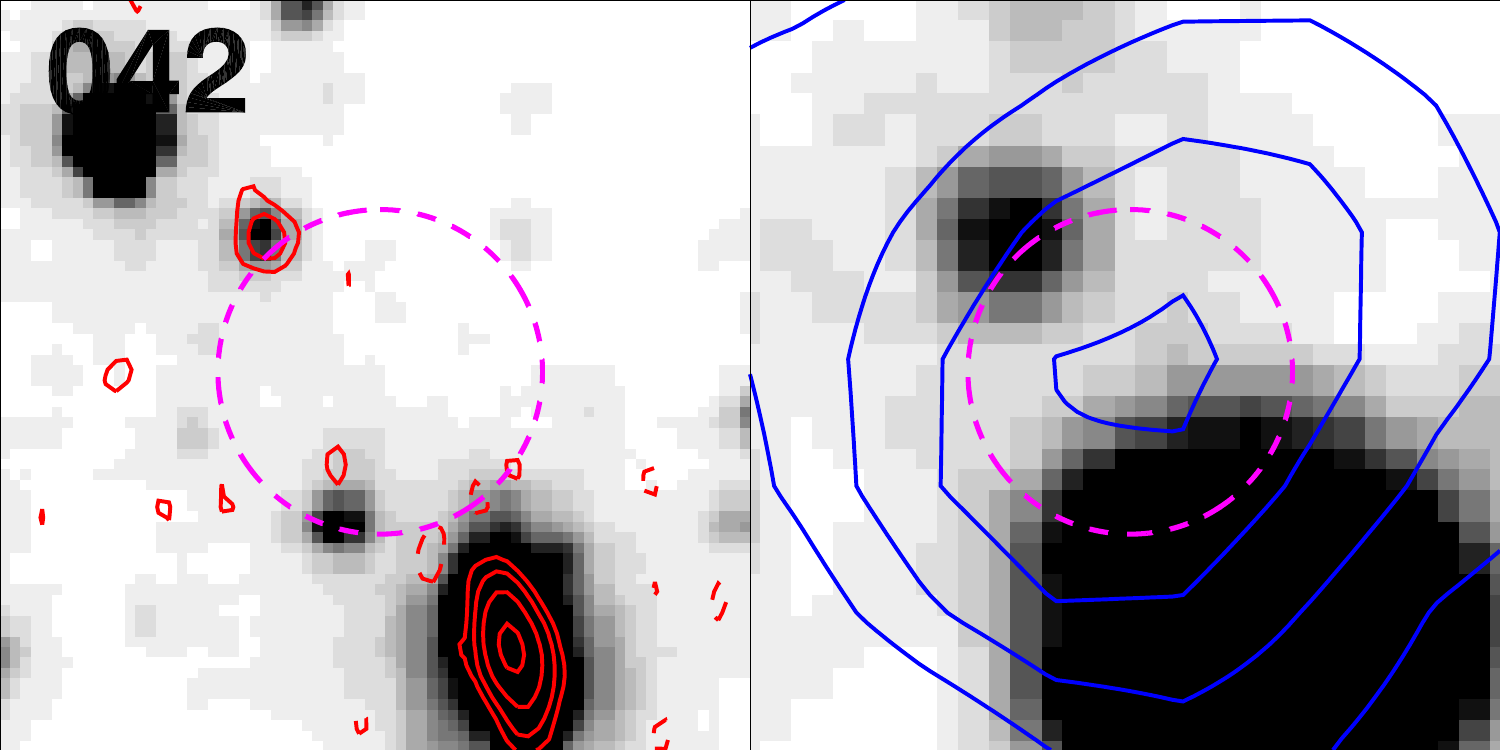}
\includegraphics[scale=0.295]{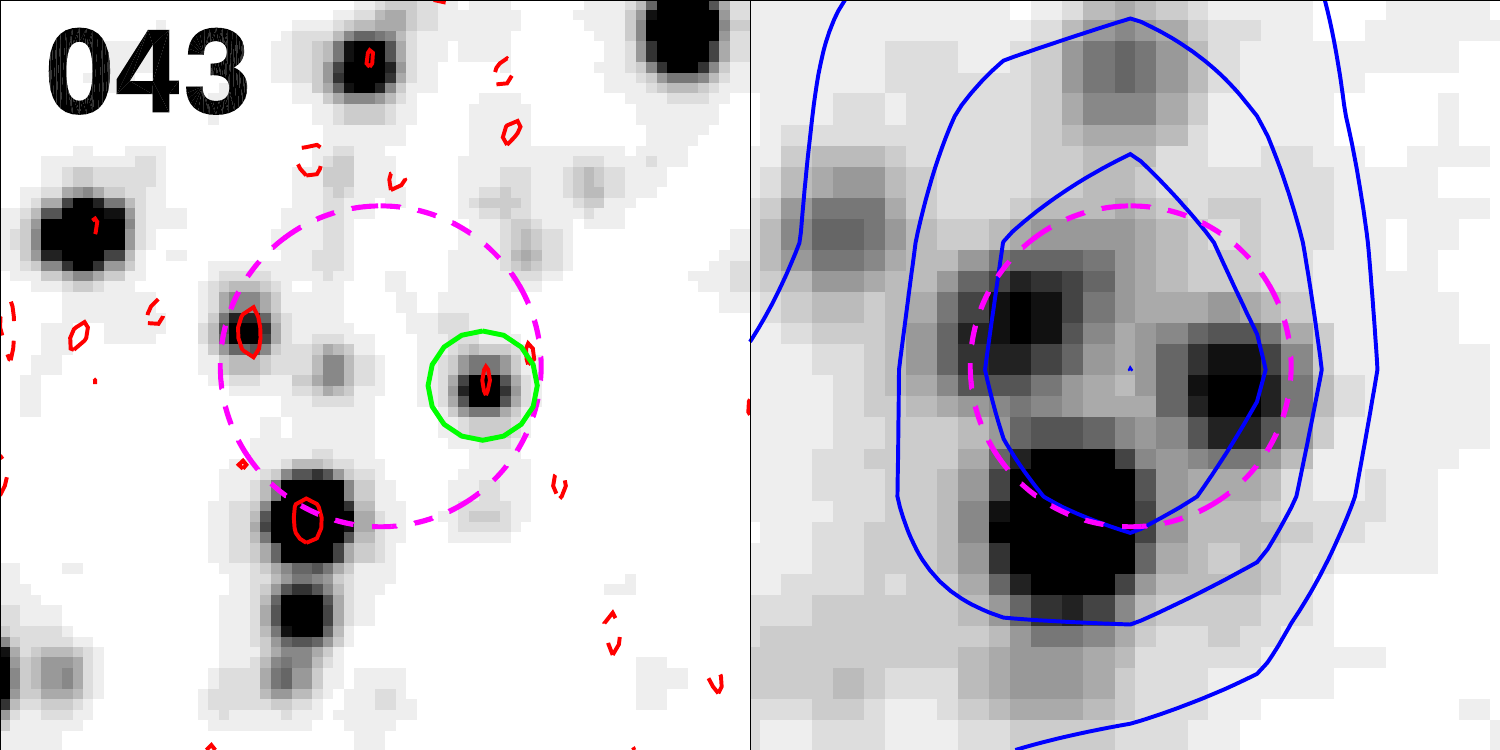}%
\hspace{1cm}%
\includegraphics[scale=0.295]{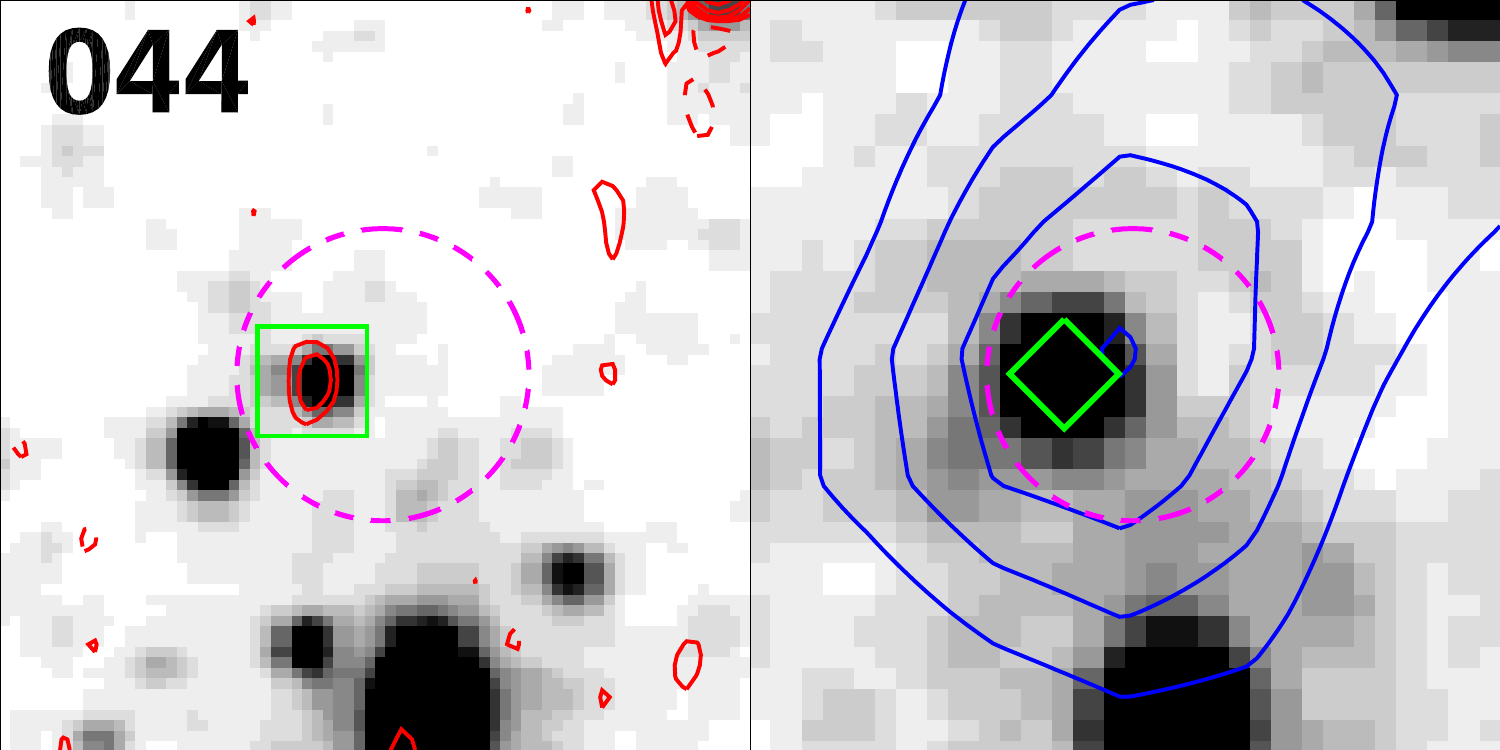}%
\hspace{1cm}%
\includegraphics[scale=0.295]{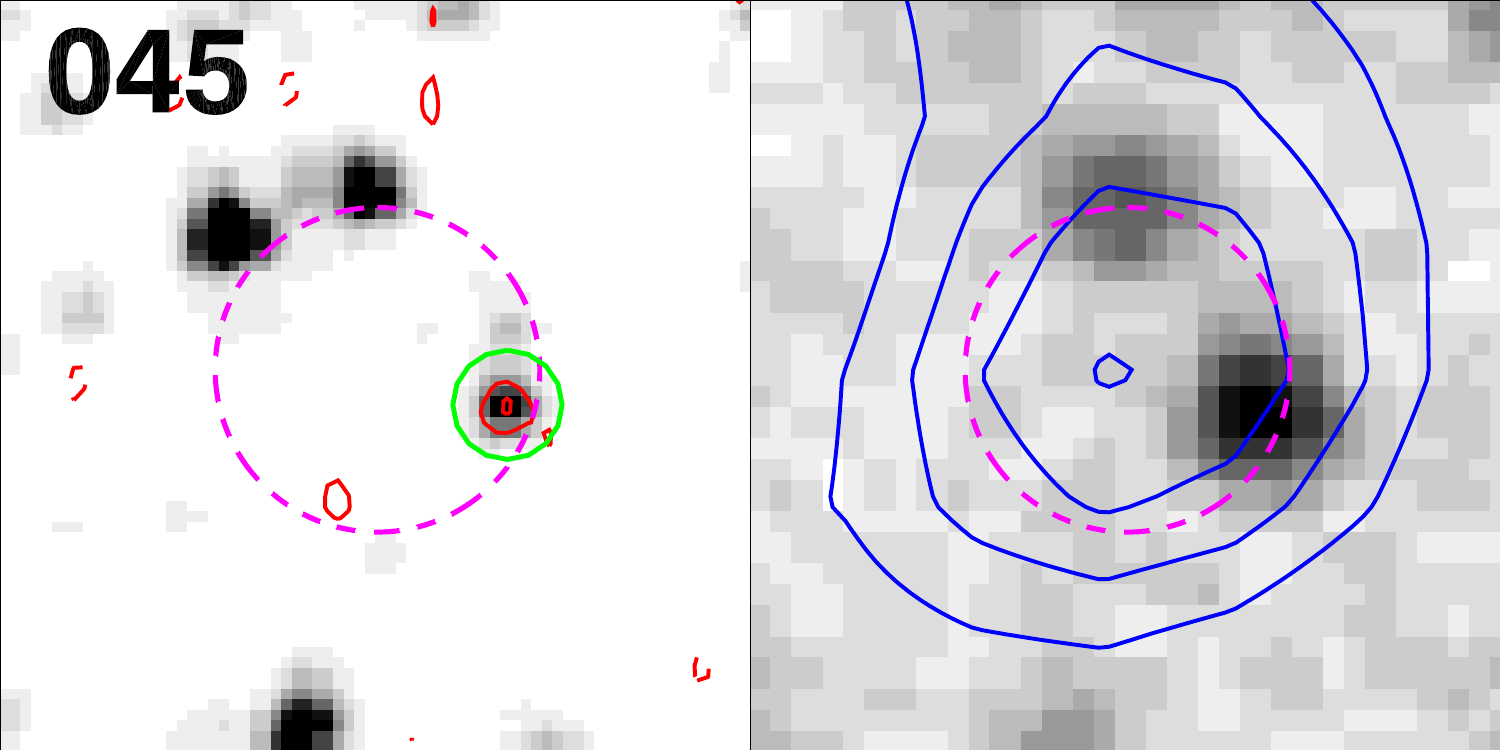}
\includegraphics[scale=0.295]{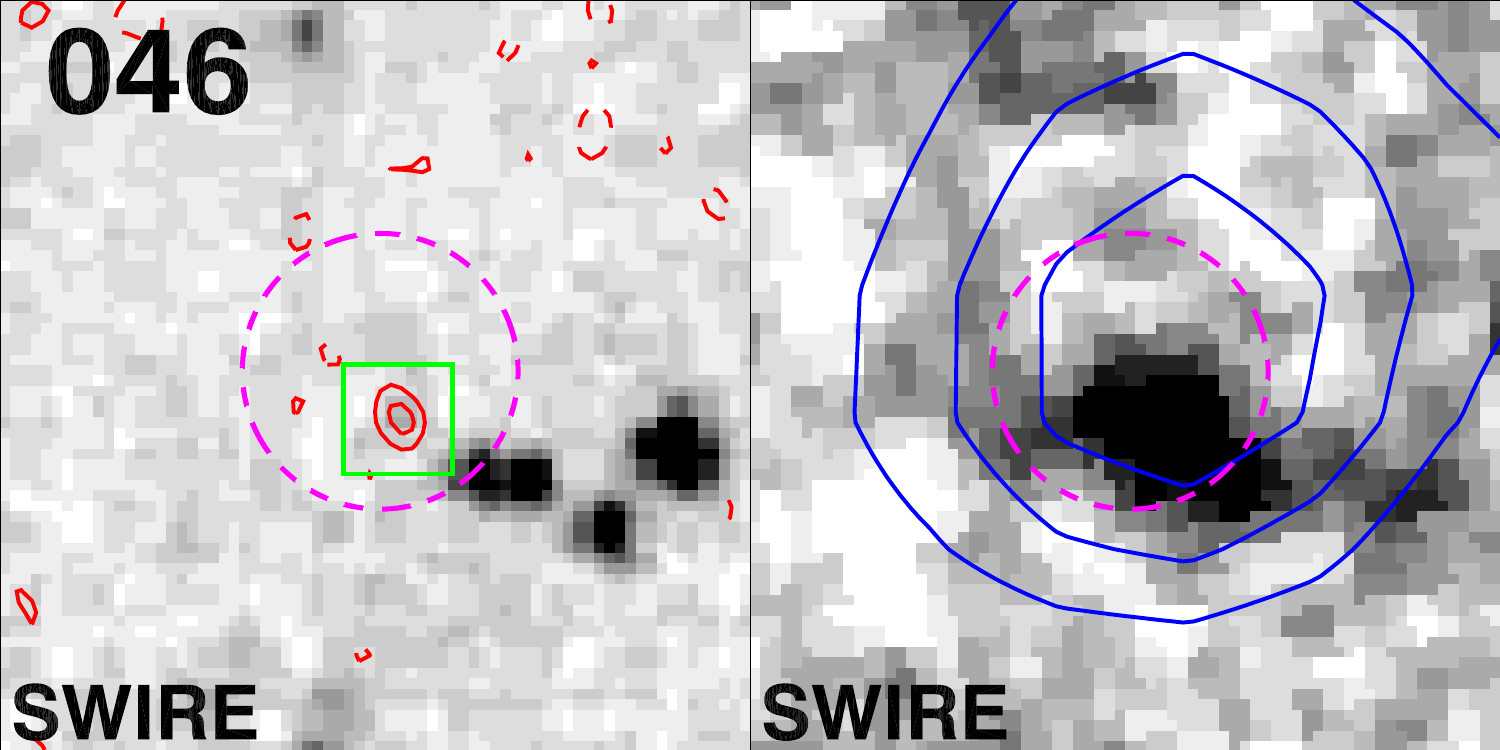}%
\hspace{1cm}%
\includegraphics[scale=0.295]{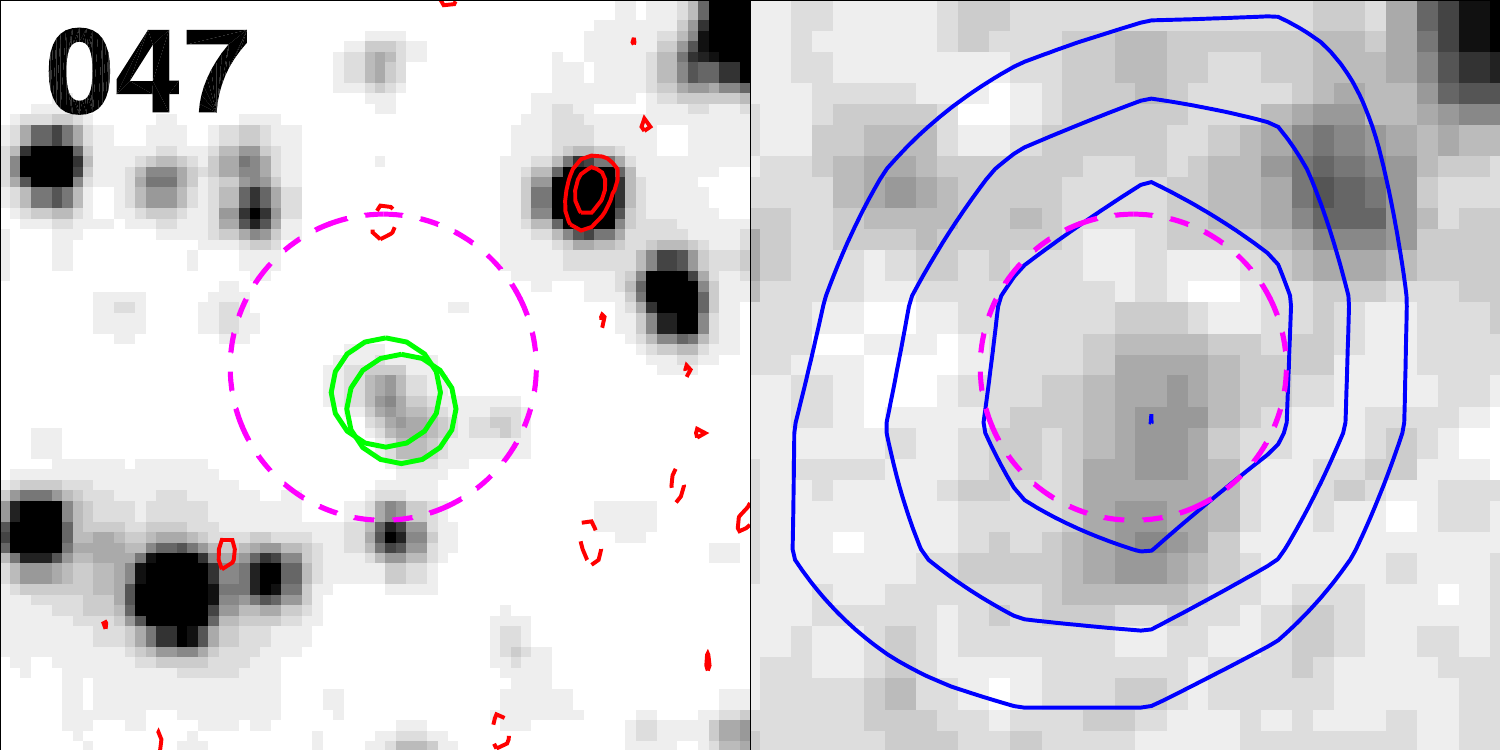}%
\hspace{1cm}%
\includegraphics[scale=0.295]{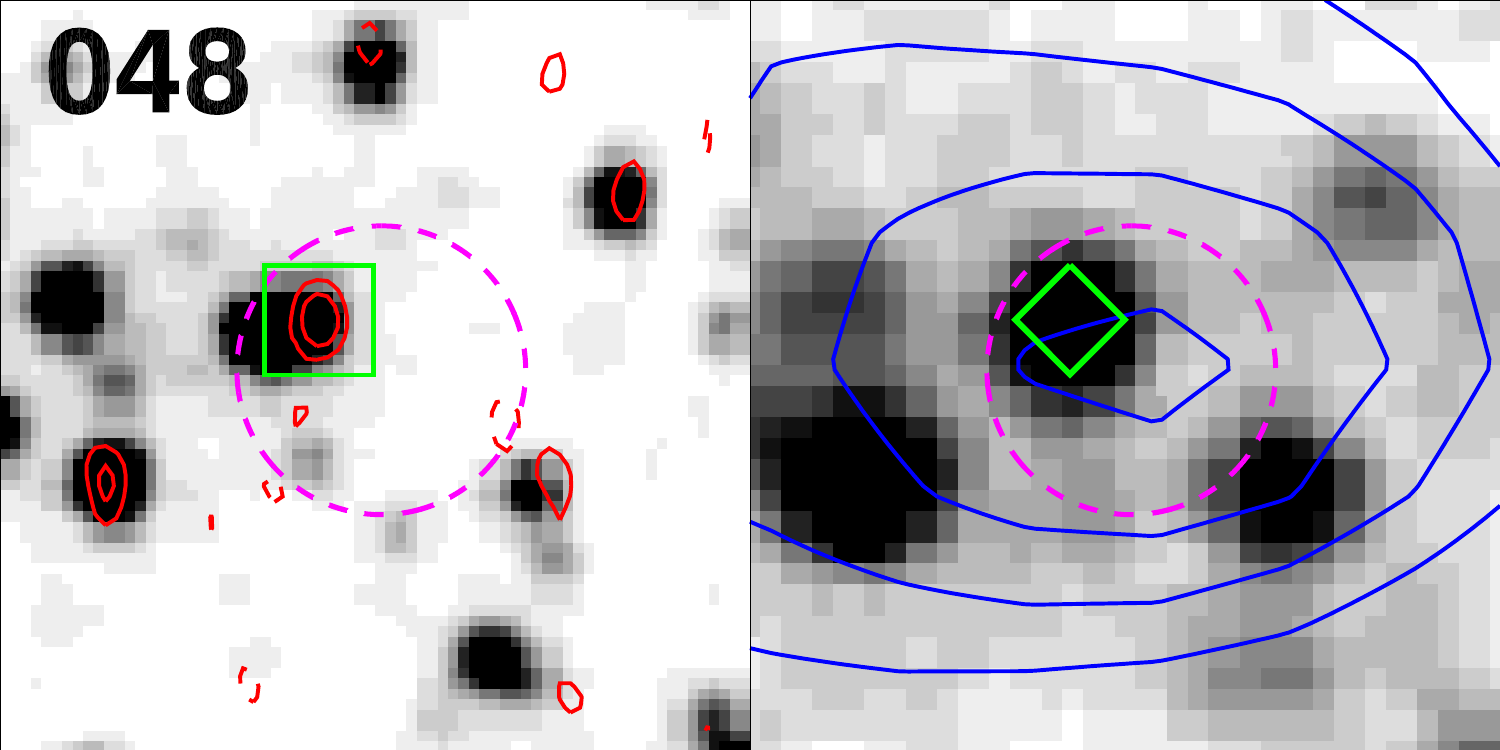}
\includegraphics[scale=0.295]{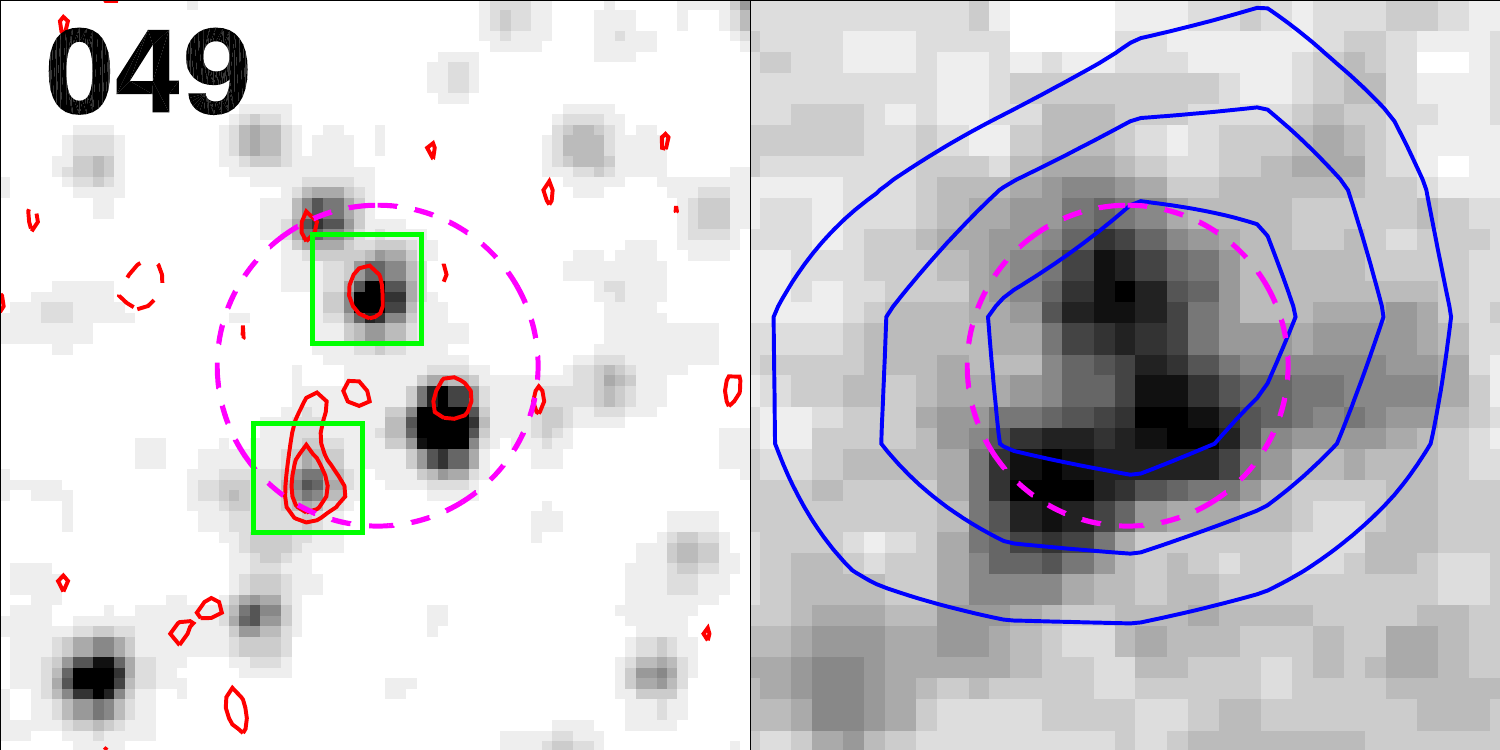}%
\hspace{1cm}%
\includegraphics[scale=0.295]{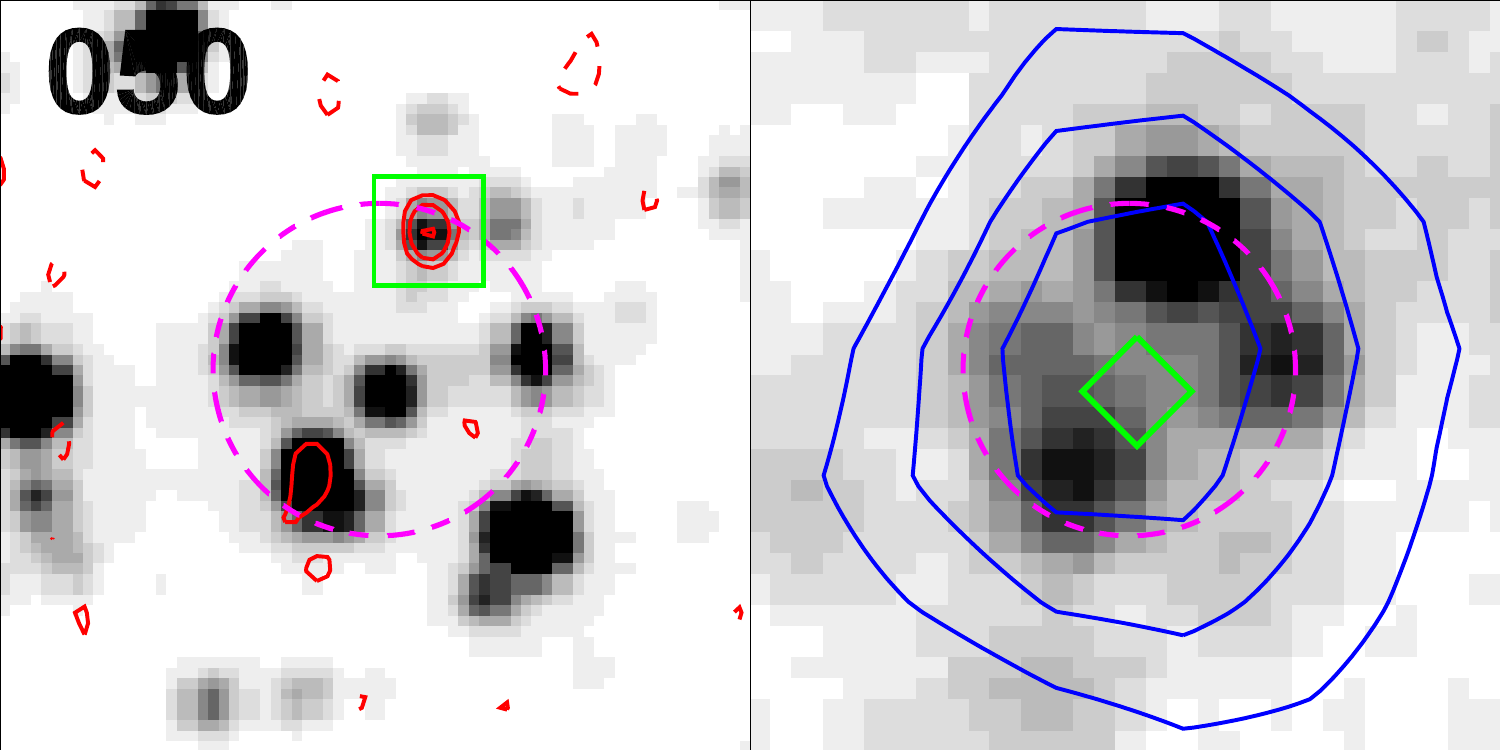}%
\hspace{1cm}%
\includegraphics[scale=0.295]{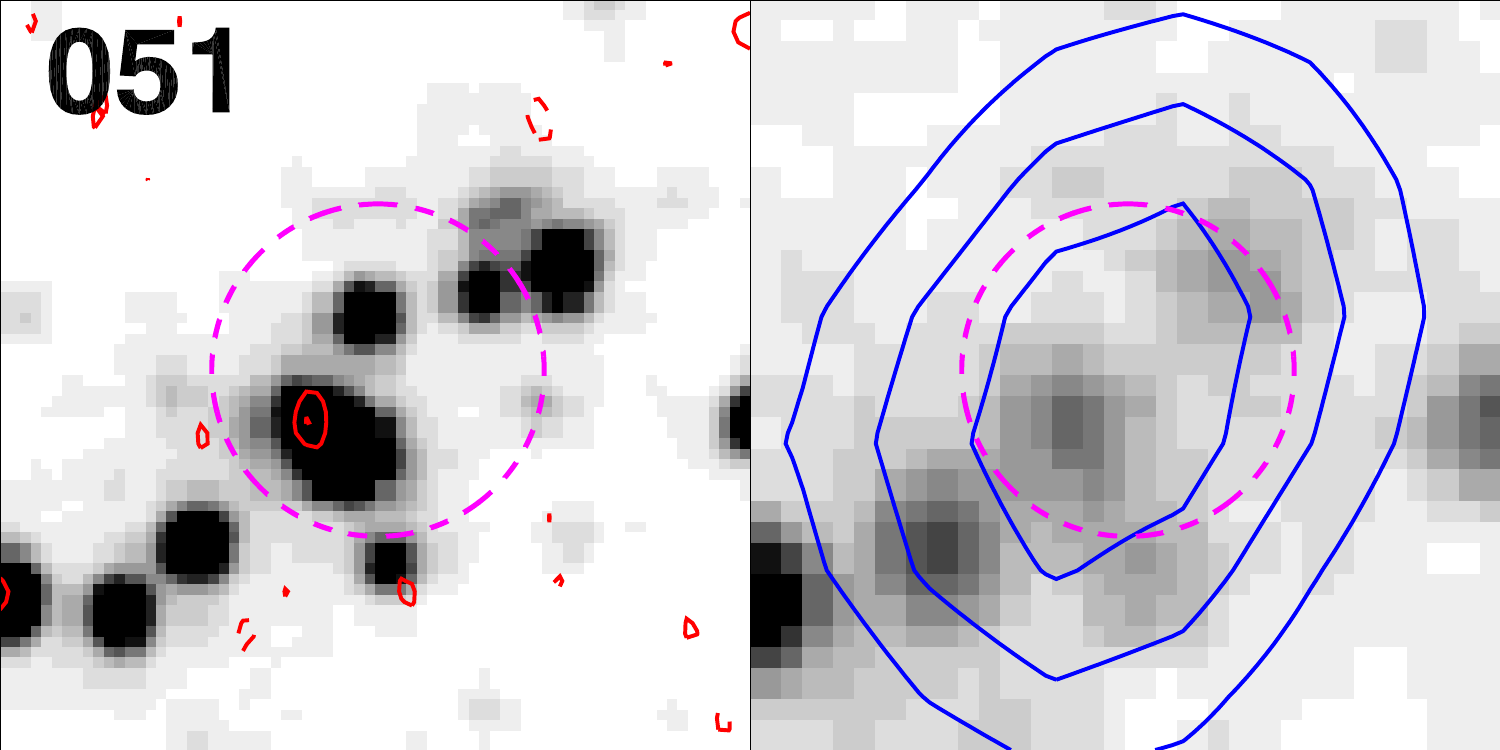}
\includegraphics[scale=0.295]{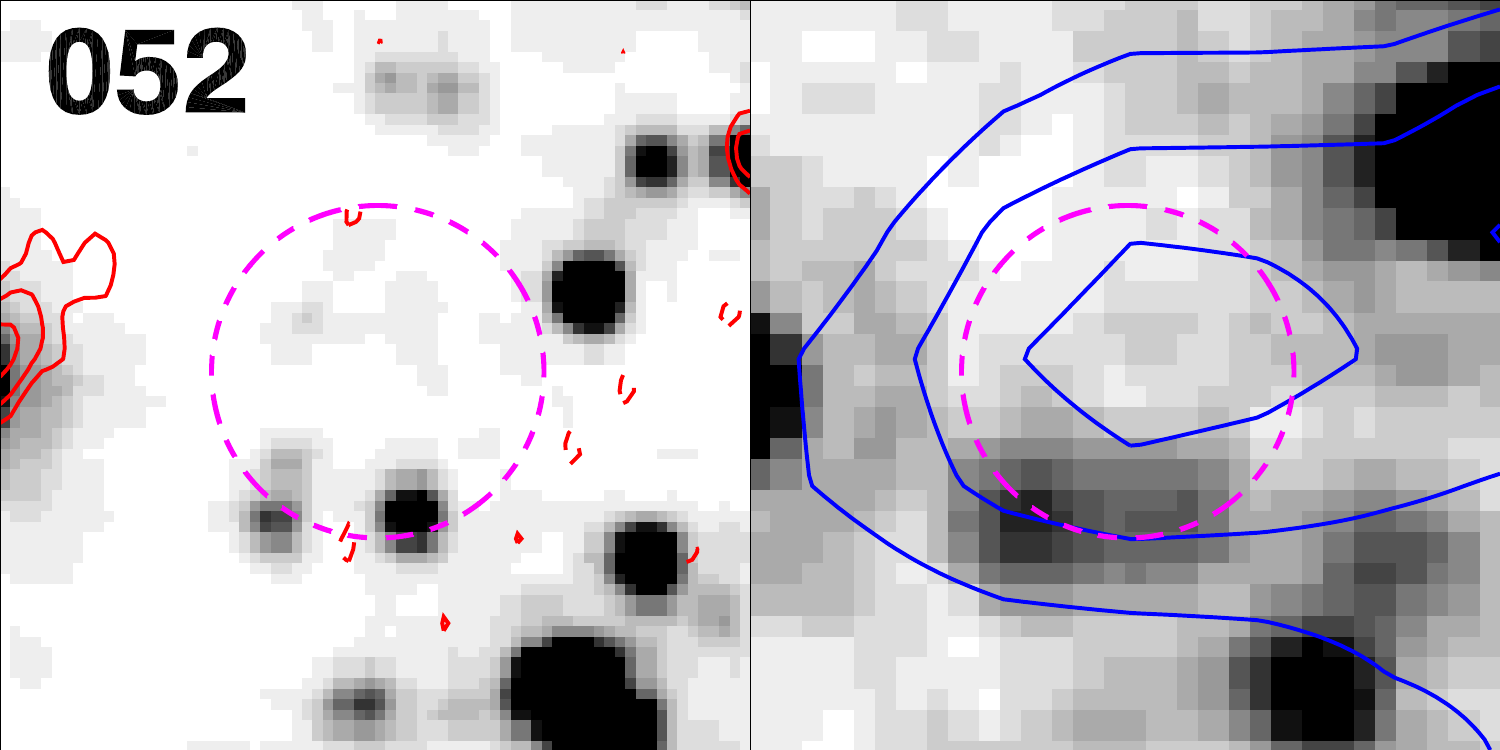}%
\hspace{1cm}%
\includegraphics[scale=0.295]{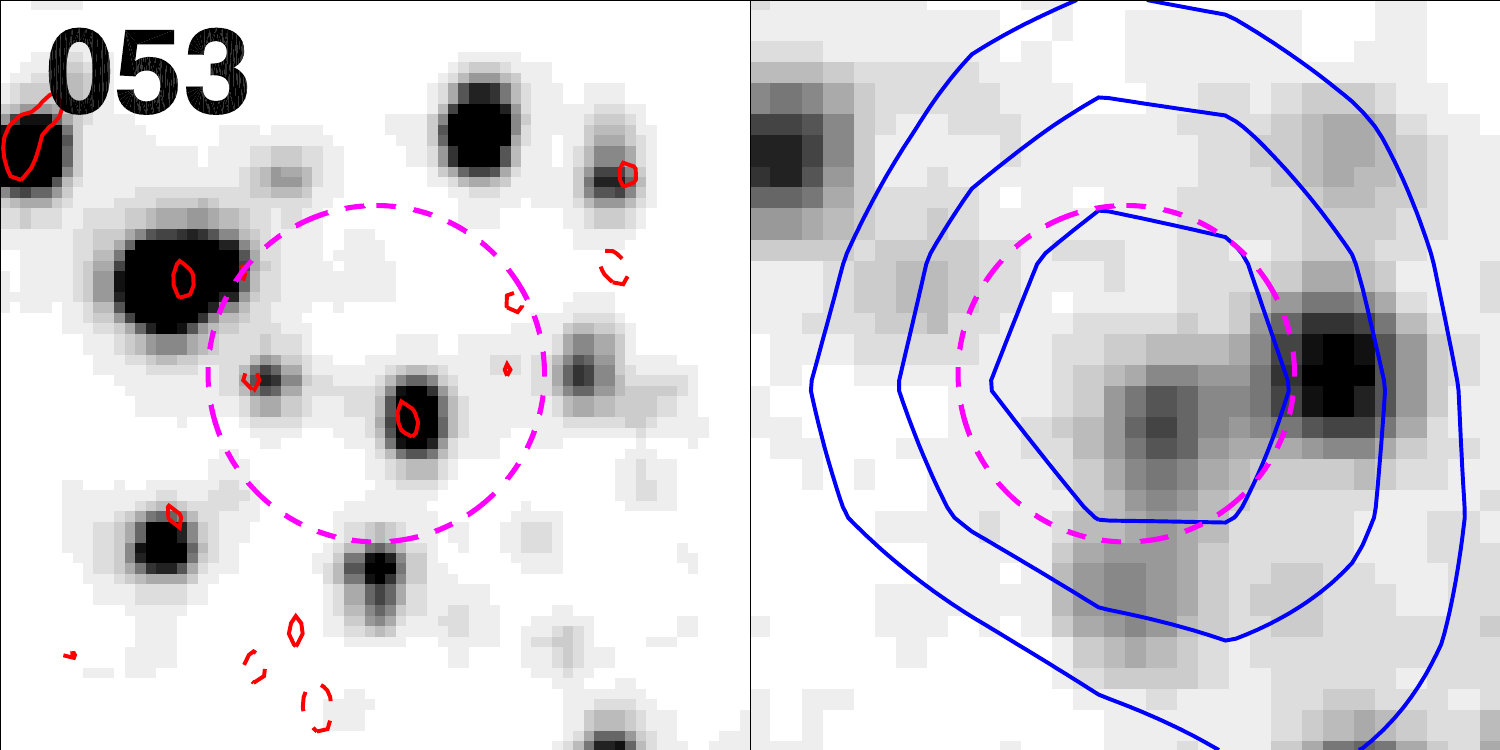}%
\hspace{1cm}%
\includegraphics[scale=0.295]{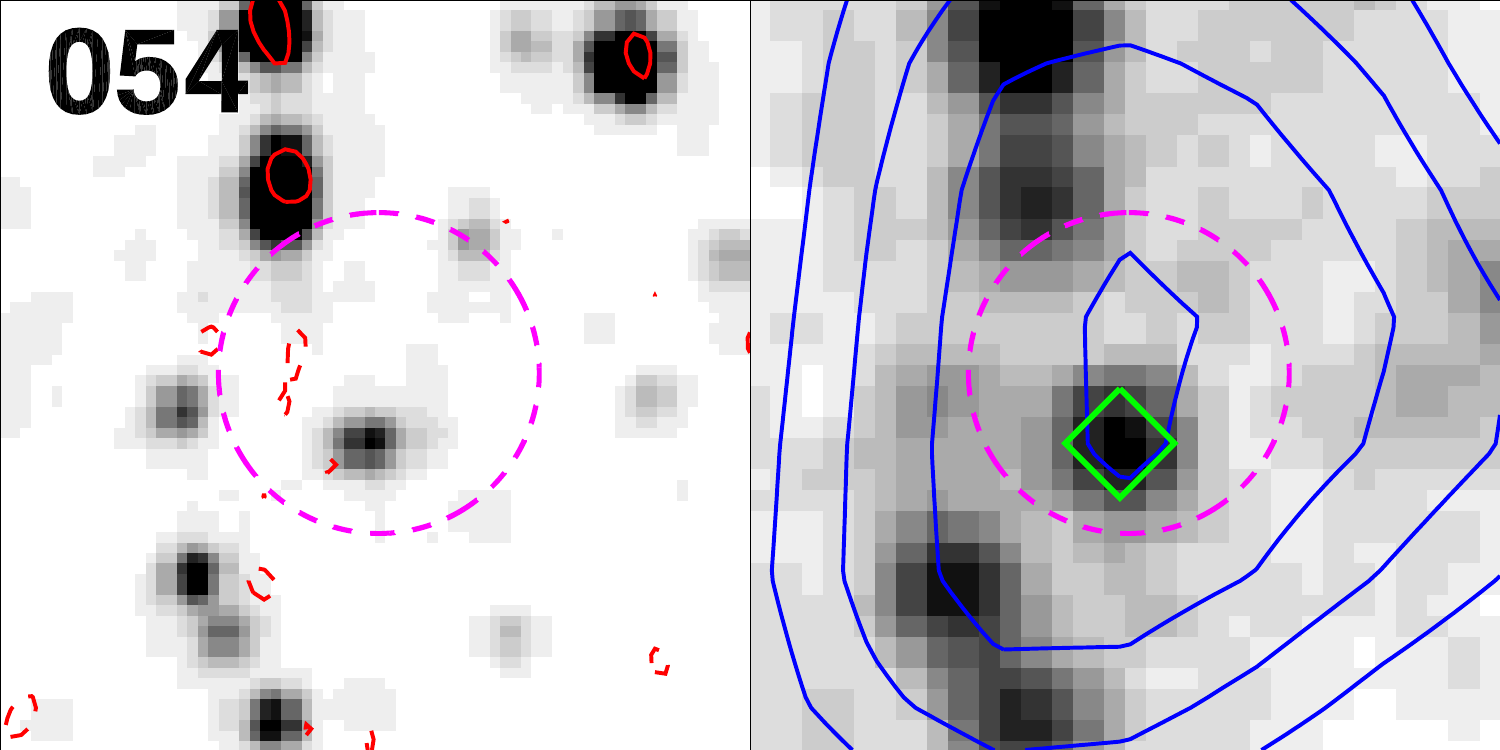}
\includegraphics[scale=0.295]{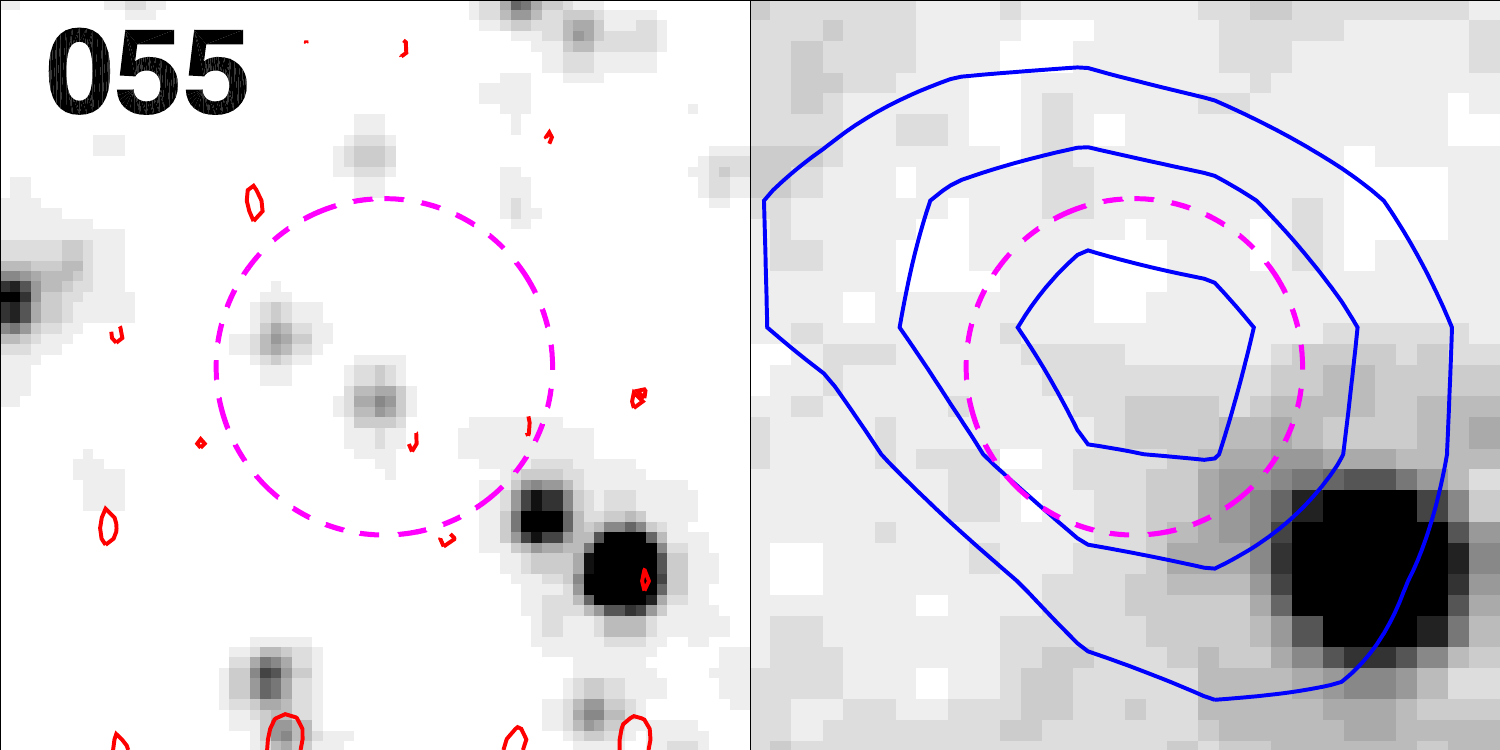}%
\hspace{1cm}%
\includegraphics[scale=0.295]{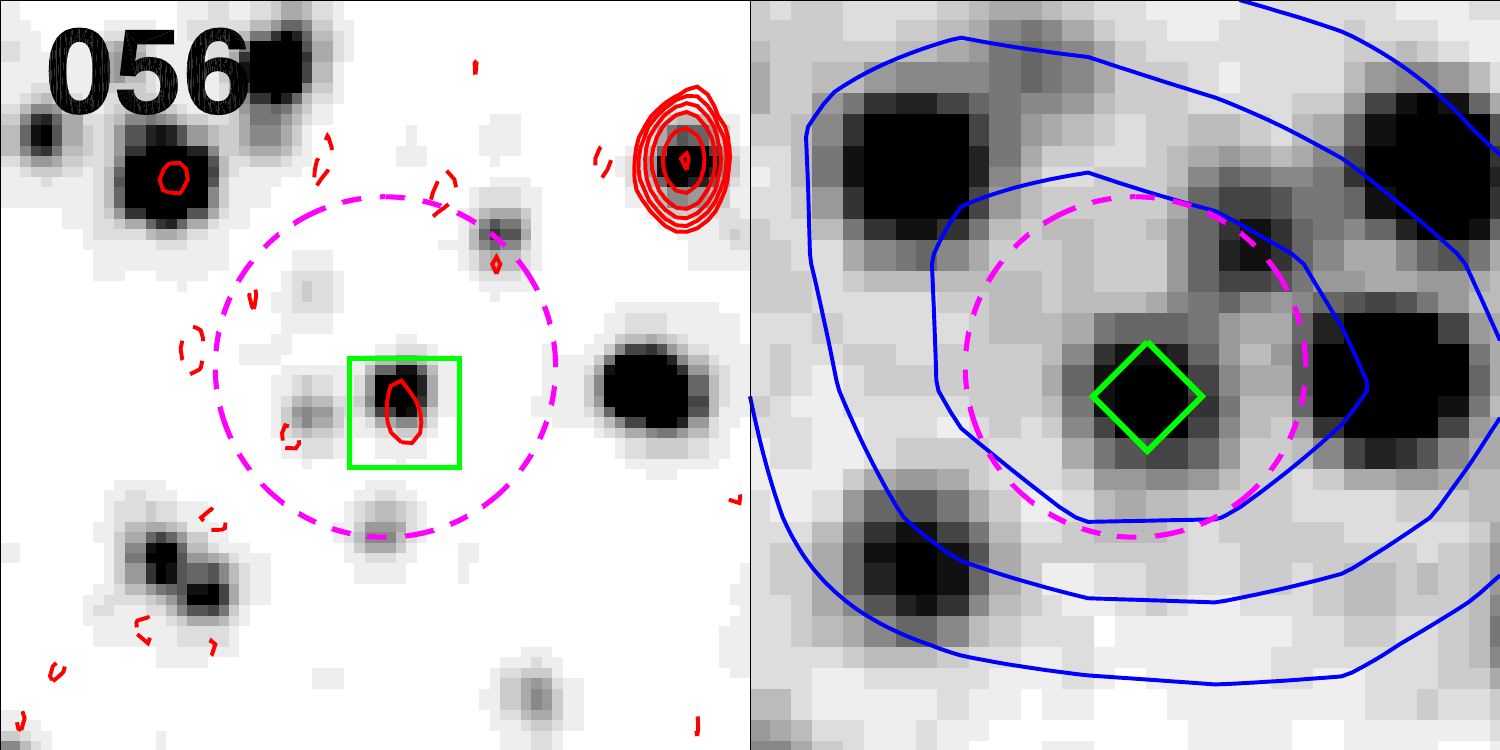}%
\hspace{1cm}%
\includegraphics[scale=0.295]{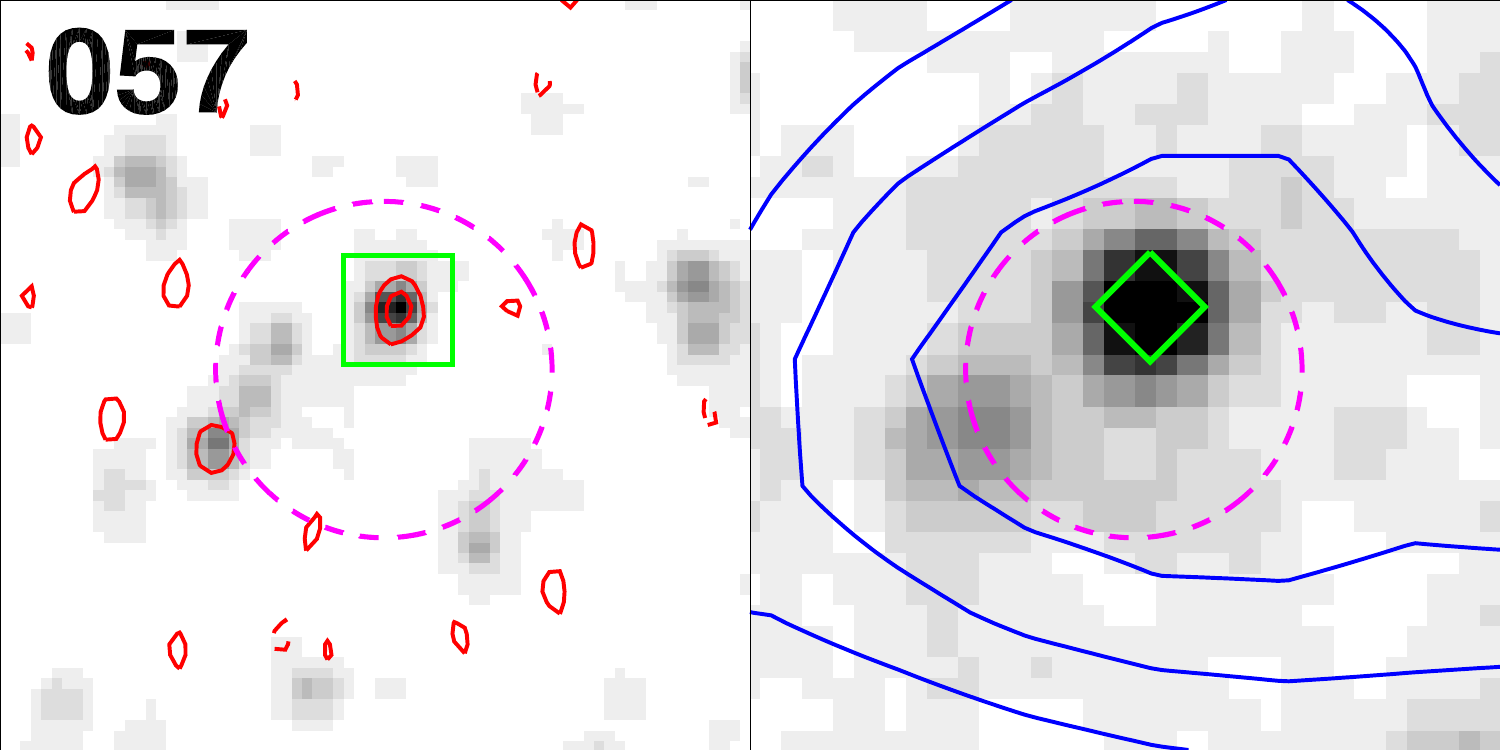}

\contcaption{}
\end{figure*}

\begin{figure*}
\begin{center}
\includegraphics[scale=0.295]{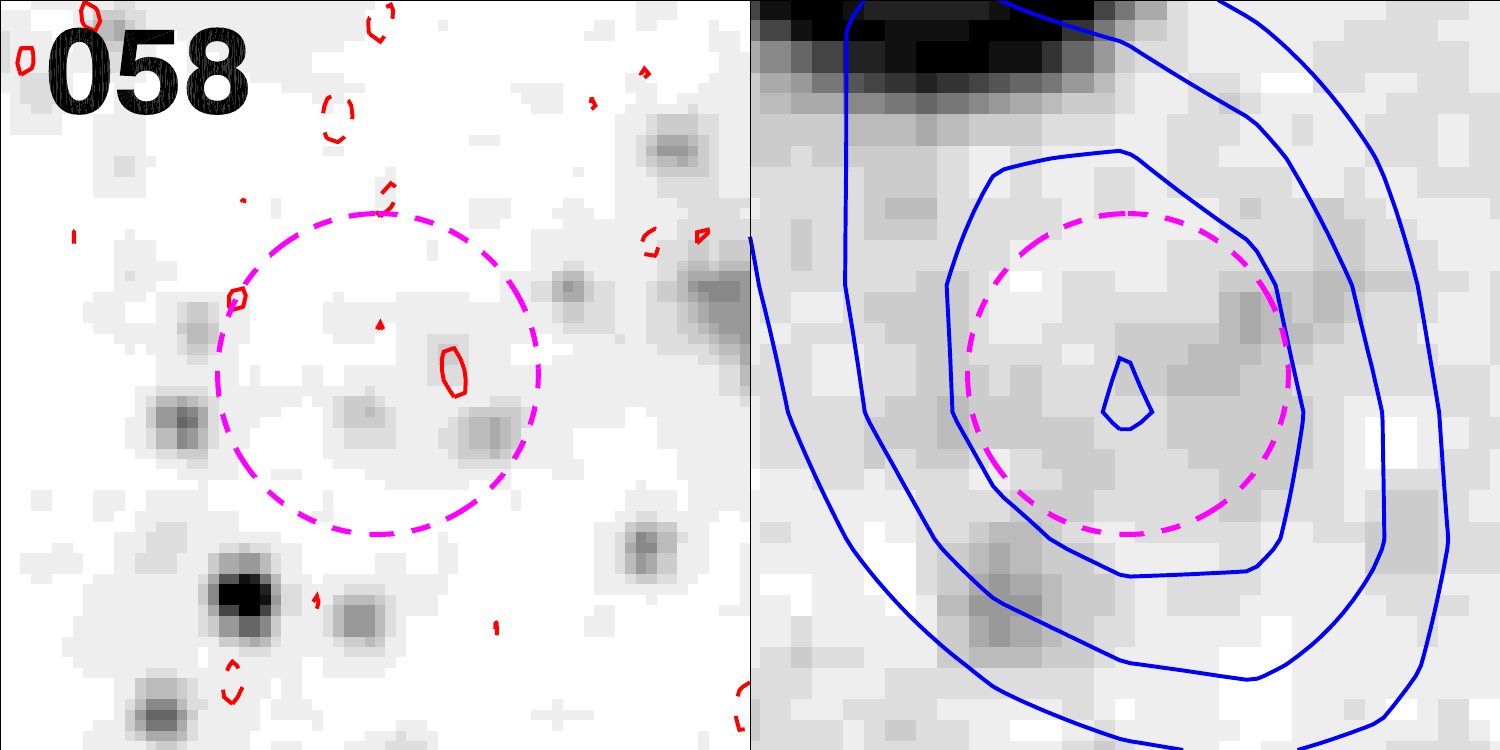}%
\hspace{1cm}%
\includegraphics[scale=0.295]{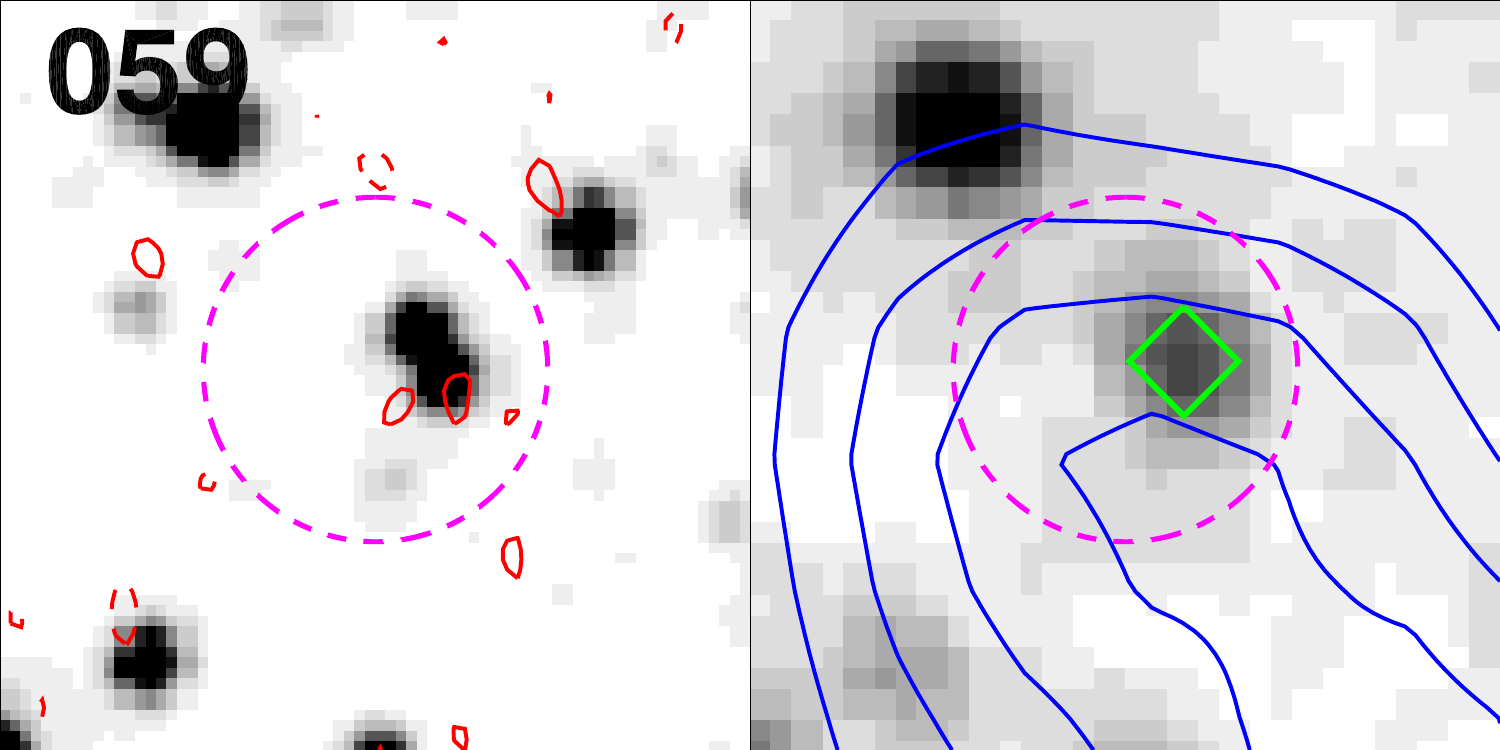}%
\hspace{1cm}%
\includegraphics[scale=0.295]{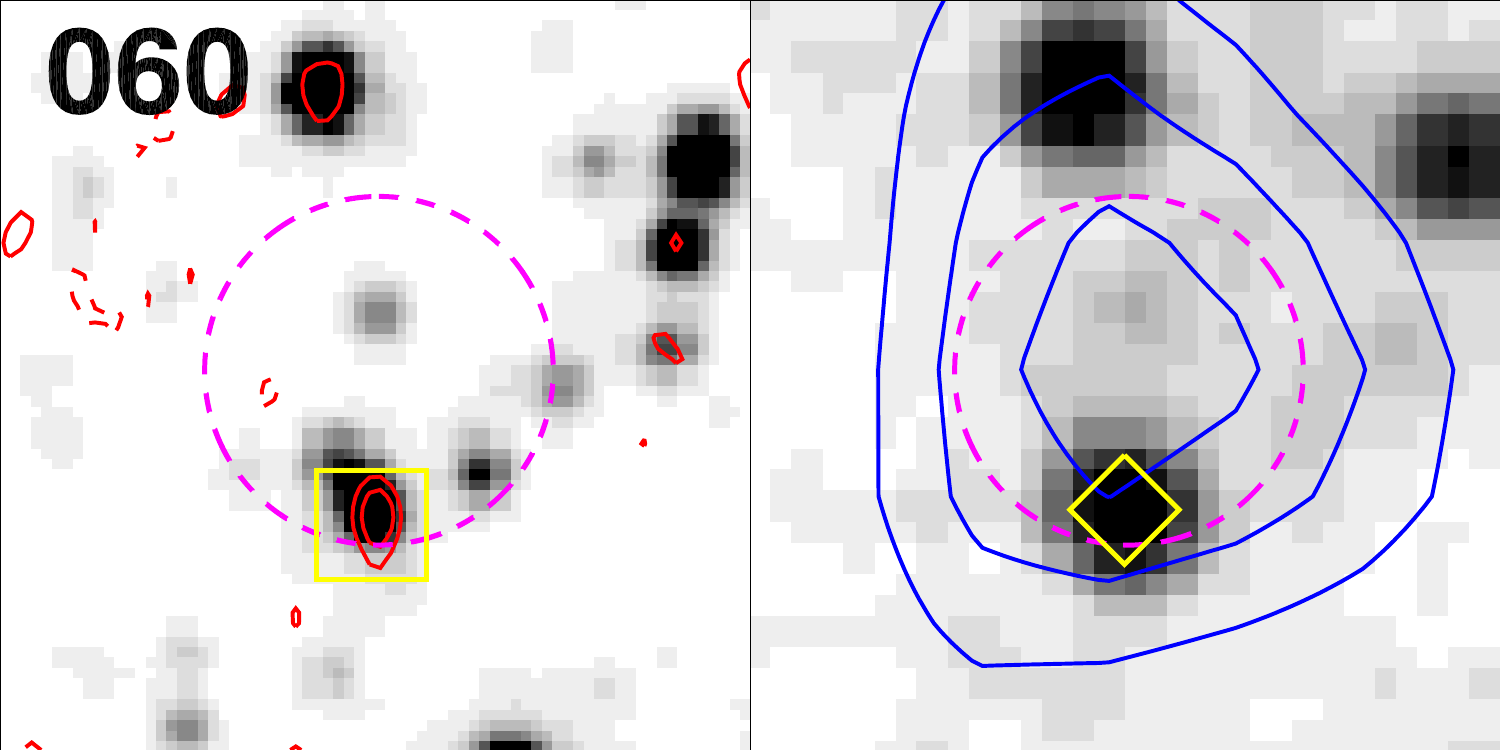}
\includegraphics[scale=0.295]{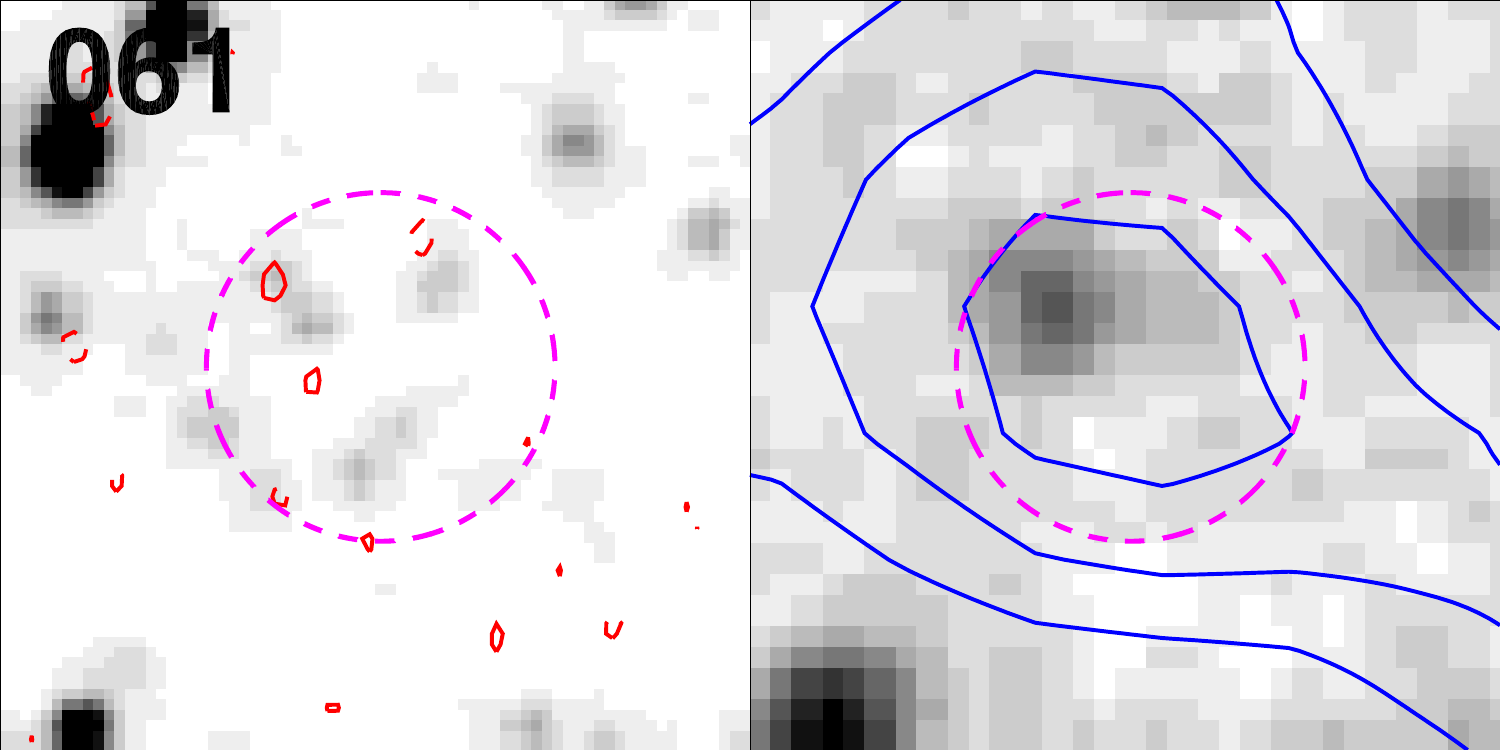}%
\hspace{1cm}%
\includegraphics[scale=0.295]{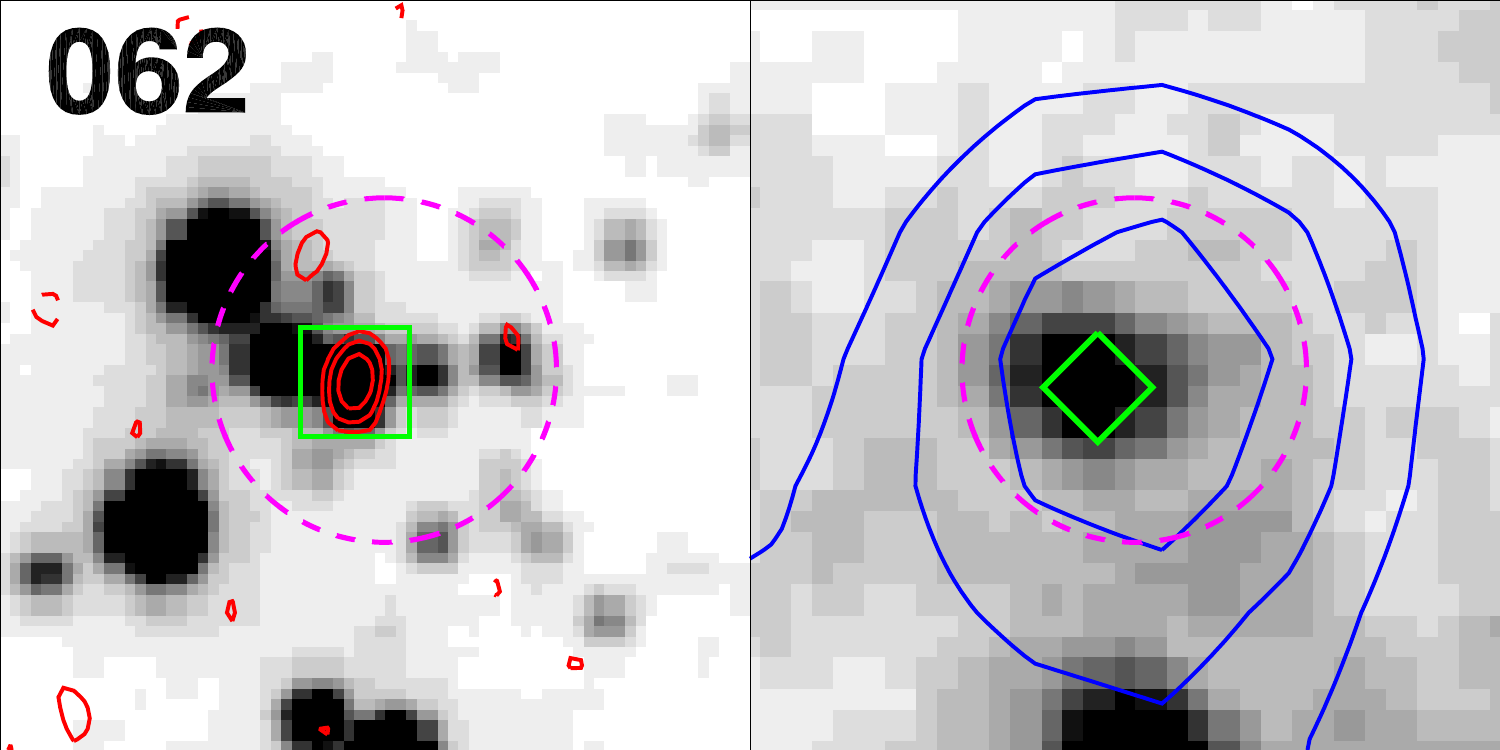}%
\hspace{1cm}%
\includegraphics[scale=0.295]{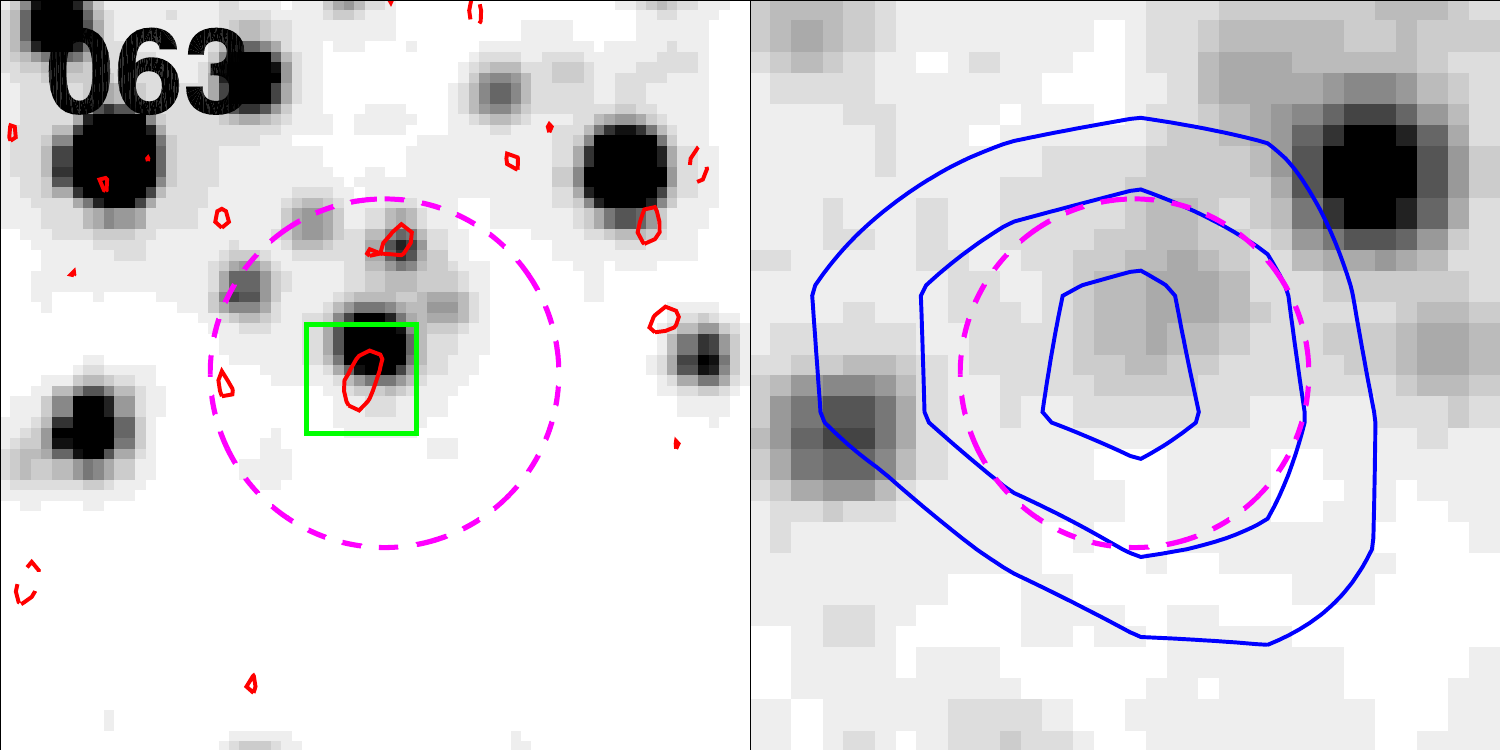}
\includegraphics[scale=0.295]{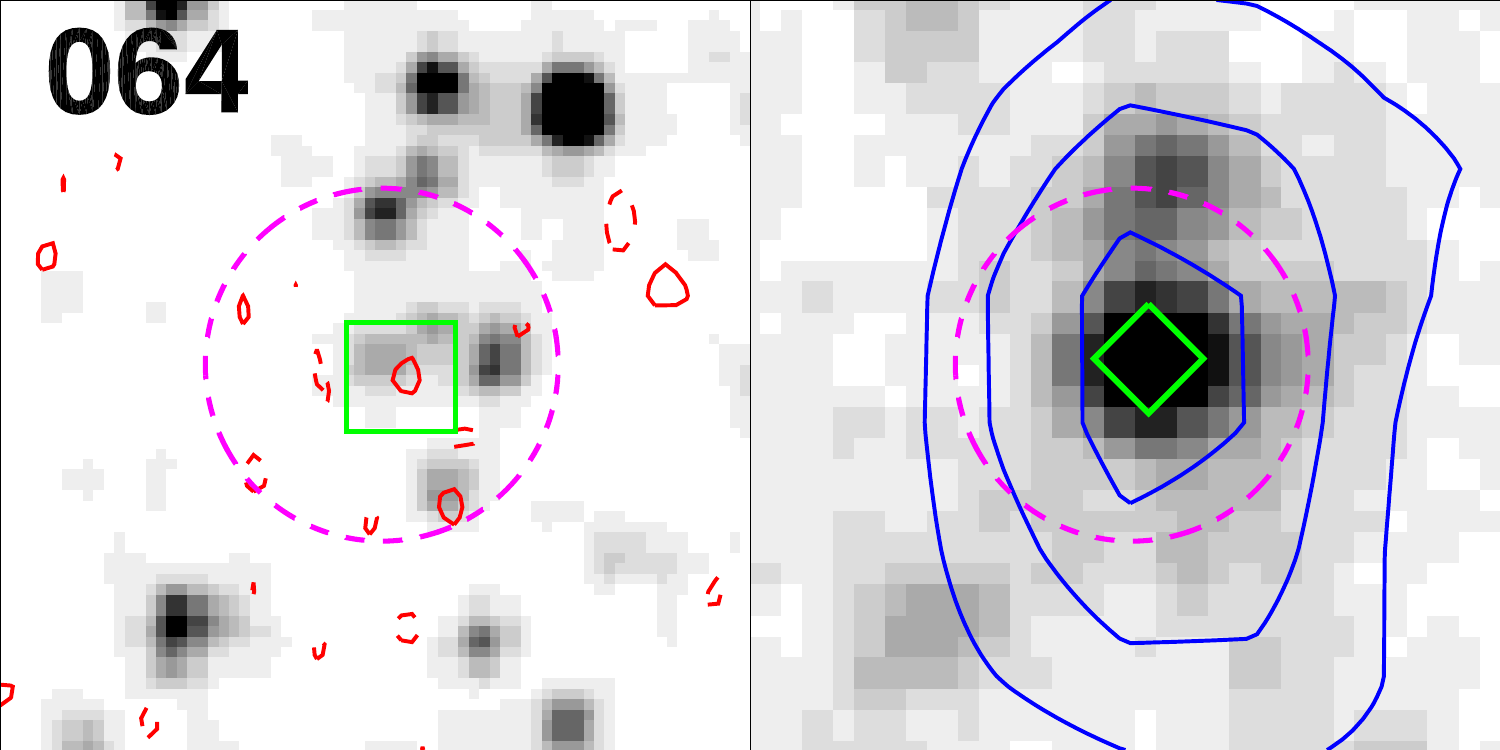}%
\hspace{1cm}%
\includegraphics[scale=0.295]{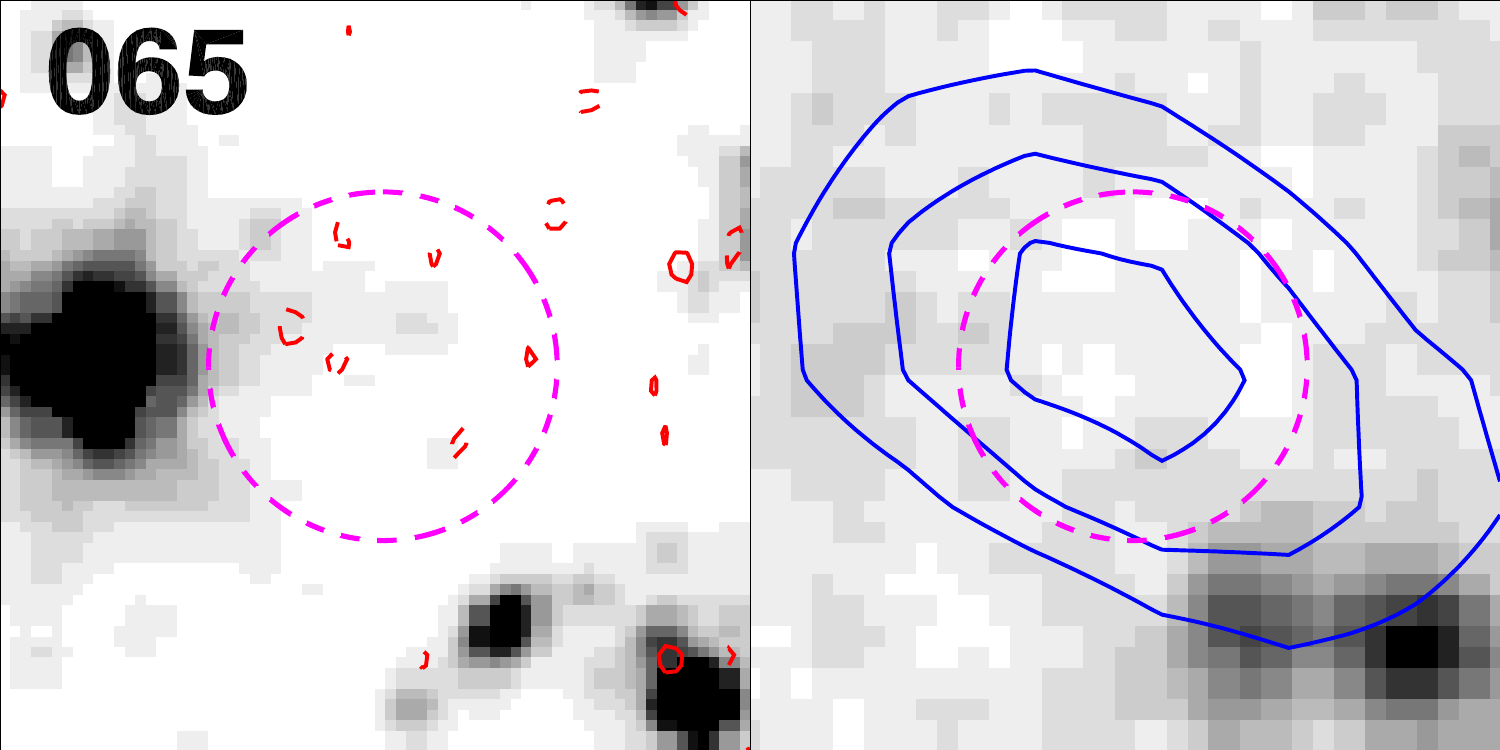}%
\hspace{1cm}%
\includegraphics[scale=0.295]{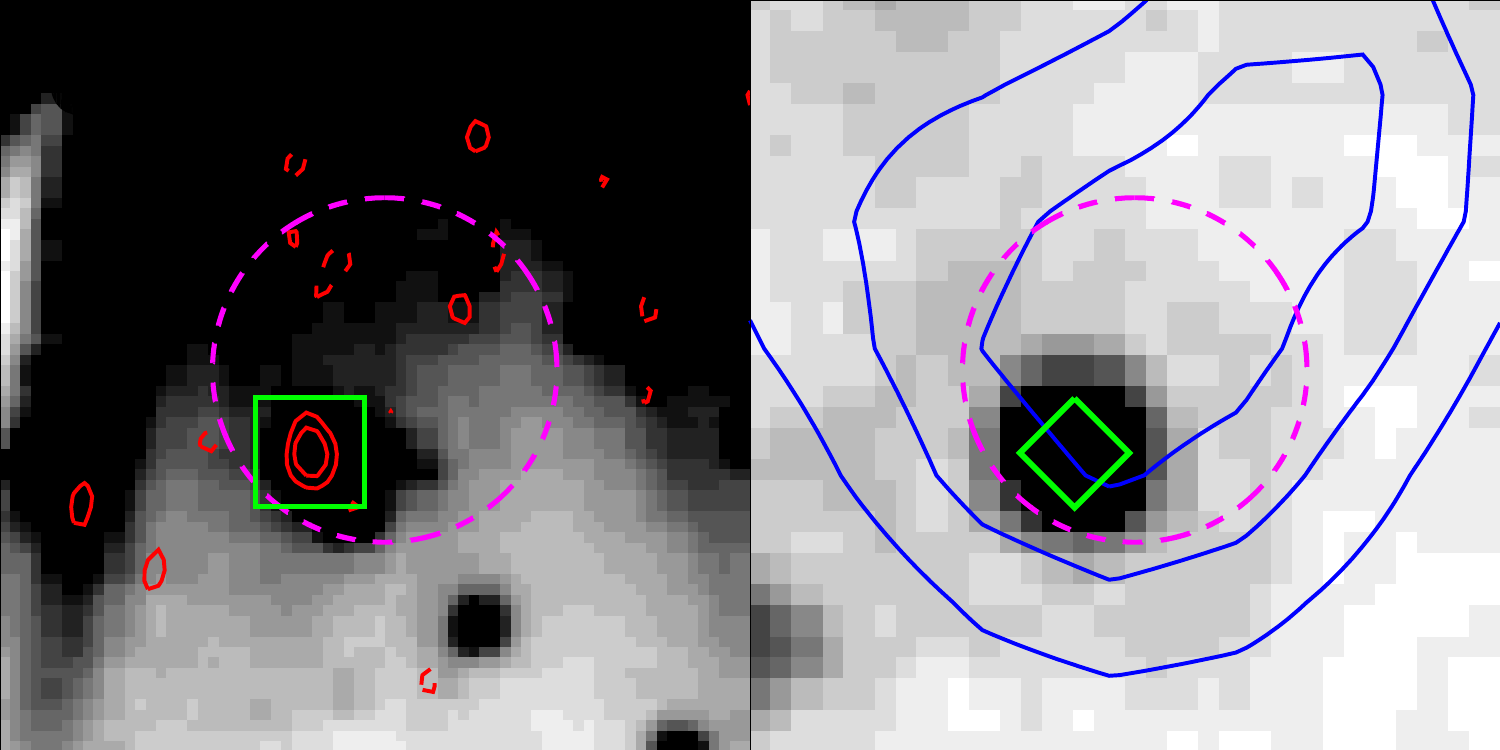}
\includegraphics[scale=0.295]{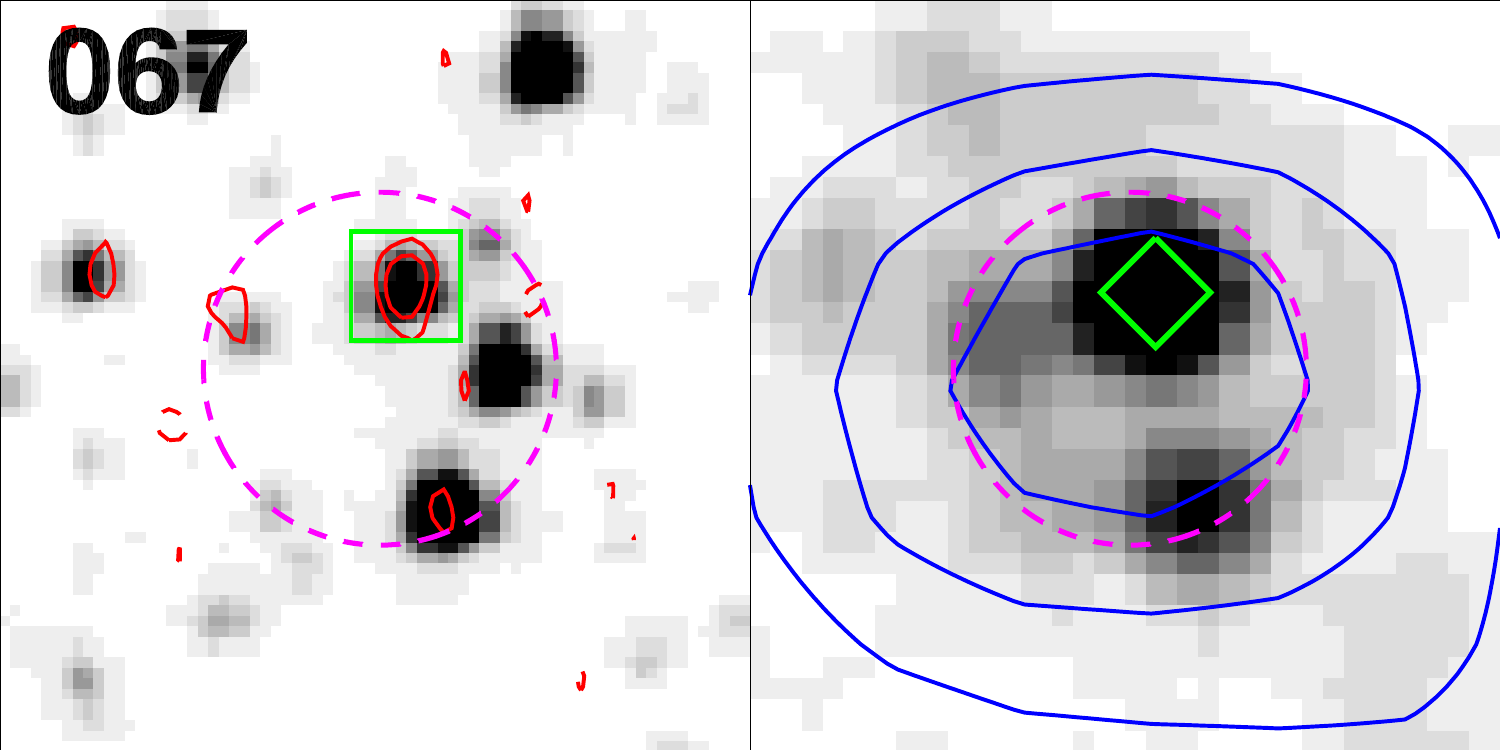}%
\hspace{1cm}%
\includegraphics[scale=0.295]{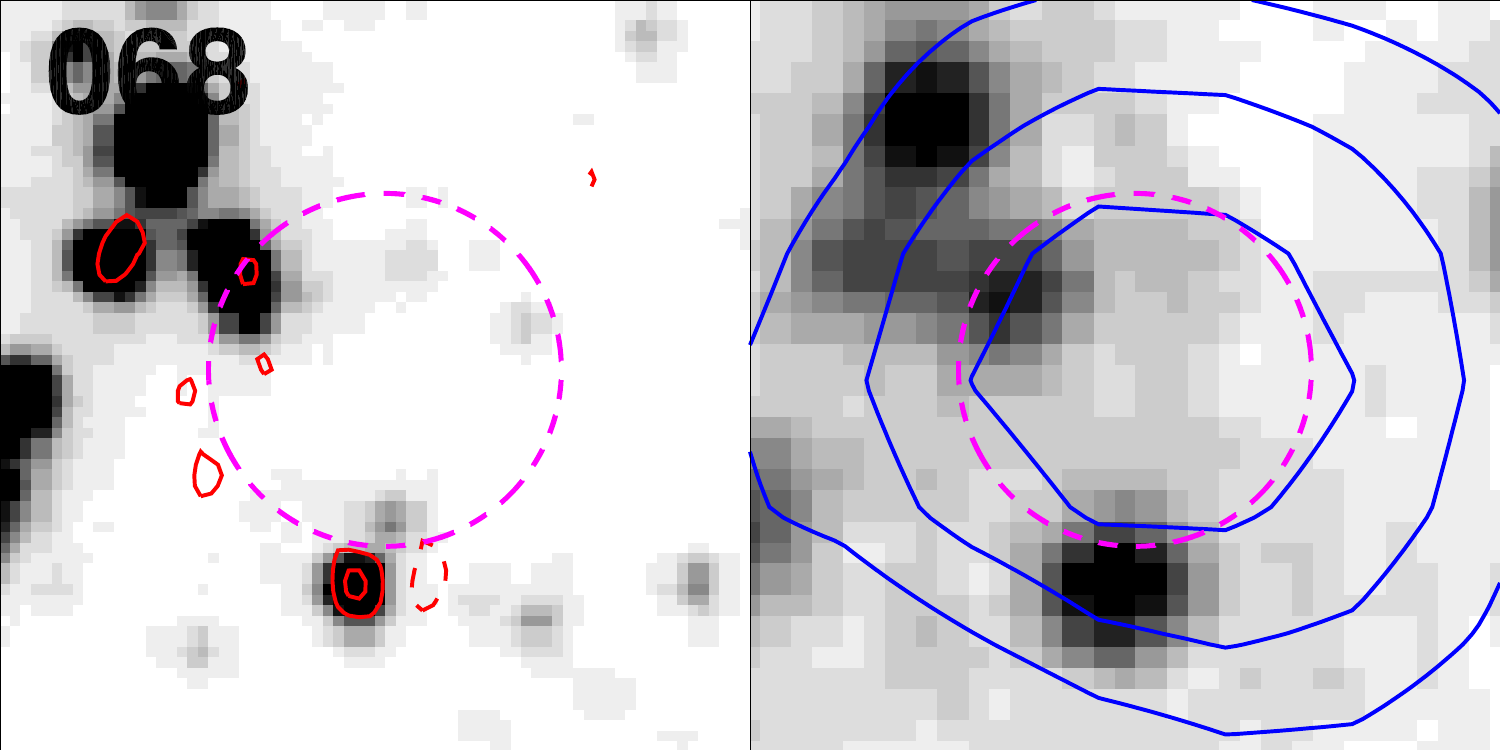}%
\hspace{1cm}%
\includegraphics[scale=0.295]{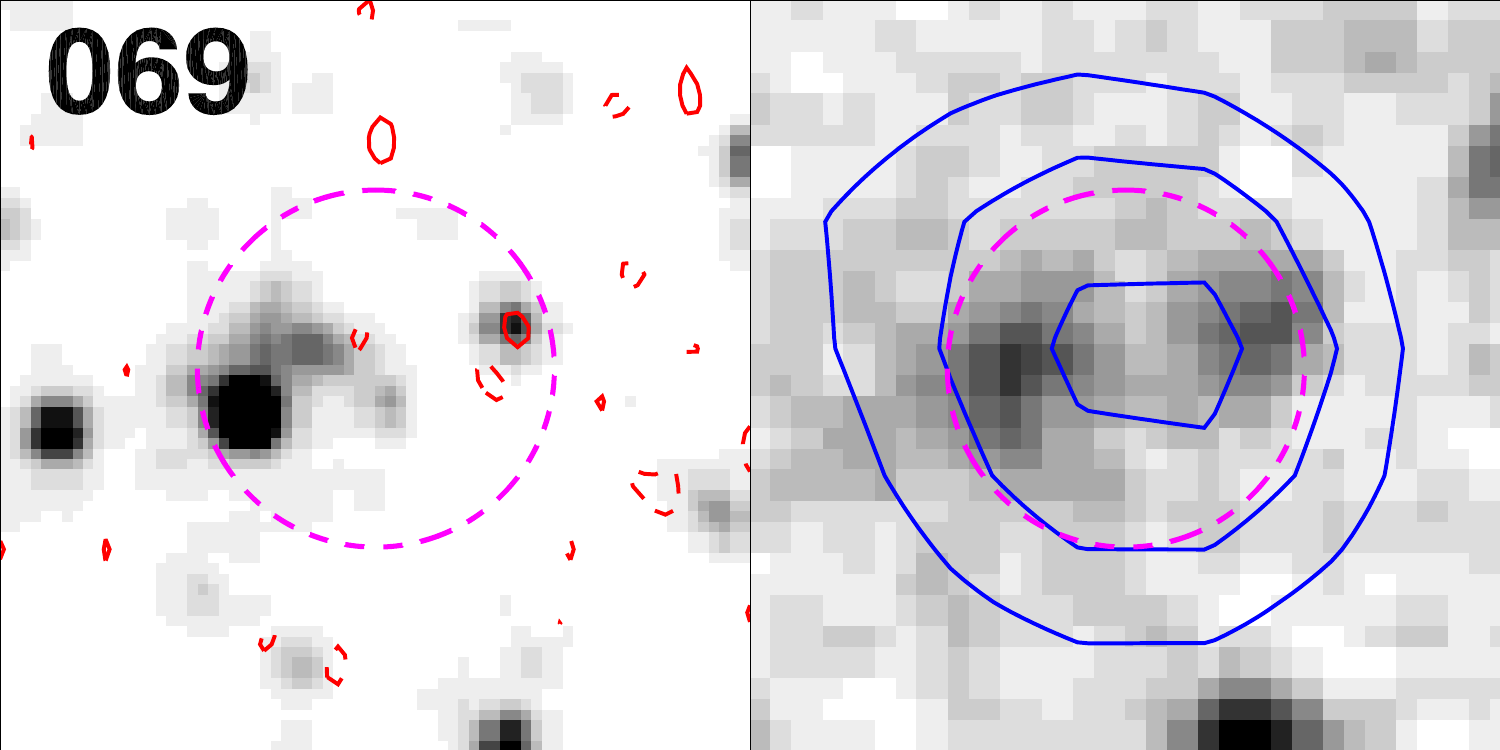}
\includegraphics[scale=0.295]{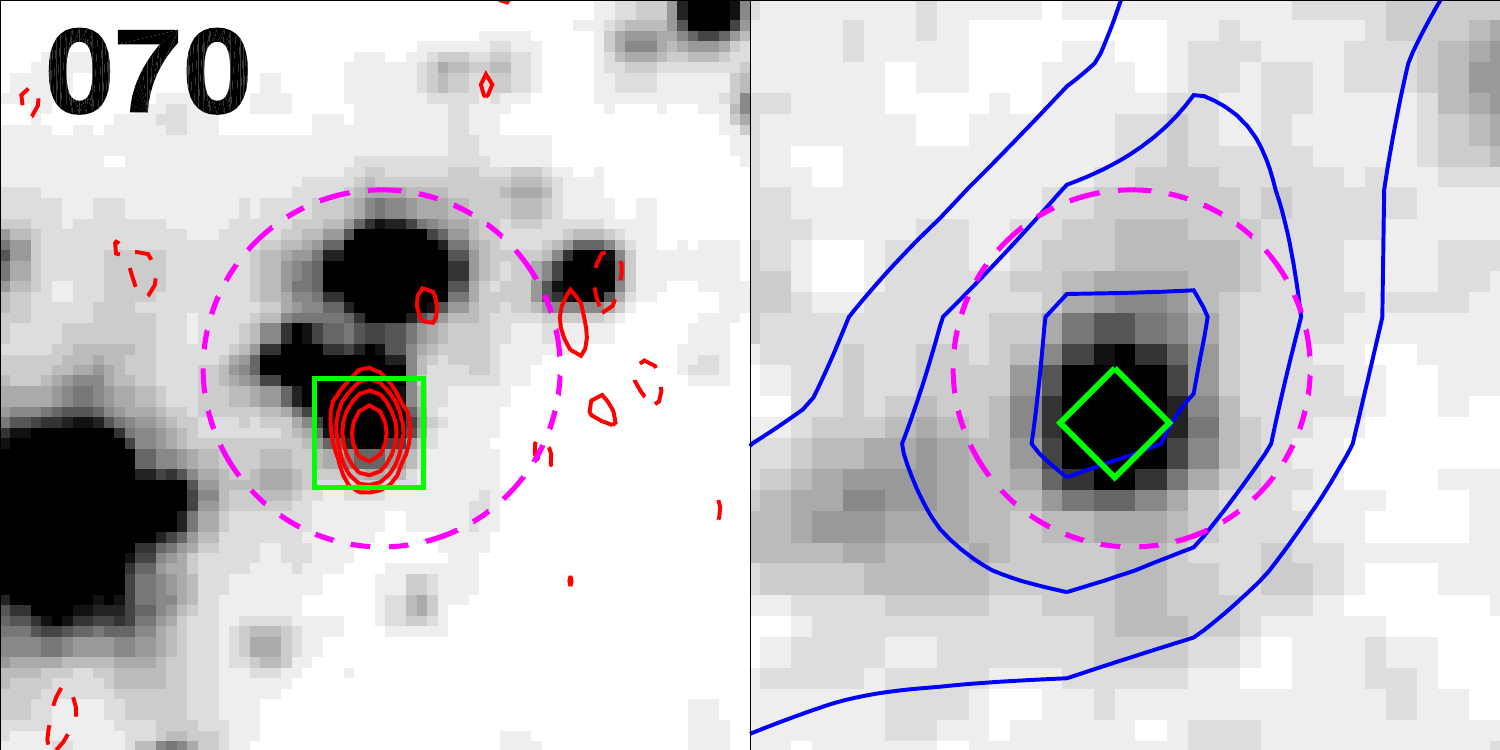}%
\hspace{1cm}%
\includegraphics[scale=0.295]{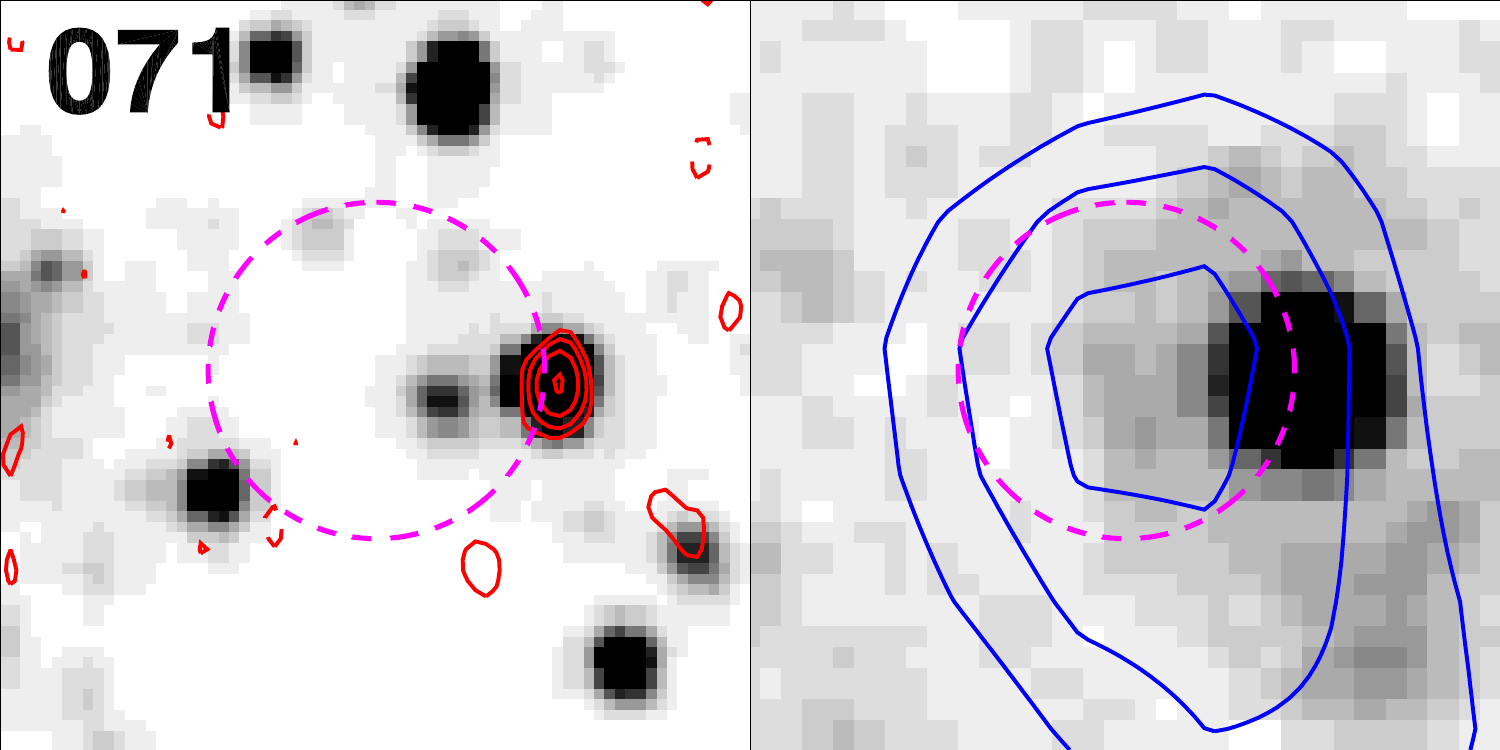}%
\hspace{1cm}%
\includegraphics[scale=0.295]{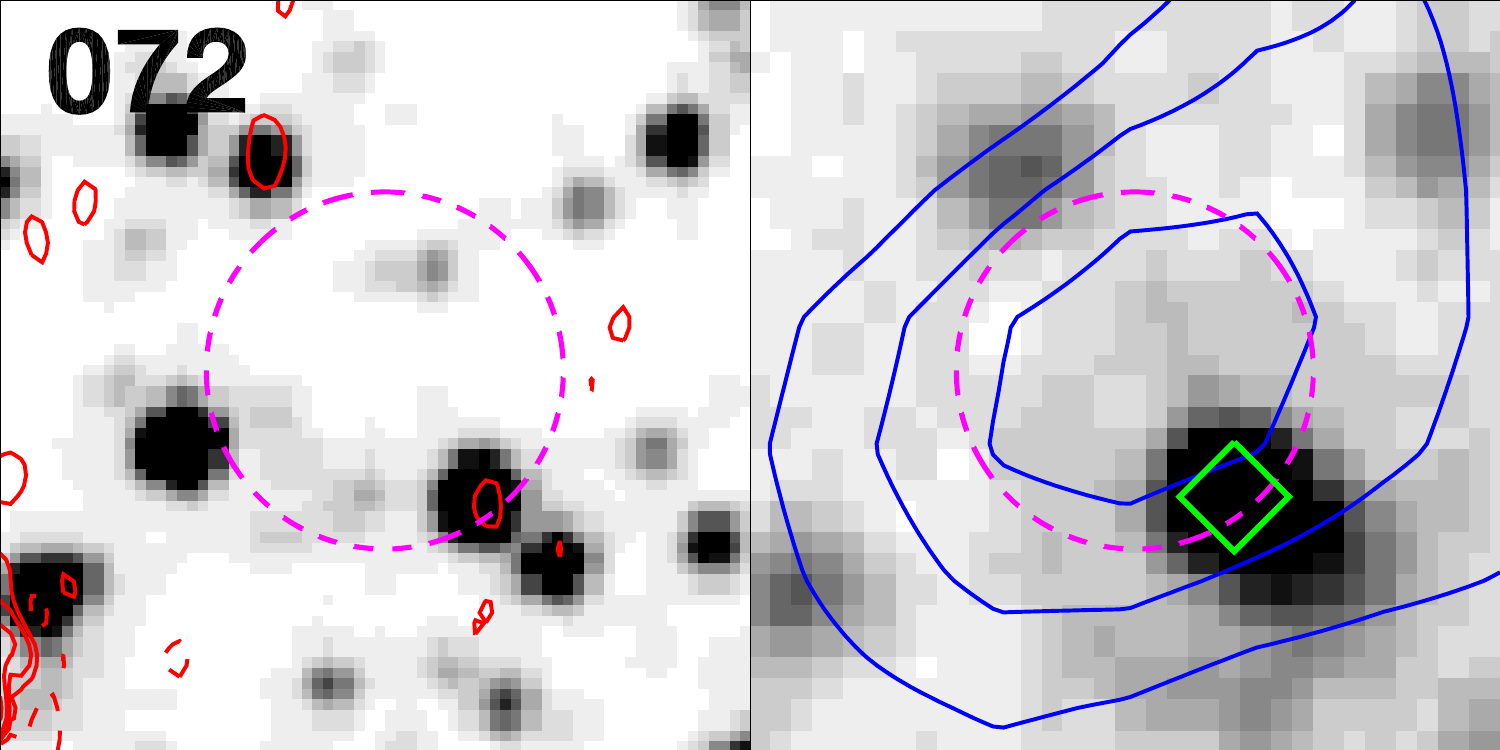}
\includegraphics[scale=0.295]{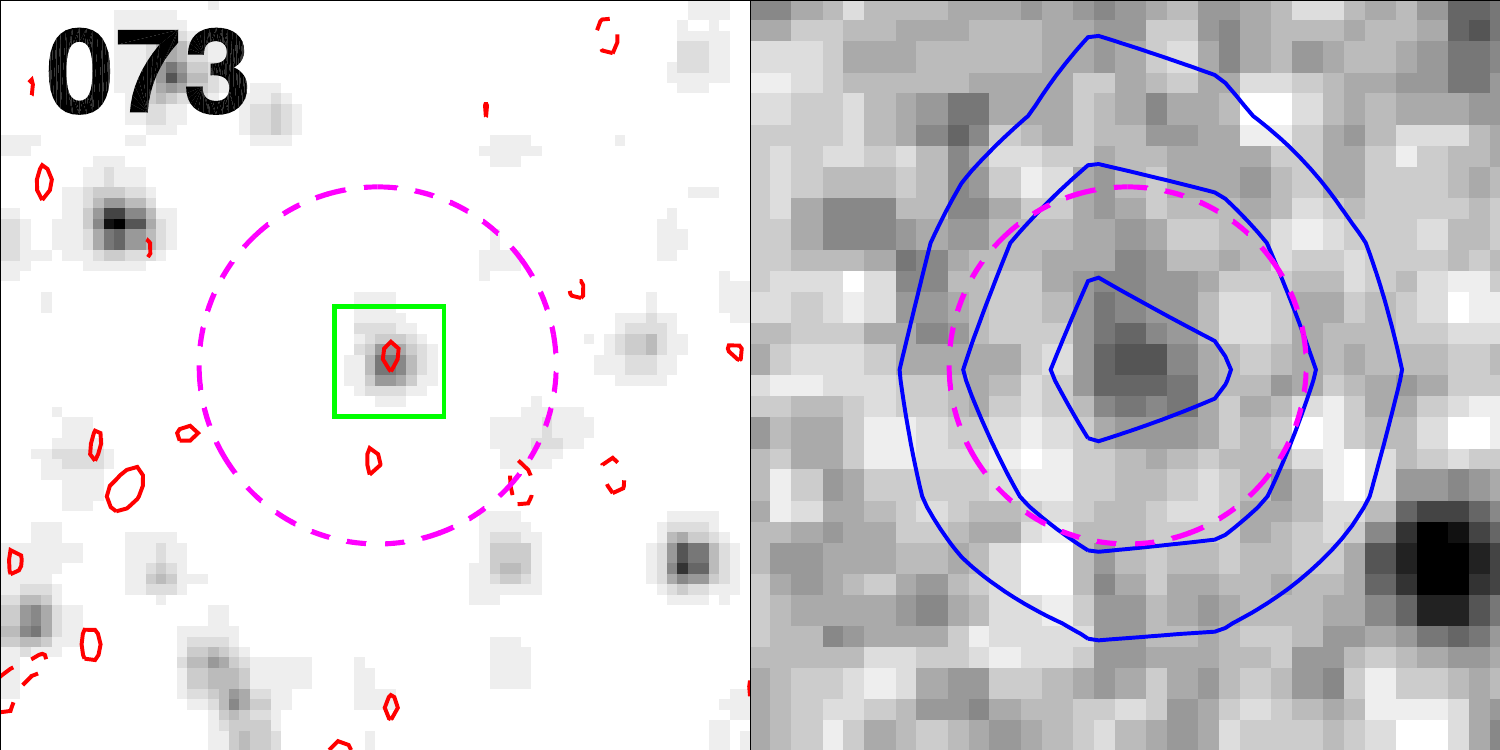}%
\hspace{1cm}%
\includegraphics[scale=0.295]{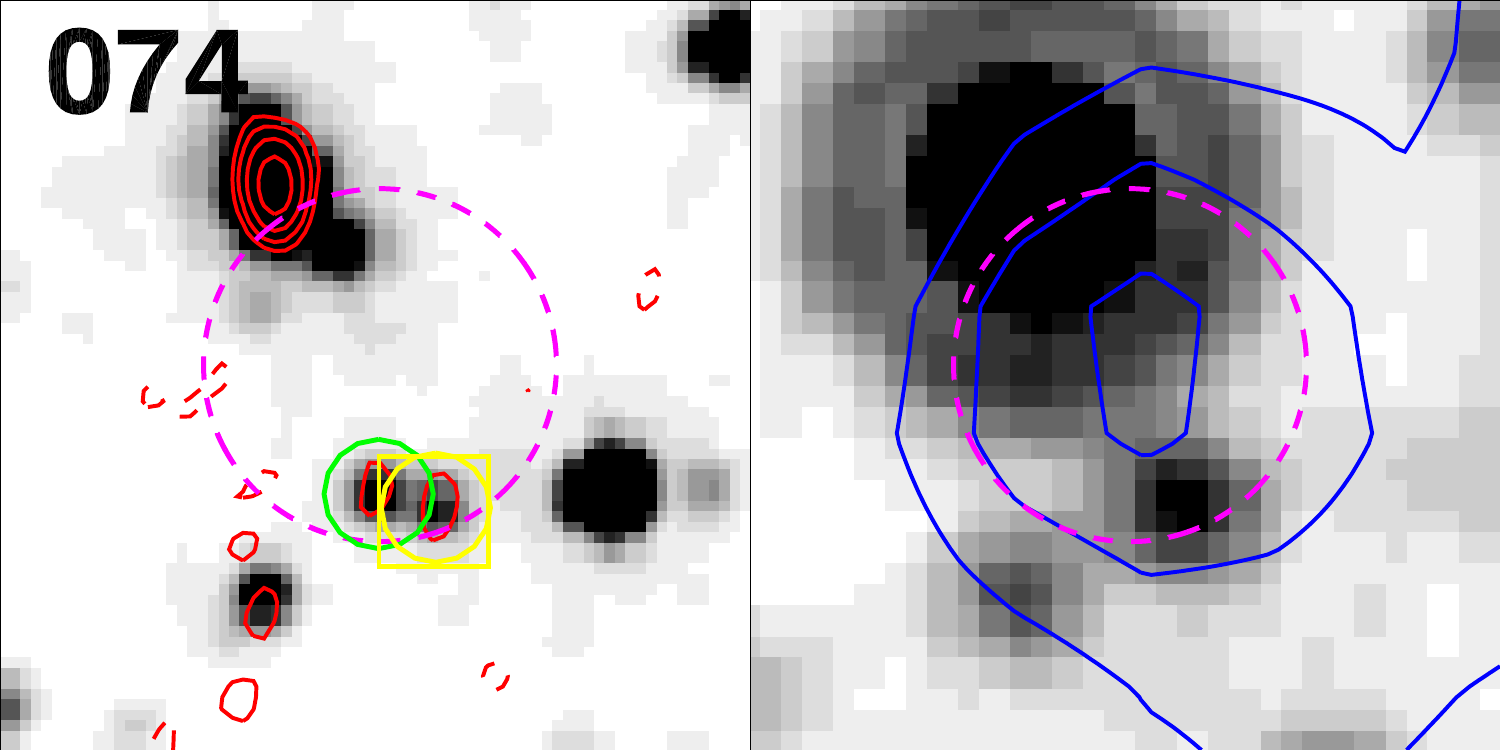}%
\hspace{1cm}%
\includegraphics[scale=0.295]{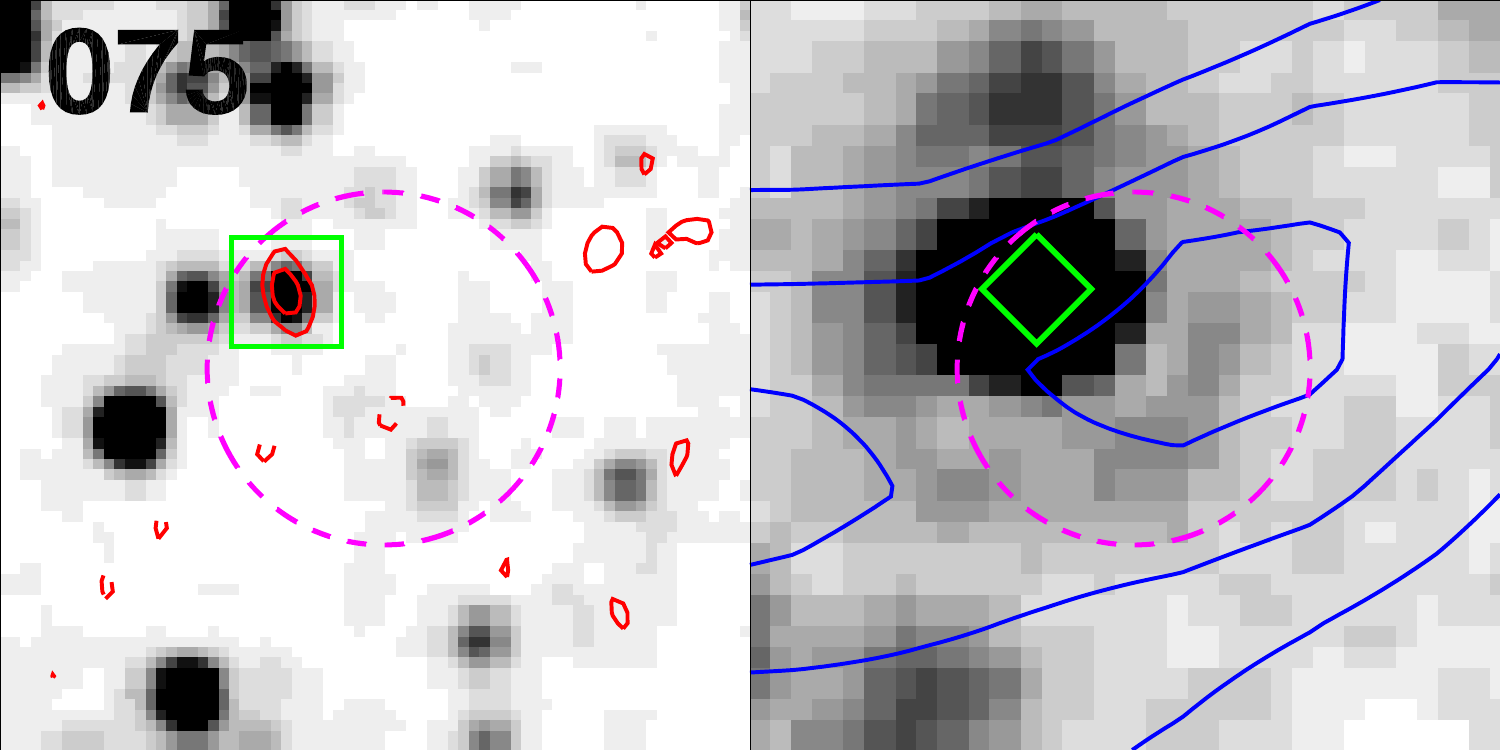}
\includegraphics[scale=0.295]{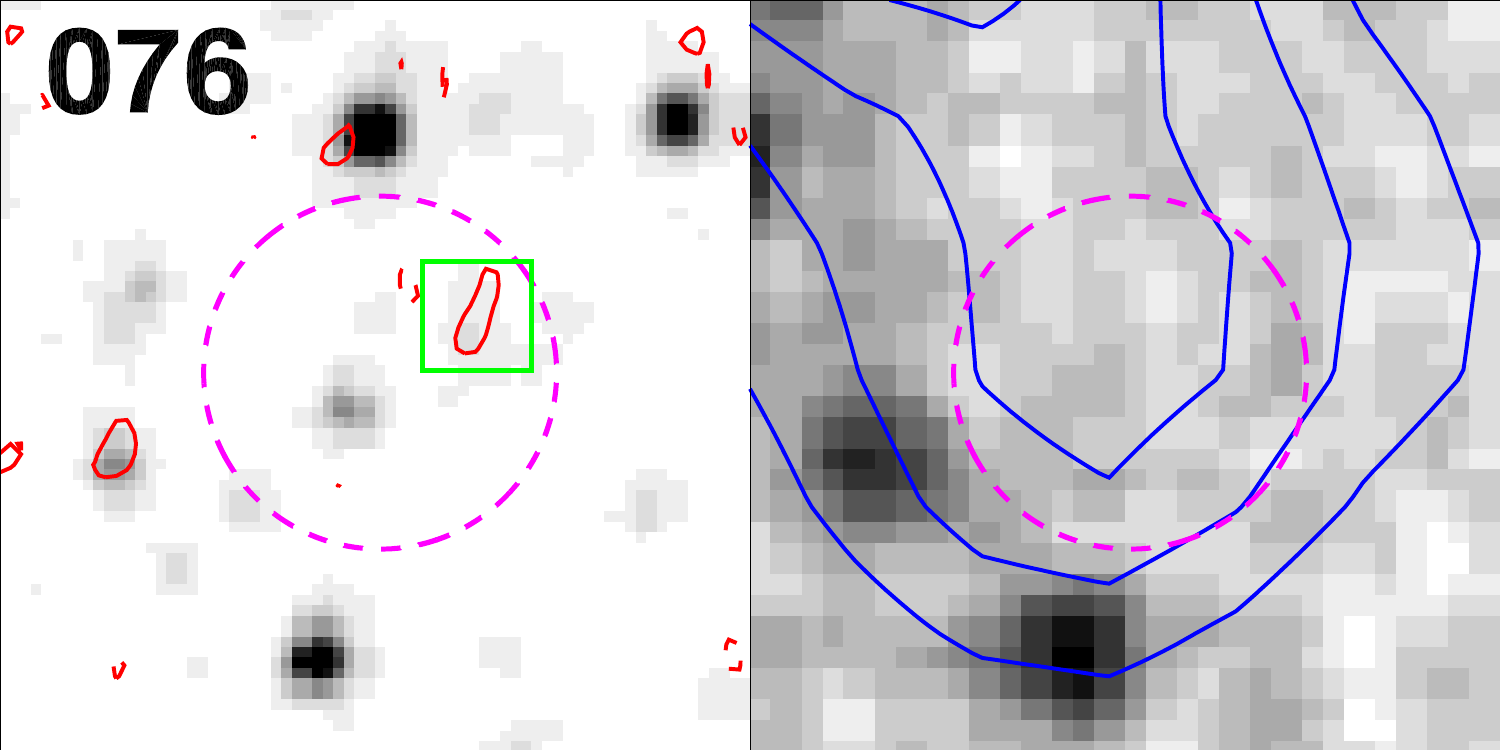}%
\hspace{1cm}%
\includegraphics[scale=0.295]{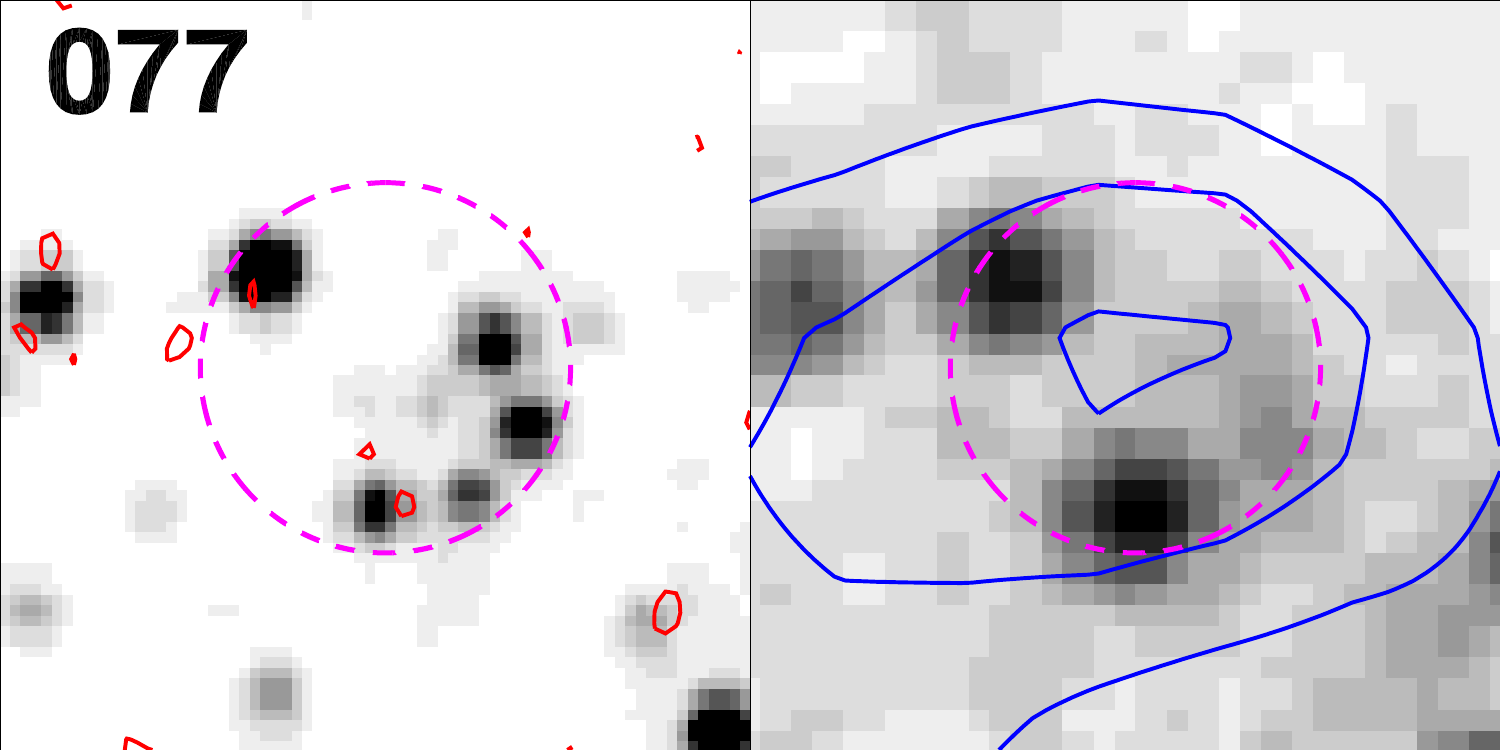}%
\hspace{1cm}%
\includegraphics[scale=0.295]{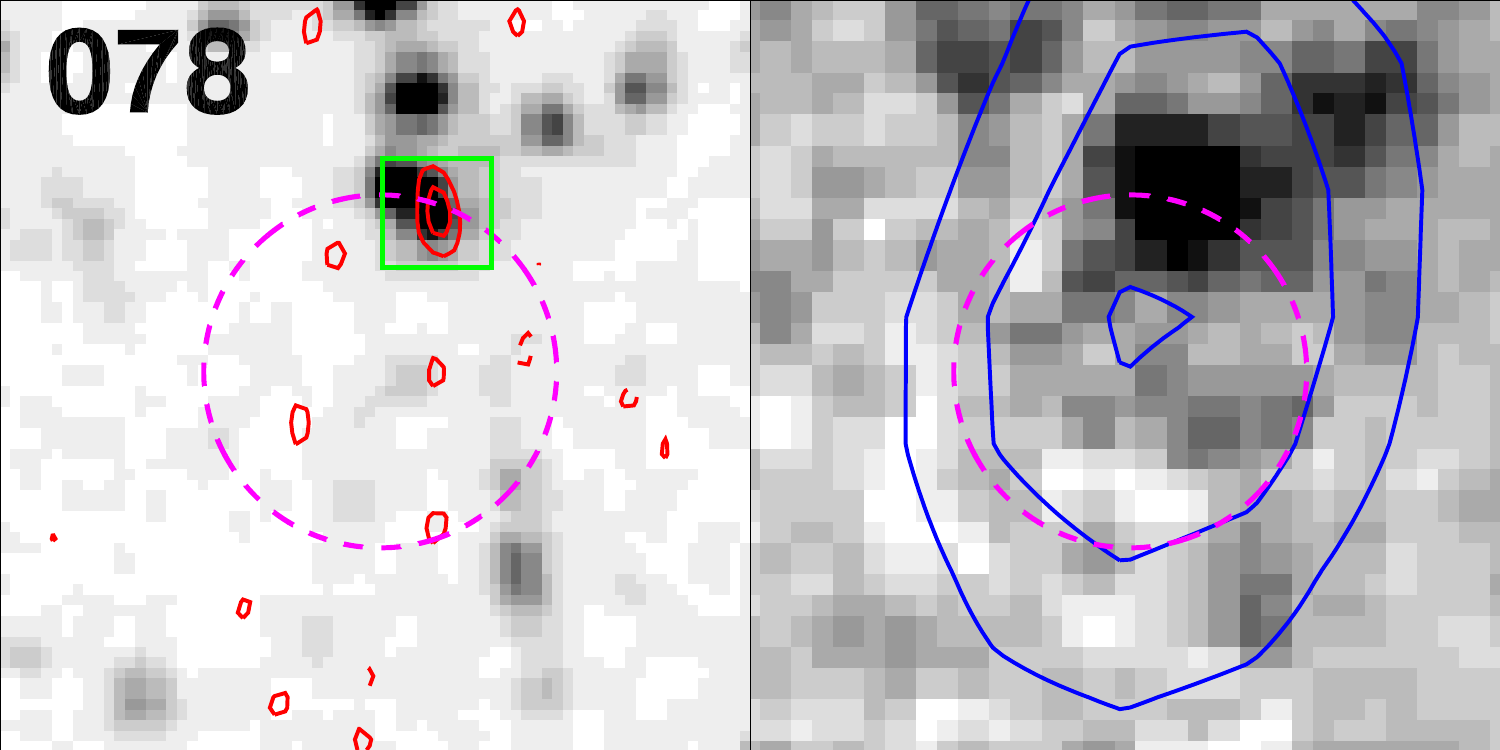}
\includegraphics[scale=0.295]{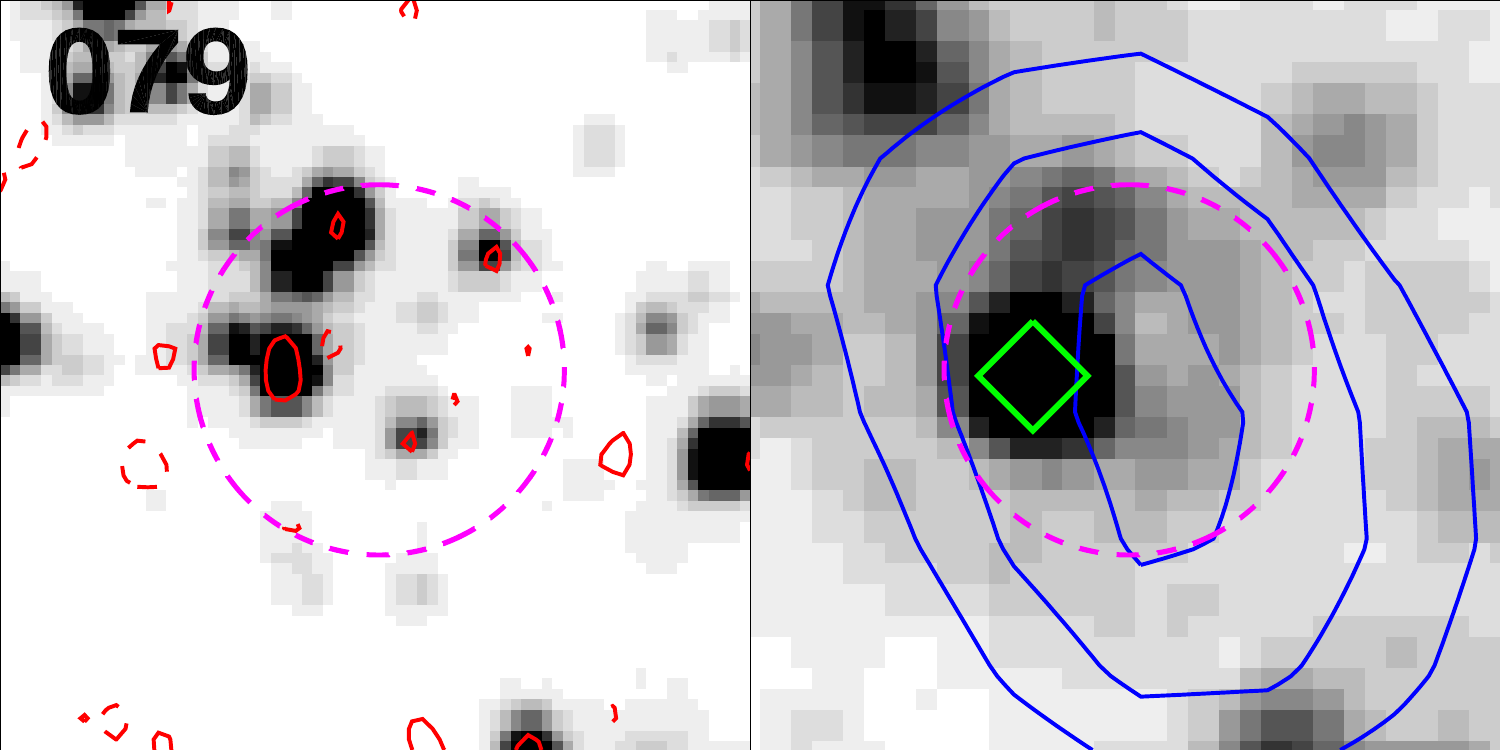}%
\hspace{1cm}%
\includegraphics[scale=0.295]{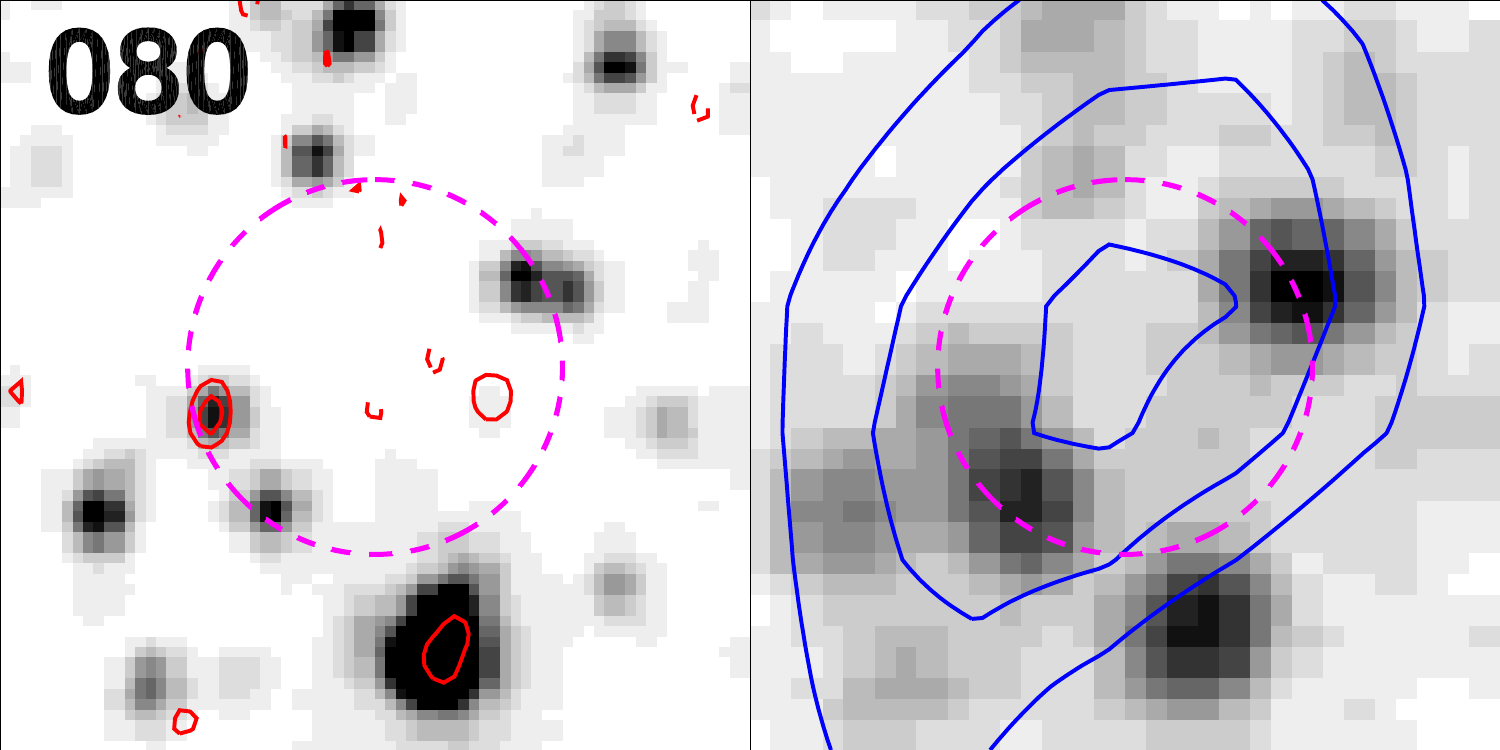}%
\hspace{1cm}%
\includegraphics[scale=0.295]{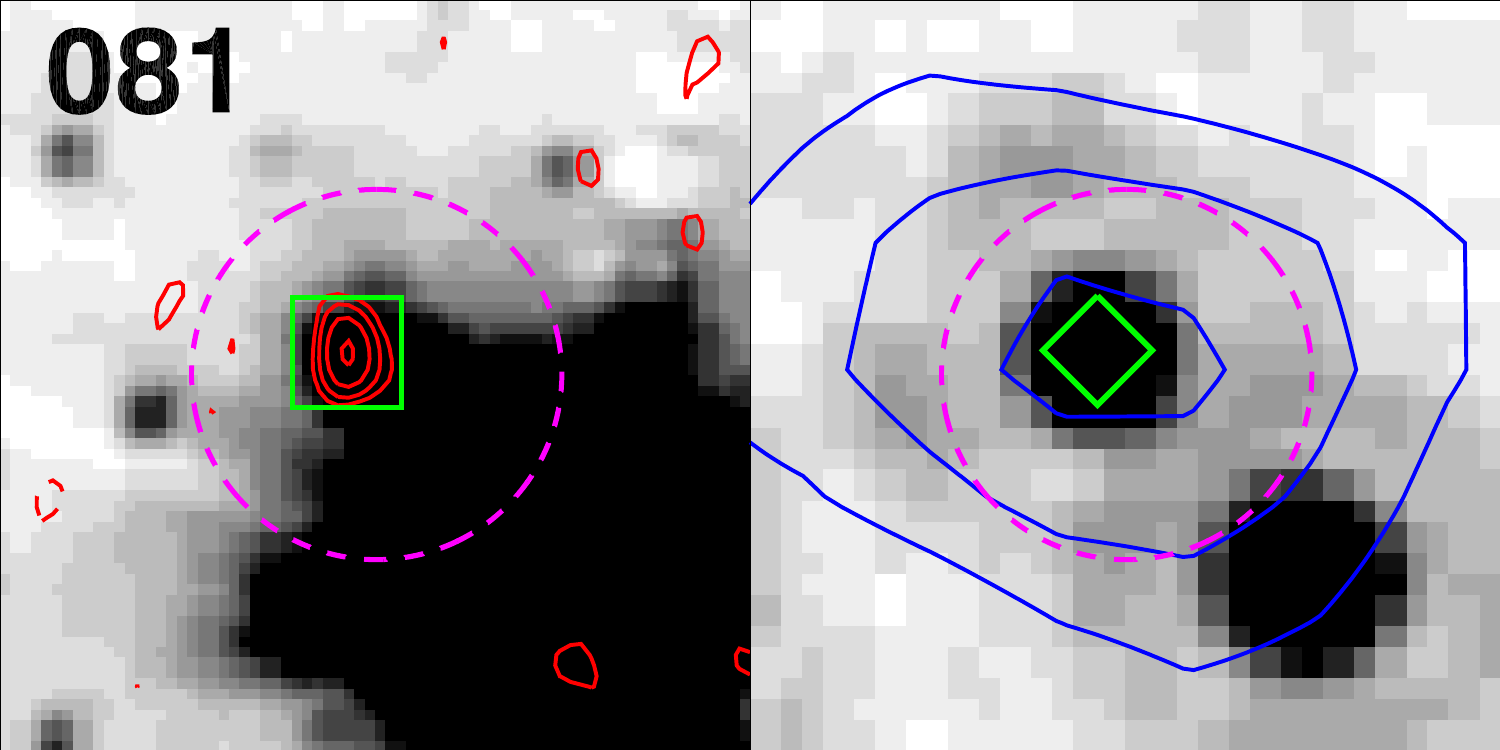}
\includegraphics[scale=0.295]{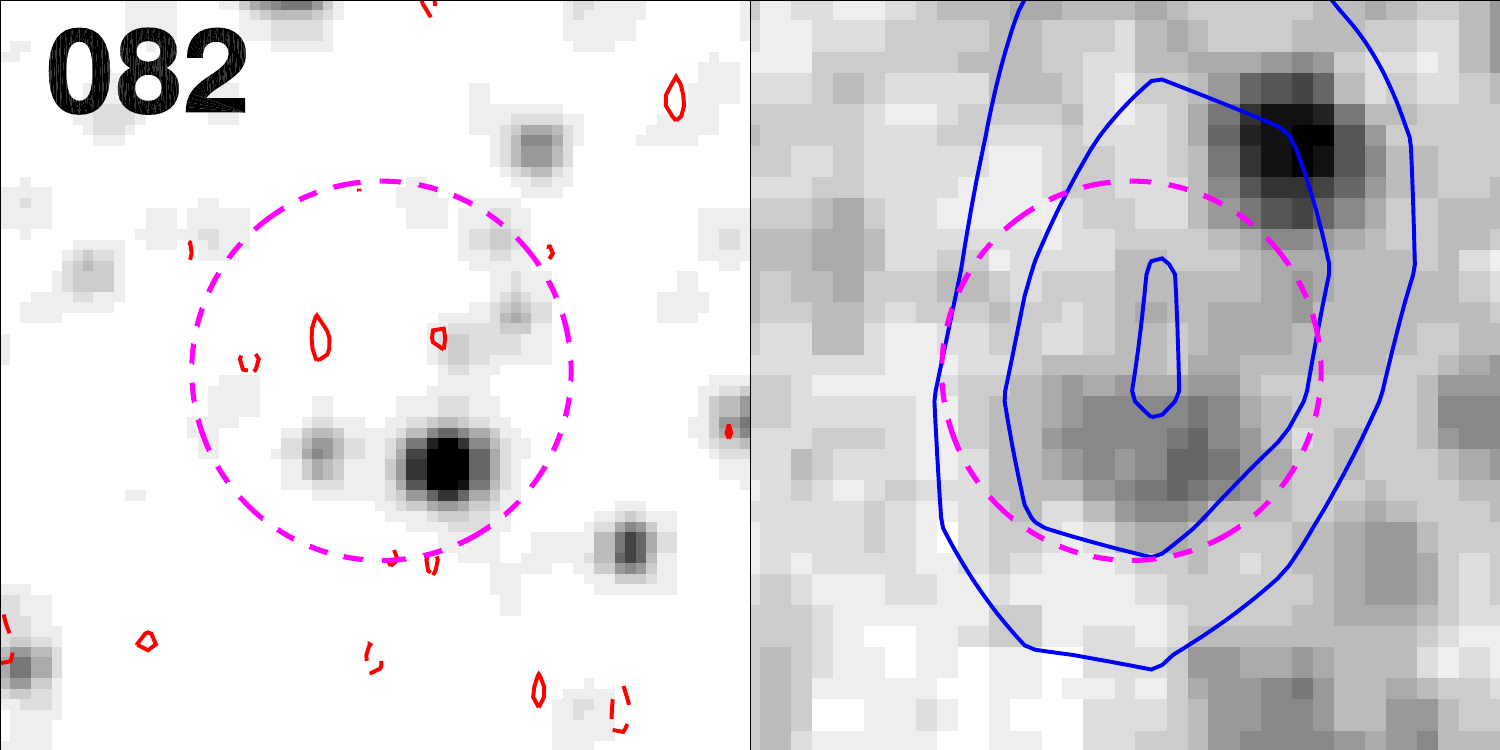}%
\hspace{1cm}%
\includegraphics[scale=0.295]{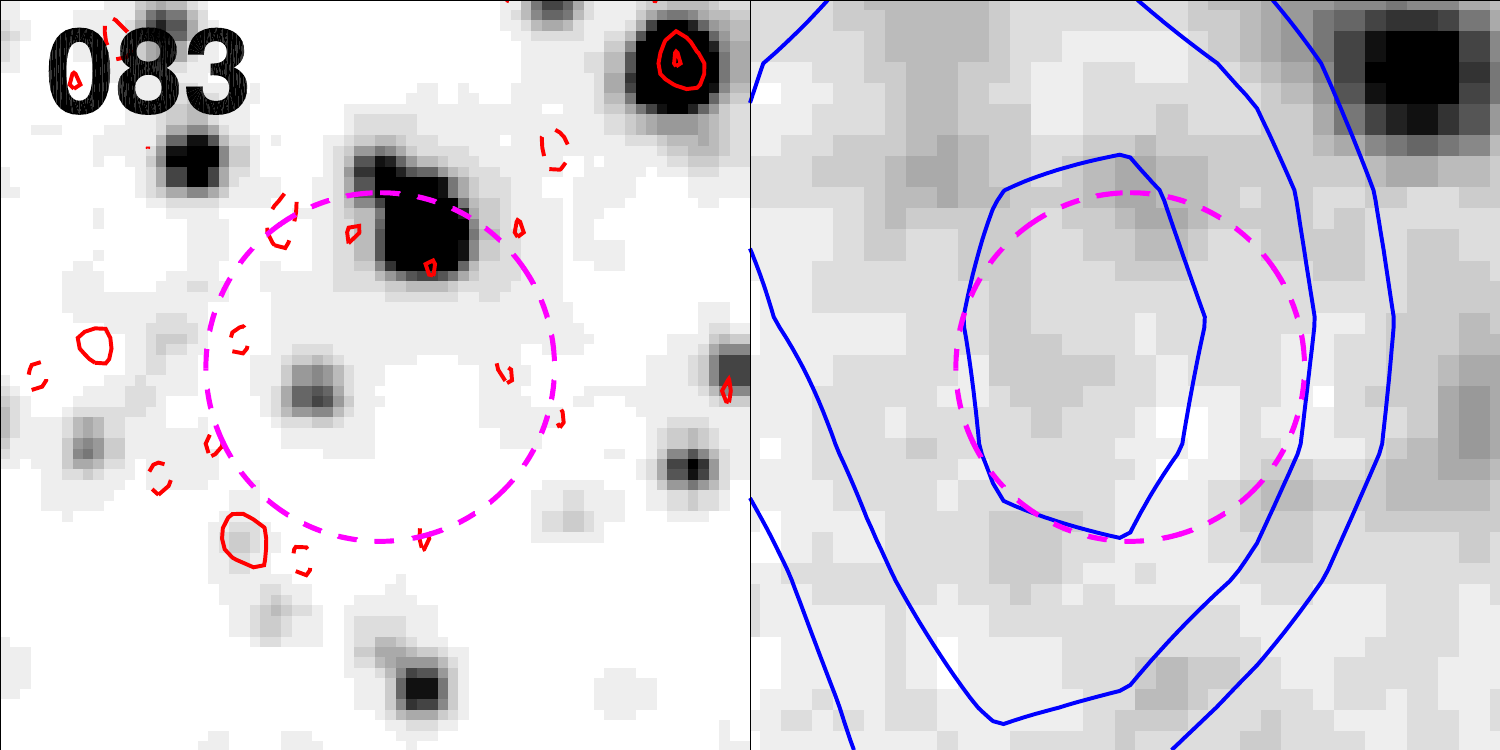}%
\hspace{1cm}%
\includegraphics[scale=0.295]{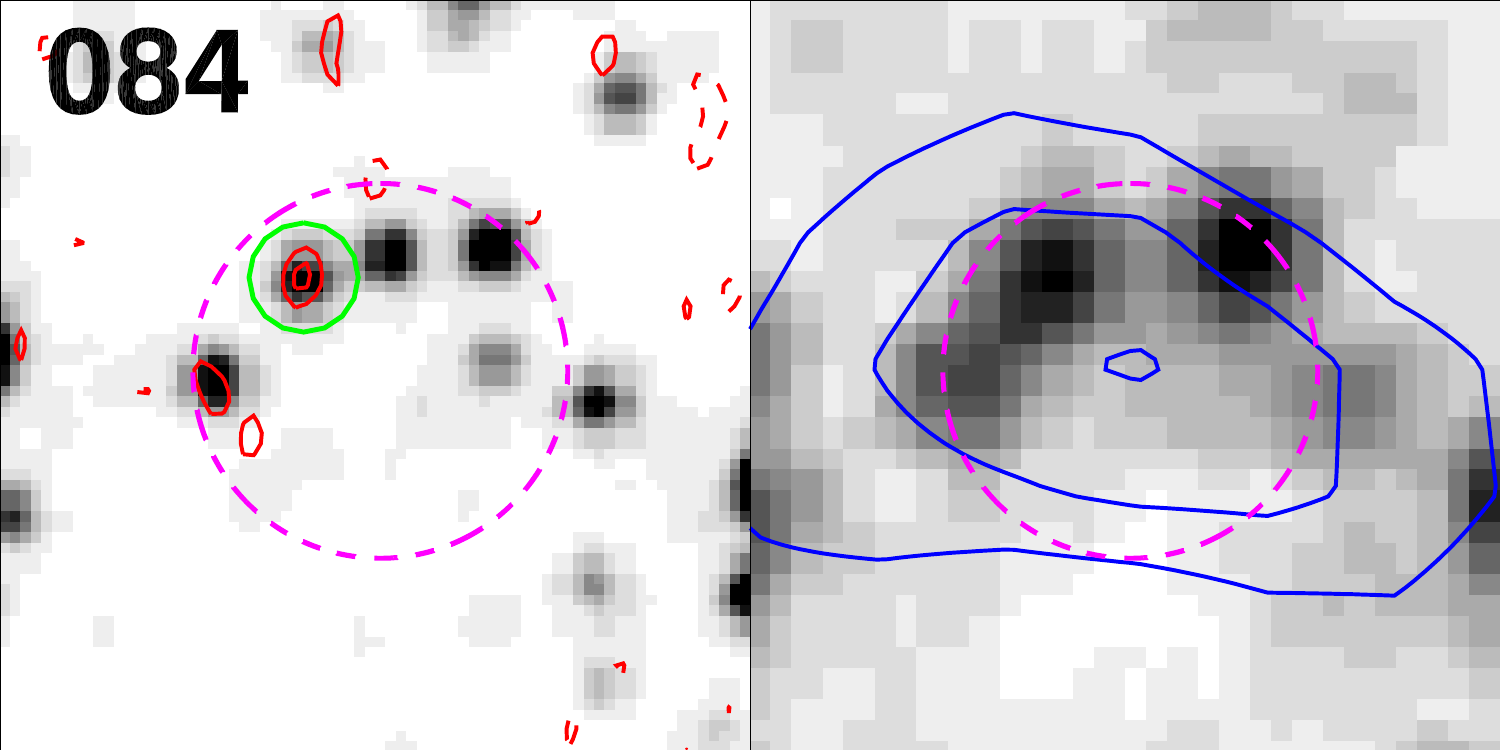}
\includegraphics[scale=0.295]{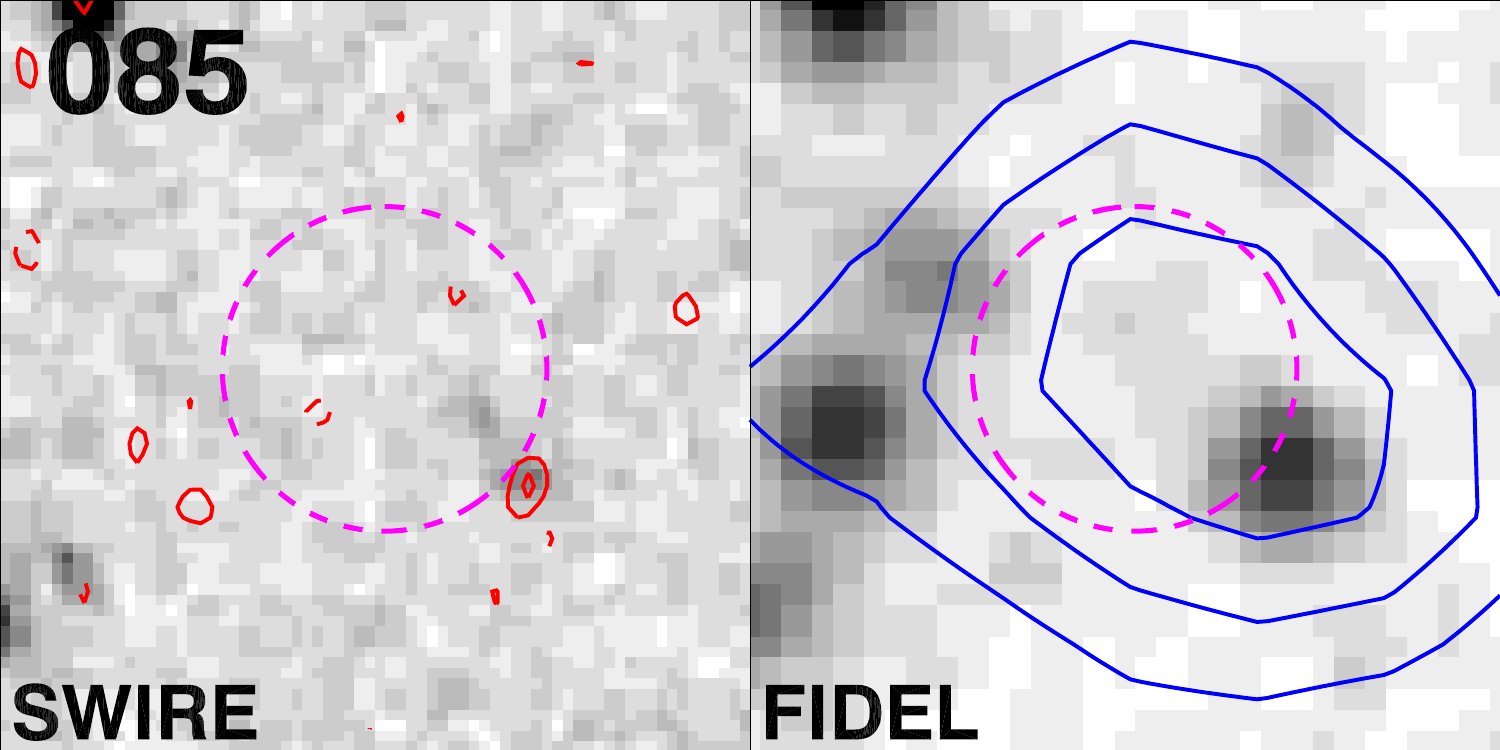}%
\hspace{1cm}%
\includegraphics[scale=0.295]{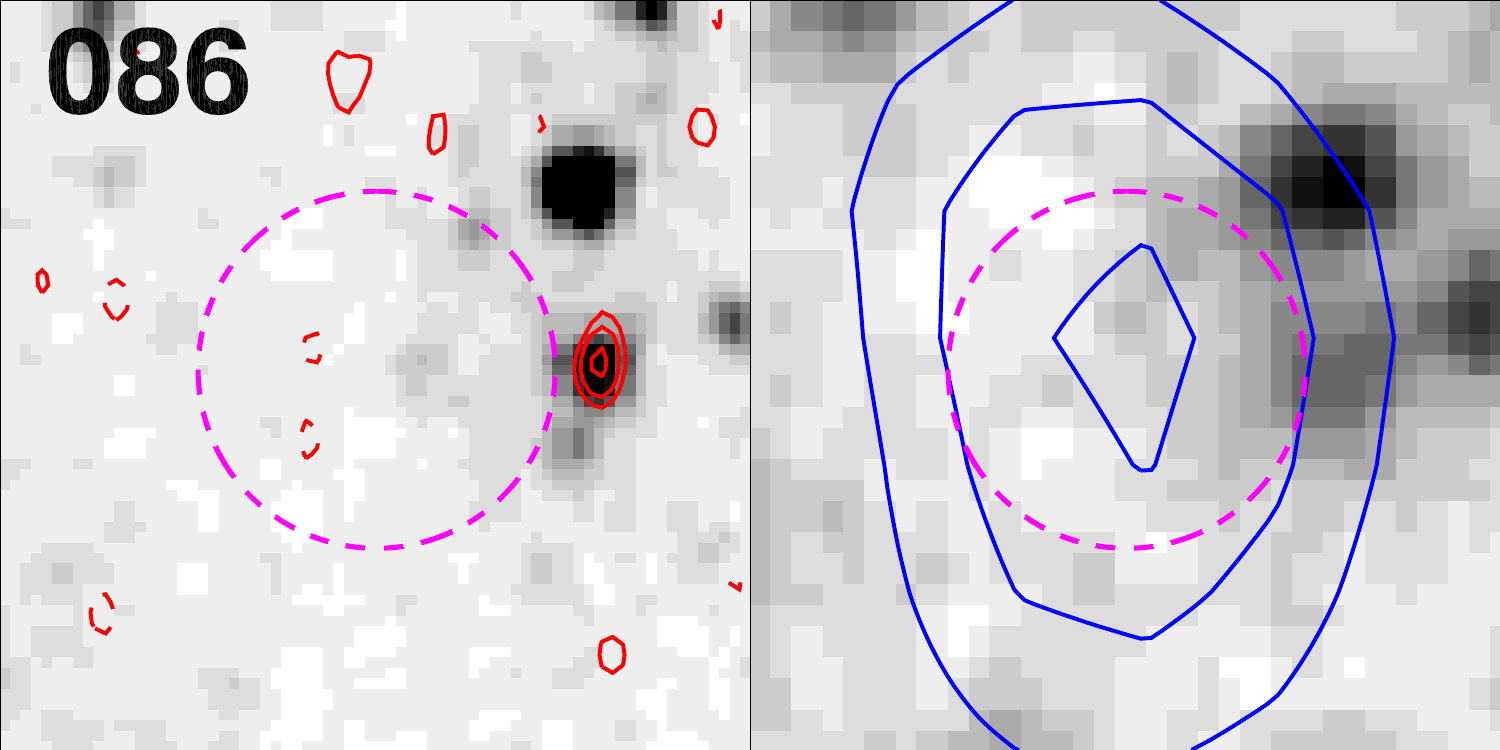}%
\hspace{1cm}%
\includegraphics[scale=0.295]{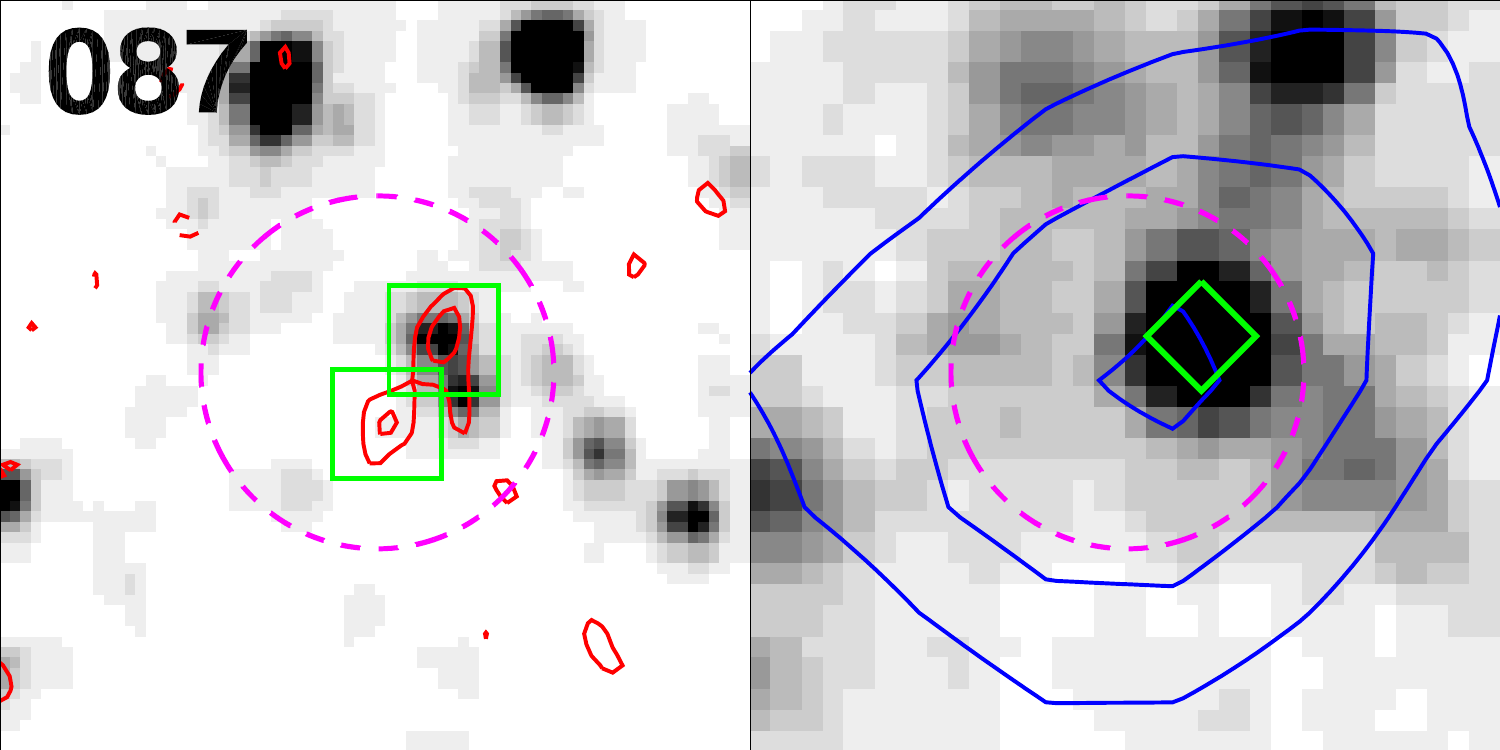}

\contcaption{}
\end{center}
\end{figure*}

\begin{figure*}
\begin{center}
\includegraphics[scale=0.295]{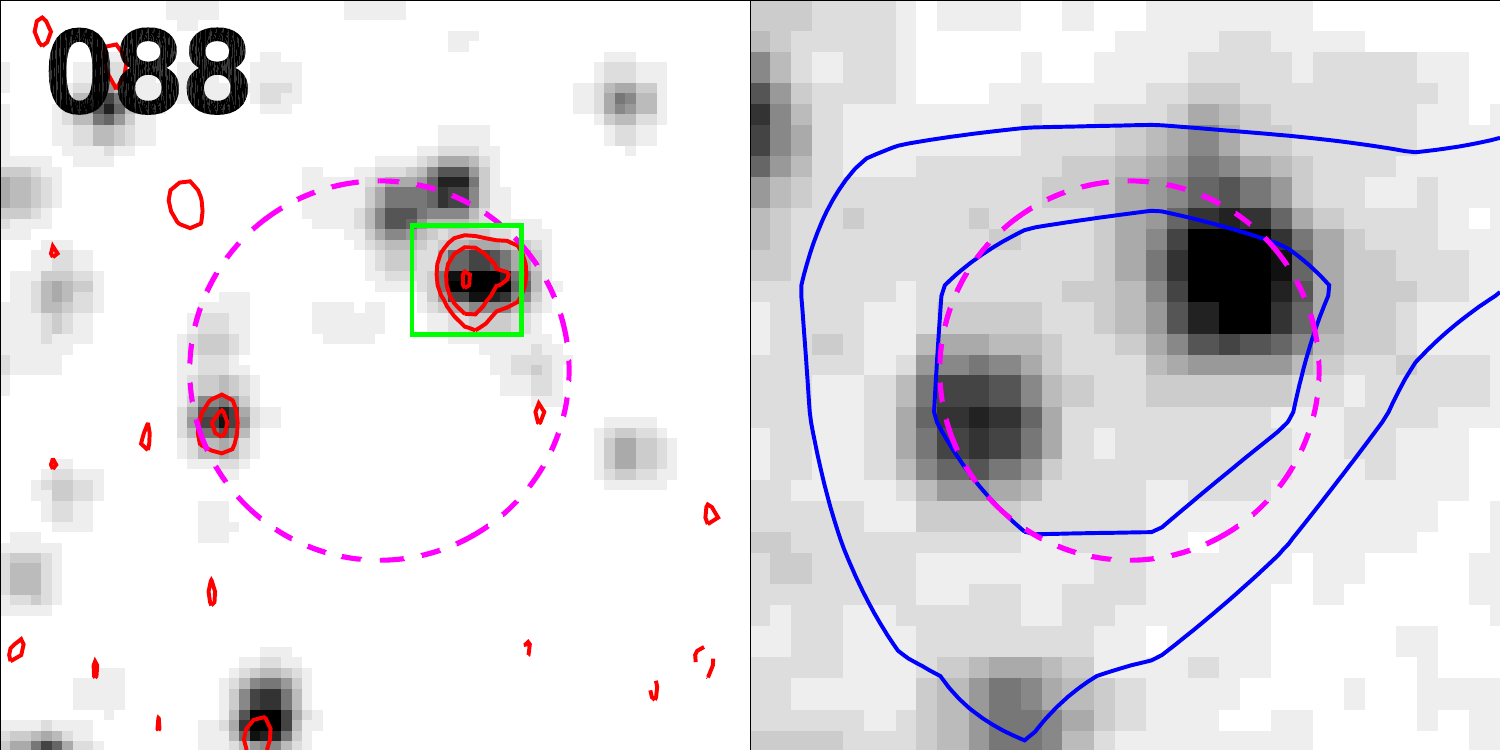}%
\hspace{1cm}%
\includegraphics[scale=0.295]{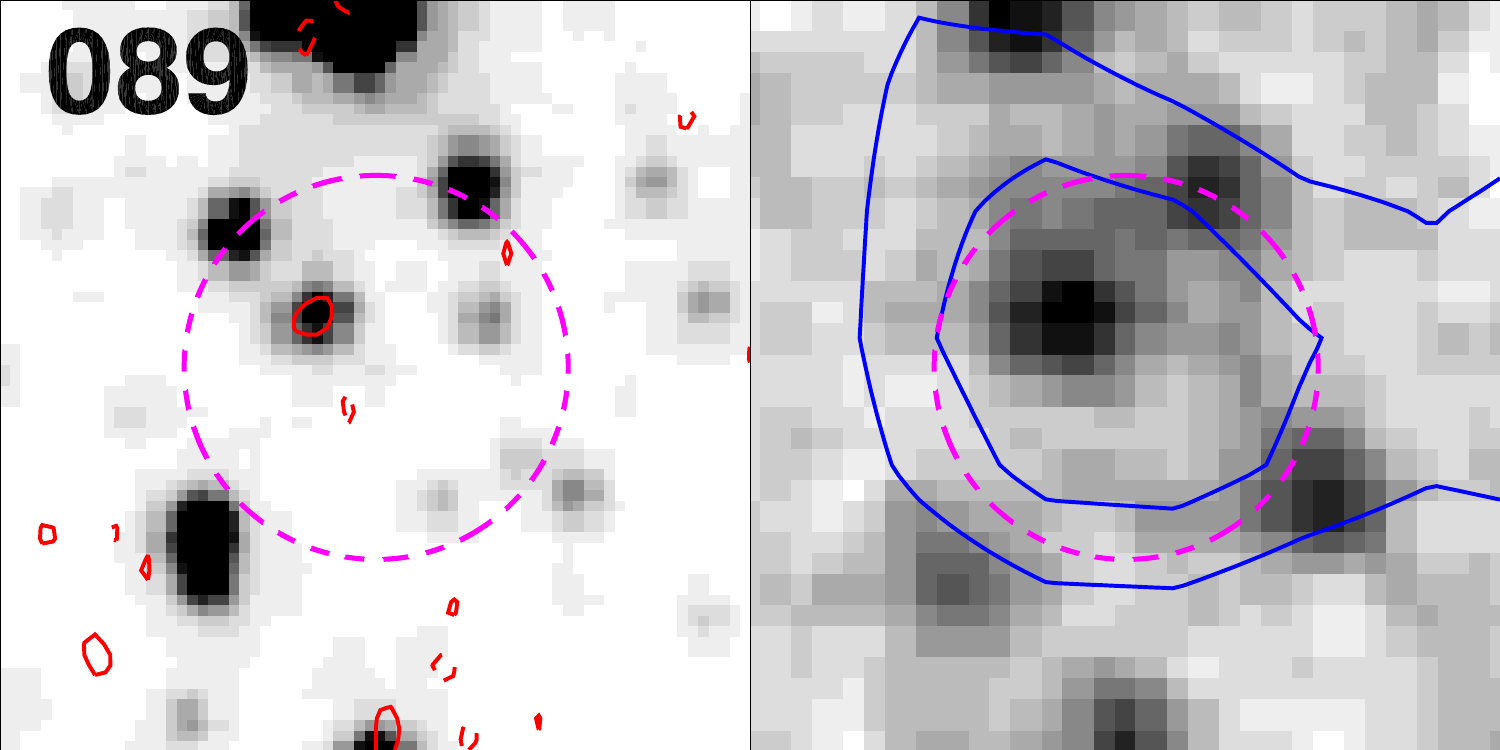}%
\hspace{1cm}%
\includegraphics[scale=0.295]{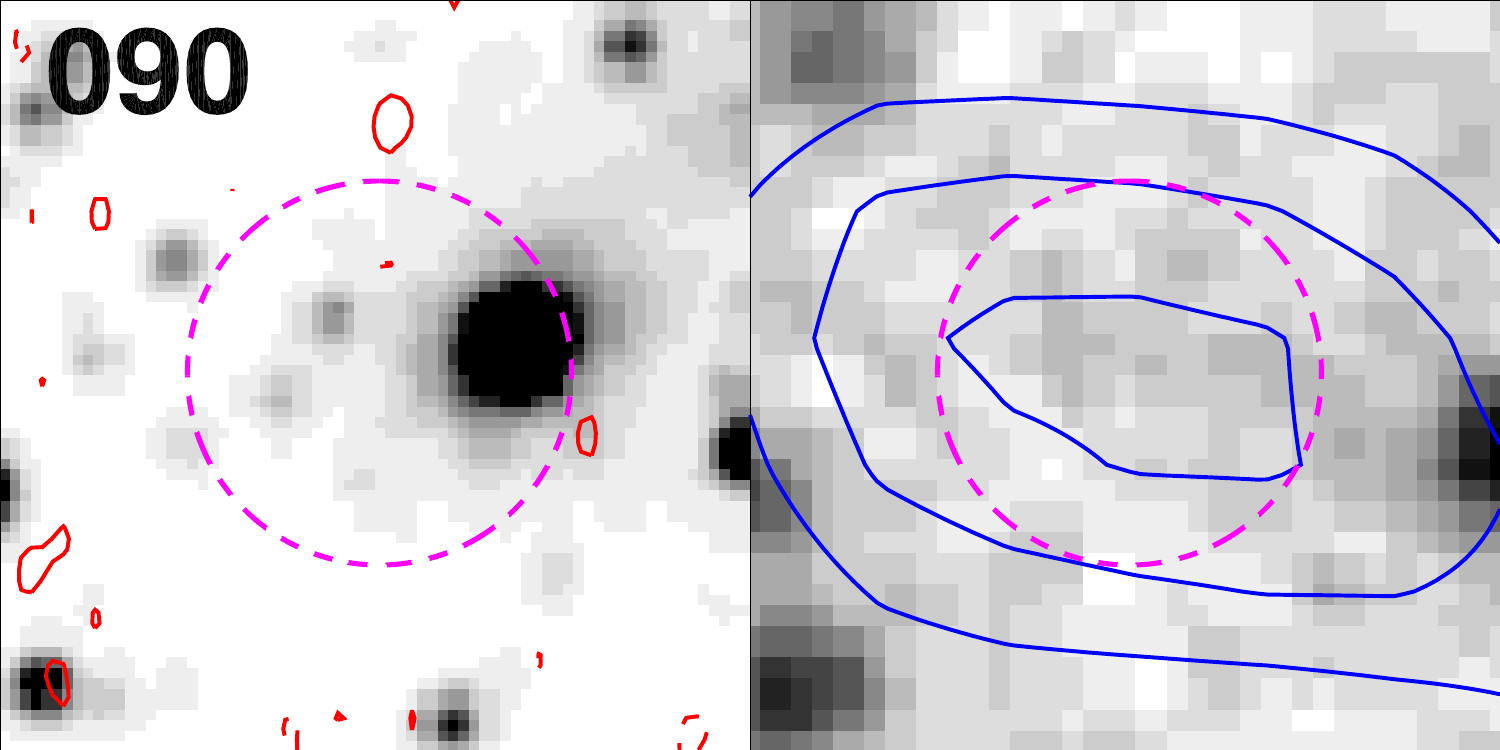}
\includegraphics[scale=0.295]{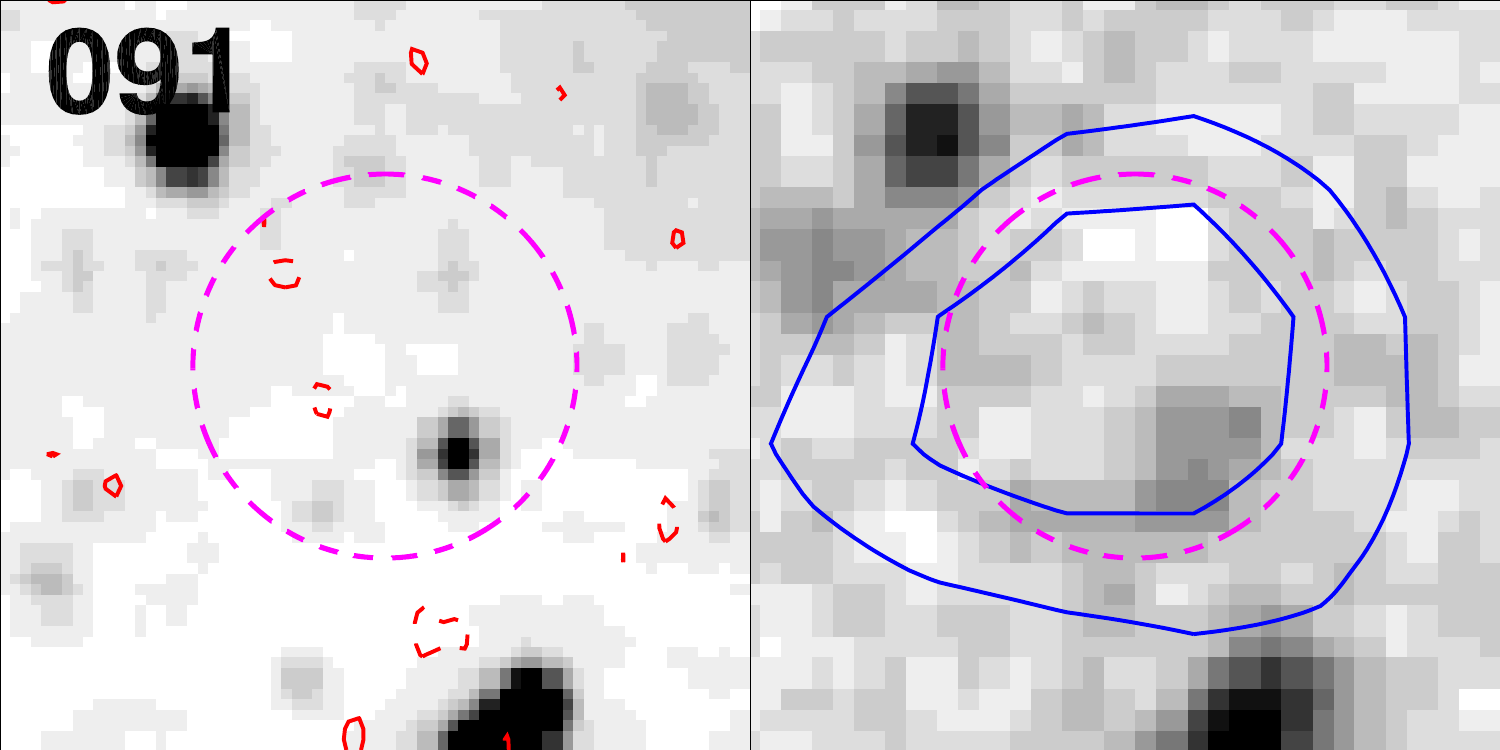}%
\hspace{1cm}%
\includegraphics[scale=0.295]{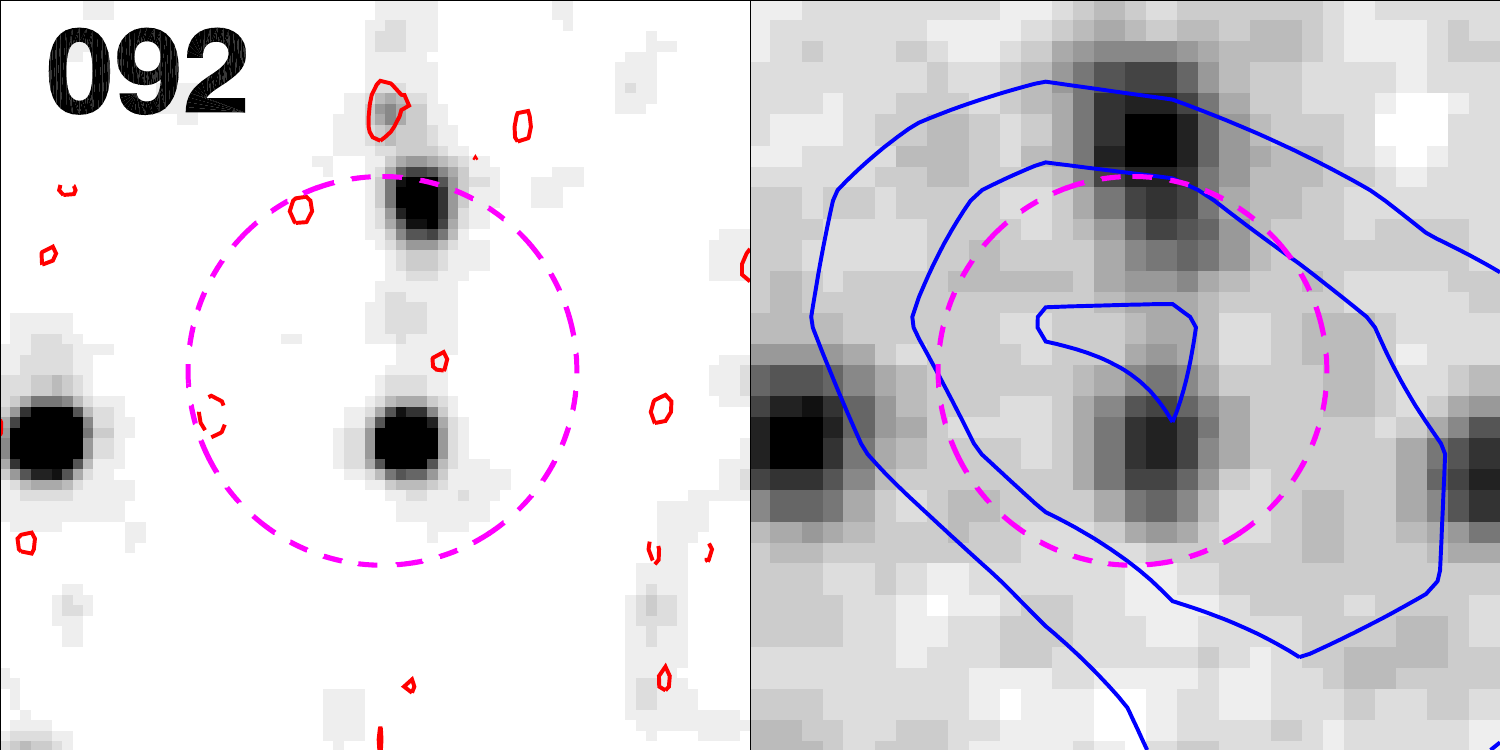}%
\hspace{1cm}%
\includegraphics[scale=0.295]{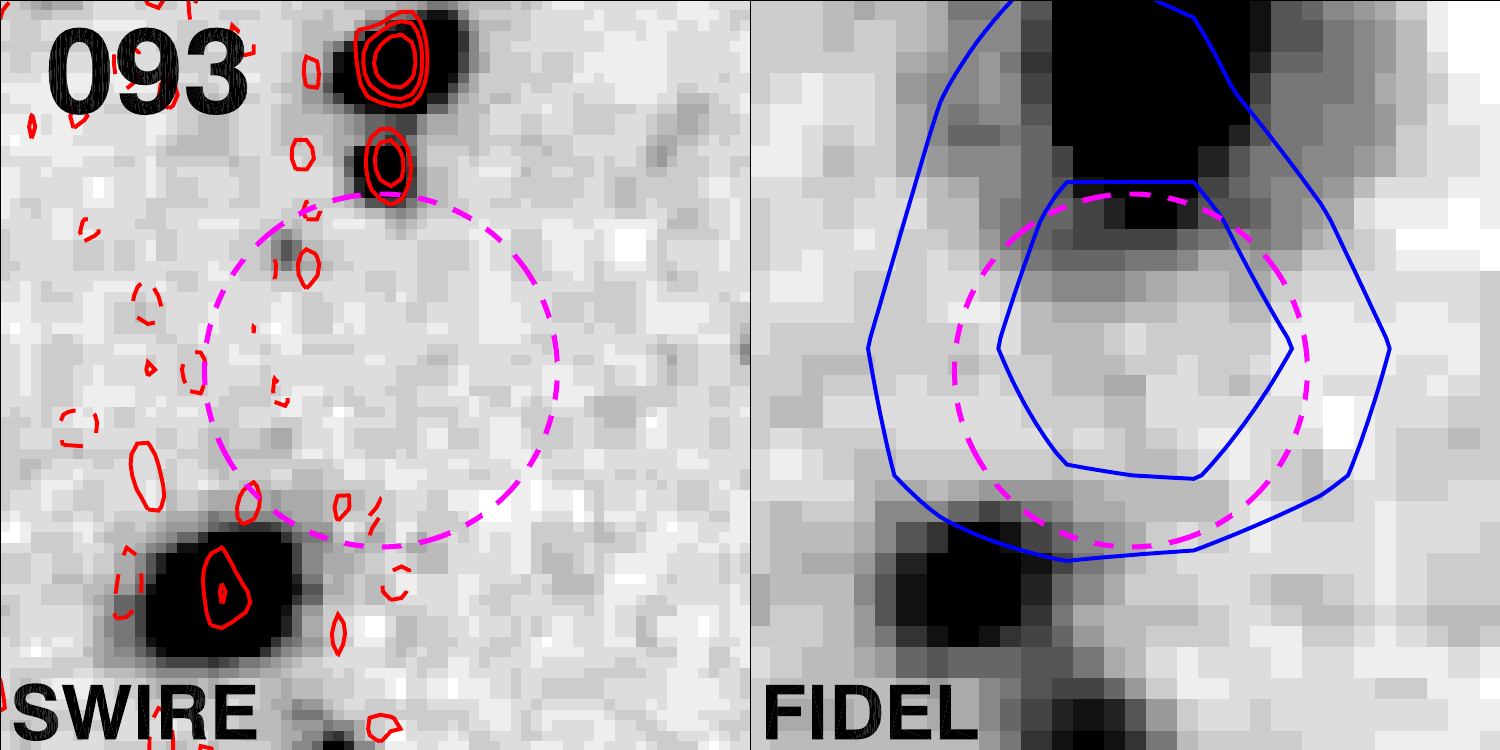}
\includegraphics[scale=0.295]{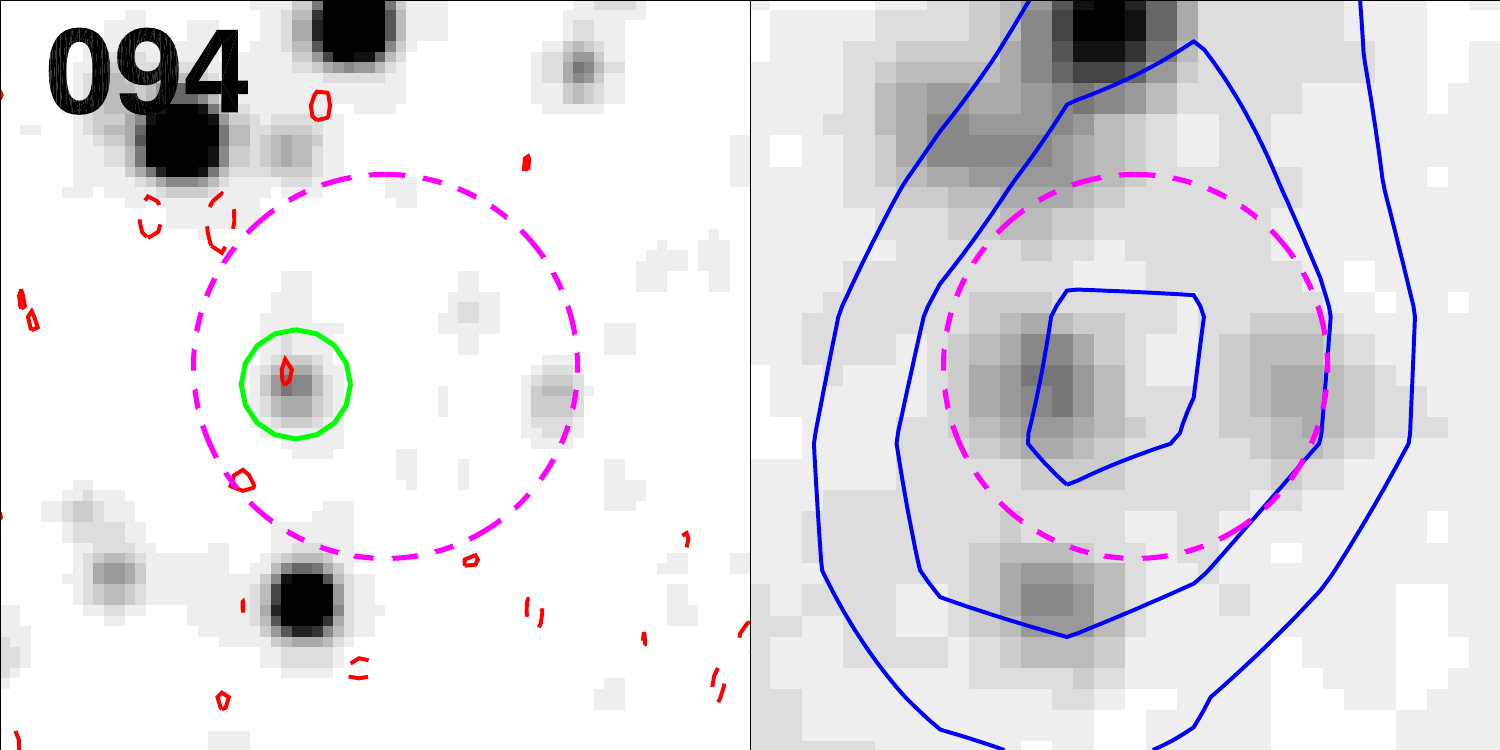}%
\hspace{1cm}%
\includegraphics[scale=0.295]{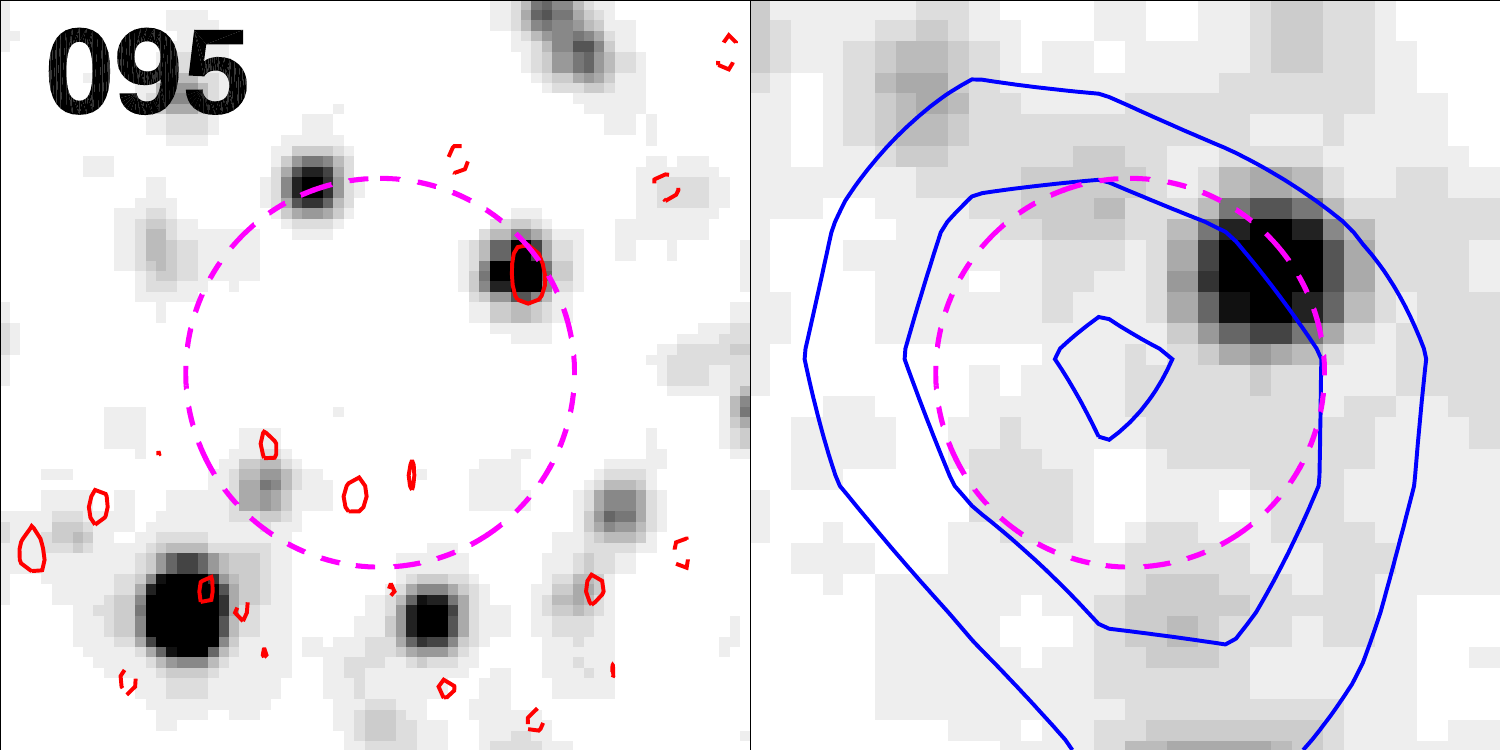}%
\hspace{1cm}%
\includegraphics[scale=0.295]{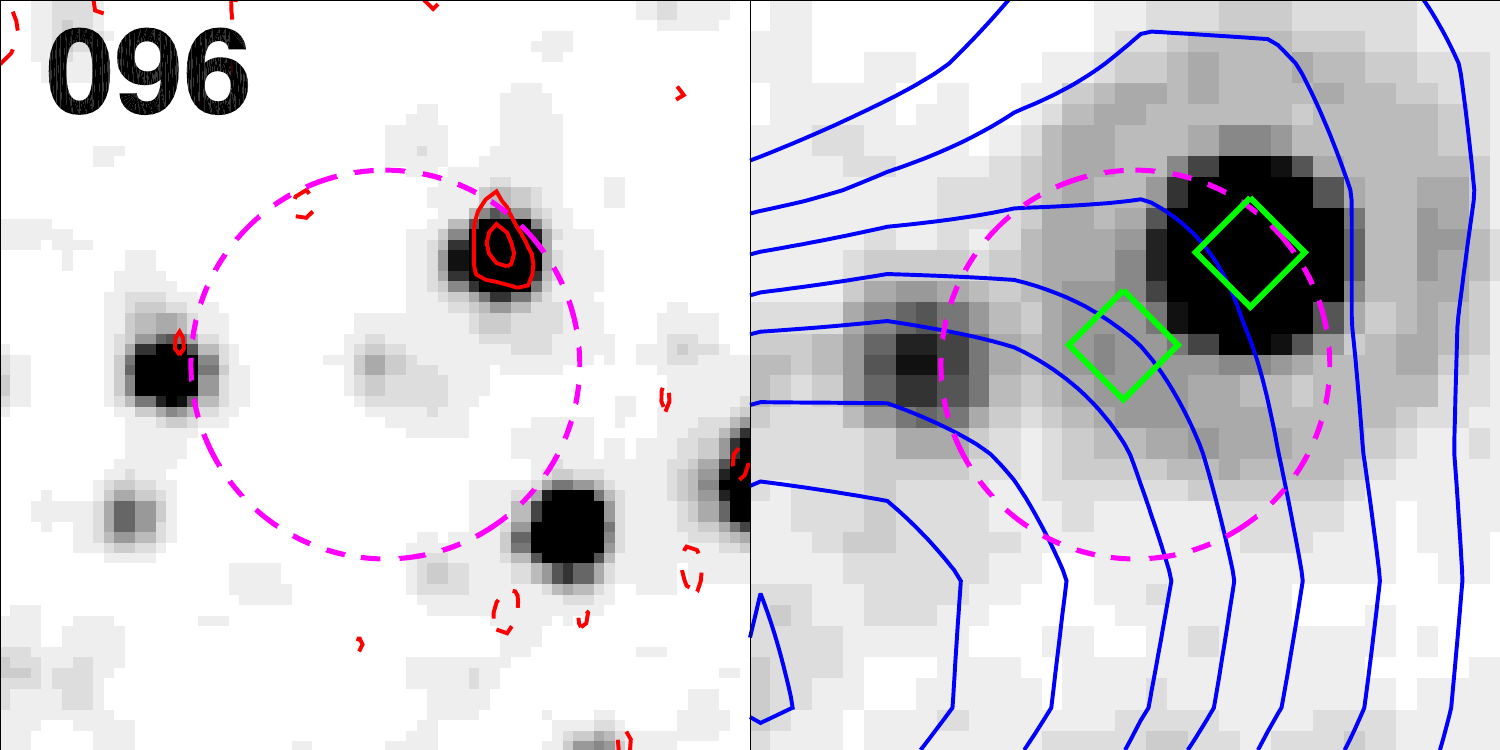}
\includegraphics[scale=0.295]{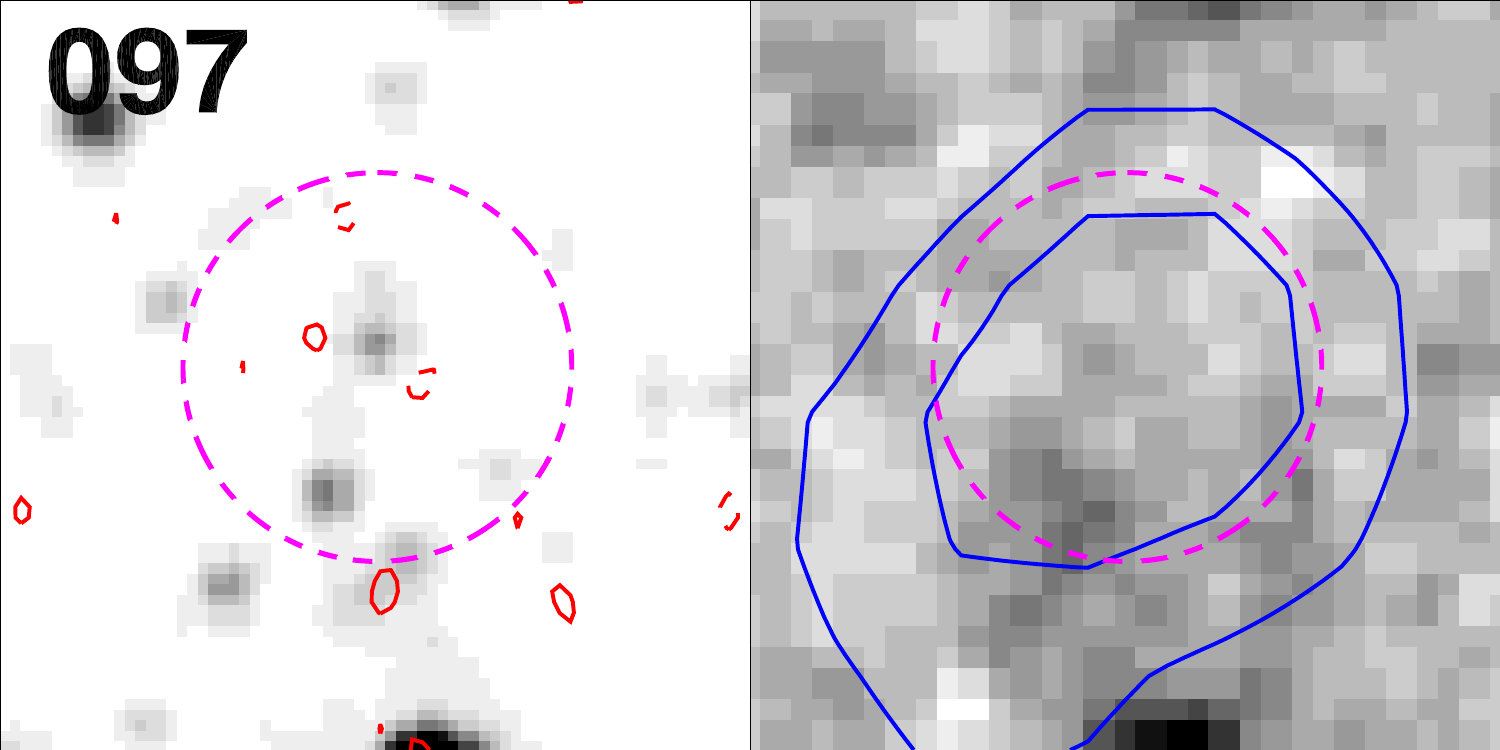}%
\hspace{1cm}%
\includegraphics[scale=0.295]{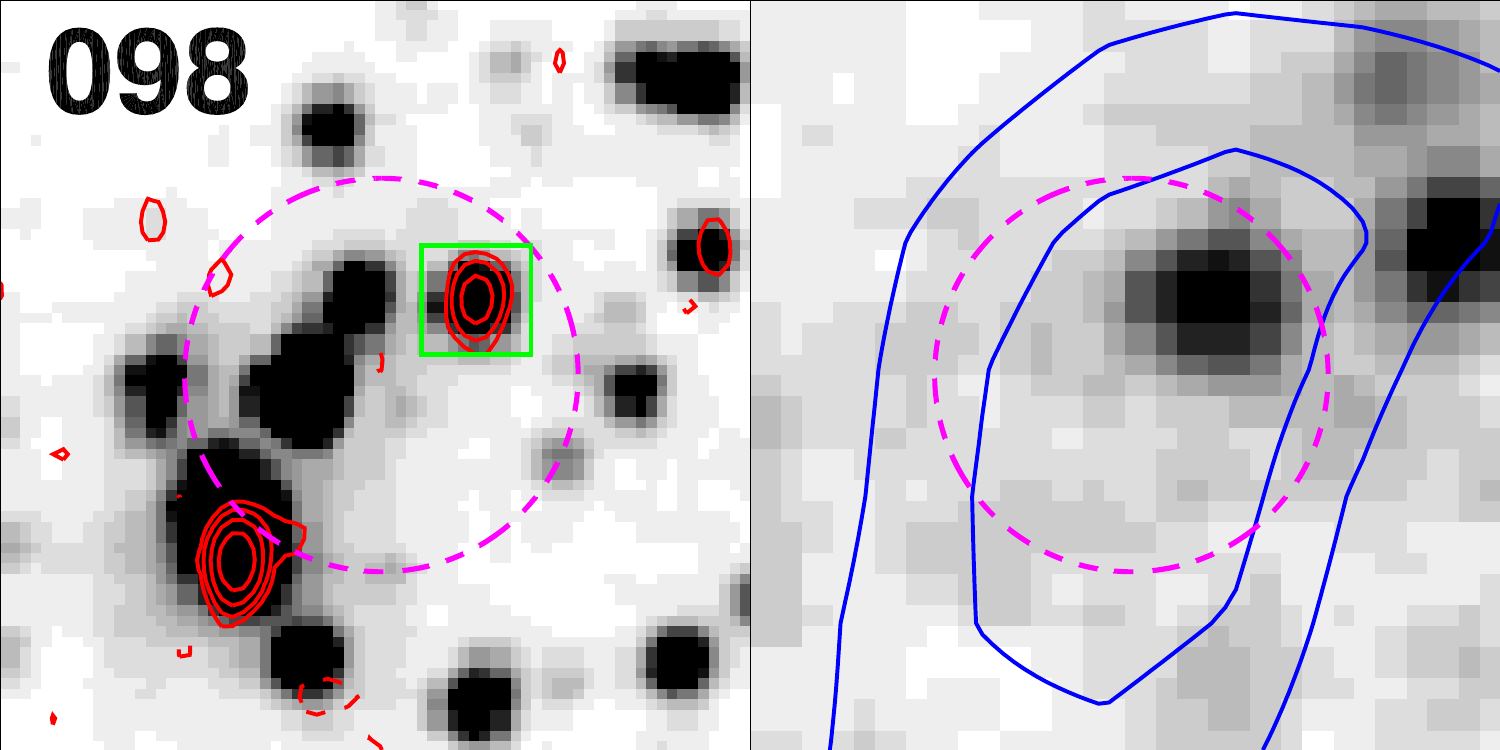}%
\hspace{1cm}%
\includegraphics[scale=0.295]{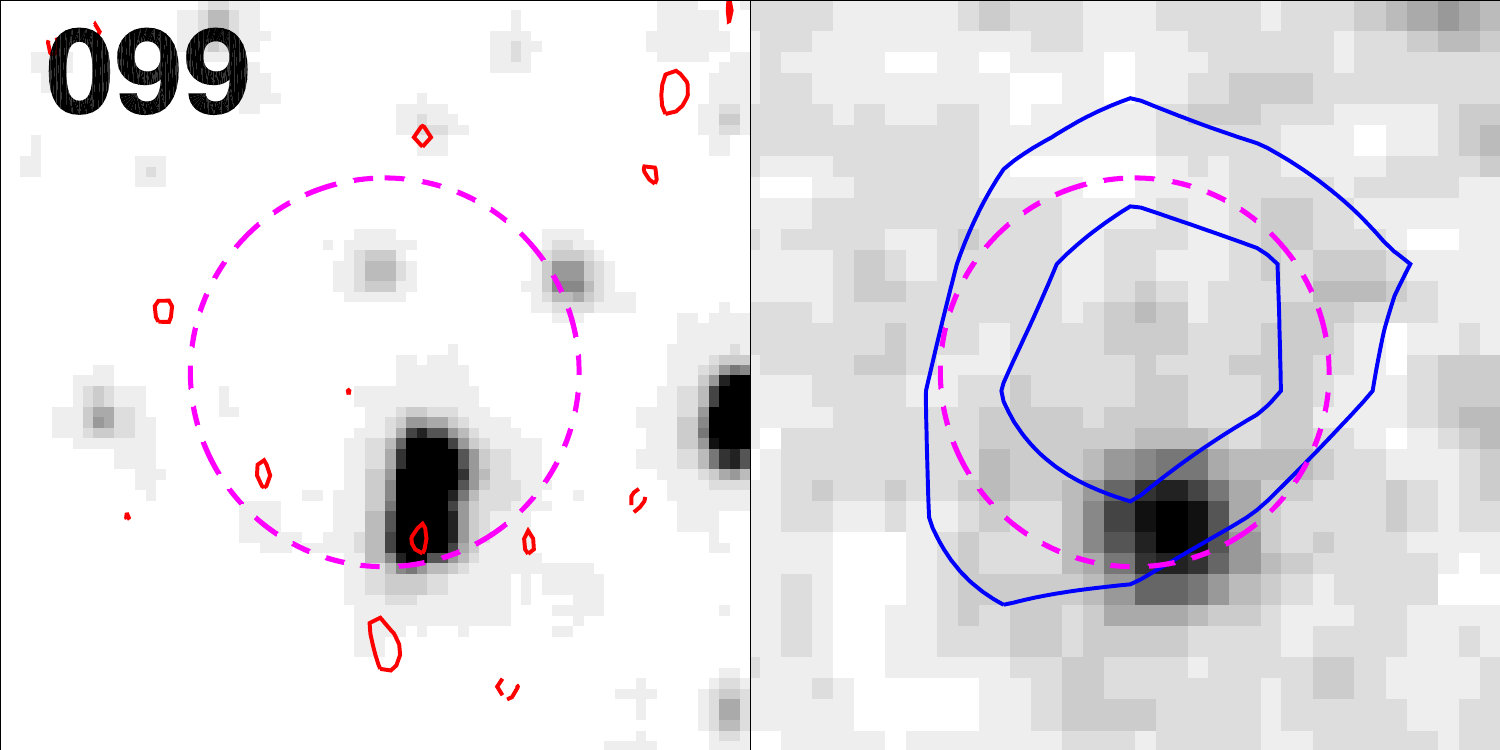}
\includegraphics[scale=0.295]{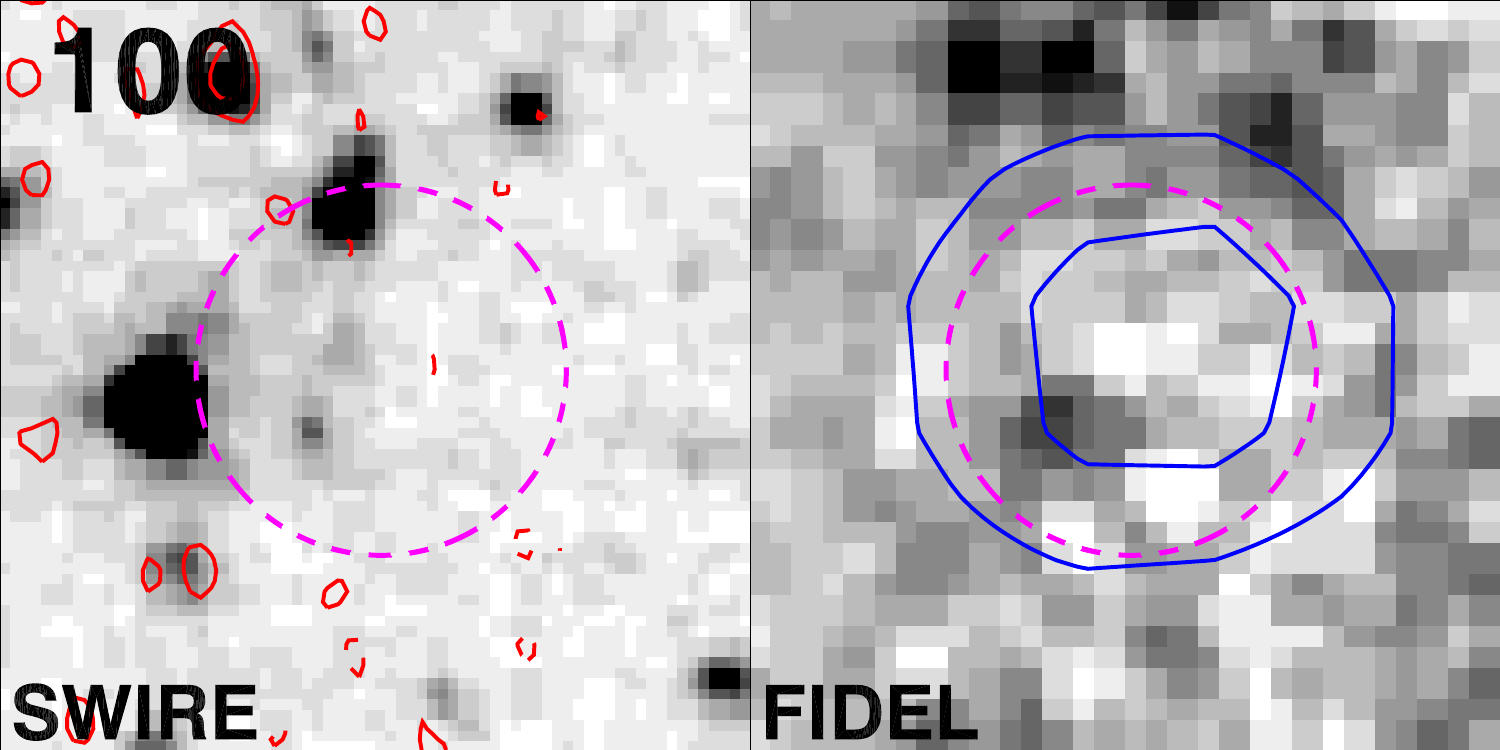}%
\hspace{1cm}%
\includegraphics[scale=0.295]{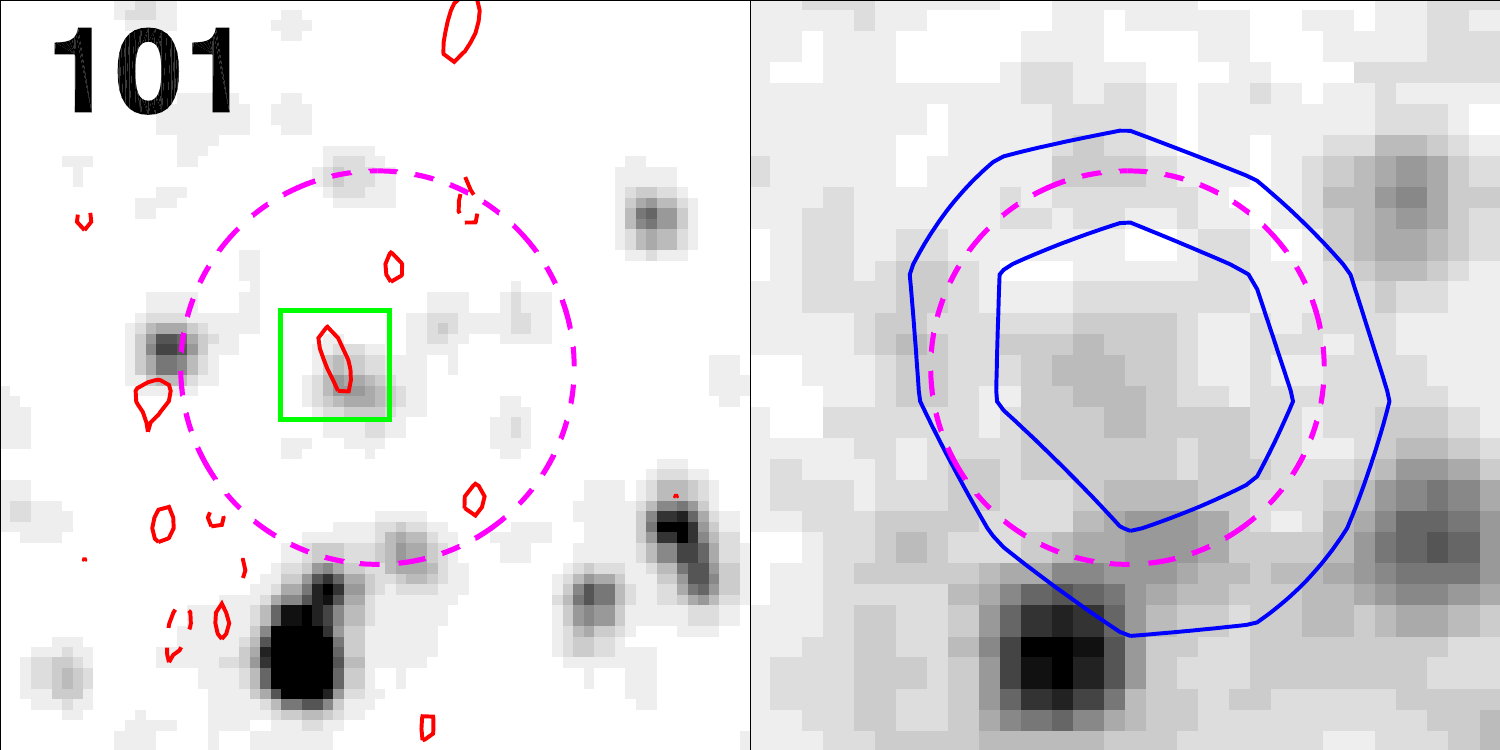}%
\hspace{1cm}%
\includegraphics[scale=0.295]{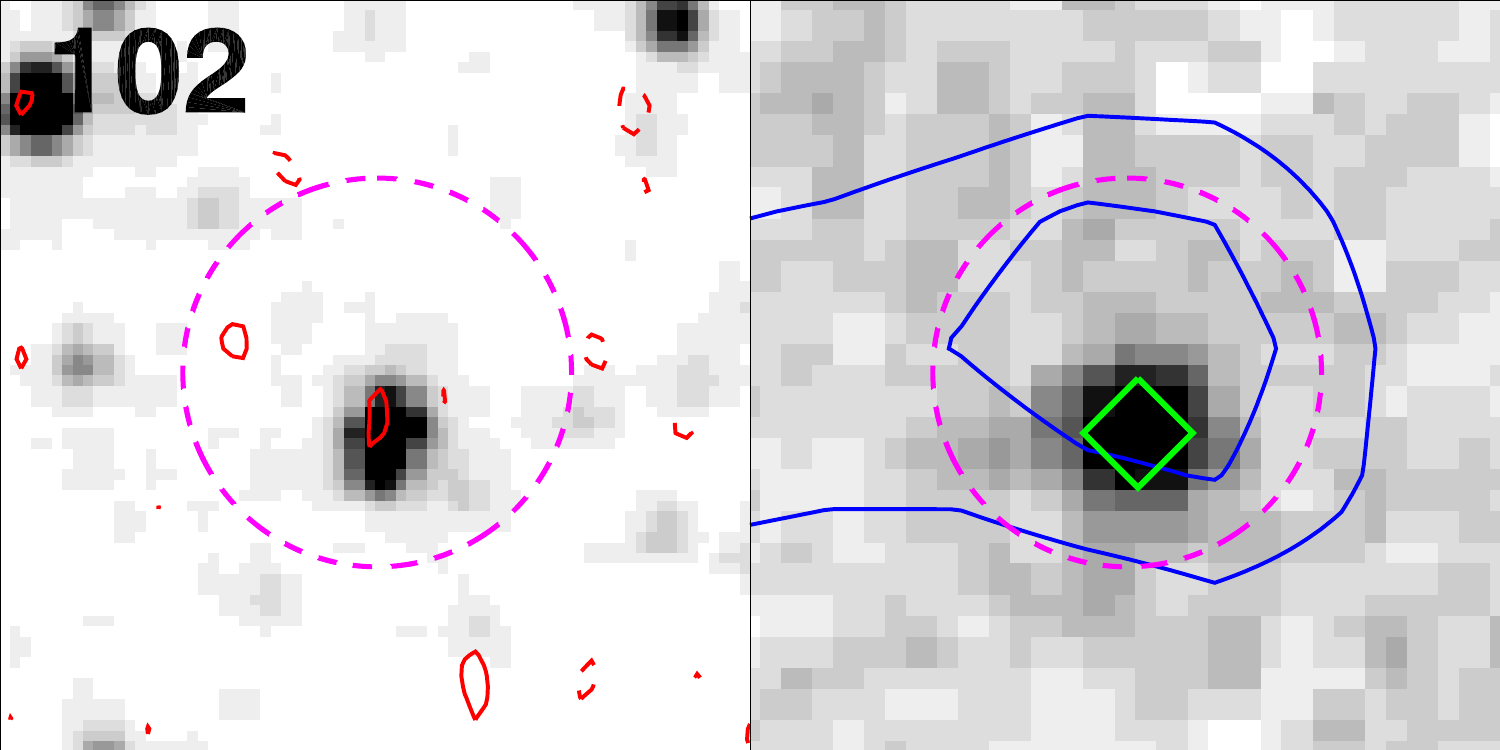}
\includegraphics[scale=0.295]{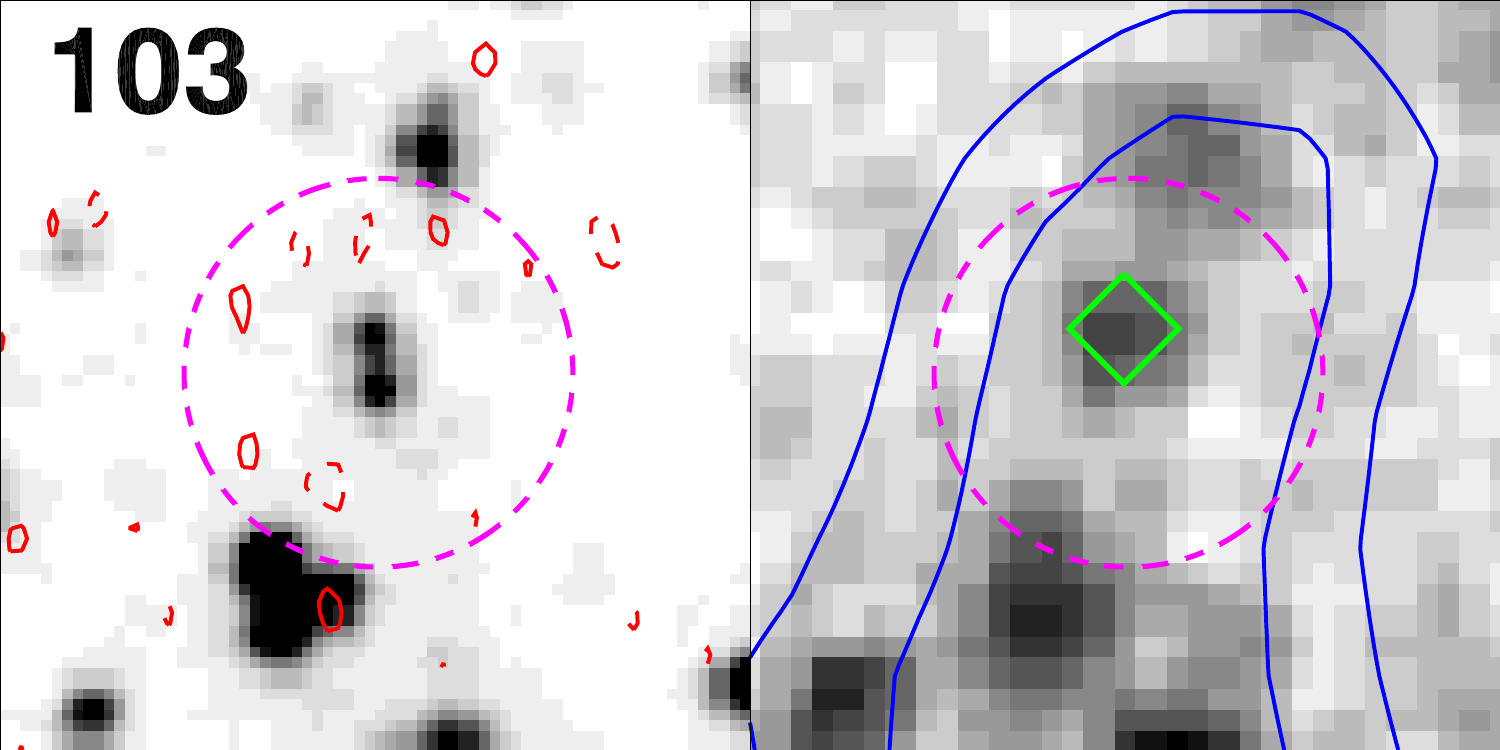}%
\hspace{1cm}%
\includegraphics[scale=0.295]{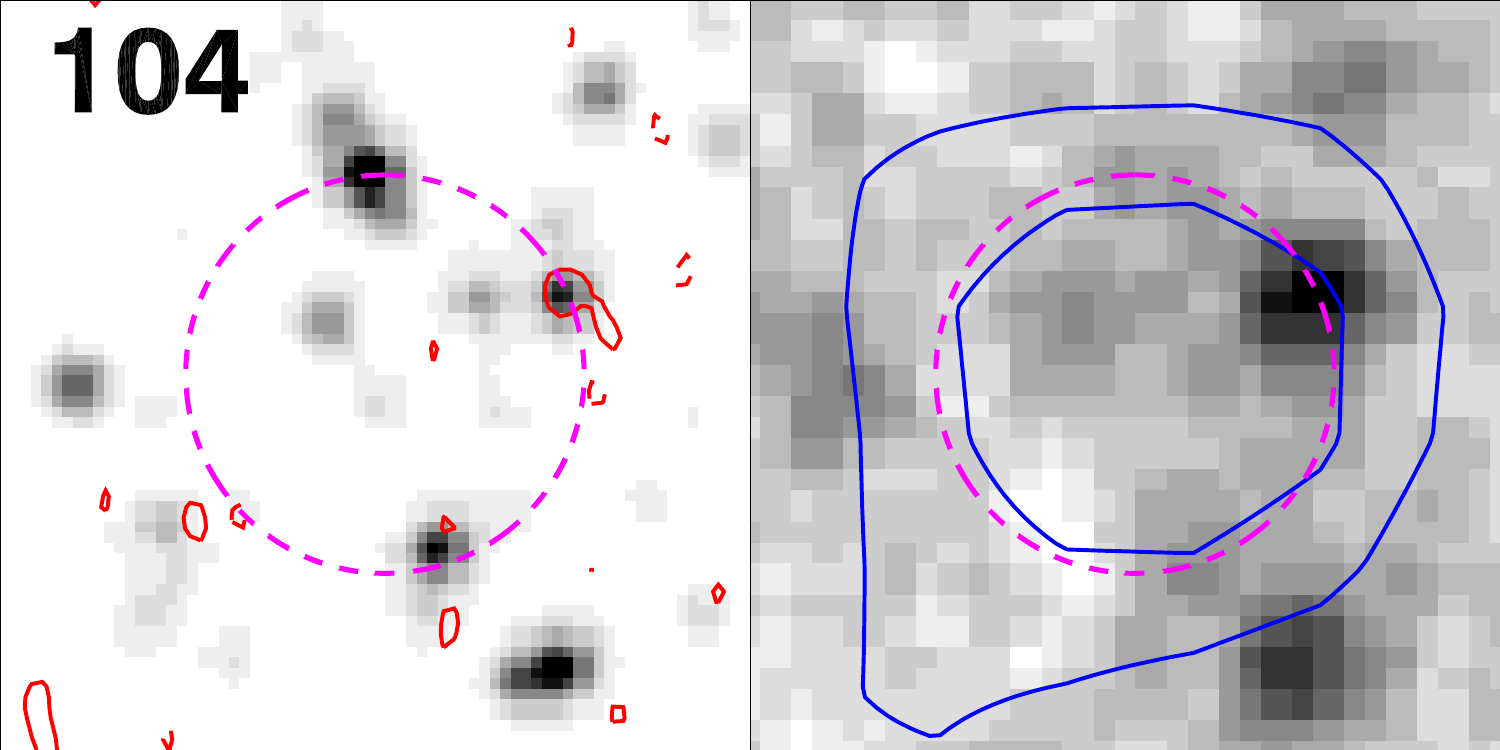}%
\hspace{1cm}%
\includegraphics[scale=0.295]{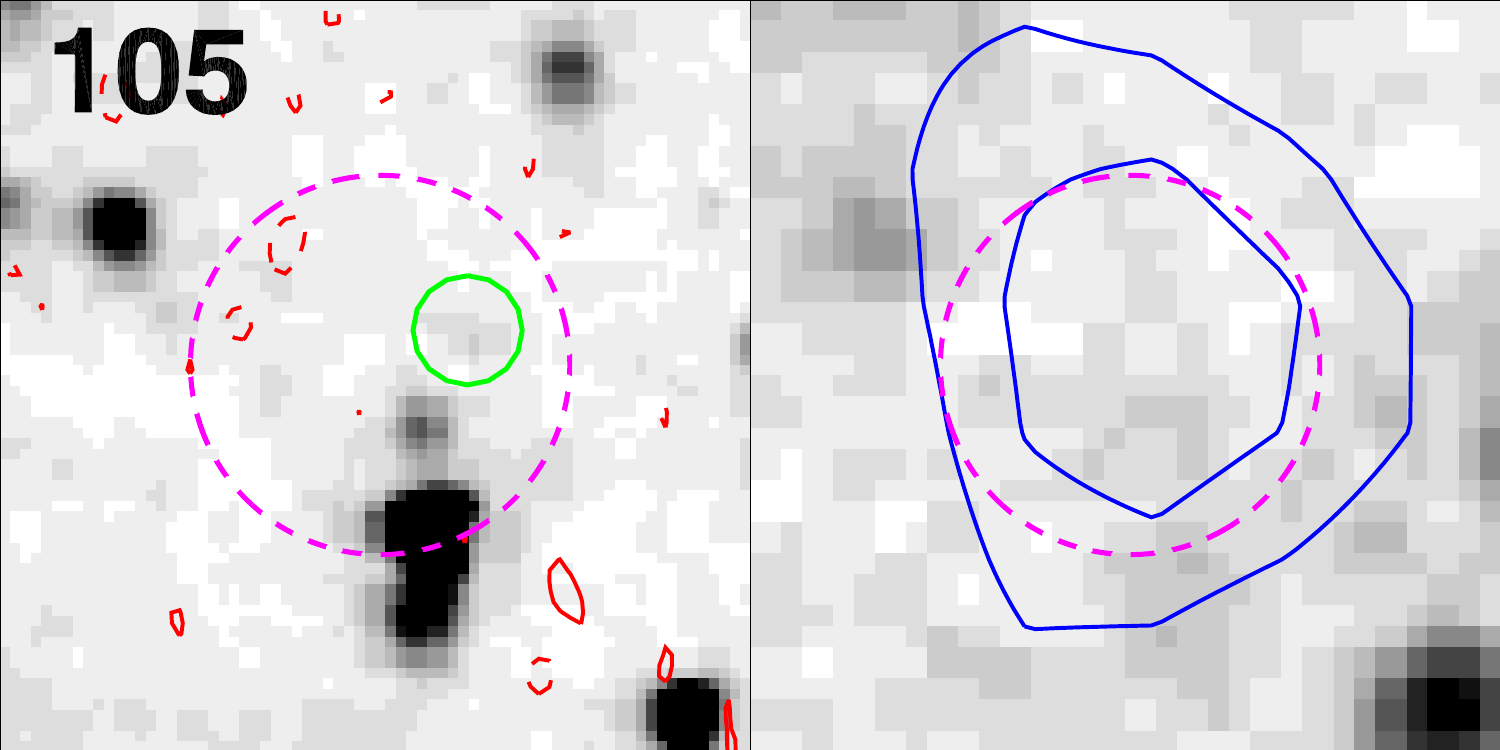}
\includegraphics[scale=0.295]{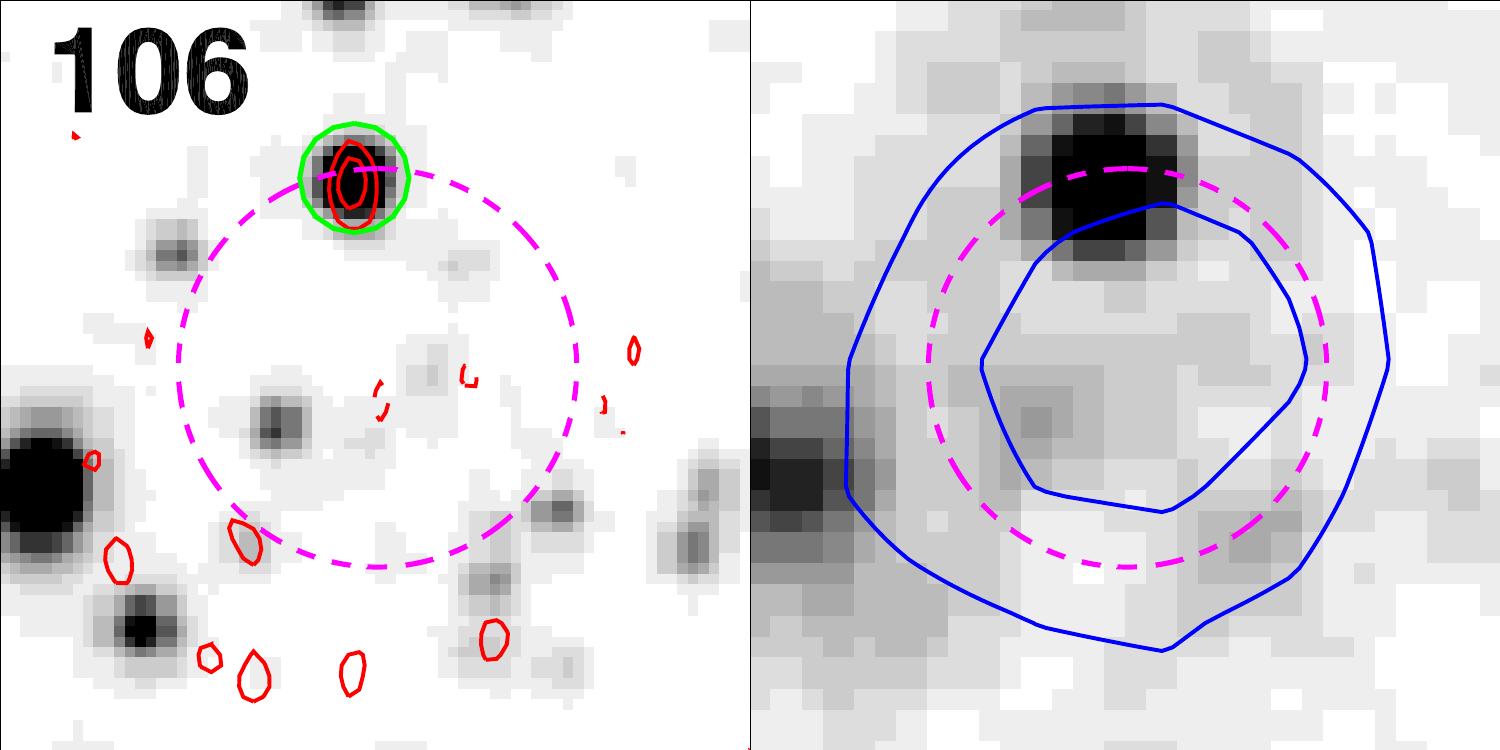}%
\hspace{1cm}%
\includegraphics[scale=0.295]{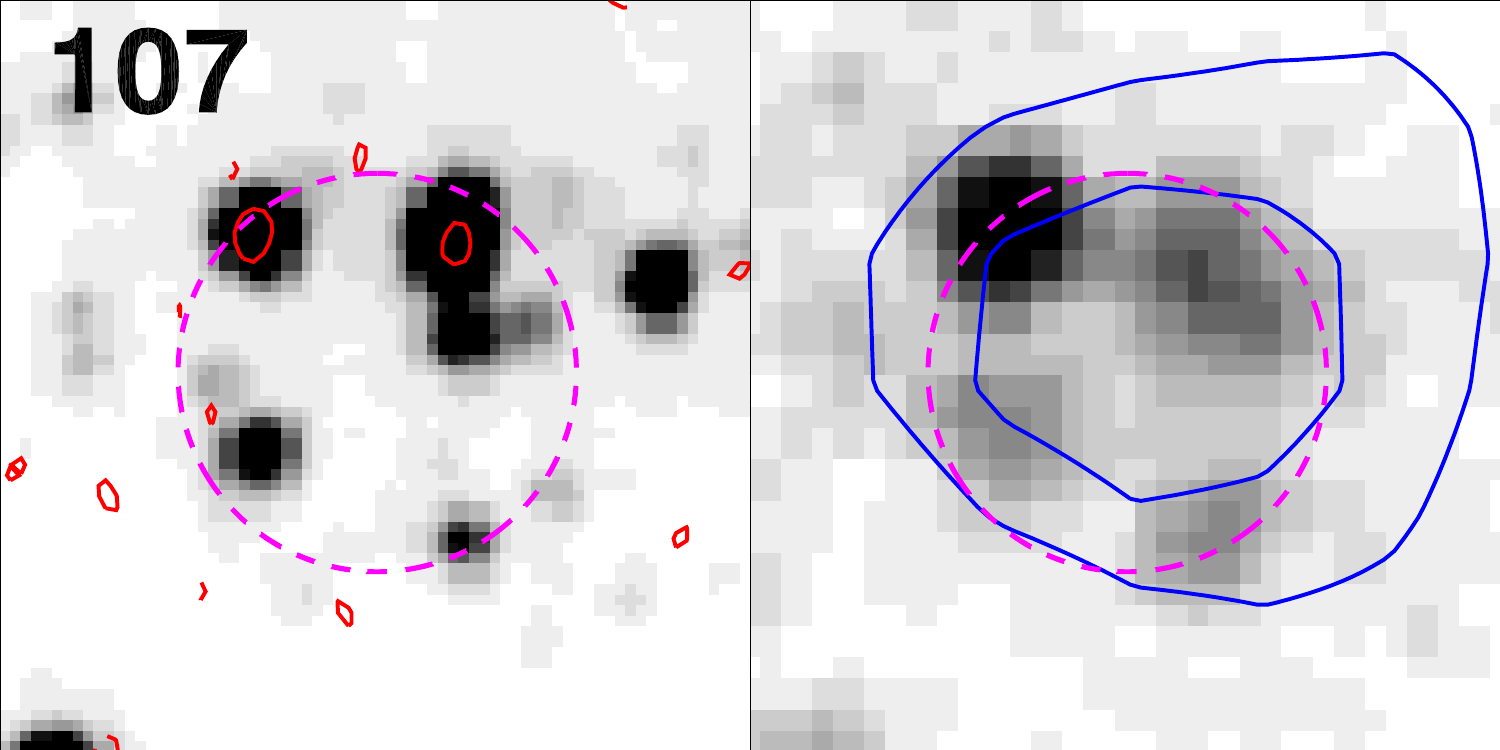}%
\hspace{1cm}%
\includegraphics[scale=0.295]{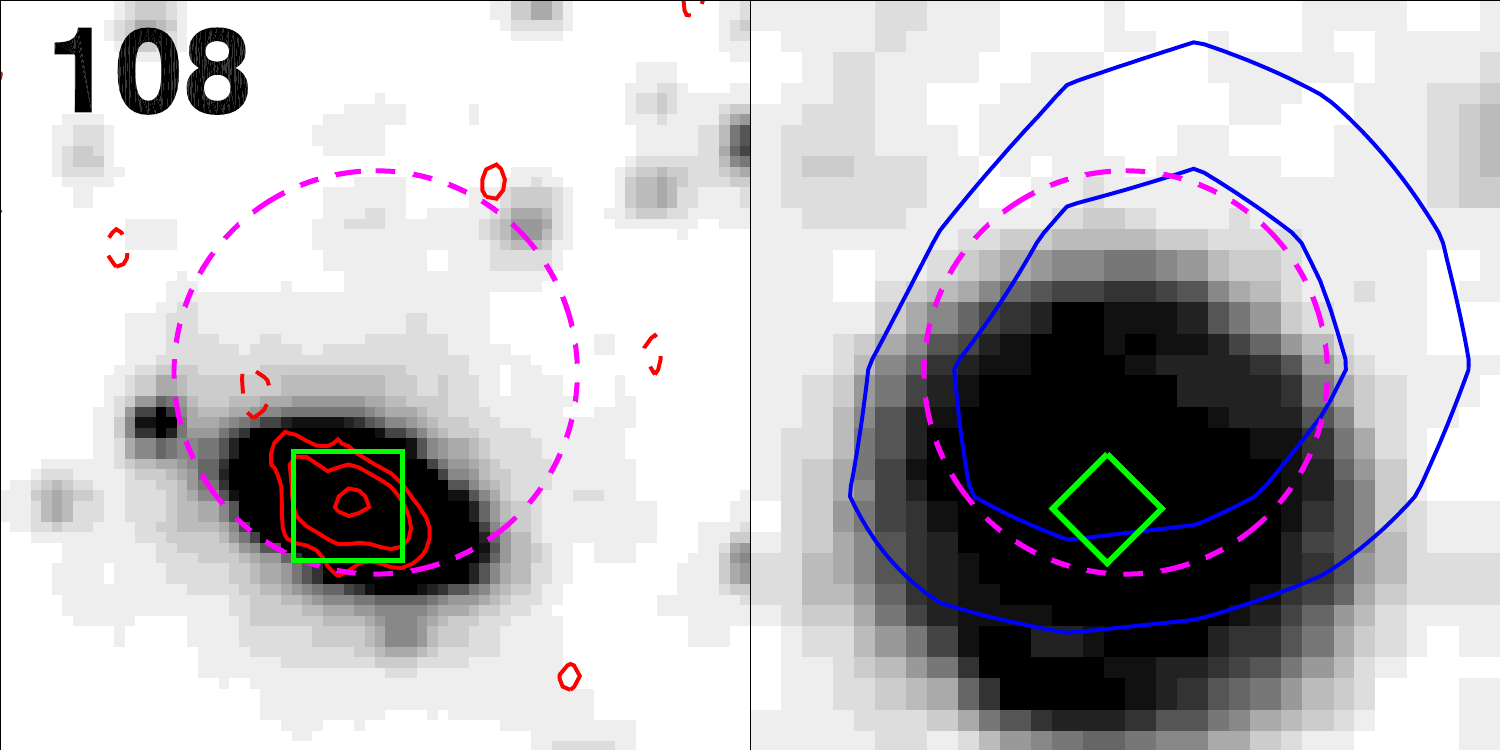}
\includegraphics[scale=0.295]{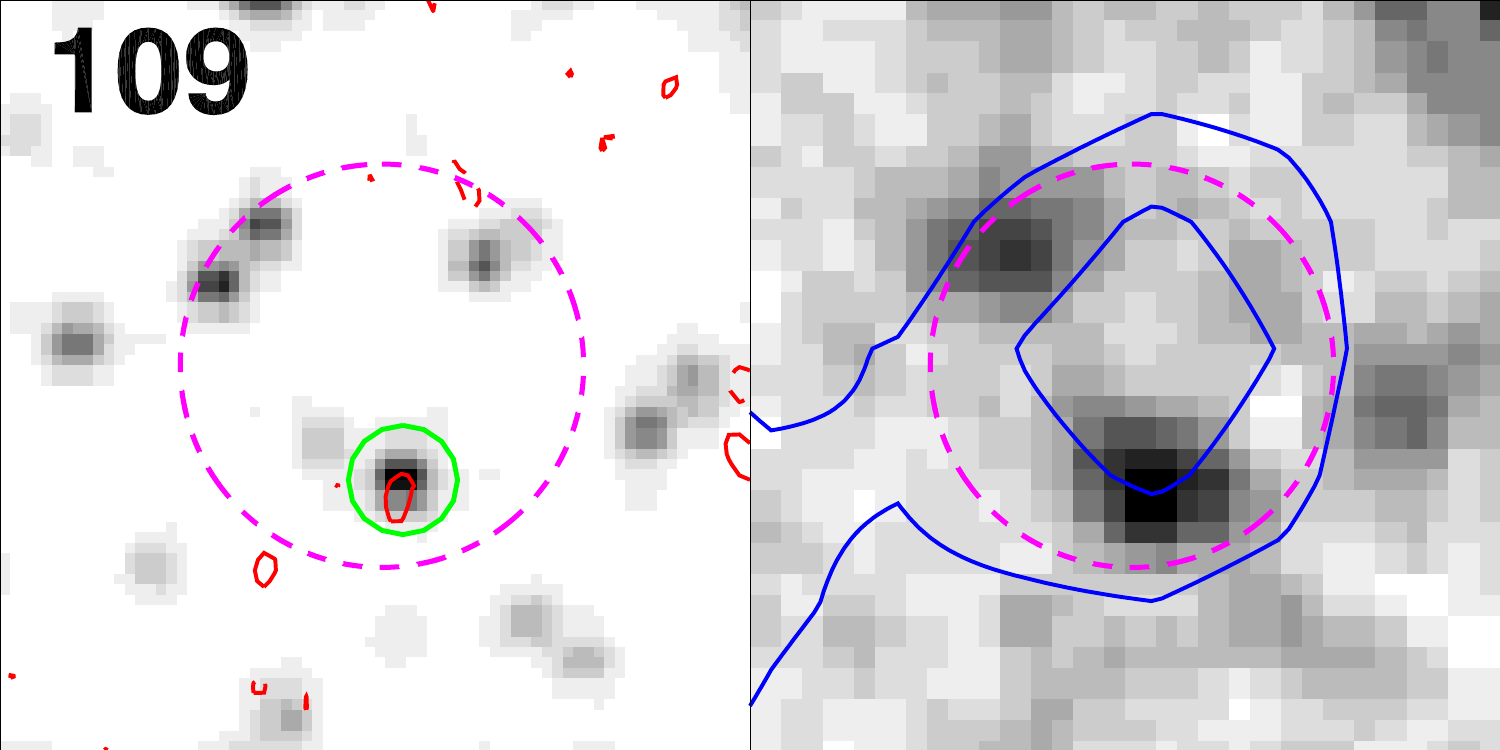}%
\hspace{1cm}%
\includegraphics[scale=0.295]{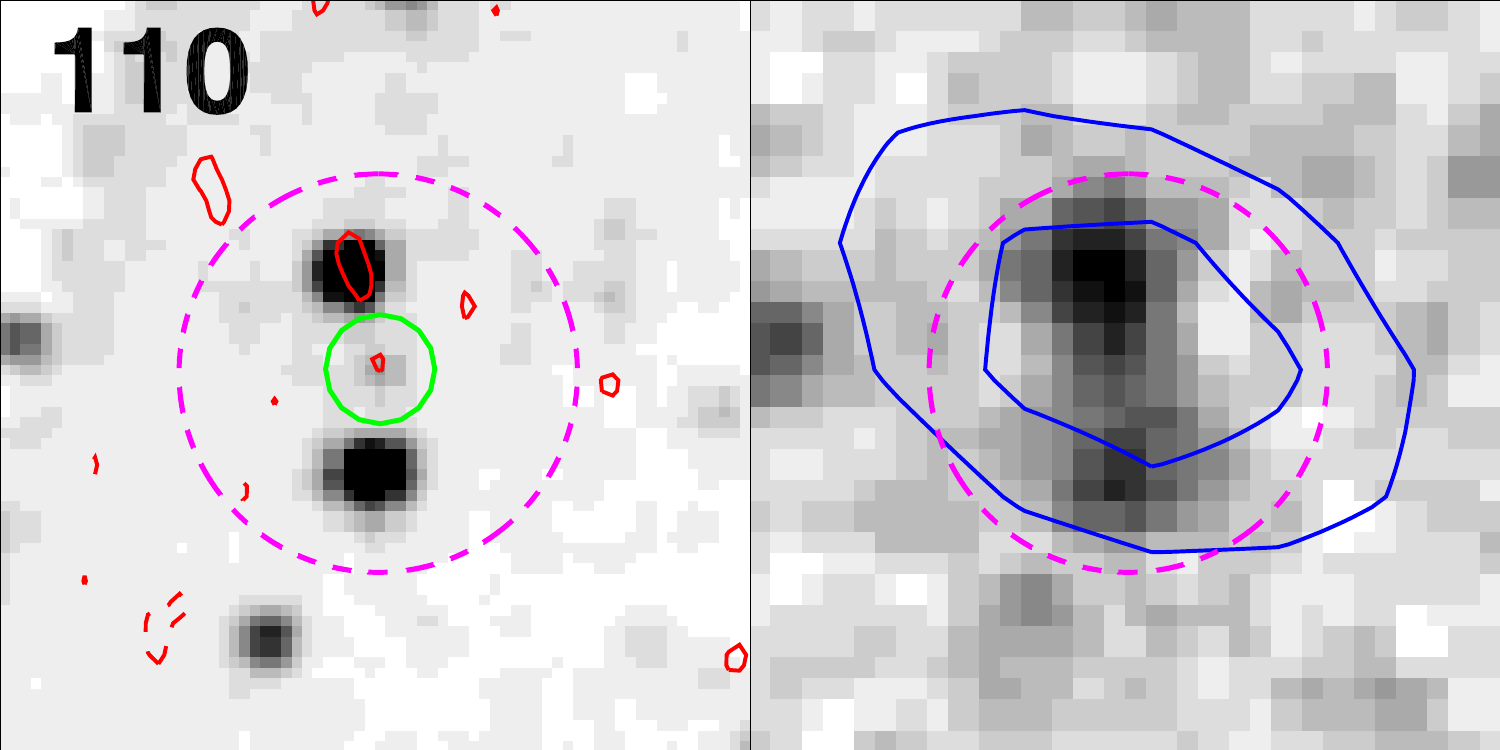}%
\hspace{1cm}%
\includegraphics[scale=0.295]{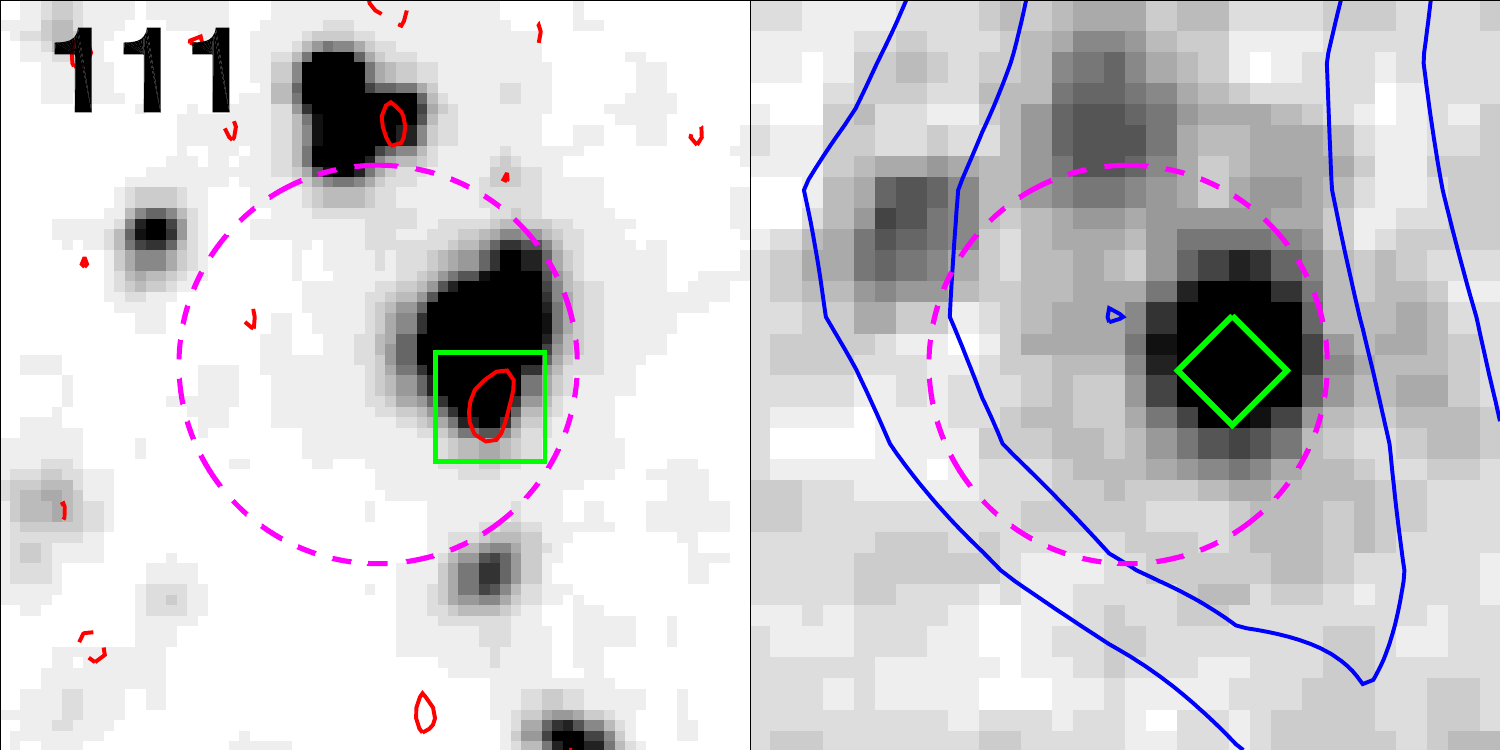}
\includegraphics[scale=0.295]{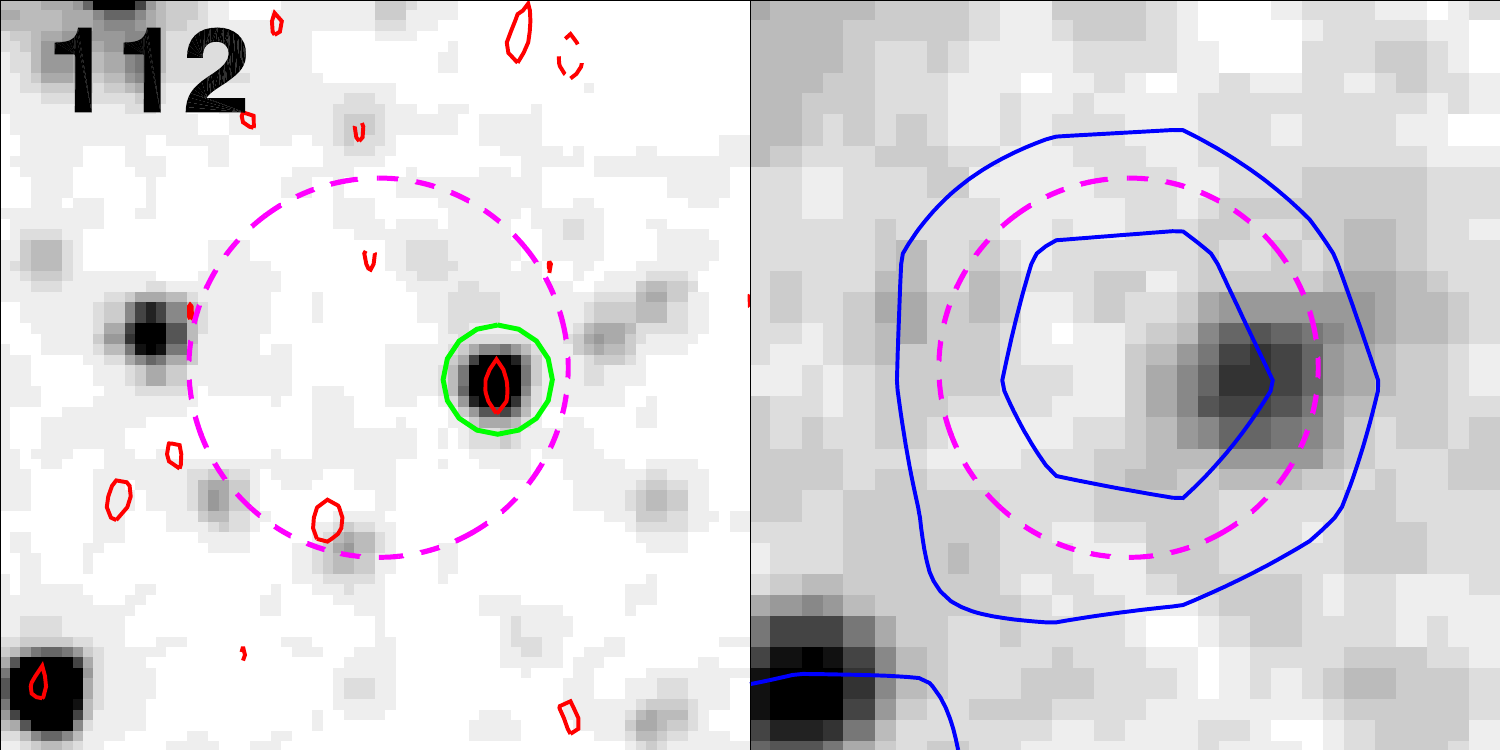}%
\hspace{1cm}%
\includegraphics[scale=0.295]{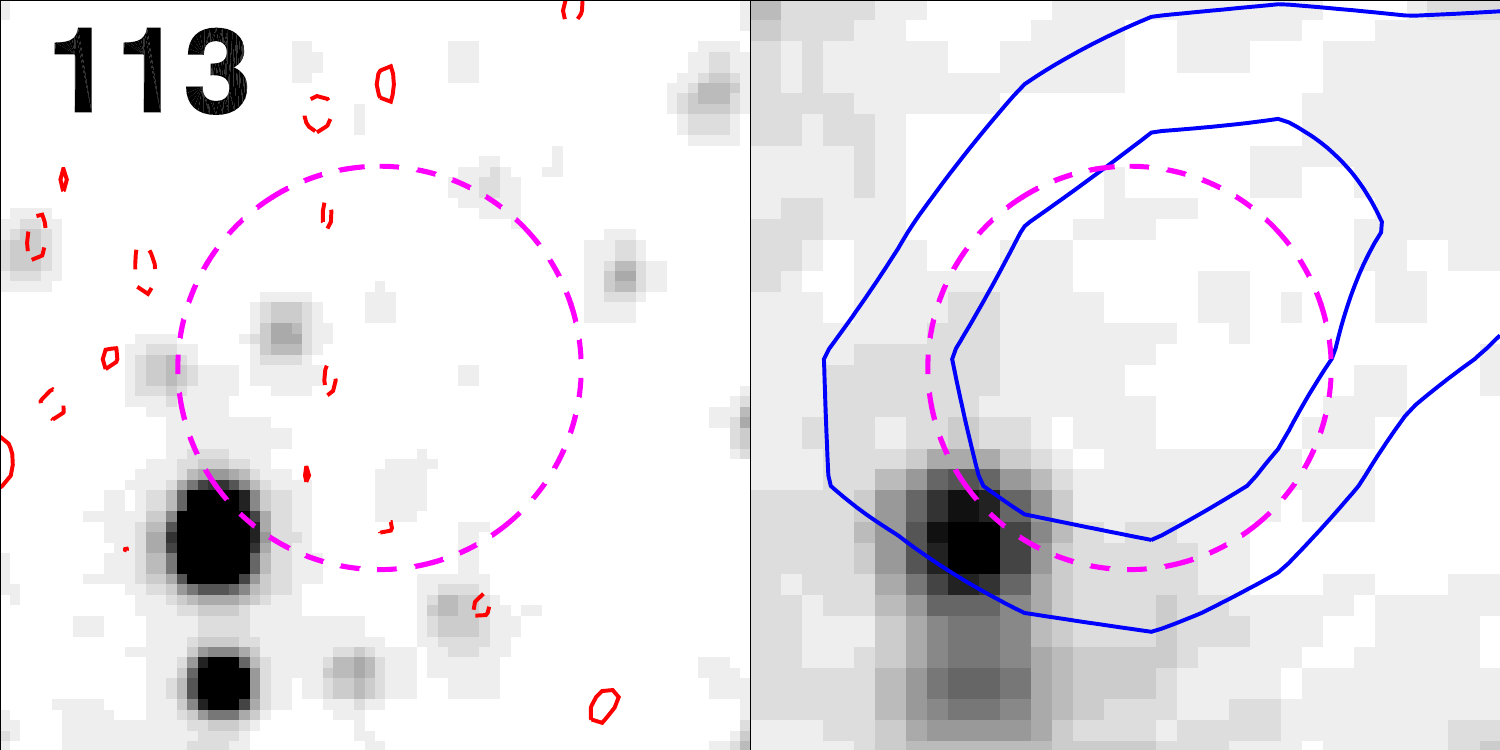}%
\hspace{1cm}%
\includegraphics[scale=0.295]{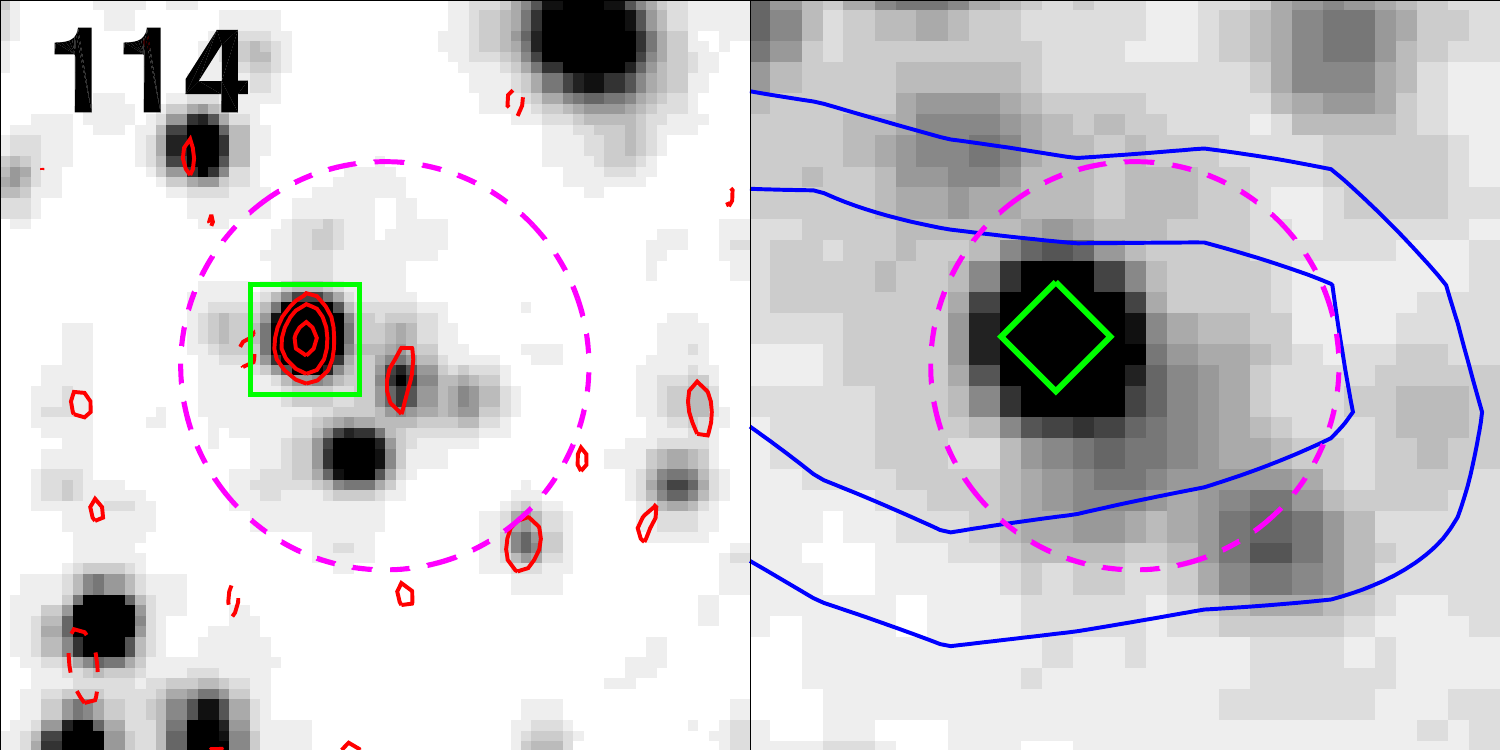}
\includegraphics[scale=0.295]{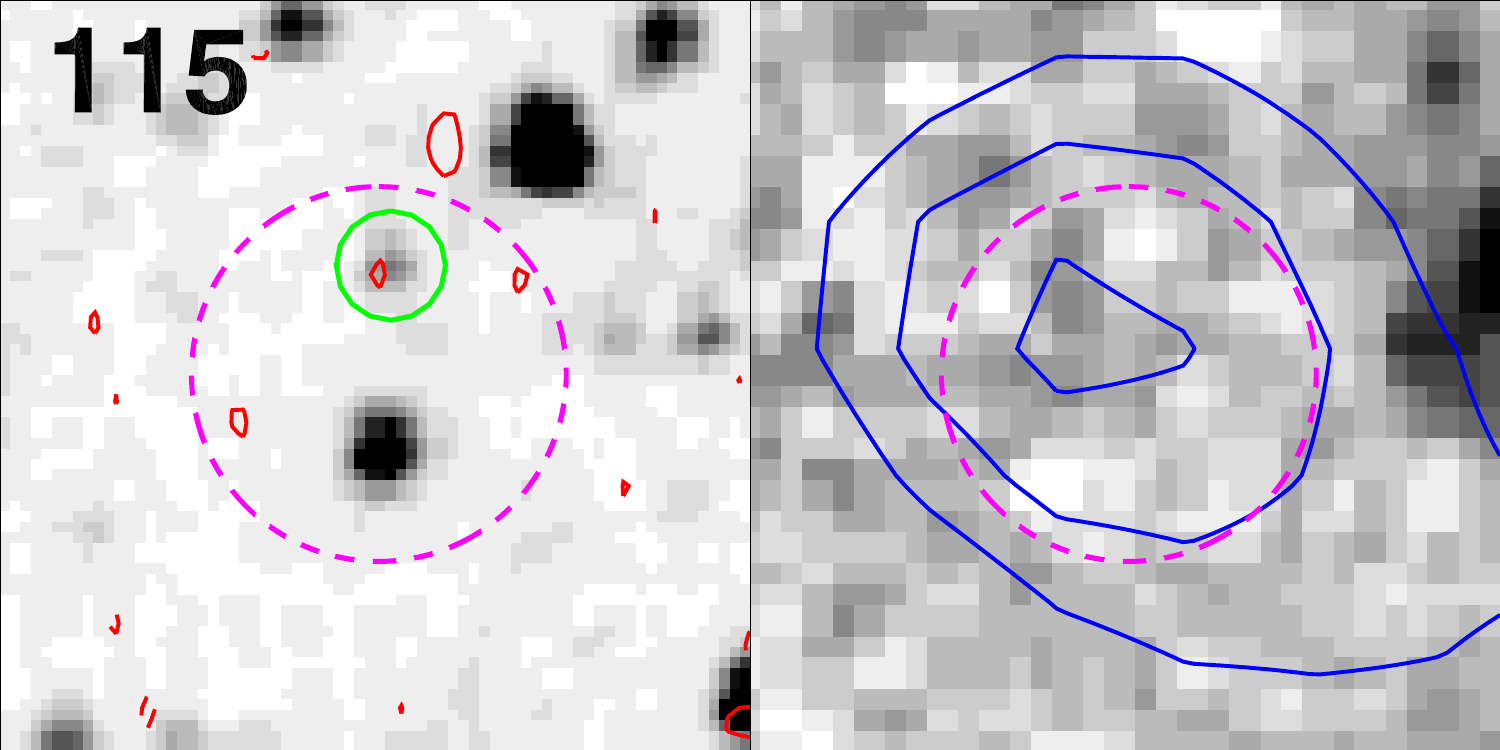}%
\hspace{1cm}%
\includegraphics[scale=0.295]{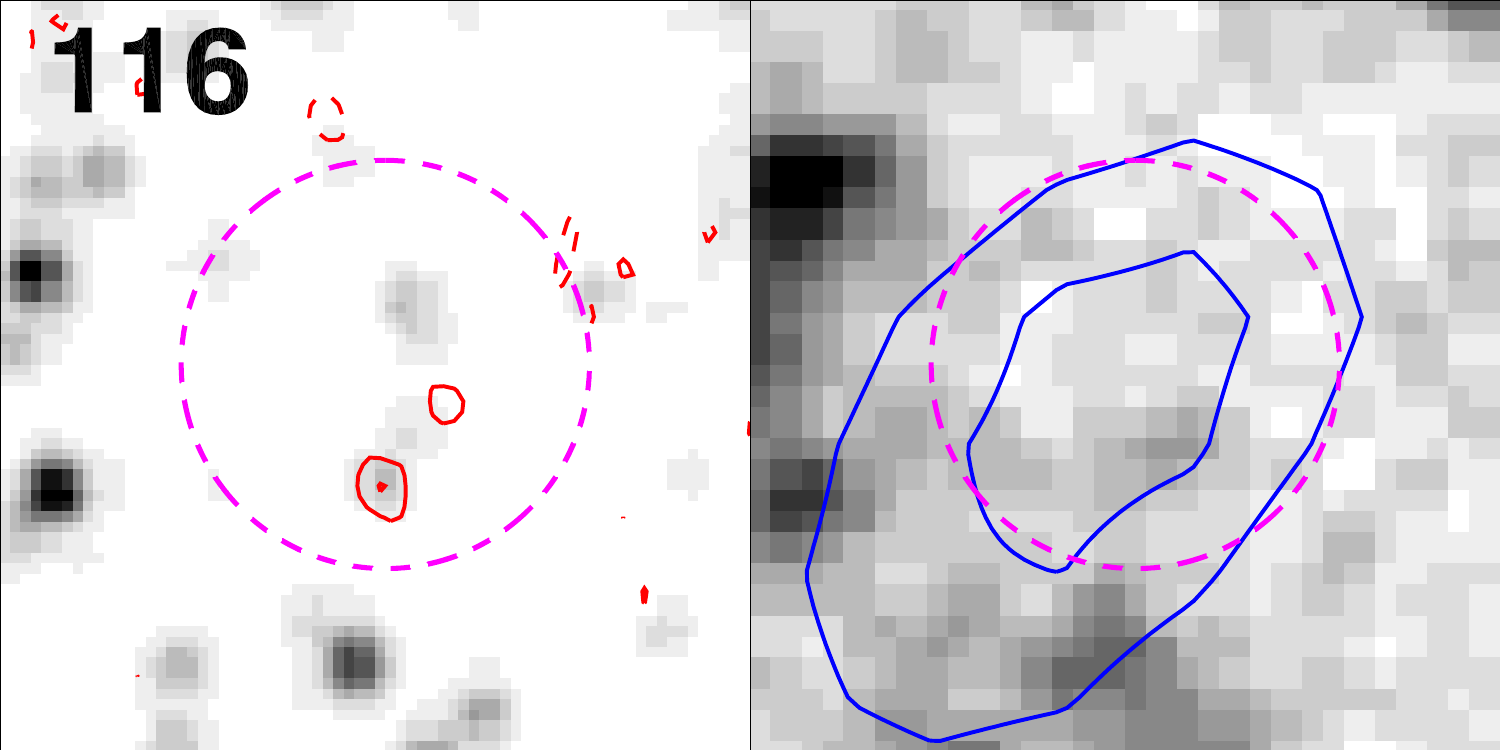}%
\hspace{1cm}%
\includegraphics[scale=0.295]{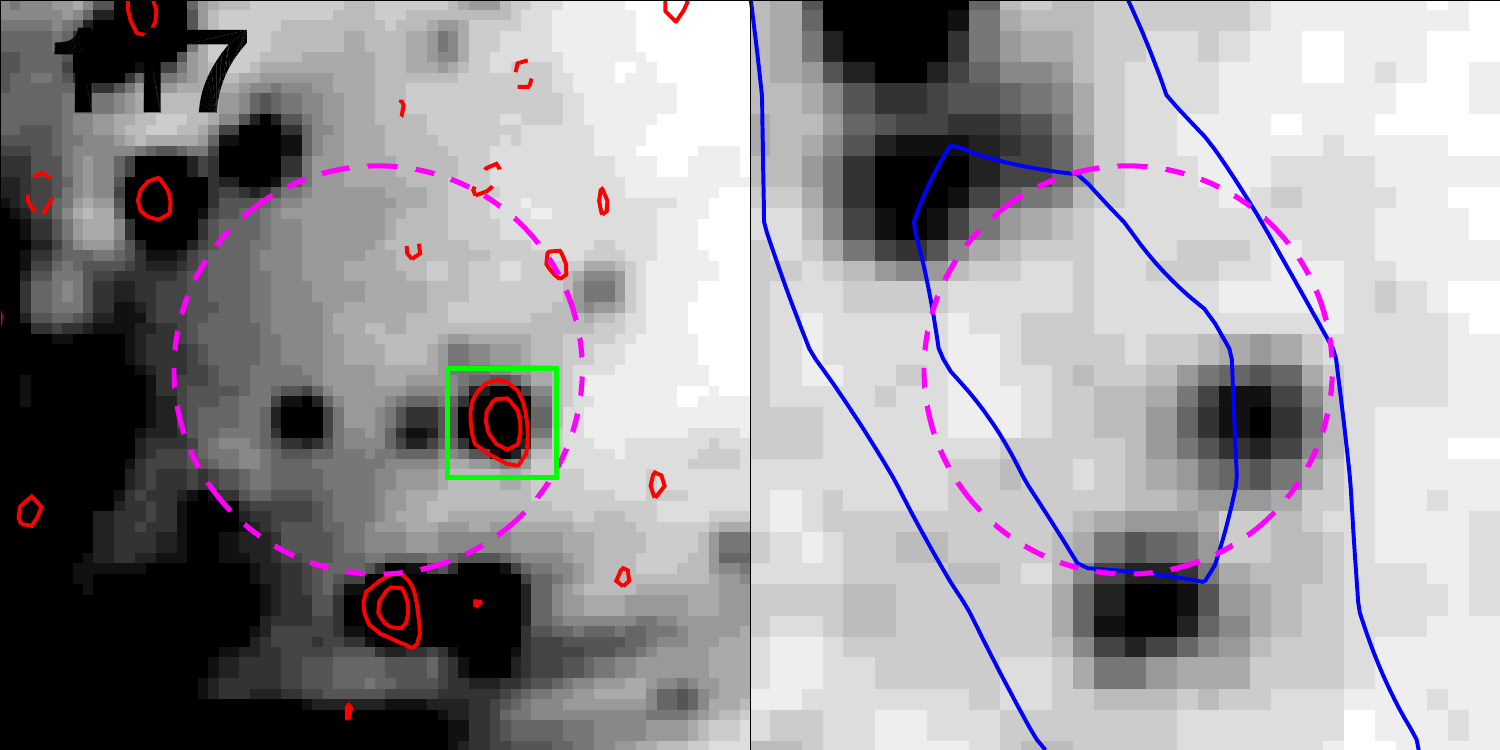}

\contcaption{}
\end{center}
\end{figure*}

\begin{figure*}
\begin{center}
\contcaption{}
\end{center}
\end{figure*}

\begin{figure*}
\begin{center}
\includegraphics[scale=0.295]{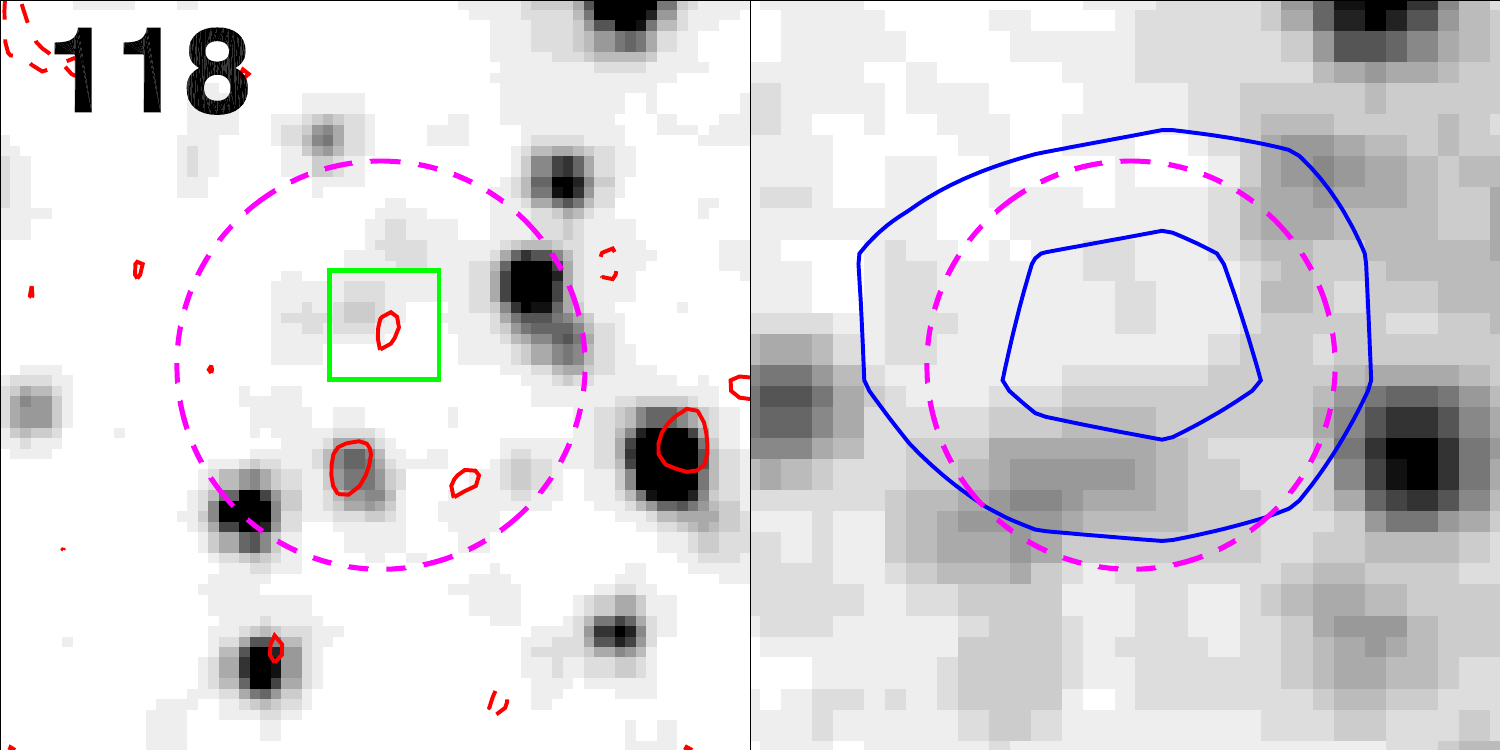}%
\hspace{1cm}%
\includegraphics[scale=0.295]{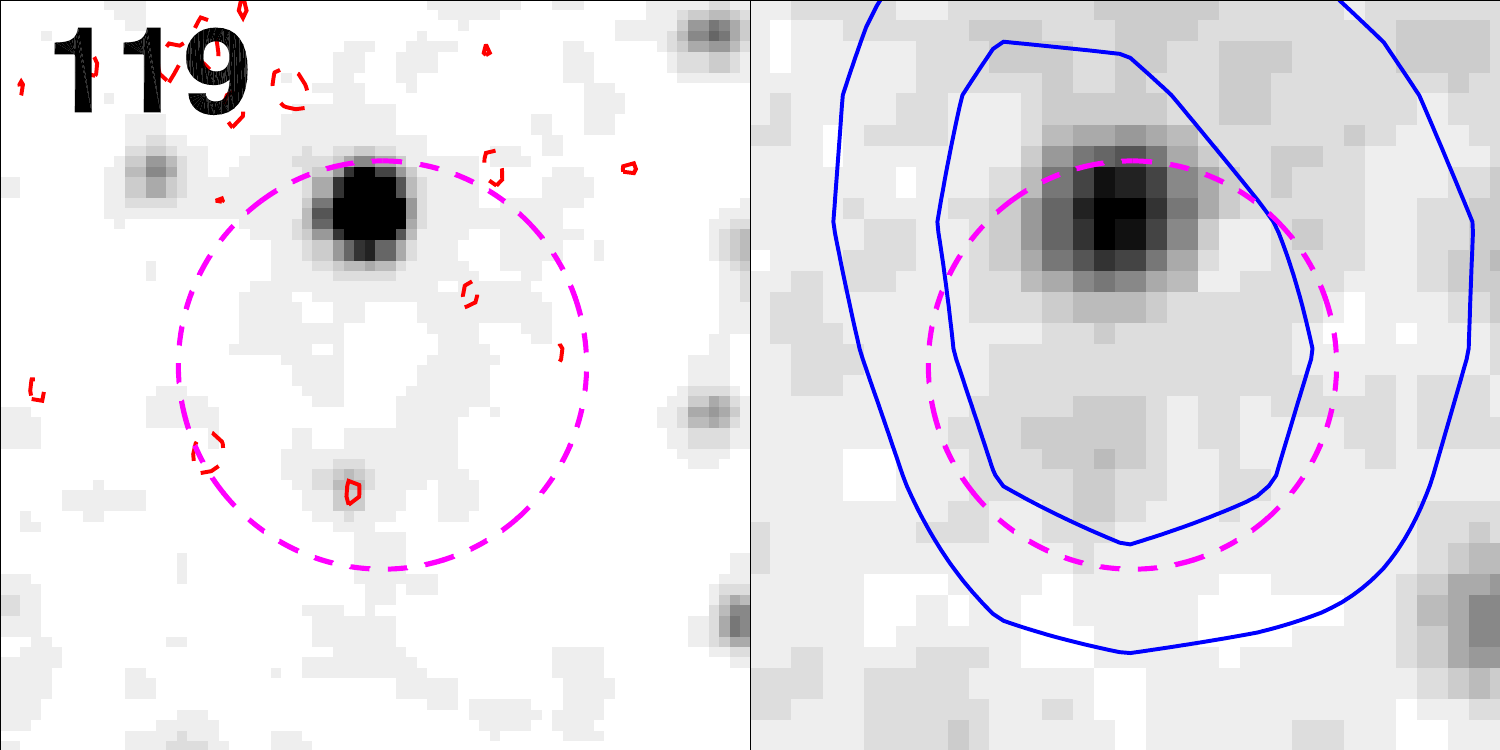}%
\hspace{1cm}%
\includegraphics[scale=0.295]{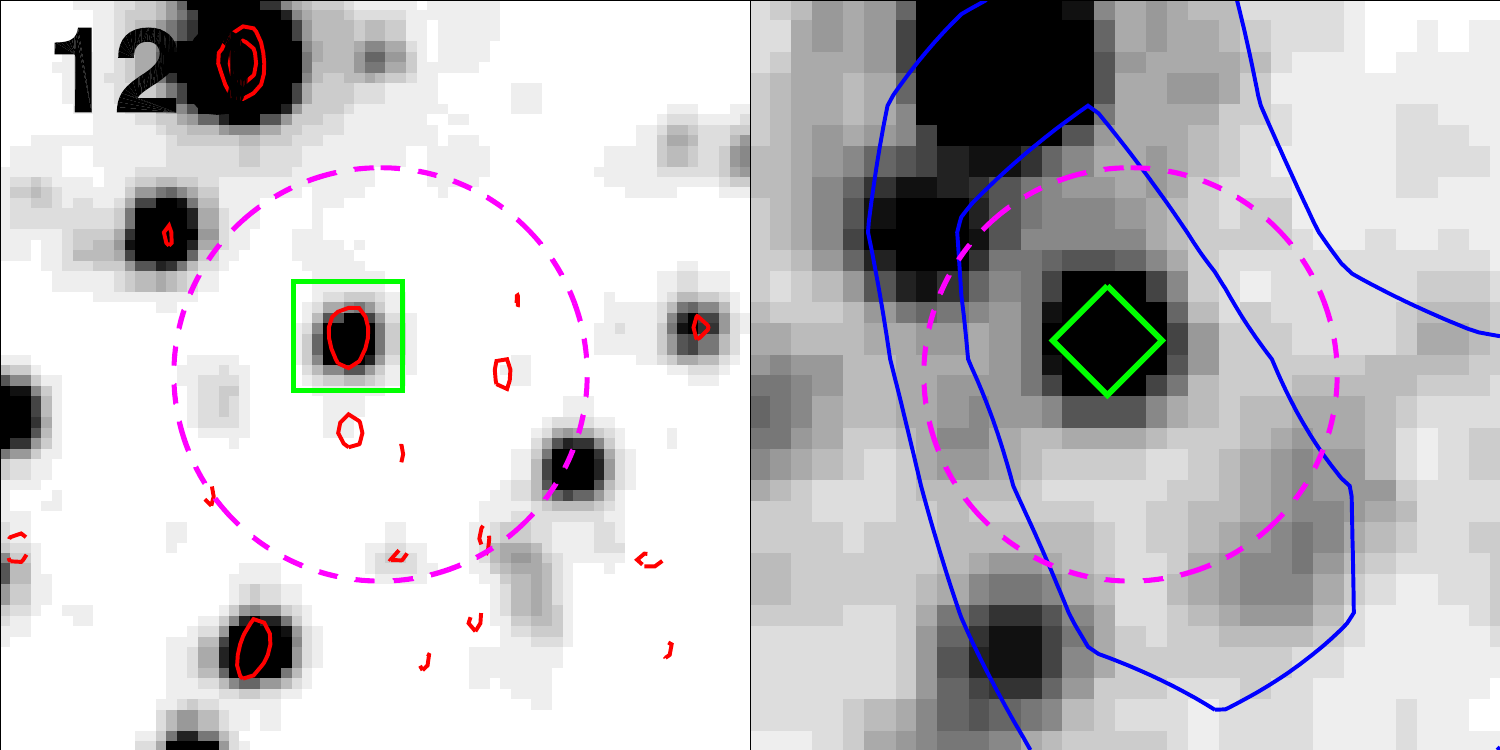}
\includegraphics[scale=0.295]{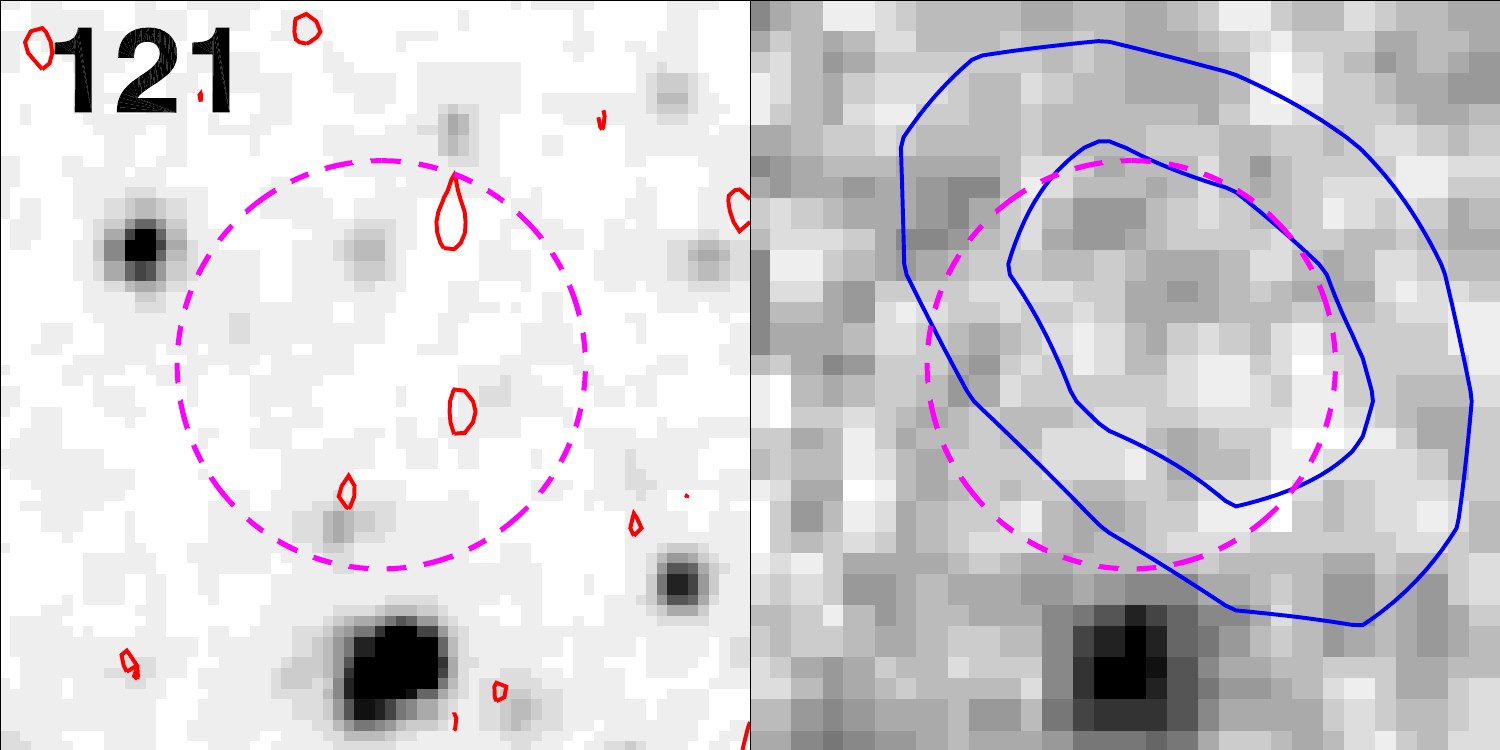}%
\hspace{1cm}%
\includegraphics[scale=0.295]{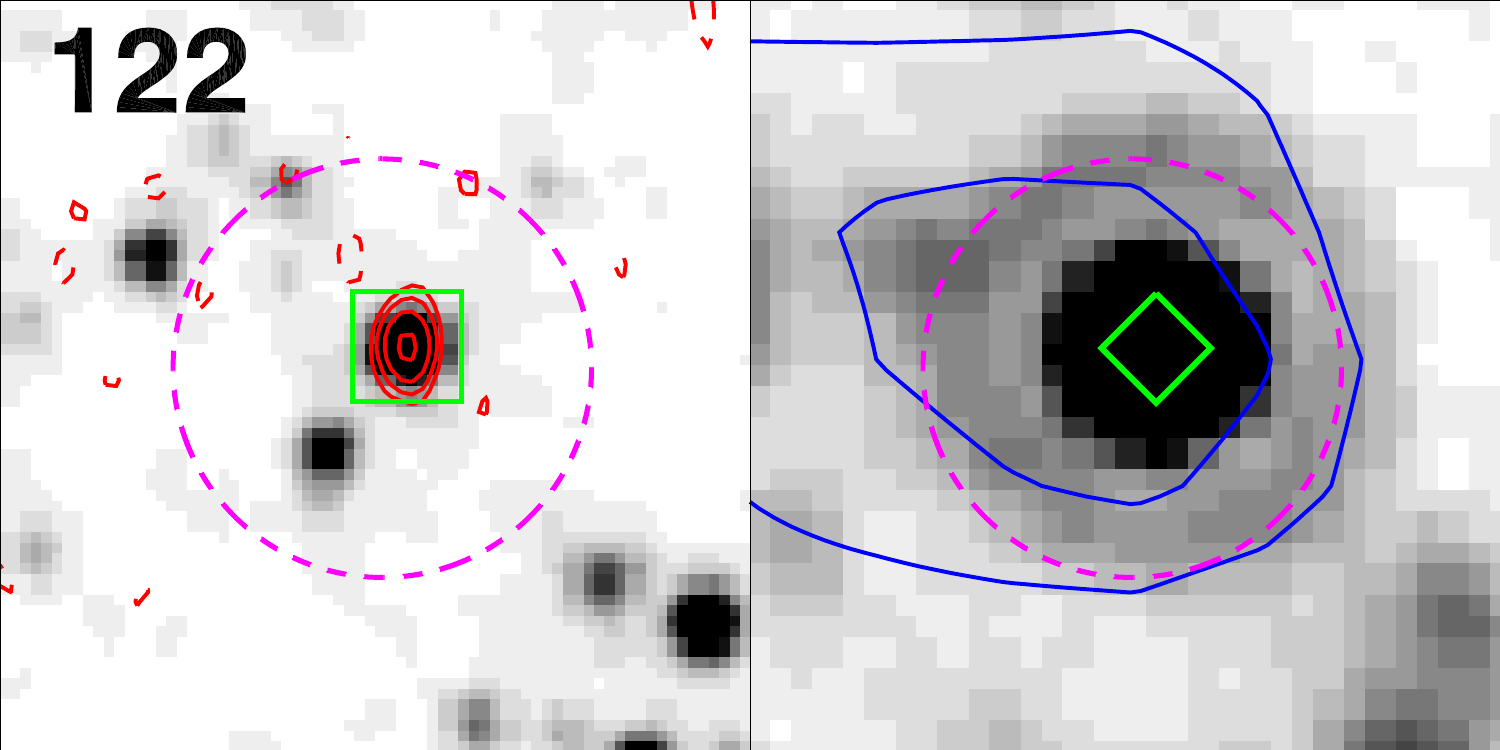}%
\hspace{1cm}%
\includegraphics[scale=0.295]{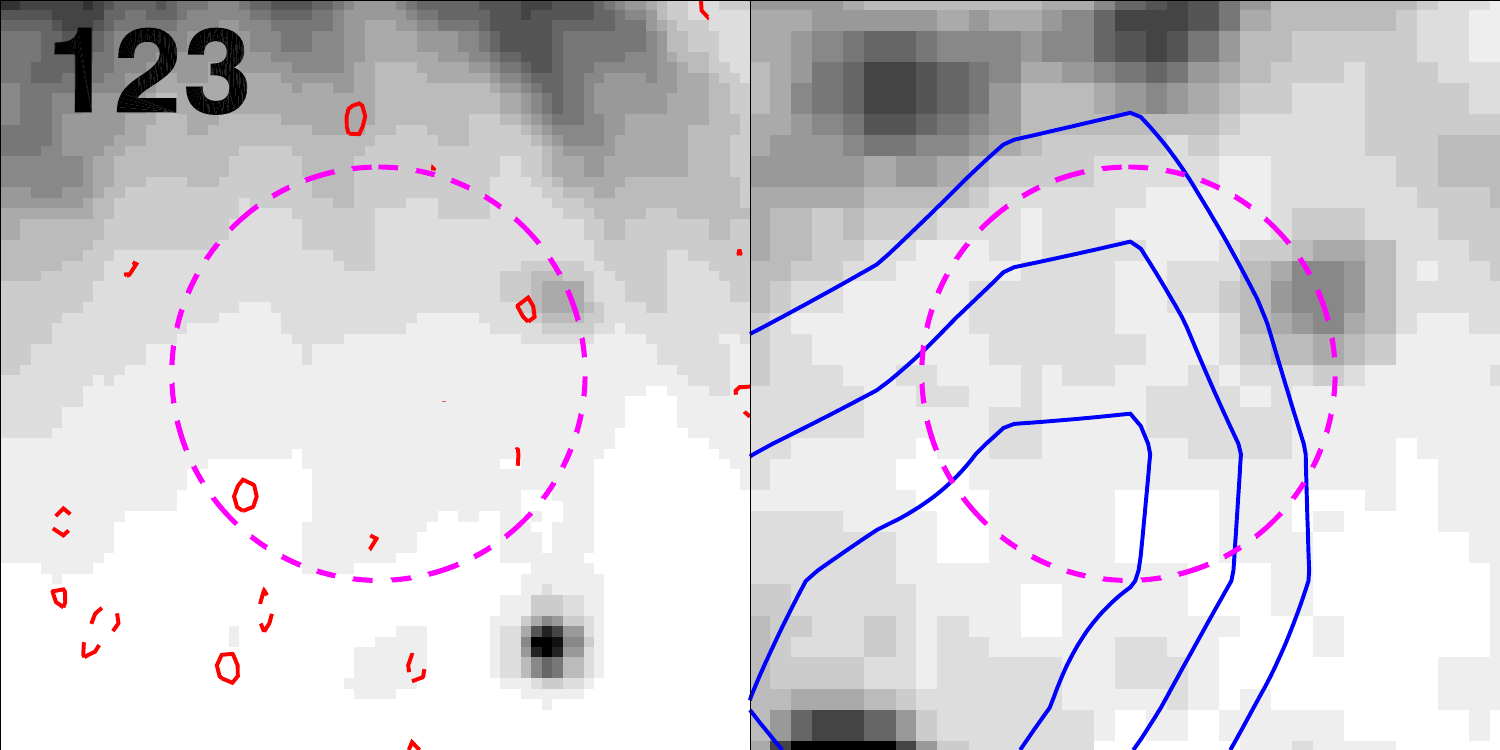}
\includegraphics[scale=0.295]{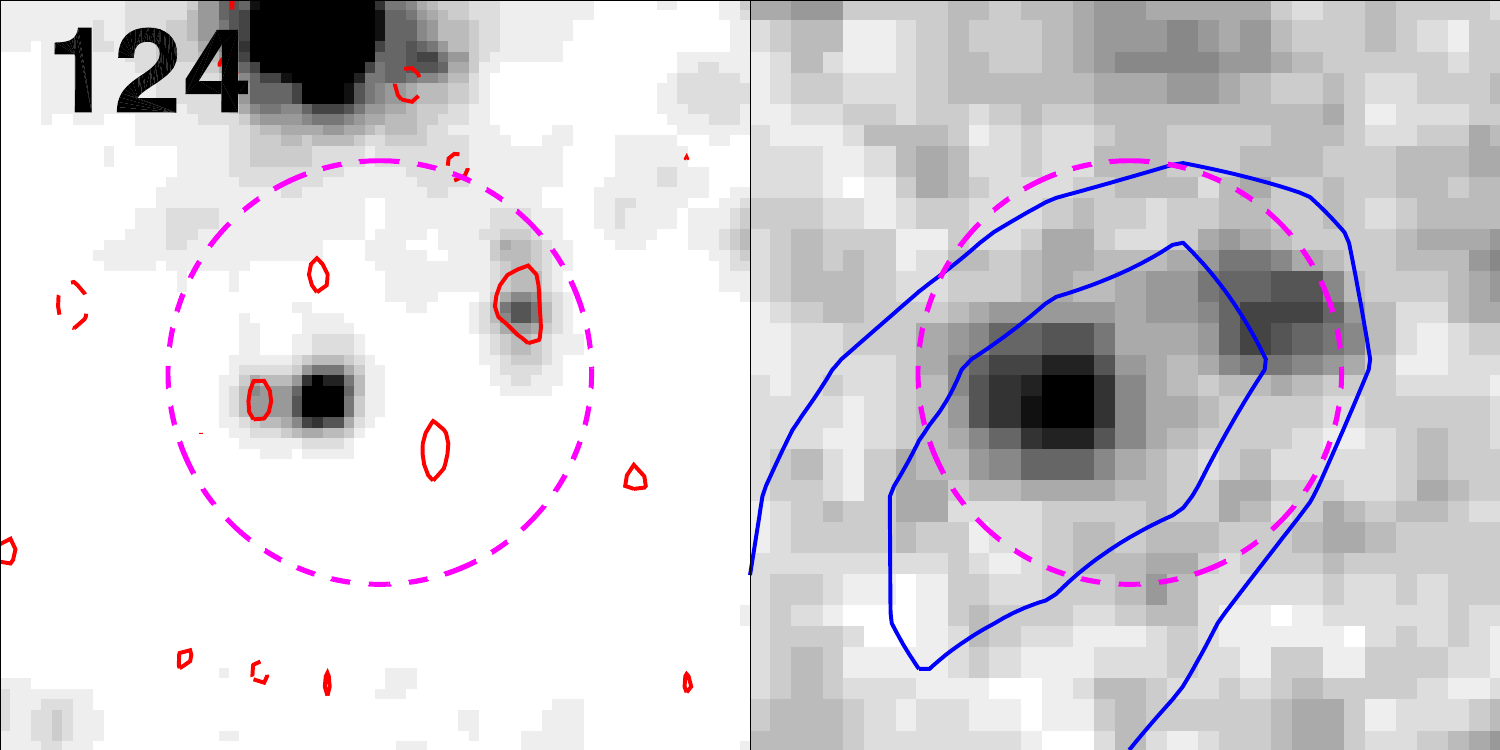}%
\hspace{1cm}%
\includegraphics[scale=0.295]{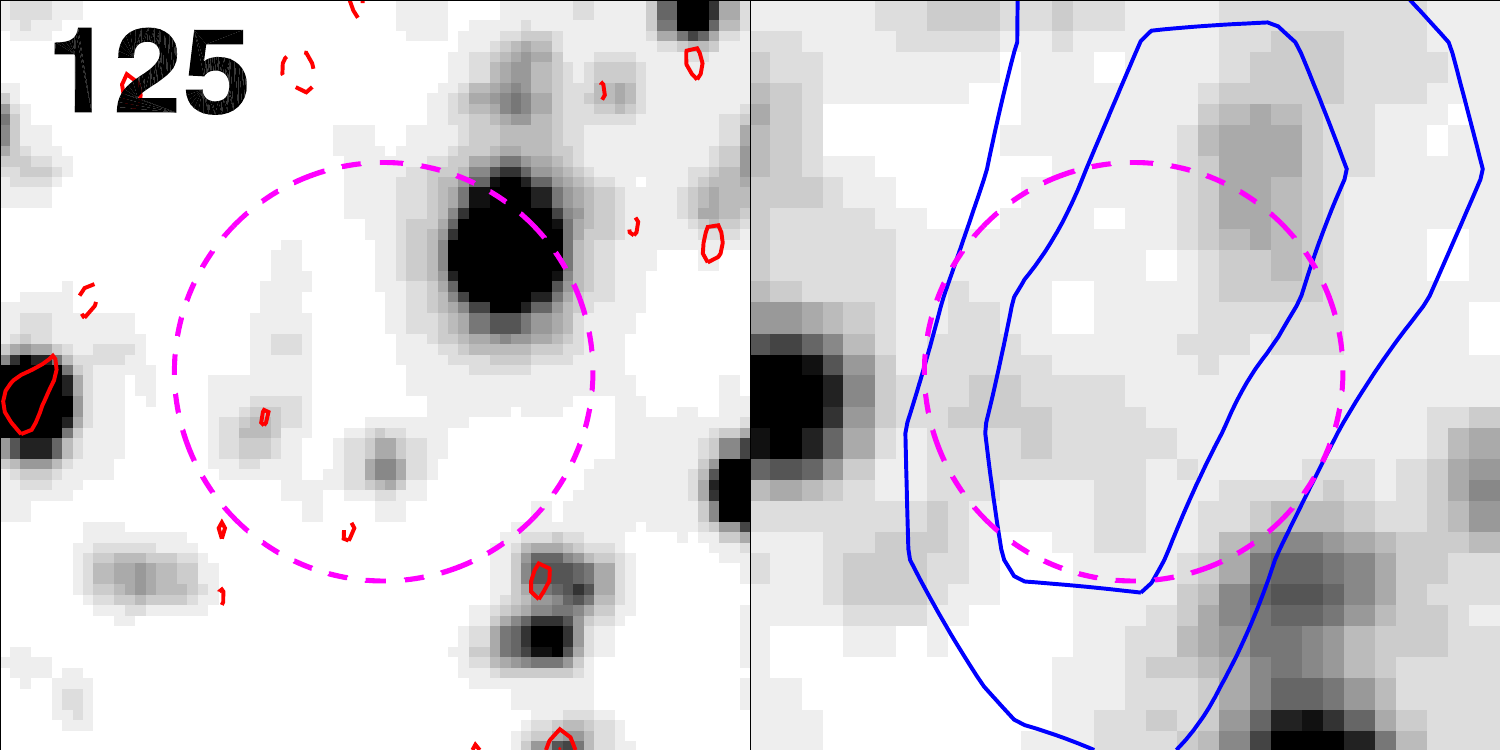}%
\hspace{1cm}%
\includegraphics[scale=0.295]{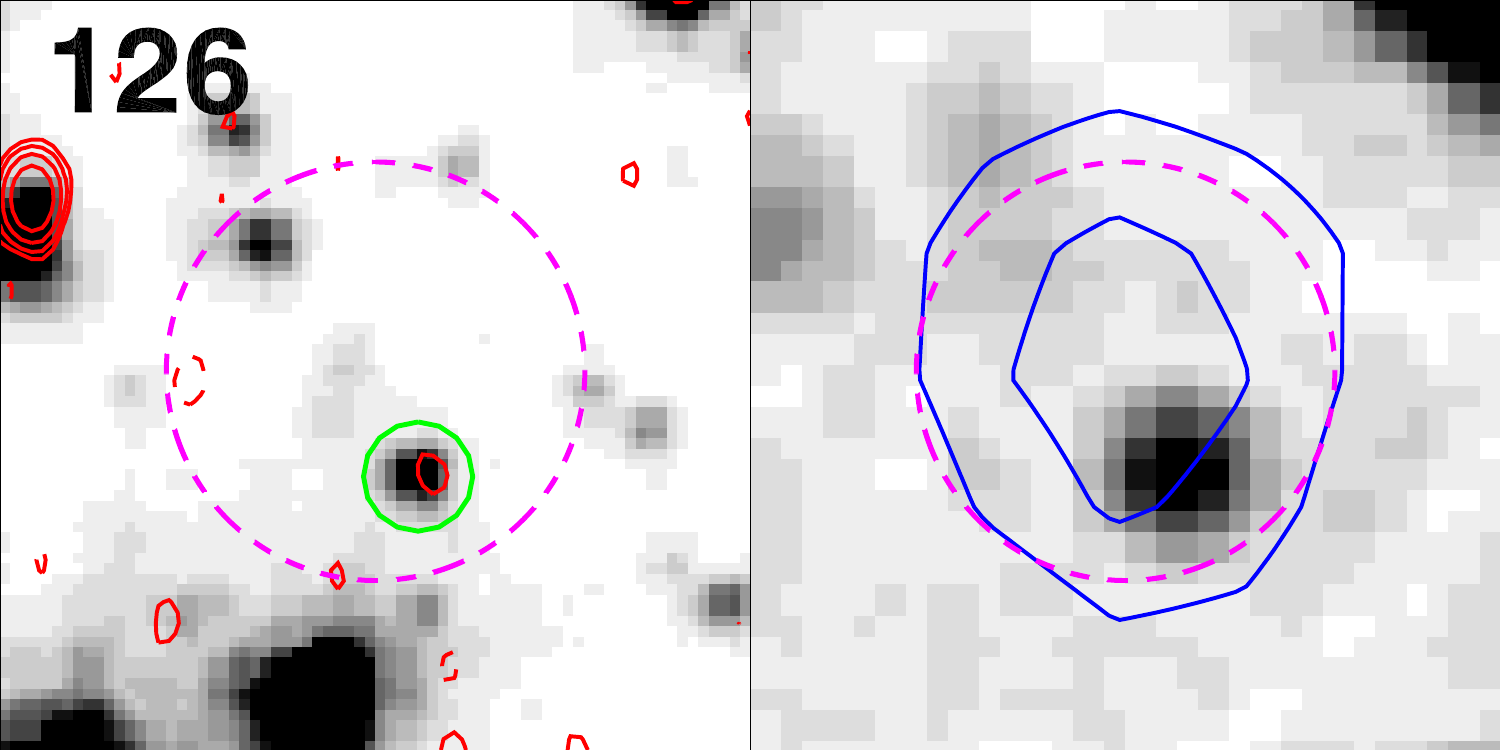}

\contcaption{}
\end{center}
\end{figure*}

Here we give short descriptions of those SMGs that merit further
discussion; postage stamp maps of each SMG are shown in
Fig.~\ref{fig:mainplots}.

\begin{description}
\item (LESS001) {\it LESS~J033314.3$-$275611} -- the brightest of the
  SMGs has no nearby radio emission, but an extremely faint MIPS
  component (with associated IRAC emission) is classed as a robust
  counterpart.

\item (LESS002) {\it LESS\,J033302.5$-$275643} -- high significance
  radio and 24-$\umu$m counterparts, but their positions are not
  coincident. An extension of the 24-$\umu$m emission passes under the
  radio source and the {\sc daophot} catalogue has detected this as a
  weak source (SNR$<$3.5); the redshift has been calculated assuming
  the weaker, but coincident, MIPS source is the correct counterpart
  to the radio. There is also an IRAC source at this position.


\item (LESS004) {\it LESS\,J033136.0$-$275439} -- the $p$-statistic
  finds nothing, but this is likely due to the catalogued SMG being a
  blend of up to four sources, three of which have similar
  brightnesses. The northernmost of these three is coincident with a
  76\,$\umu$Jy radio source. To the south, this chain of galaxies
  continues into LESS026.


\item (LESS006) {\it LESS\,J033257.1$-$280102} -- both the radio and
  MIPS robust counterparts are weak, but separated by
  $\sim$1~arcsec. The radio source appears to be extended, the eastern
  end of which lies closer to the MIPS position.


\item (LESS008) {\it LESS\,J033205.1$-$273108} -- the nearest radio
  source beyond the search radius has a position of 03:32:04.8146,
  $-$27:31:14.143, a flux of 69\,$\umu$Jy and is offset by 6.3~arcsec
  from the nominal submm position.


\item (LESS010) {\it LESS\,J033219.0$-$275219} -- the radio source
  gives the impression of having a bright core and a three-component
  jet. However, the end of the `jet' corresponds to a 24-$\umu$m
  source and another of the `jet' components is coincident with an
  IRAC source; the `core' also has 24-$\umu$m emission. The SMG
  emission continues to the south-west where it merges into
  LESS034. This rather confusing source has not had a redshift
  calculated.


\item (LESS012) {\it LESS\,J033248.1$-$275414} -- the bright MIPS
  source has a weak extension that {\sc daophot} resolves into
  separate components; one of these lies less than 1~arcsec from the
  secure radio counterpart.



\item (LESS015) {\it LESS\,J033333.4$-$275930} -- this SMG is a close
  neighbour of LESS076. The robust counterpart at 24-$\umu$m has
  associated weak radio emission.




\item (LESS019) {\it LESS\,J033208.1$-$275818} -- both IRAC
  identifications have faint emission at 24-$\umu$m.

\item (LESS020) {\it LESS\,J033316.6$-$280018} -- this has by far the
  brightest of the robust radio counterparts, $>$4~mJy, and therefore
  probably contains a radio-loud AGN component. For this reason, the
  CY00 redshift is greatly in error.



\item (LESS023) {\it LESS\,J033212.1$-$280508} -- the nearest radio
  source beyond the search radius has a flux of 65\,$\umu$Jy at
  position 03:32:12.2230, $-$28:05:16.752 and is associated with an
  obvious 24-$\umu$m source (245\,$\umu$Jy).

\item (LESS024) {\it LESS\,J033336.8$-$274401} -- the secure radio
  counterpart is coincident with a $p<0.1$ 24-$\umu$m source. A
  brighter radio source/24-$\umu$m source lies outside the search area
  at radio position 03:33:36.4418, $-$27:43:55.671.


\item (LESS026) {\it LESS\,J033136.9$-$275456} -- no radio or
  24-$\umu$m emission, but this source is a continuation of LESS004
  and hence probably a blend. The radio/24-$\umu$m source to the south
  (03:31:36.9524, $-$27:55:10.443) is a possible contributor to this
  submm complex.

\item (LESS027) {\it LESS\,J033149.7$-$273432} -- of the two robust
  IRAC counterparts, only the southern one has emission at 24~$\umu$m.

\item (LESS028) {\it LESS\,J033302.9$-$274432} -- this source is a
  close neighbour of LESS059. There is no sign of any significant
  emission within the search radius. The nearest radio source beyond
  the search radius has a flux of 52\,$\umu$Jy at position
  03:33:01.9865, $-$27:44:33.675 and is associated with an obvious
  24-$\umu$m source (102\,$\umu$Jy).


\item (LESS030) {\it LESS\,J033344.4$-$280346} -- the radio sources to
  the north have positions 03:33:44.6396, $-$28:03:38.273
  (240\,$\umu$Jy) and 03:33:44.9516, $-$28:03:43.435 (41\,$\umu$Jy).

\item (LESS031) {\it LESS\,J033150.0$-$275743} -- this weak radio
  source has a $p$ only slightly in excess of 0.05, is nearly
  coincident with a 24-$\umu$m source and has a robust counterpart
  from the IRAC analysis.


\item (LESS033) {\it LESS\,J033149.8$-$275332} -- this SMG is a close
  neighbour of LESS057.

\item (LESS034) {\it LESS\,J033217.6$-$275230} -- this SMG merges into
  LESS010 and three 24-$\umu$m sources lie along the line between the
  two SMGs. The nearby radio/24-$\umu$m source has a position of
  03:32:17.1874, $-$27:52:21.074 (93\,$\umu$Jy).

\item (LESS035) {\it LESS\,J033110.3$-$273714} -- the 24-$\umu$m
  emission is complex and the counterpart is difficult to see, but as
  the IRAC image reveals a faint source at the same position the {\sc
    daophot} extraction seems to have been successful.

\item (LESS036) {\it LESS\,J033149.2$-$280208} -- both radio and
  24-$\umu$m potential counterparts are coincidential and have
  $0.05<p\le0.1$ and therefore we consider this a secure
  identification.





\item (LESS041) {\it LESS\,J033110.5$-$275233} -- a pair of sources
  dominate the IRAC image, but only the one that is a robust
  identification has a counterpart at 24-$\umu$m.

\item (LESS042) {\it LESS\,J033231.0$-$275858} -- a radio/24-$\umu$m
  source lies just to the north of the search radius at 03:32:31.4500,
  $-$27:58:51.934.

\item (LESS043) {\it LESS\,J033307.0$-$274801} -- three 24-$\umu$m
  sources cluster towards the centre of the submm emission and all
  three have weak radio emission. Only one is classified as a robust
  counterpart, based on the IRAC data.



\item (LESS046) {\it LESS\,J033336.8$-$273247} -- this source is not
  covered by the 24-$\umu$m FIDEL or SIMPLE 3.6-$\umu$m data. The
  plots therefore show the shallower SWIRE data at each wavelength; at
  24-$\umu$m there is a clear counterpart to the robust radio
  identification.

\item (LESS047) {\it LESS\,J033256.0$-$273317} -- the two IRAC robust
  counterparts may be a single, extended source; 24-$\umu$m emission
  is centered closer to the western component.


\item (LESS049) {\it LESS\,J033124.4$-$275040} -- three 24-$\umu$m
  sources cluster towards the centre of the submm emission and all
  three have weak radio emission.

\item (LESS050) {\it LESS\,J033141.2$-$274441} -- as with the previous
  SMG, a cluster of several (at least four) 24-$\umu$m sources
  dominates the postage-stamp image and lie almost equidistant from
  the submm centroid; one of them is a secure identification based on
  its radio emission. The secure 24-$\umu$m identification at the very
  centre of the image is difficult to discern, but as a source is
  present at this position in the IRAC 3.6-$\umu$m image we believe
  that it is real.


\item (LESS052) {\it LESS\,J033128.5$-$275601} -- this SMG is a close
  neighbour of LESS075.






\item (LESS058) {\it LESS\,J033225.8$-$273306} -- very weak 24-$\umu$m
  emission that is not present in either the DAOPHOT or APEX
  catalogues is coincident with a $p=0.06$ radio source.


\item (LESS060) {\it LESS\,J033317.5$-$275121} -- the $p<0.1$ radio
  source to the south is coincident with a $p<0.1$ 24-$\umu$m source
  and we consider this a secure identification.



\item (LESS063) {\it LESS\,J033308.5$-$280044} -- the extremely weak
  radio emission that has been classed as a secure identification by
  the $p$-statistic is not seemingly associated with any 24-$\umu$m
  emission and we warn that it may be spurious. We do not include it
  in the redshift analysis.



\item (LESS066) {\it LESS\,J033331.7$-$275406} -- this is a close
  neighbour to LESS123.

\item (LESS067) {\it LESS\,J033243.3$-$275517} -- three 24-$\umu$m
  sources cluster towards the centre of the submm emission and all
  three have weak radio emission; one is a secure counterpart.

\item (LESS068) {\it LESS\,J033233.4$-$273918} -- the nearest radio
  source beyond the search radius has a position of 03:32:33.5615,
  $-$27:39:28.892, a flux of 50\,$\umu$Jy and is offset by 10.5~arcsec
  from the nominal submm position.



\item (LESS071) {\it LESS\,J033306.3$-$273327} -- increasing the
  search radius by a modest 0.4\,arcsec would lead to the
  identification of a $p<0.05$ radio counterpart (200\,$\umu$Jy at
  03:33:05.6632, $-$27:33:28.666) with a coincident 24-$\umu$m source.


\item (LESS073) {\it LESS\,J033229.3$-$275619} -- the radio $p<0.05$
  identification is very weak, but is coincident with a very weak
  15.2-$\umu$Jy 24-$\umu$m source that lies beneath the 3.5-$\sigma$
  catalogue threshold.

\item (LESS074) {\it LESS\,J033309.3$-$274809} -- we find two IRAC
  robust counterparts separated by only a few arcsec. One is robust
  based on the IRAC $p$-statistic alone whilst the other is robust due
  to $p<0.1$ for both the IRAC and the radio maps.

\item (LESS075) {\it LESS\,J033126.8$-$275554} -- this is a close
  neighbour of LESS052.

\item (LESS076) {\it LESS\,J033332.7$-$275957} -- the submm emission
  to the north of this source is LESS015.






\item (LESS082) {\it LESS\,J033253.8$-$273810} -- a pair of 24-$\umu$m
  sources align with the elongation of the submm emission.



\item (LESS085) {\it LESS\,J033110.3$-$274503} -- the radio source
  just outside the search radius has a position of 03:31:09.7733,
  $-$27:45:08.625, a flux of 46\,$\umu$Jy and is offset by 8.7~arcsec
  from the nominal submm position.

\item (LESS086) {\it LESS\,J033114.9$-$274844} -- the radio source
  just outside the search radius has a position of 03:31:14.1207,
  $-$27:48:44.229 (J2000), a flux of 102\,$\umu$Jy and is offset by
  10.3~arcsec from the nominal submm position.

\item (LESS087) {\it LESS\,J033251.1$-$273143} -- there are two radio
  counterparts, only one of which has 24-$\umu$m emission. This may
  be a radio core and jet.









\item (LESS096) {\it LESS\,J033313.0$-$275556} -- this SMG lies very
  close to LESS001. The secure 24-$\umu$m identification at the very
  centre of the image is difficult to discern and appears to be part
  of the Airy ring, but an IRAC 3.6-$\umu$m source at this position
  again confirms that the {\sc daophot} extraction is reliable.


\item (LESS098) {\it LESS\,J033130.2$-$275726} -- two radio sources
  align themselves closely with the submm elongation. The southern of
  the pair has a flux of 368\,$\umu$Jy and a position of
  03:31:30.7540, $-$27:57:35.129 (J2000).





\item (LESS103) {\it LESS\,J033325.3$-$273400} -- this SMG is a close
  neighbour of LESS111.




\item (LESS107) {\it LESS\,J033130.8$-$275150} -- a group of five
  catalogued 24-$\umu$m sources within the search radius form a ring
  around the submm position, two of which have associated weak radio
  emission.



\item (LESS110) {\it LESS\,J033122.6$-$275417} -- the very faint IRAC
  identification also has a counterpart in a complex 24-$\umu$m
  structure that {\sc daophot} disentangles into three separate
  components.

\item (LESS111) {\it LESS\,J033325.6$-$273423} -- this SMG is a close
  neighbour of LESS103. It has a robust counterpart in both the radio
  and the MIPS catalogues, but the weaker radio component is
  significantly offset (2~arcsec) from the much brighter MIPS
  detection. This perhaps suggests that the radio source is spurious,
  but its relatively high SNR ($>$5) argues that it is unlikely to be
  a false detection.






\item (LESS117) {\it LESS\,J033128.0$-$273925} -- there is a striking
  alignment of five 24-$\umu$m sources with the elongation of the
  submm emission along 30\degr, several of which have associated radio
  emission.

\item (LESS118) {\it LESS\,J033121.8$-$274936} -- the extremely weak
  radio emission that has been classed as a secure identification by
  the $p$-statistic is not seemingly associated with any MIPS or IRAC
  emission and we warn that it may be spurious. We do not include it
  in the redshift analysis.





\item (LESS123) {\it LESS\,J033330.9$-$275349} -- this is a close
  neighbour to LESS066.



\end{description}

\bsp

\end{document}